\documentclass[aps,rmp,reprint,amsmath,amssymb,graphicx,longbibliography,preprintnumbers]{revtex4-1}

\usepackage{dcolumn}
\usepackage{bm}
\usepackage{color}
\usepackage{url}
\usepackage{epsfig}

\begin{document}


\preprint{DESY 14-032}
\preprint{KYUSHU-RCAPP 2015-01}
\preprint{LAL 15-235}

\title{Experimental Tests of Particle Flow Calorimetry}

\author{Felix Sefkow}
\affiliation{Deutsches Elektronen-Synchrotron DESY, Hamburg, Germany}

\author{Andy White}
\affiliation{University of Texas at Arlington, Arlington, TX, USA}

\author{Kiyotomo Kawagoe}
\affiliation{Kyushu University, Fukuoka, Japan}

\author{Roman P\"oschl}
\affiliation{Laboratoire de l'Accel\'erateur Line\'eaire LAL, Orsay, France}

\author{Jos\'e Repond}
\affiliation{Argonne National Laboratory ANL, Argonne, IL, USA}



\begin{abstract}
Precision physics at future colliders requires highly granular calorimeters to support the Particle Flow Approach for event reconstruction. This article presents a review of about 10 - 15 years of R\&D, mainly conducted within the CALICE collaboration, for this novel type of detector. The performance of large scale prototypes in beam tests validate the technical concept of particle flow calorimeters. The  comparison of test beam data with simulation, of e.g.\ hadronic showers, supports full detector studies and gives deeper insight into the structure of hadronic cascades than was possible previously.  
\end{abstract}


\maketitle

\tableofcontents{}


%
		\section{Introduction%
\label{sec:Intro}}


\subsection{Role and limitations of calorimeters in high energy physics}

Progress in elementary particle physics has been driven by advances in detector technology as much as by the increased reach of accelerators. 
The transition from photographic event collection to electronic data recording has given access to rare processes in high rate experiments, and silicon strip detectors made the picosecond lifetimes and oscillations of heavy flavoured particles visible. 
In the past 10 years, calorimeters with imaging capabilities have been developed, which opens a new era of precision in the measurement of particle jet energies.

Today's particle physics experiments, mostly at colliders, use multi-purpose detectors aiming at capturing as precise and complete information as possible for the reconstruction of all particles in the final state of each collision -- their type and kinematic properties. 
Whilst trackers infer the momenta of charged particles from the curvature of ionisation trails in magnetic fields, calorimeters obtain the energies of particles only indirectly via the debris emerging from their interaction with a block of matter, so-called showers. 
Without calorimeters, the reconstruction of jets, which contain many types of particles, is not complete. 


Future particle physics projects, with hadron, lepton or neutrino beams, place ever-increasing demands on the detailed reconstruction of the beam interaction final states.
For example, the next generation of $e^+e^-$ linear colliders, the ILC~\cite{Baer:2013cma} with centre-of-mass (c.m.)\ energies of 250~GeV -- 1~TeV
and its detectors ILD and SiD~\cite{Behnke:2013lya},
and CLIC~\cite{Linssen:2012hp} (250~GeV -- 3~TeV) with adapted detectors,
will enter a domain of precision measurements with heavy bosons ---
W, Z and H --- being copiously produced.
They must be reconstructed in multi-jet final states and identified on the basis of their invariant mass. 
The W - Z mass separation is about 10~GeV.
Since the dijet invariant mass is given by 
$M^2  = 2 E_1 E_2 (1-\cos \theta_{12})$, where $E_{1,2}$ are the energies of the jets and $\theta_{12}$ the angle between them, a jet energy resolution of $\sigma_E/E$ translates into a mass resolution 
$\sigma_M/M = (1/\sqrt{2}) \sigma_E/E$, 
if angular uncertainties can be neglected. 
Taking the natural width of about 2.7\% into account, a $3 \sigma$ separation 
then requires a jet energy resolution of 3-4\% over a wide range of jet energies, from 50 to 500 or 1000~GeV, for the ILC or CLIC, respectively~\cite{Thomson:2009rp}.  

Physics channels particularly sensitive to the jet energy performance are those with heavy bosons to be identified in dijet final states. 
The process $e^+e^-\rightarrow$WW$\nu_e\bar{\nu}_e$, which probes the WW scattering amplitude and which is to be separated 
from its irreducible ZZ$\nu_e\bar{\nu}_e$ background, is an often quoted example. 
Other examples are the measurement of the H~$\rightarrow$~WW$^*$ branching ratio, which together with the cross section for Higgs production in the WW fusion channel enters into the determination of the Higgs total width. Here both W and Higgs masses are reconstructed.
For ZH final states, where the Z decays into neutrinos  only the Higgs decaying into a pair of jets is visible in clean conditions, which give access 
to the Higgs coupling to charm quarks. 
Channels with large jet multiplicity, like $t\bar{t}$H, are less affected, since there jet finding ambiguities dominate the mass resolution. On the other hand, the ZH$\rightarrow 4$~jets final state provides a Higgs mass resolution comparable to the recoil mass technique, using leptonic Z decay modes, and is sensitive to the calorimeter performance, even though kinematic constraints can be applied. 

The classical way of measuring jet energies is to sum up the energy depositions of all charged and neutral particles of the jet in the calorimeter system, generally composed of an electromagnetic section (ECAL) followed by a hadronic part (HCAL). 
In this case, the resolution dependence on the energy $E$ approximately follows a form 
$\frac{\sigma_E}{E} = \frac{a}{\sqrt{E(\rm{GeV})}} \oplus b$
where $a$, called the stochastic term, reflects statistical fluctuations in the shower evolution and measurement, and $b$, called the constant term, arises from imperfections in detector homogeneity, stability and calibration. 
At higher energies, there are additional contributions from the fluctuations of non-contained energy, or {\it leakage}. 
Typically, the term $a$ attains values of $50 - 100\%$, while $b$ is a few percent, which altogether is not sufficient to meet the linear collider goals. 

The limitation of the classical approach stems from the fact that most of the jet energy ($\sim 70\%$) is carried by hadrons~\cite{Knowles:1997dk,lepphys} and the jet measurement thus inherits the intrinsically poor performance of traditional hadronic calorimetry, if no tracking information is used.
In contrast to the cascades of bremsstrahlung and pair creation initiated by electromagnetic particles, hadronic showers are characterised by a much smaller number of subsequent nuclear collisions and smaller number of produced particles, and thus suffer from much larger fluctuations. 
In addition, a large number of different fundamental processes, in general give rise to a different detector response for the same deposited energy~\cite{Fabjan:1976da}. 
Target nucleus recoil and nuclear excitations do not or only partially contribute to the signal ("invisible energy") at all. 
Particles like $\pi^0$ and $\eta$ decay into photons which initiate electromagnetic showers and do not participate further in the hadronic cascade. 
Most calorimeters are non-compensating, which means that the response ($e$) to this electromagnetic fraction differs from that to the hadronic part, $h$, i.e.\ $e/h \neq 1$~\cite{Wigmans:1986pe}.  
The response to hadrons thus fluctuates with the fractions of electromagnetic and invisible energy. 
Since the electromagnetic fraction increases with energy~\cite{Gabriel:1993ai}, the hadron response also becomes non-linear. 
The jet response then in addition fluctuates with how the energy is shared between particles, and with the electromagnetic content of the jet fragmentation itself. 

As a result of these factors, one direction of research on hadronic calorimeters has concentrated on compensation methods to restore $e/h = 1$, for instance, by identifying electromagnetic
showers inside hadronic showers~\cite{Andrieu:1993tz}.
The best single hadron resolution at a collider detector is achieved by the ZEUS calorimeter~\cite{Bernstein:1993kj}, with $a = 35\%$. 
The performance for jets, however, generally does not reach this value, since it is subject to additional degrading effects. 
ZEUS quotes a hadronic Z mass core resolution of 6~GeV~\cite{Abramowicz:2012qy}, which is considerably worse than one might naively expect on the basis of the single hadron resolution, and not sufficient for the separation of W and Z final states. 
More recently, {\it dual readout} techniques~\cite{Akchurin:2005an} have been explored, which can in principle reduce the effect of fluctuations in the electromagnetic fraction by measuring it independently.

\subsection{Particle flow approach}

The particle flow (Pflow) approach~\cite{Brient:2002gh,Morgunov:2001cd} starts from the observation that most particles in a jet  --  charged particles and photons -- can in principle be measured with much better precision than generally provided by the calorimeter for hadrons.
In the range considered here, charged particle energies are best measured with tracking systems, which offer relative resolutions of about $10^{-4}E(\rm{GeV})$, and individual photon energies can be measured with relative precision of about 
$15\%/\sqrt{E(\rm{GeV})}$ or better in electromagnetic calorimeters. 
The PFlow method aims at optimising the jet energy resolution by reconstructing each particle individually and use the best available measurement for each. 
In a typical jet 60\% of the energy is carried by charged particles, 30\% by photons and only 10\% by long-lived neutral hadrons ($K^0_L$ and $n$), for which hadronic calorimetry is unavoidable. 
Assuming the above resolutions for tracks and photons, and $55\%/\sqrt{E(\rm{GeV})}$ for hadrons, then, in the ideal case, where each particle is resolved, a jet energy resolution of $19\%/\sqrt{E(\rm{GeV})}$ could be obtained.
Here the dominant part ($17\%/\sqrt{E(\rm{GeV})}$) is still due to the neutral hadrons.
The jet composition fluctuates from event to event; so for jets with a smaller neutral hadron fraction the precision is higher, and {\it vice versa}. 
Particle flow-like techniques were first applied in the ALEPH detector~\cite{Buskulic:1994wz}, which achieved a jet energy resolution of $60\%/\sqrt{E}$, or 6.2~GeV for hadronic Z decays. 

\begin{figure}[htb]
\includegraphics[width=0.95\hsize]{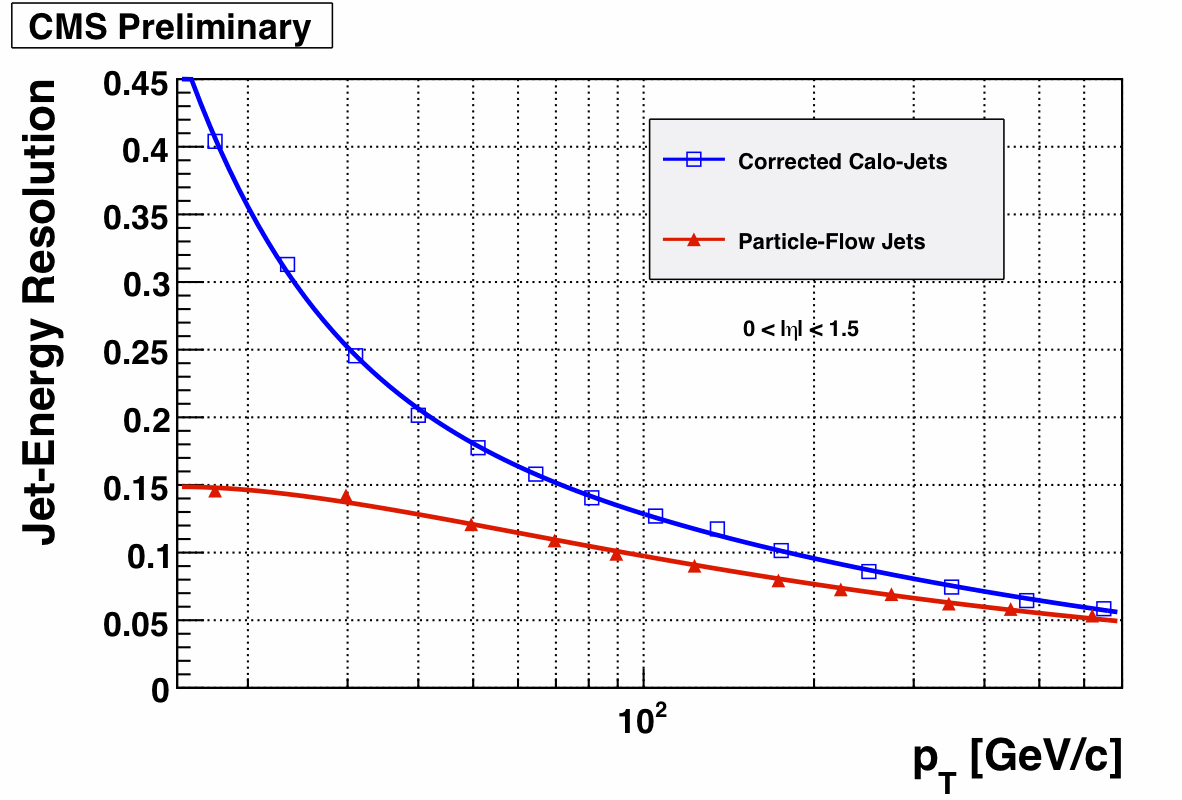}
\caption{\label{fig:Int:cmsjet} Simulated jet energy resolution as a function of transverse momentum $p_T$ for calorimeter jets and using particle flow techniques. From~\cite{CMS:2009nxa,Beaudette:2010zz}.}
\end{figure}

More recently particle flow techniques are  successfully used in the CMS experiment~\cite{CMS:2009nxa}, which is well suited for this purpose. 
CMS has a large silicon tracker in a uniform solenoidal field of 3.8T, and a finely segmented electromagnetic calorimeter made of 75,000 lead tungstate crystals surrounds the tracker. 
The hadronic calorimeter segmentation is 25 times coarser, such that, in jets above 100 GeV/c, neutral hadrons cannot be separated from charged ones, but are rather detected as an excess of the calorimeter energy over the tracker momentum. 
Using a careful and accurate reconstruction of the particle content of the event, the detector performance is significantly improved with respect to that using the calorimeter alone. 
This is illustrated for the simulated jet energy resolution as a function of transverse momentum in Fig.~\ref{fig:Int:cmsjet}.
Similar improvements are achieved for the reconstruction of missing transverse energy and for hadronic $\tau$ decays.  

\begin{figure}[htb]
\includegraphics[width=0.95\hsize]{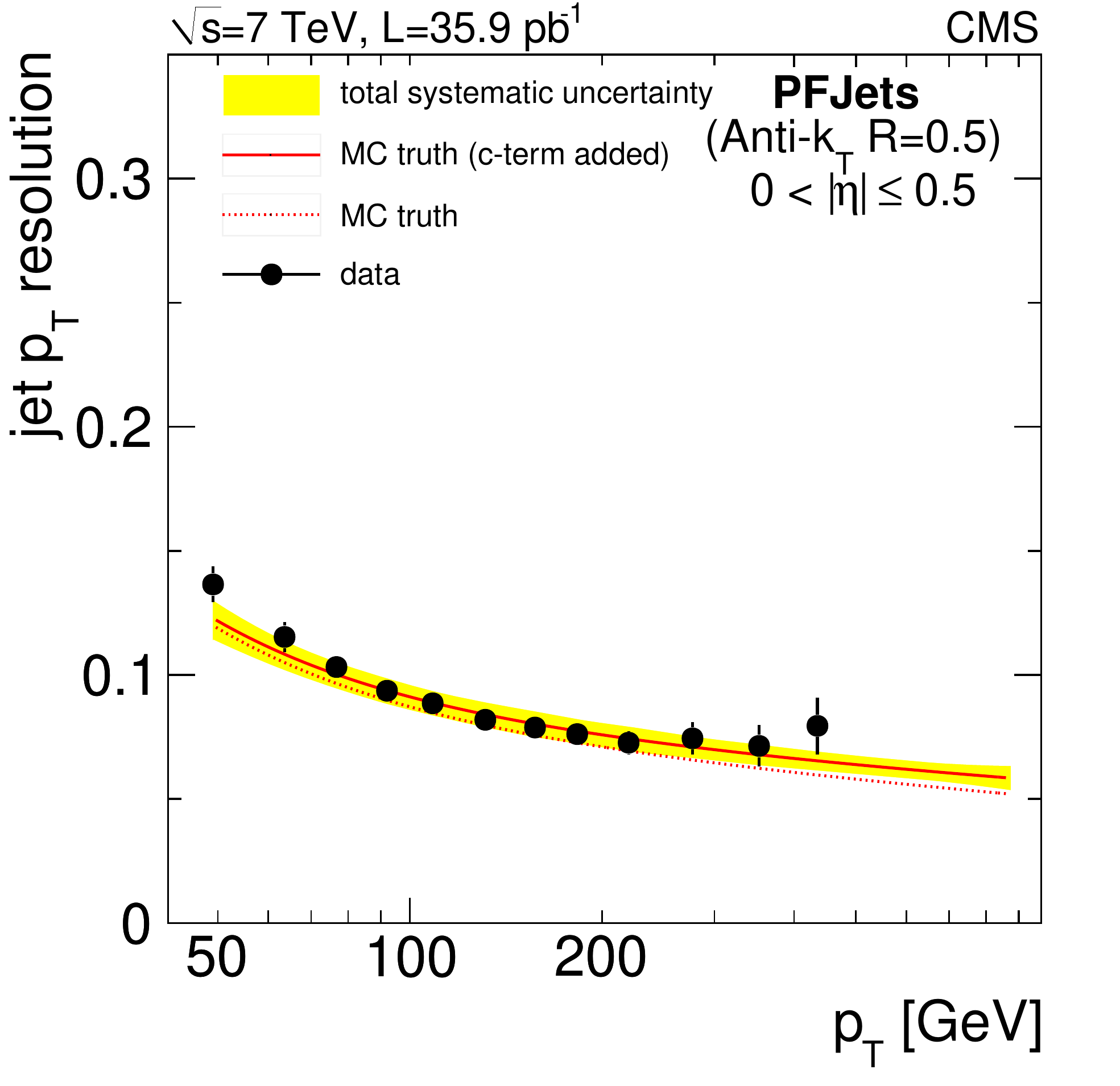}
\caption{\label{fig:Int:cmsdata} Relative jet energy resolution, extracted from the momentum asymmetry of dijet events measured with the CMS detector in proton-proton collisions. The results are compared to the width of the response for simulated jets, corrected for an additional constant term. From~\cite{Chatrchyan:2011ds}.}
\end{figure}

The CMS collaboration has verified the particle flow performance using proton-proton collision 
data~\cite{Chatrchyan:2011ds,Chatrchyan:2011tn}. 
The transverse momentum balance of dijet and $\gamma +$~jet events has been utilised to extract the resolution from the width of the observed jet asymmetry distributions. 
Soft radiation is accounted for by performing the measurement as a function of activity in the events in addition to the two jets and extrapolating to zero. 
The results are shown in Fig.~\ref{fig:Int:cmsdata} and compared to the expectation based on the relative width of the simulated jet energy response, corrected for an additional constant (c) term. 
This term was extracted from applying the same procedure to data to the simulated  events and is attributed to calibration imperfections. 

\begin{figure}[htb]
\includegraphics[width=0.95\hsize]{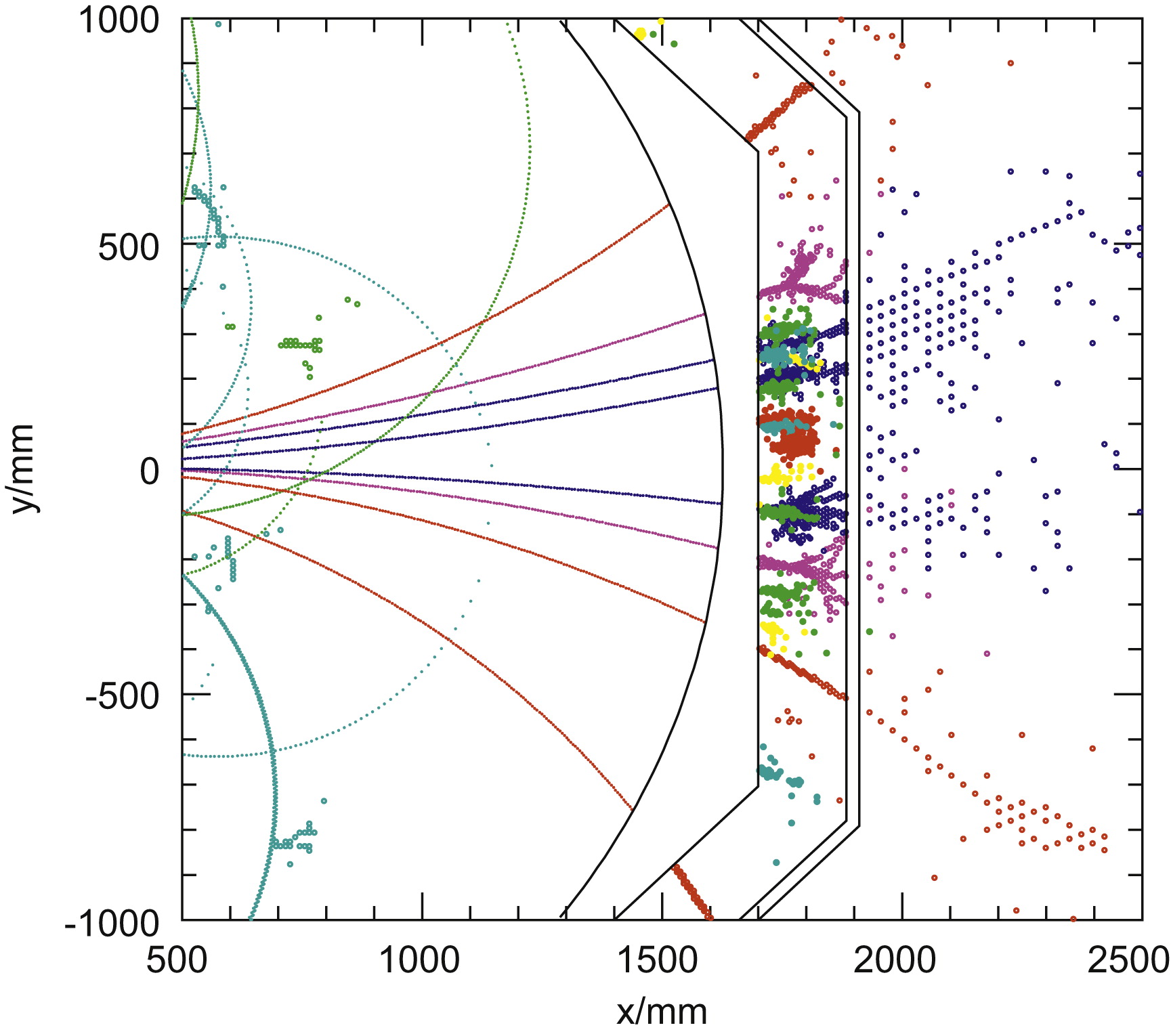}
\caption{\label{fig:Int:ildjetevt} Simulated jet in the ILD detector, with particle flow objects reconstructed by the Pandora algorithm shown in different colours. From~\cite{Thomson:2009rp}.}
\end{figure}

The CMS data confirm the expected gain in performance using the particle flow techniques. 
The net performance, however, is comparable to that of the ATLAS detector using calorimetric methods~\cite{Aad:2012ag} 
and does not yet match the goals formulated for the linear collider experiments. 
The particle flow method relies on the ability to properly assign the calorimetric energy depositions to individual particles, 
which places high demands on the imaging capabilities of the calorimeters, and on the pattern recognition performance of the reconstruction software. 
The steps towards a detector fully optimised for particle flow consist in extending the detailed topological reconstruction and high granularity into the hadron calorimeter section, in further optimising the segmentation of the ECAL and the efficiency of the tracking system, and in developing more sophisticated reconstruction algorithms. 
For illustration, a jet simulated in the highly granular ILD detector and the colour-coded result of the reconstruction algorithm is shown in Fig.~\ref{fig:Int:ildjetevt}.
Only the deposits left over, after removing those associated with charged particles or identified as photons, will be interpreted as neutral hadrons. 
In practice, this cannot always be done unambiguously, if the particles impinge too close to each other. 
Mis-assignments give rise to additional measurement uncertainty, which is called {\it confusion}. 
For example, a neutral particle shower overlapping with that of a charged one could remain unresolved and mis-interpreted as part of the charged hadron shower, which is replaced by the track measurement; so the neutral energy would be lost. 
On the other hand, a detached fragment of a charged particle shower could be mis-identified as a separate neutral hadron, and the fragment energy could be double-counted. 
Therefore, it is not {\it a priori} guaranteed that the particle flow reconstruction yields a better resolution than the calorimetric measurement alone. 

The Pandora particle flow algorithm (PFA)~\cite{Thomson:2009rp} is the most developed and best performing today. 
Recently developed alternatives like ARBOR~\cite{bib:arbor-chef13} are still less performant than Pandora but have the merit of delivering an independent validation of the Particle Flow Concept. 
For the assignment of the energy depositions to particles, Pandora makes use of topological information, including the sub-structure of showers, as well as the compatibility of calorimetric and track-based measurements.  
In this way, it is ensured that at higher energies, as jets get more collimated and particles become harder to separate, a smooth transition is made to a classic  {\it energy flow} like reconstruction, in which neutral particles are rather identified as excess in energy above the track-based expectation, and the classical, purely calorimetric performance for the jet is either retained or improved where the track assignment is unambiguous. 

\begin{figure}[hb]
\includegraphics[width=0.95\hsize]{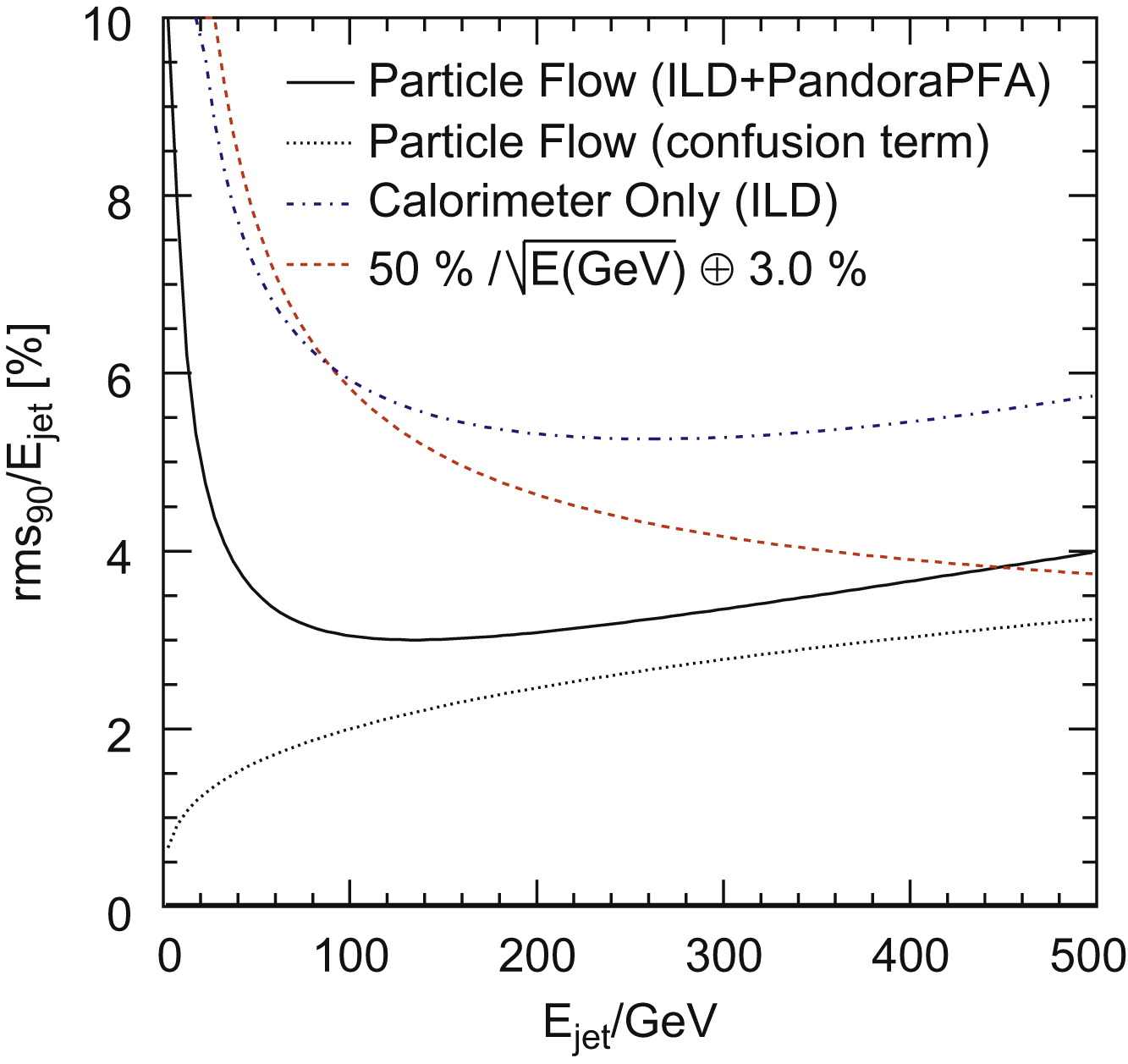}
\caption{\label{fig:Int:pflowvsenergy} Empirical form of the simulated ILD jet energy resolution as a function of energy. Also shown is the contribution from confusion, and, for comparison, the resolution obtained from the ILD calorimeter alone, and that of an ideal calorimeter  with given parameters. From~\cite{Thomson:2009rp}.}
\end{figure}

The development of the algorithm has proceeded along with a detailed understanding of the relative roles which resolution and confusion effects play in different energy regimes, and which properties of the detectors drive the performance. 
For the use of energy momentum match in the assignment of energy depositions, and for energy flow treatment of dense jets, particle flow calorimeters with their emphasis on imaging must still feature a good energy resolution. 
Furthermore, the neutral hadron energy uncertainty is the dominant contribution to the jet resolution for low energy jets, where particles are well separated, while at higher energies the confusion effects take over. 
For the detectors envisioned for the ILC, the transition is around 100~GeV,
as can be seen in Fig.~\ref{fig:Int:pflowvsenergy}.
Empirically, this can be parametrized as
\begin{equation}
\frac{\rm{rms}_{90}}{E} = \frac{21}{\sqrt{E}} \oplus 0.7 \oplus 0.004 E \oplus  \left( \frac{E}{100}\right)^{0.3} \% \;\;\; ( E \; \rm{in \; GeV} )
\end{equation}
where rms$_{90}$ is the r.m.s.\ of the smallest range containing 90\% of the events.%
\footnote{The particle flow resolution function is not Gaussian. Its statistical power 
was shown~\cite{Thomson:2009rp} to be  equivalent to that of a Gaussian with a standard deviation of 1.1$\times$rms$_{90}$.}
The individual terms represent contributions from the intrinsic calorimetric resolution,
tracking imperfections, leakage and confusion.  
The performance is compared with 
the resolution obtained with a traditional approach from calorimetric information alone. 
Even at jet energies as high as 500~GeV, where confusion, also shown separately, 
becomes dominant, particle flow brings in a significant improvement. 
Note that the degradation at high energies is also due to leakage, which affects the 
purely calorimetric measurement much more severely. 
A particular strength of high granularity is the possibility to use topological information 
such as the reconstructed starting point of the shower for the estimation of leakage.
This has not yet been exploited here and has the potential to mitigate the effects further. 

In the framework of studies for CLIC~\cite{Linssen:2012hp}, it was shown that the required 
jet energy resolution can be achieved with the PFlow technique for jet energies up to 1500~GeV. 
The studies~\cite{Marshall:2012ry} also demonstrate that missing transverse momentum can be measured with 
a similar precision as jet energy, and that fake missing momentum (in one coordinate) 
is limited to 1 -- 2\% of the event energy.

Both detector concepts developed for the ILC, ILD and SiD, ~\cite{Behnke:2013lya}, and in modified versions for CLIC~\cite{Linssen:2012hp}, have their design based on the particle flow paradigm.
PFlow demands a highly efficient tracking system.
In order to separate the particles, it calls for extremely compact electromagnetic calorimeters -- to keep the Moli\`ere radius small -- and for unprecedented fine calorimeter granularities. 
To isolate photons and resolve the sub-structure of hadron showers, transverse and longitudinal cell sizes in ECAL and HCAL must be of the order of a radiation length $X_0$, resulting in channel counts of $10^7 - 10^8$.
Both ECAL and HCAL must fit inside the magnetic coil, in order not to looe continuity in tracking the shower evolution. 
The radial and longitudinal distance of the calorimeter from the interaction point and the magnetic field should be large to allow separation of shower components. 
For the same relative change, the dependence on radius is stronger than on the field, but then cost considerations need to be folded in, as well. 
The main difference between ILD and SiD is that ILD has chosen to favour a larger radius tracking system with a time projection chamber (TPC) and a smaller field, while SiD follows a more compact design with an all-silicon tracker and a higher field. 
At CLIC energies, leakage becomes more important. Since the radius of the coil is limited by technical and cost considerations, tungsten is chosen as the HCAL barrel absorber material to ensure sufficient shower containment. 
The calorimeter technologies, however, are very similar for all cases and have motivated a common R\&D and validation effort. 

\subsection{Validation approach}

The detector requirements imposed by the particle flow principle  -- high field, large size, dense materials, fine segmentation --  drive the cost of the resulting detector systems far beyond that of previous $e^+e^-$ collider experiments. 
A careful optimisation is thus mandatory, and the quest for an experimental validation of the performance potential held by the particle flow approach is highly motivated. 

It has been suggested to directly test the jet energy performance in test beams by creating bundles of particles from a primary beam hadron impinging on a thin target. 
Leaving aside the differences in particle momenta and multiplicity, or energy density,  between these "jets" and those generated in quark fragmentation at the same primary energy, such an experiment would have prohibitive cost, as simulation studies have shown~\cite{jmpe}.
For particle flow reconstruction magnetic momentum spectroscopy and large acceptance are indispensable. 

Consequently the experimental strategy must be to validate the critical ingredients of particle flow calorimetry individually. 
First of all, the need for high granularity has spurred the development of novel calorimeter read-out technologies, such as large area silicon diode arrays, silicon photo-multipliers (SiPMs) for optical read-out, 2D-segmented resistive plate chambers and micro-pattern gas amplification structures. 
Large scale prototypes have been built and high statistics test beam data have been collected over several years in a worldwide effort organised by the CALICE collaboration~\cite{Adloff:2012dla}. 
The goals were to test the new technologies and demonstrate their performance, to validate the simulation models for hadronic shower evolution in the detail necessary for particle flow reconstruction and, finally, to test the particle flow algorithms on real data.  
This must be done for a number of different absorber materials and read-out media, since, for example, the role of neutron production is different for light and heavy absorbers, and gas and scintillator have different sensitivity to the various sub-components of the showers. 
The sensitivity to, e.g.\ neutrons depends on the  hydrogen content of the active medium, and on the timing of the read-out electronics. 
 
In detail, the issues to be addressed are:
\begin{itemize} 
\item {\bf Technical performance:}
The novel technologies must be tested with prototypes large enough to contain electromagnetic or hadronic showers, respectively, and demonstrate the expected performance in terms of noise, linearity, resolution, uniformity and stability.   
\item {\bf Detector understanding:} 
The detector performance must be modelled in simulations in sufficient detail, and the models must be validated with muons and electrons, for which the interactions with the detector materials can be reliably simulated. 
\item {\bf Software compensation:}
The potential offered by fine read-out segmentation to apply software compensation methods~\cite{Abramowicz:1980iv,Andrieu:1993tz,Cojocaru:2004jk} for restoring linearity and improving the hadronic energy resolution should be realised. 
\item {\bf Calibration:}
It must be demonstrated that the unprecedented number of channels individually read out can be monitored and calibrated to the required precision. 
\item {\bf Hadron shower simulation models:}
Particle flow performance is driven by hadronic energy resolution and confusion. The hadronic shower evolution, detector response and resolution as well as lateral and longitudinal extension must be correctly modelled and confronted with high precision test beam data. 
\item {\bf Shower sub-structure:}
The reconstruction of individual particle showers makes detailed use of the shower topology, and sub-structures like tracks linking different fragments of the same shower, which should be seen in beam data and quantitatively tested. 
\item {\bf Particle flow algorithms:}
Particle flow algorithms should be applied to real test beam data, and their capability to resolve the topologies and to separate particle showers from each other should be measured and compared with predictions. 
\end{itemize}

This paper deals with the issues above and thus lays the experimental foundation for the ingredients that go into full detector simulation studies with realistic and complex collider event topologies. These are documented in the context of the Pandora development~\cite{Thomson:2009rp}, or in the 
reports on detector concepts~\cite{Behnke:2013lya,Linssen:2012hp}.
We do not recapitulate these studies here, but we would like to point out that a complete exploration of the particle flow approach consists of both, the full simulation studies and the experimental tests reported here.

The paper will first introduce the candidate technologies for highly granular calorimeters, sketch their implementation in the linear collider detector concepts ILD and SiD, and describe the large prototypes built and tested by the CALICE collaboration.  
Their performance in test beams will be presented and confronted with expectations, and the required simulation details will be discussed. It should be noted that not all prototypes are at the same stage of implementation or testing.
To represent the current situation we present the full range of prototypes at their various stages of development.
Improvements achieved with different approaches to software compensation are described in a separate section, together with the results obtained with a combination of ECAL and HCAL prototypes. 
A calibration section deals with 
the extrapolation to a full detector system,  and the challenges and benefits of high granularity in this respect will be discussed. 
The section on shower model validation starts with an overview of current state-of-the-art simulation programs and presents comparisons of their predictions with data for a number of global observables as well as for shower sub-structure. 
A central result of the validation effort is the application of the Pandora particle flow algorithm~\cite{Thomson:2009rp} to CALICE data and the test of the two particle separation power with test beam events. 
The article closes with a near-term outlook and possible future directions. 
\section{Particle flow detectors and calorimeter technologies%
\label{sec:Technologies}}
%

\subsection{The ILD and SiD calorimeter systems}

\begin{figure}[htb]
\centering		    
       \includegraphics[scale=0.8]{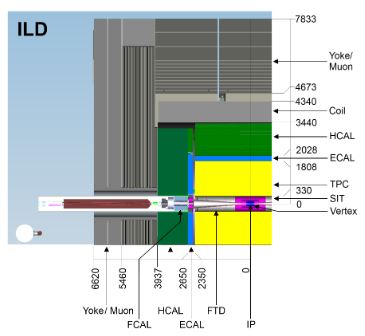}
		
\caption{ \label{fig:ILD_Quadrant} The ILD Detector for the ILC. From~\cite{Behnke:2013lya}.}
\end{figure}

\begin{figure}[htb]
\centering
       \includegraphics[scale=0.2]{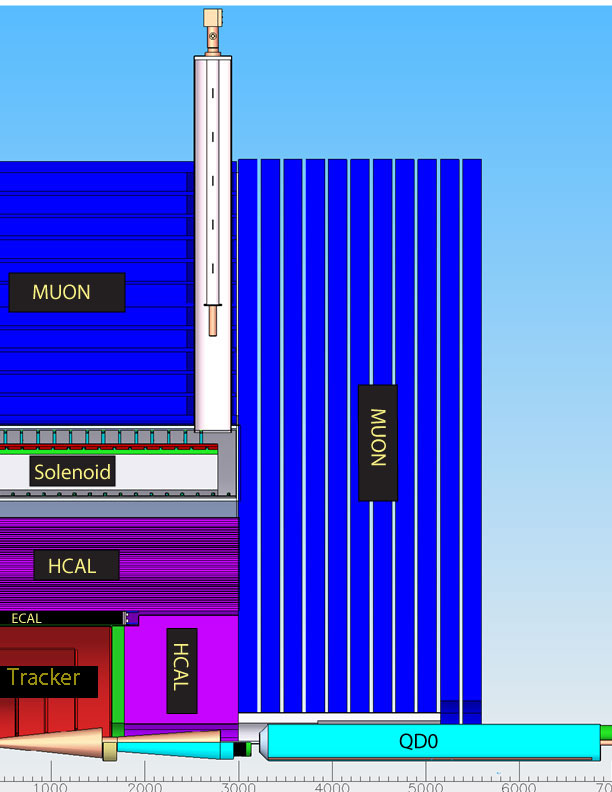}
     
\caption{  \label{fig:SiD_Quadrant} The SiD Detector for the ILC. 
From~\cite{Behnke:2013lya}.}
\end{figure}

The ILD and SiD calorimeter systems are central features of these 
detector concepts~\cite{Behnke:2013lya,Linssen:2012hp}.
The calorimeters are part of integrated detector designs to take advantage of the 
PFA approach to achieving the excellent jet energy resolutions required by experiments 
at future lepton colliders. Quadrant views of the ILD and SiD detectors are shown in 
Fig.~\ref{fig:ILD_Quadrant} and Fig.~\ref{fig:SiD_Quadrant} respectively.

Somewhat modified designs have also been developed for experiments at the higher energies
of CLIC, the main differences being greater calorimeter depths, and a more demanding timing
requirement.

The most significant difference between the ILD and SiD designs lies in the tracking systems:
ILD uses a time projection chamber (TPC) supplemented with layers of silicon, while the SiD
tracker is an all-silicon system. As a result, the inner radius of the ILD calorimeters is
larger than that for the more compact SiD design. The central, solenoidal, magnetic fields
have values of 3.5T and 5T for ILD and SiD respectively. ILD benefits from greater track
separation at the entrance to the calorimeter, due to the larger radius, while SiD has high 
precision tracking in a limited number of layers in its compact design using a larger field.
For both ILD and SiD, the calorimeter systems are separated into electromagnetic (ECAL) and hadronic
(HCAL) depth sections. The SiD HCAL features a long barrel with a plug forward HCAL, while ILD
has a shorter barrel HCAL with endcaps. 
In both concepts the iron flux return yoke is foreseen to be instrumented for muon tracking and to 
act as a tail catcher to the calorimeters. 

The calorimeter implementations for use with a PFA are highly granular in nature - as required
 to record very detailed images of showers for subsequent identification and separation of energy
deposits, and the measurement of the energies of neutral particles. Early studies showed that 
for small enough cell sizes
there is an approximately linear relation between the number of calorimeter cells recording a hit
for a shower and the energy of the particle(s) causing the shower~\cite{Ammosov:2004xb}. 
At very high hit densities saturation effects occur and corrections need to be made. Several variations on this 
approach have been developed, digital and semi-digital, in addition to the more conventional analog approach.
The fully digital method applies a predetermined threshold to each cell as the data is taken. 
The semi-digital method is the same except that it allows for more than one threshold to be applied.
The analog approach records cell signal magnitudes and stores the information for offline reconstruction.
In the digital methods based on hit counting the energy resolution depends on the longitudinal and transverse 
granularity, while in the analogue case it depends on the longitudinal sampling only. 
Therefore the digital methods require smaller cell sizes. 

The ECAL serves to identify electrons and photons and to measure photon
energies. Approximately 60\% of hadrons interact in the ECAL featuring a depth of about one interaction length. In the ECAL overlapping photon-hadron energy deposits need to be disentangled and electron-hadron charged tracks need to be separated. To minimise overlaps, electromagnetic showers
must be confined as much as possible, favouring absorber materials with a small Moli\`ere radius.
Additionally, to facilitate the separation and identification of electromagnetic and hadronic showers,
it is helpful to have a large ratio between the interaction length and radiation length.
Distinguishing electromagnetic showers is facilitated by using a high transverse granularity,
and fine depth segmentation with many layers. This latter feature also assists with following
charged particles through the ECAL to the HCAL. 

The ECAL's for ILD and SiD have tungsten as the absorber. Tungsten has a Moli\`ere radius of 9.7mm
satisfying the requirement to narrowly confine electromagnetic showers. 
Silicon is the material of choice for the ECAL active layers, as such layers are easily segmented, 
although there is an alternative design for the ILD ECAL that uses orthogonal scintillator strip layers. 
The combination of silicon sensor layers and tungsten absorber allows for compact active layer designs 
and for division into small (${\cal O}$(5~mm)) cells in the transverse plane. Longitudinally the ILD and SiD
ECAL's have 30 layers, with, for instance, SiD having the first 20 layers with 2.5 mm tungsten thickness
and 1.25 mm readout gap, and the last 10 layers having 5.00 mm tungsten - a compromise between cost,  
the sampling frequency, and the containment of showers. Providing full information for a particle flow
algorithm demands that every cell in every layer is read out. 

Two different designs have been developed and prototypes constructed for ECAL's using silicon and
tungsten, one by CALICE and a second independently by SiD. 

\begin{figure}[htb]
\centering
       \includegraphics[width=1.0\hsize]{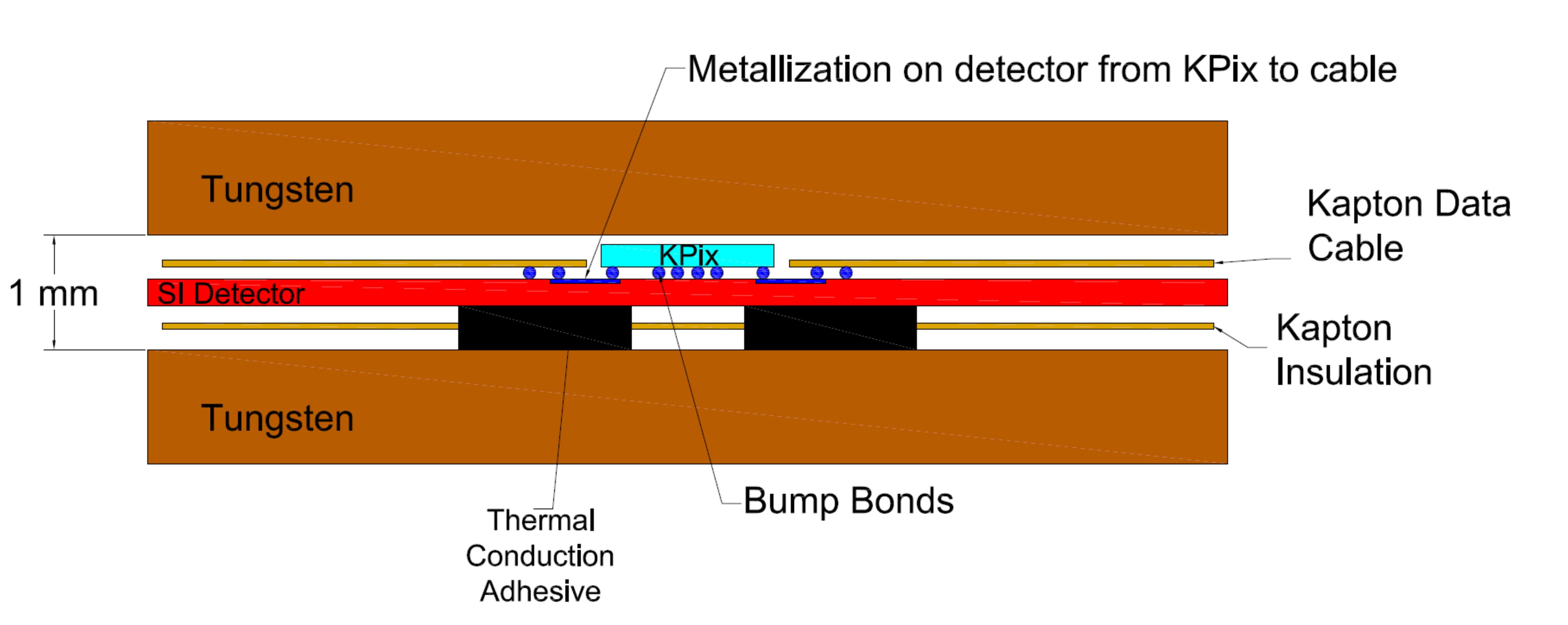}
          
\caption{ \label{fig:SiD_ECAL} The layer structure of the SiD Silicon-Tungsten Electromagnetic Calorimeter.}
\end{figure}

The SiD ECAL design is based on a self-supporting structure of tungsten plates interconnected with
screw and insert-spacer assemblies. Wedge-shaped modules, for the barrel ECAL, are assembled by stacking 
alternating layers of tungsten and silicon - with the sensor layers permanently 
installed in the structure. The current baseline design of the sensor layers is a tiling scheme using 15 cm 
hexagonal sensors, each subdivided into 1024 hexagonal pixels of 13~mm$^2$ area. All pixels are connected
to a single KPiX 1024-channel ASIC mounted directly in the center of the sensor. The digitised data are read out
via Kapton flex cables, which also carry power. The overall power requirement is significantly reduced by
employing power pulsing, a central feature of ILC detectors made feasible by the long interval between
bunch trains.
A view of the layer
structure
for the SiD ECal is shown in Fig.~\ref{fig:SiD_ECAL}.
%

An alternative for the ECAL is based on Monolithic Active Pixel Sensors with 50 micron x 50 micron 
silicon pixels. This is a digital approach to electromagnetic calorimetry (DECAL).
Such sensors could be manufactured in a commercial
mixed-mode CMOS process using standard 300 mm wafers.  
First-generation DECAL sensors with $168\times 168$ pixels 
have been manufactured and tested. 
The sensor
supports single-bunch time stamping with up to 13 bits and power-pulsing.

\begin{figure}[htb]
\centering
\includegraphics[width=0.3\textwidth]{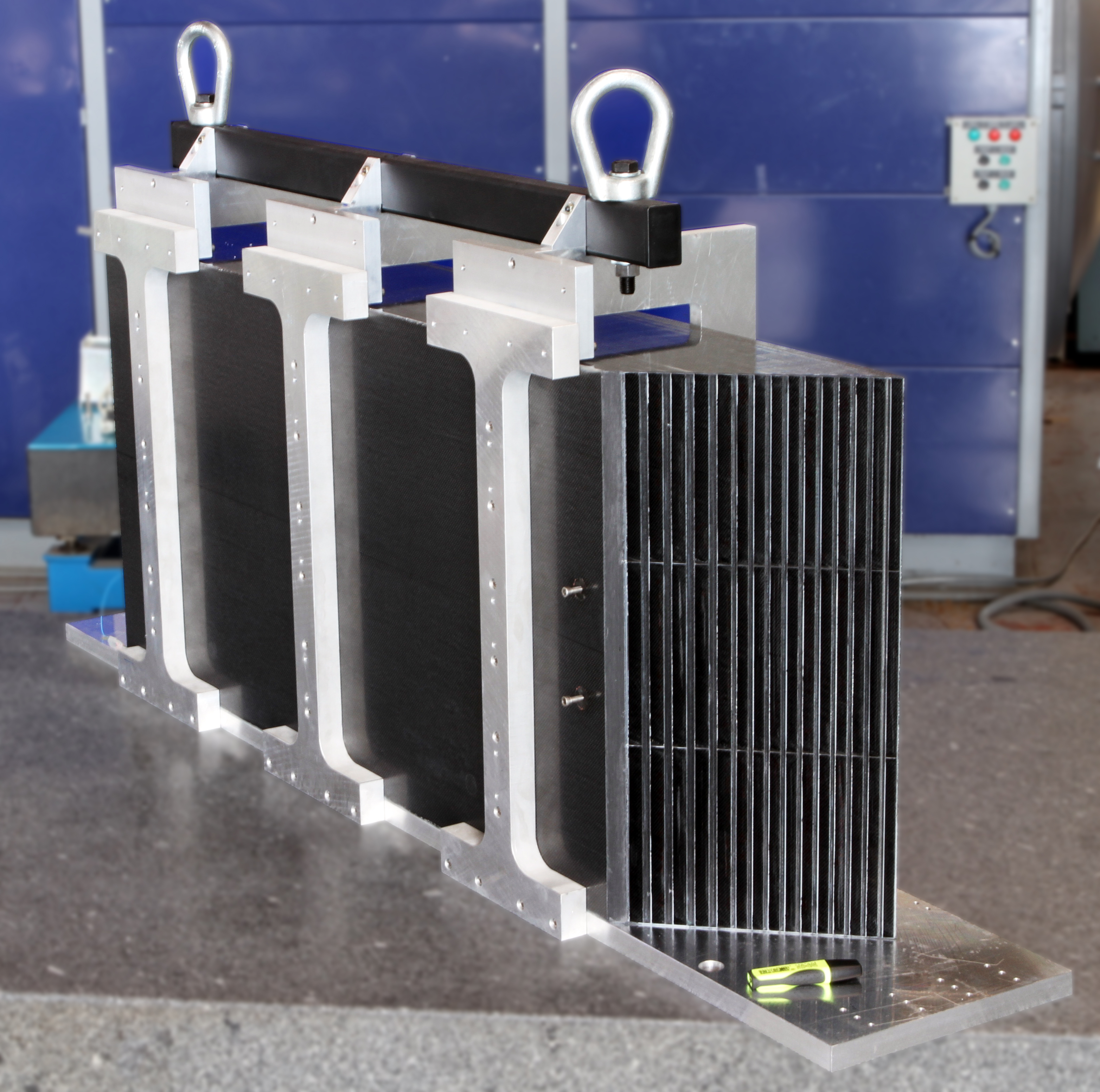}
\caption{\label{fig:ILD_ECAL} Prototype ILD Silicon-Tungsten Electromagnetic Calorimeter. From~\cite{Behnke:2013lya}}
\end{figure}

The design retained so far by ILD is pursued by the CALICE collaboration. It uses an alveolar structure of carbon fibre, into which are inserted
bare tungsten plates and tungsten plates carrying sensor, readout elements and control services on both
sides. 
The silicon sensors are segmented into square pads of 5~mm size. A view of a prototype for this structure is shown in Figure~\ref{fig:ILD_ECAL}. The front-end electronics for 
this calorimeter are provided by ASICs, called SKIROC2~\cite{Callier:2011zz} that are integrated into the layer structure. In the ILD baseline the ASICs will be bonded onto a very thin multilayer printed circuit board that is part of the assembly that is inserted into the alveolar structure. The SKIROC ASICs as well as others of the same 'family' mentioned in this section combine signal amplification, shaping, triggering and digitisation. The ASICs can be power pulsed and address thus all aspects needed for an experiment at the LC. In parts 
their performance are subject of this review.

%
An alternative CALICE ECAL design, to reduce the channel count and cost, uses 5~mm wide and 45~mm long orthogonal scintillator strips with silicon photo-multiplier (SiPM) readout in alternating layers between tungsten absorber plates. The readout electronics is provided by ASICs adapted to the readout of silicon photomultipliers, SPIROC2 in this case~\cite{ConfortiDiLorenzo:2013vka} bonded to a thin printed circuit board.

The HCAL must allow efficient separation and identification of energy deposits from charged particles,
and provide an adequate measurement of the energies of neutral hadrons. These requirements again
argue for fine transverse and longitudinal segmentation. The transverse segmentation should be small
compared to the typical size of a hadronic shower, while longitudinally there should be a large number 
of layers for shower pattern recognition, while being consistent with sufficient thickness to contain
a high fraction of the energies of the most energetic particles/jets. The whole calorimeter
system must be contained within the inner radius of the central solenoid, whose size is limited
by achievable technology and cost, and whose thickness corresponds to about 1.5 nuclear interaction lengths.
The possibility exists to use the first layers of the muon system,
radially outside the coil, as a tail-catcher, to identify and measure the energy of any small components 
of hadronic showers that propagate through the calorimeters and coil (for the barrel section of the
calorimeter).

\begin{figure}[htb]
\centering
       \includegraphics[width=0.8\hsize]{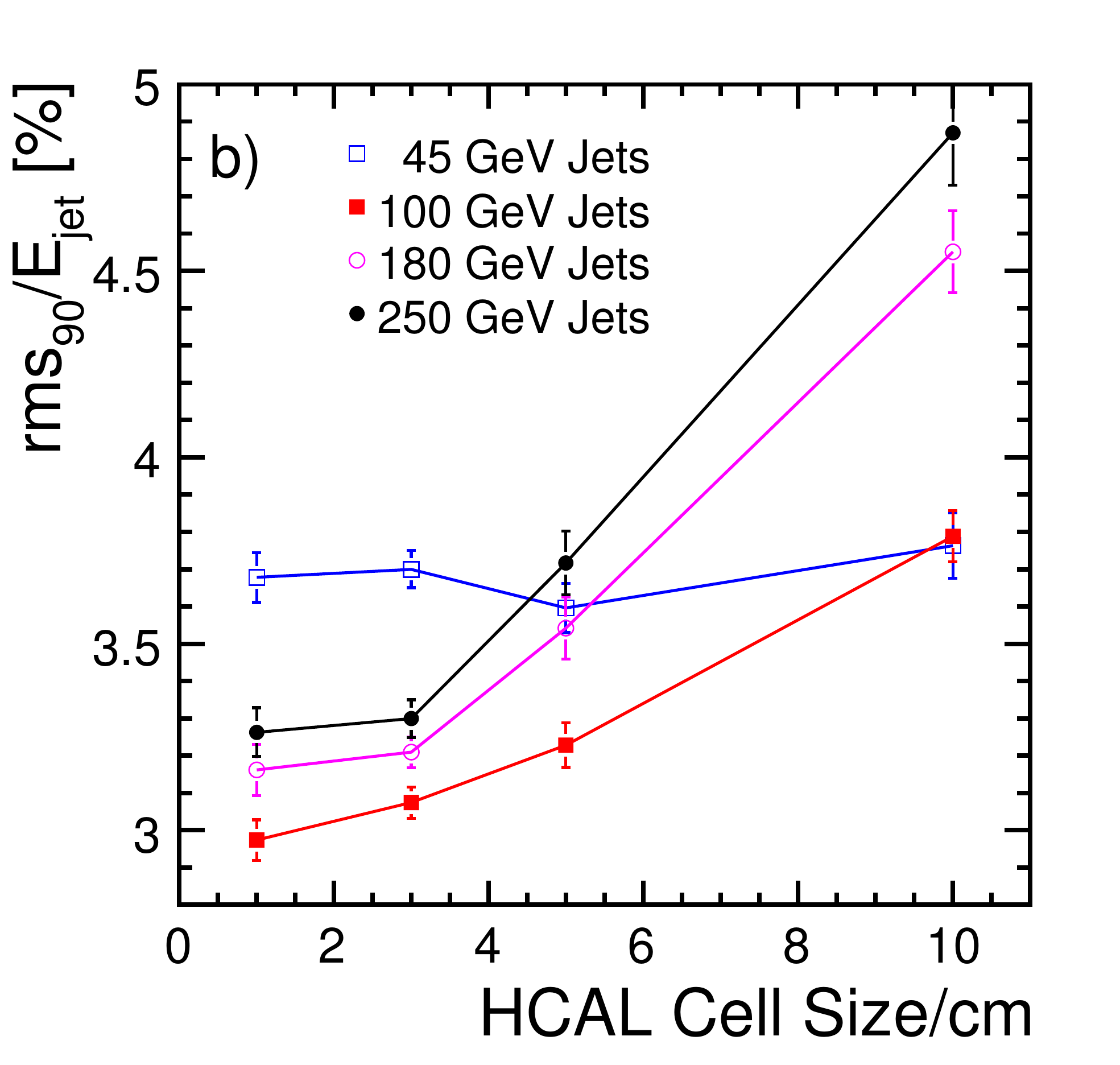}

\caption{\label{fig:AHCAL_Cell_size} Simulated jet energy resolution of the ILD detector as a function of the 
AHCAL cell size. From~\cite{Thomson:2009rp}.}
\end{figure}

The analog HCAL uses layers of small scintillator tiles and iron or tungsten absorber. 
Each small tile, with 3 cm x 3 cm size, has an embedded wavelength shifting fibre which is coupled to a SiPM. 
Studies have shown that there is little to be gained from using a cell size smaller than
this - see Fig.~\ref{fig:AHCAL_Cell_size}. The scintillator tiles provide energy and position 
resolution and have been shown to give a uniform response across layers and high efficiency for minimum ionising particles (MIPs) - 
essential ingredients for a successful particle flow calorimeter. Ongoing work has indicated that it
may be possible to have the SiPMs directly coupled to the scintillator tiles with the latter being
shaped to achieve good uniformity of response across their 
faces~\cite{Blazey:2009zz,AbuAjamieh:2011zz,Simon:2010hf}.
The analog HCAL for ILD is divided into 48 
longitudinal layers, corresponding to six interaction lengths to retain good calorimeter performance 
up to 1 TeV center-of-mass energy. The front-end readout electronics consists of 
SPIROC ASICs~\cite{ConfortiDiLorenzo:2013vka} mounted on
a printed circuit board and connected to the SiPMs carried with the scintillating tiles, with the
complete assembly being the active layer between absorber plates. Interface boards for signal, calibration 
and power are located at the end faces of AHCAL modules.
%
%
%
%

\begin{figure}[htb]
\centering
       \includegraphics[width=0.98\hsize]{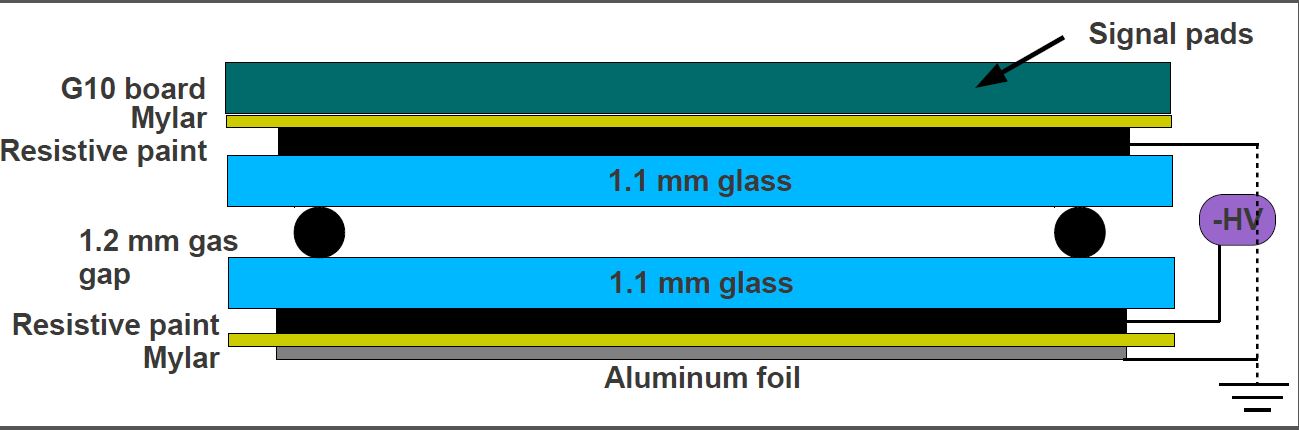}

\caption{\label{fig:RPC_layer} A schematic of the layer structure of the RPC Digital Hadron Calorimeter}
\end{figure}

In contrast to the scintillator-based HCAL, a number of gas-based calorimeters have been proposed.
Two variations of calorimeters using Resistive Plate Chambers (RPC) have been built and tested. The first is
a fully digital RPC system with glass plates acting as resistive plates. In the original design, two
glass plates, coated on their outer surfaces with resistive paint, define an active layer with a
typical thickness of less than 8 mm. High voltage is applied to the resistive layers. A charged particle
crossing the gas gap between the plates causes an avalanche which induces a signal on readout pads
adjacent to the plates. A schematic of the layer structure is shown in Fig.~\ref{fig:RPC_layer}. 
A pad size of 1 cm$^2$ is used and a pad is counted as either hit or not hit depending on the result
of a comparison between the signal size and a pre-determined threshold, downloaded to the front-end
electronics. The 1 cm$^2$ pad size was chosen to mitigate the effects of saturation on energy resolution. 
A pad multiplicity of less than two has been achieved for a hit efficiency of 95\%. 
The ASIC used has been the DCAL chip developed specifically for this purpose.
This glass RPC design is the baseline choice for the SiD HCAL system.
In recent developments, a one-glass (the one with the resistive paint) RPC has been tested and offers the possibility of a thinner active
layer, together with a hit multiplicity close to one independent of hit efficiency.

A multiple threshold variation of the glass RPC design for the ILD HCAL has also been developed. This 
semi-digital RPC HCAL, called SDHCAL hereafter,  uses two-bit readout, implementing three thresholds. This allows mitigation of
 saturation effects provoked by large energy deposits, and the determination of 
whether one, a few, or many particles crossed a given cell - providing additional information to the
particle flow algorithm. The front-end electronics is provided by HARDROC ASICs~\cite{{Dulucq:2010ssa}} mounted on one side of
a printed circuit board, the other side of which carries the inductive signal pickup pads.

\begin{figure}[htb]
\centering
       \includegraphics[scale=0.4]{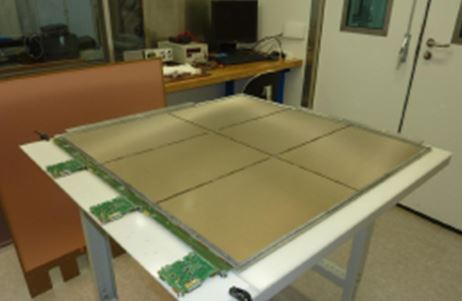}

\caption{ \label{fig:MM_plane} A view of a plane of the micromegas-based Digital Hadron Calorimeter}
\end{figure}

Gas-based HCAL designs have also been developed to take advantage of micro-patterned gas detector technology,
using both micromegas and gas electron multiplier (GEM) approaches. In the micromegas design a commercial 20 micron 
woven mesh separates the 3mm drift gap from a 128 micron amplification gap. Signals are acquired on 1 cm$^2$
pads on one side of a printed circuit board, the other side of which contains the readout ASIC's. This approach
is proposed as an alternative to the RPC semi-digital design and also uses several signal thresholds. A view of 
a large area, assembled micromegas digital hadron calorimeter plane is shown in Fig.~\ref{fig:MM_plane}.

\begin{figure}[htb]
\centering
     \includegraphics[width=0.5\hsize]{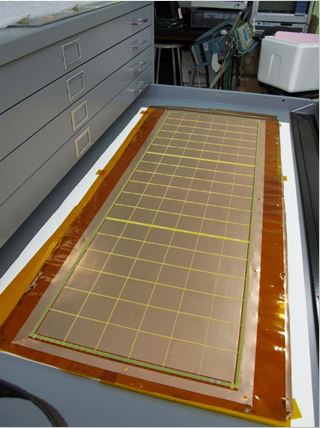}

\caption{\label{fig:GEM_plane} A plane under assembly for the GEM-based Digital Hadron Calorimeter}
\end{figure}

Finally, two approaches using gas-electron-multiplier technology (GEM) have been proposed.
The first GEM-based design uses two layers of GEM foils separated by 1 mm, a 3 mm drift region in front of the
foils, and 1 mm induction region beyond the foils. As for the micromegas case, signals are collected on
1 cm$^2$ anode pads on a printed circuit board also containing the readout ASICs. The ASIC used with this
technology has been the KPiX 1024-channel analog device, allowing the possibility of recording the signal
level on each pad for later processing with offline thresholds. A view of a large GEM chamber under construction 
is shown in Fig.~\ref{fig:GEM_plane}.

\begin{figure}[htb]
\centering
       \includegraphics[width=0.8\hsize]{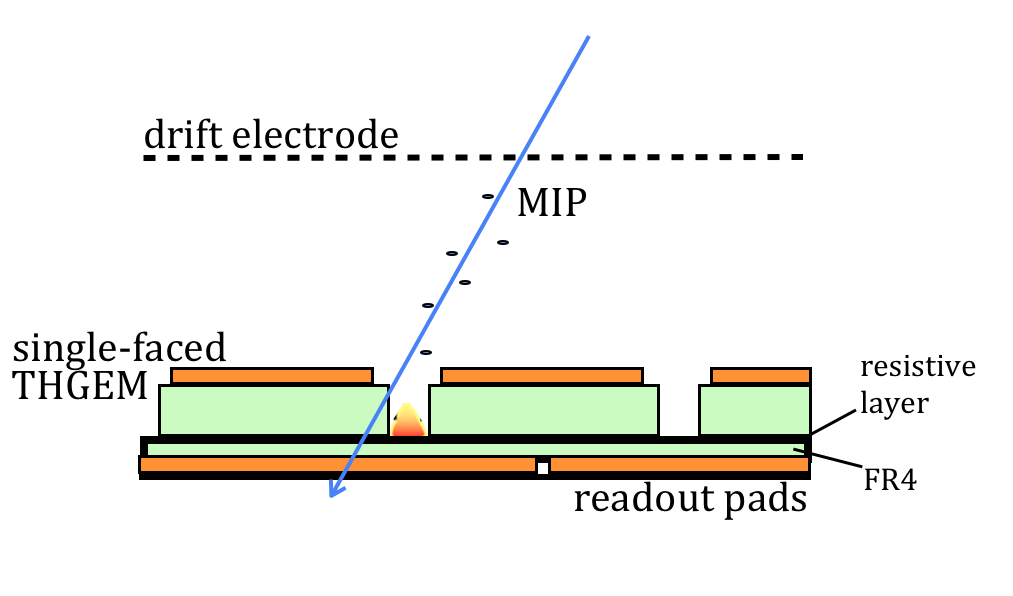}
\caption{\label{fig:THGEM_design} A thick-GEM structure which provides discharge protection.}
\end{figure}

The second GEM-based design uses thick-GEM (THGEM) technology with holes in 400-500 micron thick circuit boards.
Several single and double THGEM structures have been developed. Promising solutions for thin active layers
use a single THGEM which includes a resistive layer to prevent sparking and shield the front-end electronics
from the effects of any residual discharges. A schematic of a possible THGEM structure is shown in
Fig.~\ref{fig:THGEM_design}.

 \subsection{The large CALICE beam test prototypes}
 
The candidate absorber materials and read-out technologies for highly granular electromagnetic and hadronic calorimeters are summarised in Table~\ref{Tab:Tech:tree}. 
\begin{table}[htb]
\begin{ruledtabular}
\begin{tabular}{lcc}
 & ECAL & HCAL  \\
\colrule
Absorber & tungsten & stainless steel (ILC) \\
& & tungsten (CLIC barrel) \\
Analogue  & {\bf silicon, scintillator} & {\bf scintillator} \\
Digital  & MAPS & {\bf RPC}, GEM, Micromegas \\
\end{tabular}
\end{ruledtabular}
\caption{\label{Tab:Tech:tree} Absorber materials and read-out technologies considered for particle flow calorimeters. Those realised in large prototypes are printed in bold face.}
\end{table}

 Test beam experiments played a key role in establishing these technologies and validating the associated simulations. 
 A large international effort has been carried out in the framework of the CALICE collaboration, where all major technologies have been exposed to particle beams, and large sample of data has been collected at DESY, CERN and Fermilab facilities.  
Given the scale of the effort, this only became possible by maximising the use of common infrastructure, such as mechanical devices and absorber stacks, a common data acquisition system and a family of front end ASICs with common building blocks, computing, storage and reconstruction software.  
In the following we present the large prototypes, capable of providing full shower physics data, and a few smaller set-ups for technical tests. 

 
%

 

\begin{figure}
\begin{center}
\includegraphics[width=0.4\textwidth]{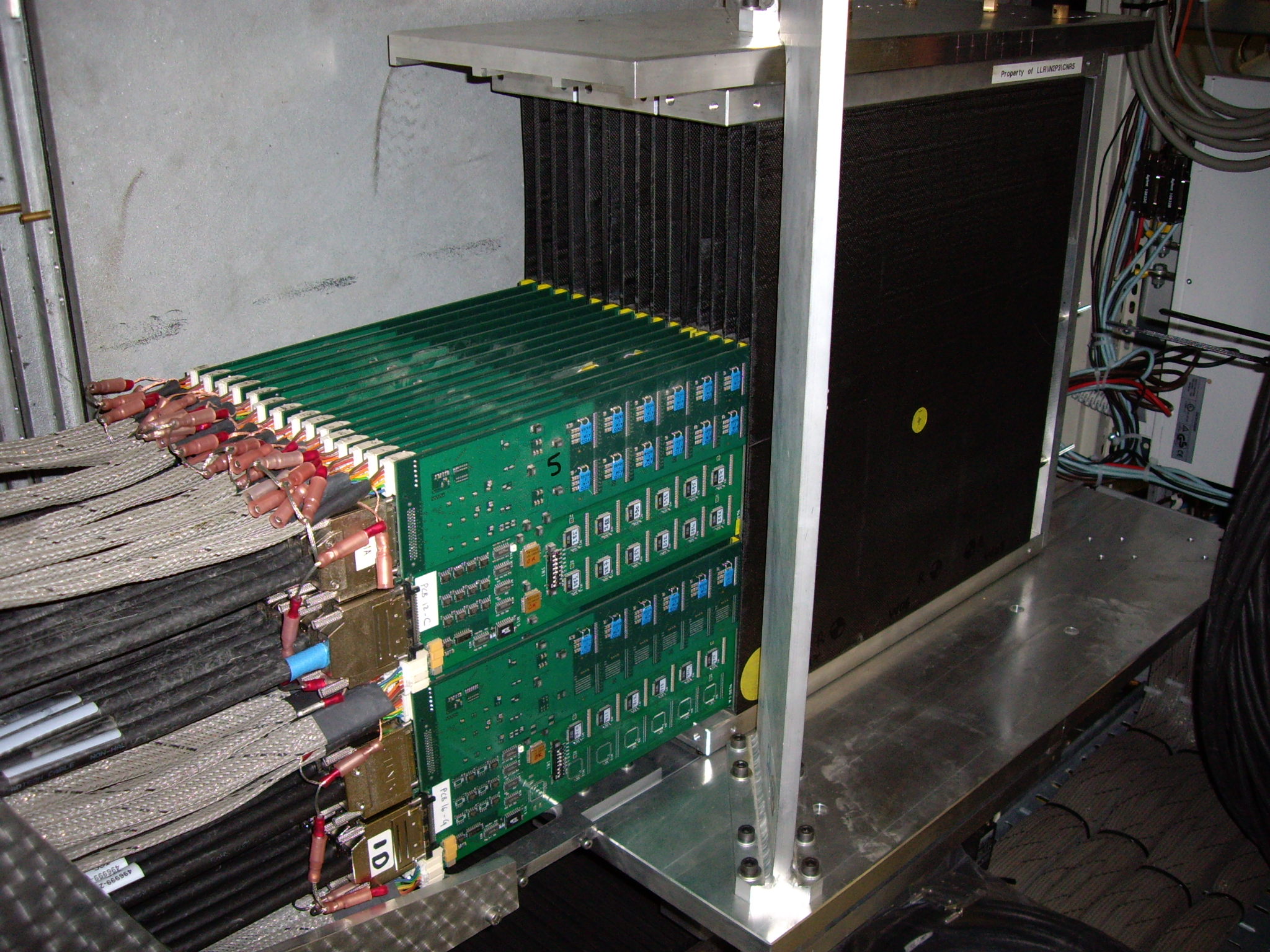}
\end{center}
\caption{The CALICE SiW ECAL physics prototype detector.}
\label{siw:fig:physProto}
\end{figure}

\paragraph {\textit{\textbf{SiW ECAL:}}}\label{par:siwecal-intro}
The CALICE SiW ECAL group has developed a first so-called ``physics prototype``~\cite{Anduze:2008hq},
shown in Fig,~\ref{siw:fig:physProto}, whose aim was to demonstrate the ability of such an ECAL to meet the
performance requirements. For details of the layout please consult Ref.~\cite{Anduze:2008hq}. Here only the gross features are recapitulated.
The physics prototype has an active area of  ${\rm 18 \times 18\,cm^2}$ in width and approximately 20\,cm in depth, subdivided longitudinally into 30 layers. 
The layers are composed alternately of W absorber plates and a matrix of PIN diode sensors on a silicon wafer substrate. The active part of a layer consists of
${\rm 3 \times 3}$ silicon wafers featuring a matrix of ${\rm 6 \times 6}$ PIN diodes. Altogether the SiW ECAL comprises thus a total of
of 9720 1x1\,${\rm cm^2}$ calorimeter cells. 
Electronic read-out proceeds via an 18 channel ASIC, FLC\_PHY3~\cite{Anduze:2008hq} followed by an
off-detector digitisation and data acquisition system~\cite{Dauncey:2002vf}.
At normal incidence, the prototype has a total depth of 24\,$X_0$ achieved using 10 layers of 0.4\,$X_0$ tungsten absorber plates, followed by 10 layers of 0.8\,$X_0$, and another 10 layers of 1.2\,$X_0$ thick plates. Each layer is subdivided into a central part featuring a $3\times2$ array of silicon wafers and a bottom part consisting of a $3\times1$ 
array of silicon wafers. In 2006 the bottom part of the first six layers was missing. The detector was progressively developed until completion in 2008.

%

\paragraph {\textit{\textbf{SciW ECAL:}}}\label{par:sciwecal-intro} 
The CALICE scintillator strip-based ECAL (ScECAL) achieves the required granularity for Particle Flow 
with a scintillator strip structure.
Scintillator layers interleaved with absorber plates
are placed in two alternative orientations,
with horizontally and vertically aligned strips,
to achieve effectively fine square segmentation.
Each strip is individually read out by a Multi Pixel Photon Counter~\cite{Gomi:2007zz}
(MPPC, a silicon photo-multiplier produced by Hamamatsu Photonics) with 1600 pixels.
Signals from the MPPCs are read out by the same front-end electronics developed for the CALICE analog hadron calorimeter prototype~\cite{collaboration:2010hb}.

The first ScECAL prototype, consisting of 468 channels, was constructed and tested at DESY~\cite{Francis:2013uua}.
It consisted of 26 pairs of 3~mm thick scintillator and 3.5~mm thick absorber layers. 
The absorber material was composed of 82\% tungsten, 13\% cobalt and about 5\% carbon. 
Each scintillator layer consisted of two $45\times 90$~mm$^2$ ``mega-strip'' structures consisting of nine $45 \times 10$~mm$^2$ strips.
The total active volume was about $90 \times 90 \times 200$~mm$^3$.
The total thickness of the prototype was 17.3 radiation lengths.
%
The mega-strips were produced by machining holes and grooves in a 3~mm-thick Kuraray SCSC38 plastic scintillator plate. 
Two types of detection layers were produced: one with a 1 mm diameter Kuraray Y-11 wavelength shifting (WLS) fibre running along the length of the strip (type-F), and the other without the WLS fibre or its associated hole (type-D).The presence of the WLS fibre improves the response uniformity along the strip length.

\begin{figure}[htb]
\includegraphics[width=0.5\hsize]{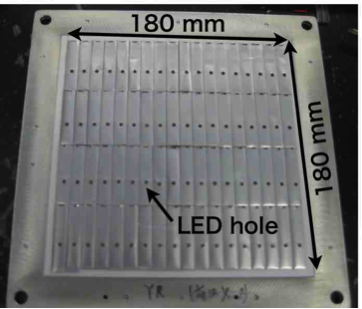}
\includegraphics[width=0.45\hsize]{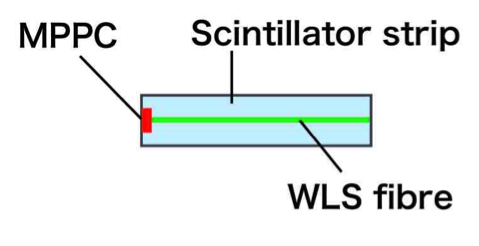}
\caption{\label{fig:tech:scecalphoto} {\it Left:} An active layer of the SciECAL prototype, consisting of 72 scintillator strips, where each strip is wrapped with reflective foil having a hole for calibration LED light. {\it Right:} Schematic view of a scintillator strip with a WLS fibre and a MPPC.}
\end{figure}

Following the experience of the first prototype tested at DESY, 
the physics prototype of ScECAL, with 30 layers and  a transverse size of $180\times 180$~mm$^2$ was constructed, with a total of 2160 channels. 
Rather than the mega-strips used in the first prototype, the physics prototype used individual small scintillator strips, wrapped with reflective foil to increase photon yield and reduce optical cross-talk between 
strips (Fig.~\ref{fig:tech:scecalphoto} {\it left}).
Each active layer consisted of 72 scintillator strips with a size of  $45\times 10 \times 3$~mm$^3$.
Each strip, produced by an extrusion method, had a hole along its length into which 
a 1~mm diameter Kuraray Y-11 WSF fibre was inserted along the strip (Fig.~\ref{fig:tech:scecalphoto} {\it right}).
The physics prototype was equipped with an improved calibration system, based on LEDs and notched clear fibers, to monitor the gain of the MPPCs in the physics prototype.
It was tested at Fermilab~\cite{CALICE:2012st}, mounted in front of
the CALICE AHCAL physics prototype.
%

\begin{figure}[htb]
\includegraphics[width=0.45\hsize]{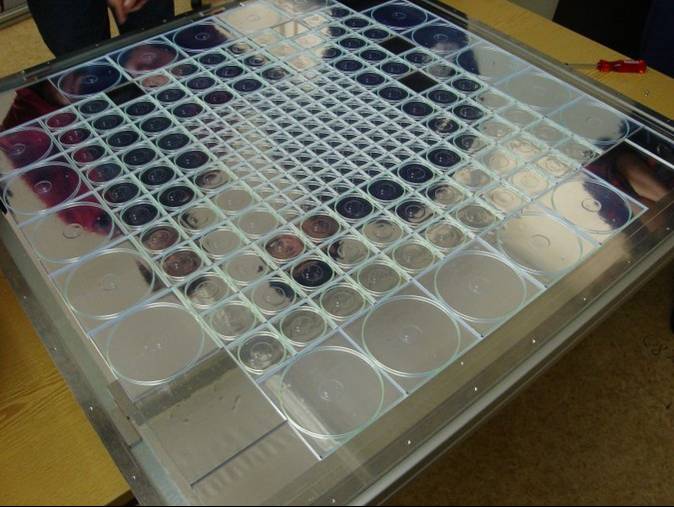}
\includegraphics[width=0.45\hsize]{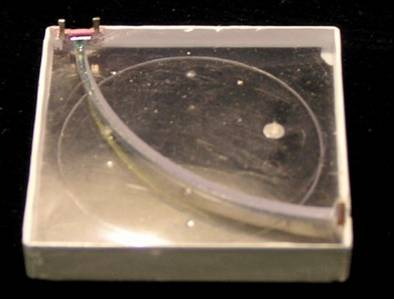}
\caption{\label{fig:tech:ahcalproto} Active layer of the AHCAL prototype, scintillator tile with SiPM.}
\end{figure}

 
 \paragraph {\textit{\textbf{Sci Fe and W AHCAL:}}}\label{par:ahcal-intro} 
 The scintillator-based analogue HCAL (AHCAL) prototype~\cite{collaboration:2010hb} is a sandwich structure made of 38 layers of scintillator tiles (5~mm thick) interleaved with steel absorber plates of 17~mm thickness, which were later replaced by 10~mm thick tungsten plates~\cite{Adloff:2013jqa}.
 The active layers are housed in steel cassettes with two times 2~mm thick cover plates, which contribute to the absorber structure. 
 The total thickness corresponds to 5.3~nuclear interaction lengths. The transverse dimensions of the active part are $90\times 90$~cm$^2$. 
 The tile size is  $3\times 3$~cm$^2$ in the central region; for the outer and rear parts larger sizes are used, see Fig.~\ref{fig:tech:ahcalproto}.
In total there are 7608 tiles, each individually read out via a wavelength-shifting fibre by a silicon photo-multiplier (SiPM) produced by MEPhI / PULSAR in Russia~\cite{Buzhan:2003ur}.

The CALICE AHCAL prototype was the first application of SiPMs on a large scale. 
SiPMs (see~\cite{Renker:2009zz} for a review) are pixelated avalanche photo-diodes operated in Geiger mode, with a typical gain of $10^6$. 
The pixels, 1156 on a 1.1$\times$1.1~mm$^2$ surface, have individual quenching resistors and operate on a common substrate.  
The output charge signal is proportional to the number pixels fired by photo-electrons and measures the light intensity. 
Electronic read-out proceeds via an 18 channel ASIC, FLC\_SiPM, 
which is based on the FLC\_PHY3 chip~\cite{Anduze:2008hq} used for the ECAL prototype and  therefore compatible 
with the same digitisation and data acquisition system~\cite{Dauncey:2002vf}.
A LED based calibration system was used to measure the gain of each SiPM, and their temperature was monitored by five sensors in each layer. 

Test bench characterisation of SiPMs and tiles plays a vital role.
For each SiPM, the over-voltage (reverse bias voltage excess over breakdown voltage) was adjusted 
to equalise the light yield to about 12 pixels per MIP normally traversing a tile, 
and the non-linear response as a function of light intensity was recorded for use in the offline reconstruction. 
 
 
 \paragraph {\textit{\textbf{RPC Fe and W DHCAL:}}}\label{par:dhcal-intro} 
 The Digital Hadron Calorimeter or DHCAL uses Resistive Plate Chambers (RPCs) as active elements~\cite{Drake:2007zz}. The chambers are read out with $1\times 1$~cm$^2$ pads and 1-bit (digital) resolution. A small-scale prototype was assembled and tested in the Fermilab test beam in 2007 to validate the 
 concept~\cite{Bilki:2008df,Bilki:2009ym,Bilki:2009xs,Bilki:2009wp,Bilki:2009gc}.

A large prototype with up to 54 layers and close to 500,000 readout channels was built in 2008 - 2011. Each layer measured approximately $96\times 96$~cm$^2$ and was equipped with three chambers, stacked vertically on top of each other. 

In the test beams the DHCAL layers were inserted into a main stack of 38 or 39 layers, followed by a tail catcher with up to 15 layers. For the tests performed at Fermilab the main stack contained steel absorber plates ~\cite{collaboration:2010hb}. At CERN the absorber plates contained a tungsten based alloy. In both cases the tail catcher featured steel absorber plates.

\begin{figure}[htb]
\includegraphics[width=0.9\hsize]{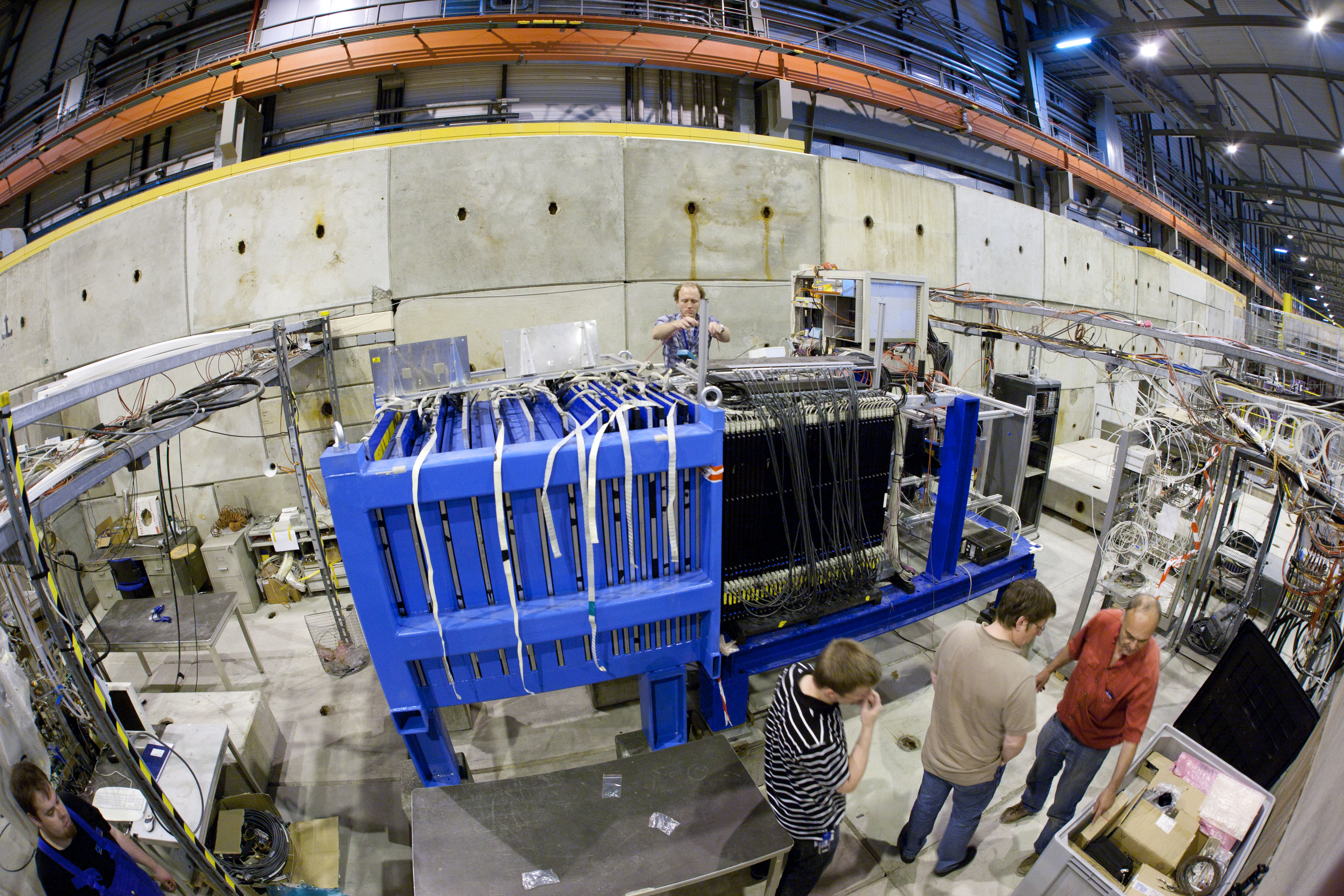}
\caption{\label{fig:DHCAL_CERN} Photograph of the DHCAL setup in the PS test beam at CERN.}
\end{figure}

In the various test beam campaigns combined, spanning the years 2010 - 2012, the DHCAL recorded several ten million muon and  secondary beam events, where the latter contained a mixture of electrons, muons, pions, and protons. A photograph of the setup in the CERN test beam is shown in 
Fig.~\ref{fig:DHCAL_CERN}.



\begin{figure}[htb]
\includegraphics[width=0.9\hsize]{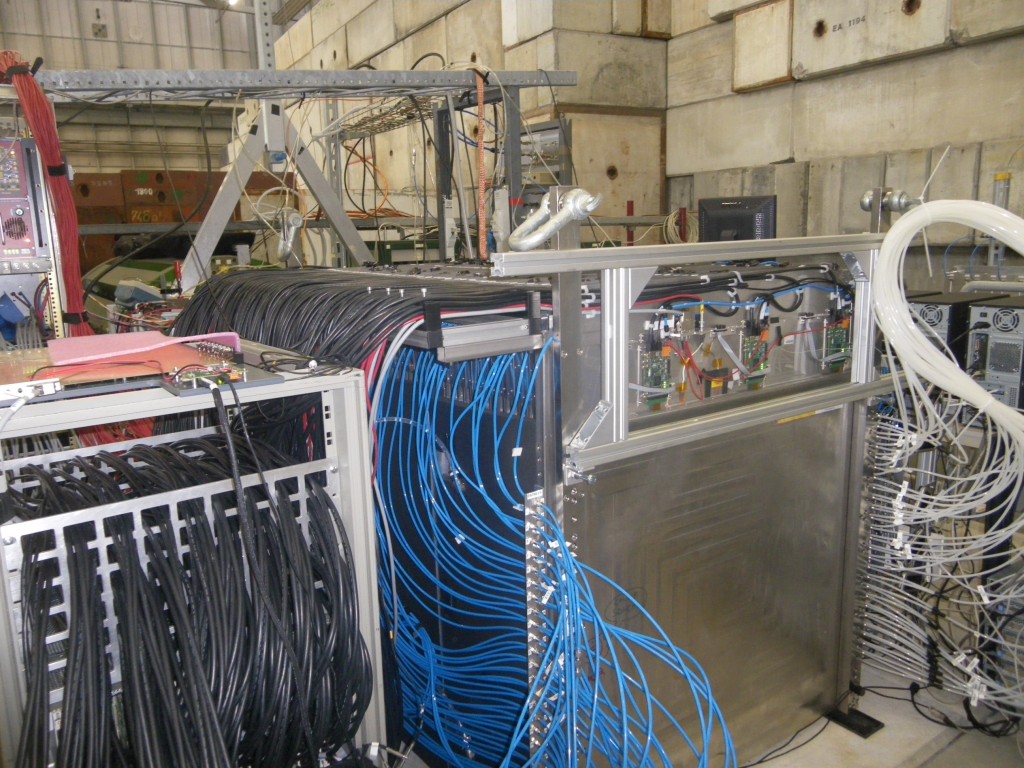}
\caption{\label{fig:grpc-sdhcal} CALICE's Semi-Digital Hadronic Calorimeter prototype during the April 2012 test beam at CERN.}
\end{figure}

 \paragraph {\textit{\textbf{RPC Fe SDHCAL:}}}\label{par:sdhcal-intro} 
Similar to the DHCAL described above the CALICE RPC-SDHCAL~\cite{Mannai:2013qka} uses glass resistive plate chambers as the sensitive medium. By virtue of a  pad board a $100\times100\,\rm{cm^2}$ large RPC 
chamber is subdivided into cells of $1\times1\,\rm{cm^2}$  The chambers are coated with a novel mixture of colloidal graphite allowing for the application of the silk screen print method that ensures a uniform effective surface resisitivity. The chambers are integrated into a stainless cassette with an overall thickness of 11\,mm  that constitutes therefore a part of the absorber medium.  This steel cassette hosts also the readout components consisting of the HARDROC ASICs mounted on PCBs. The steel cassettes are inserted into a mechanical structure that can host up to 51 cassettes.  The mechanical structure features stainless steel plates of 1.5\,cm thickness as the main absorber medium. A layer composed of the cassette and an absorber plate has a depth of about 0.12 hadronic interaction lengths $\lambda_I$.
For beam tests 50 RPC-chambers were built within six months at the beginning of 2012 and the prototype has been operated with 48 chambers. A water cooling system, together with the capabilities of the HARDROC ASICs to cycle the power supply synchronously to the duty cycle of the beam, provided an important noise reduction.  The prototype was commissioned in 2012 and an extensive beam test program has been conducted at the PS and SPS beam test facilities of CERN. A picture of the prototype is shown in Fig.~\ref{fig:grpc-sdhcal}.
 
\paragraph{\textit{\textbf{Alternative SDHCAL technology:}}}\label{par:alt-intro} 
A 1~m$^2$ micromegas chamber comprises 9216 cells. The PCBs and the readout ASICs are integrated into the gas volume. This prevents a large energy deposition in the mesh which in turns protects the ASICs from being damaged in case of a spark. In total 4 of these chambers were built and operated in beam tests in 2012. They were tested in standalone mode as well as interleaved with the RPC chambers described before. The latter configuration allowed for the study of the response to hadronic showers without a large number of chambers.

 
 \paragraph {\textit{\textbf{Sci and RPC Fe TCMT:}}}\label{par:TCMT-intro} 
Hadron test beam set-ups for runs at higher energies have been complemented by a tail-catcher muon-tracker (TCMT) system~\cite{CALICE:2012aa}. 
 The TCMT steel structure consists of two sections, a first one with nine 21~mm thick absorbers, and a second, coarser one  with 104~mm thick absorbers. 
The plates are interleaved with, in total, 16 read-out layers, each made of 20 scintillator strips, 5~cm wide and arranged in alternating horizontal and vertical orientation. 
The strips are read out via wavelength-shifting fibres coupled to SiPMs and use the same electronics as the AHCAL. 
In most DHCAL runs, the TCMT was instrumented with RPCs.   
 

 \subsection{Test beam overview}
 \label{sec:tbeam-intro} 
\begin{figure}[htb]
\includegraphics[width=0.95\hsize]{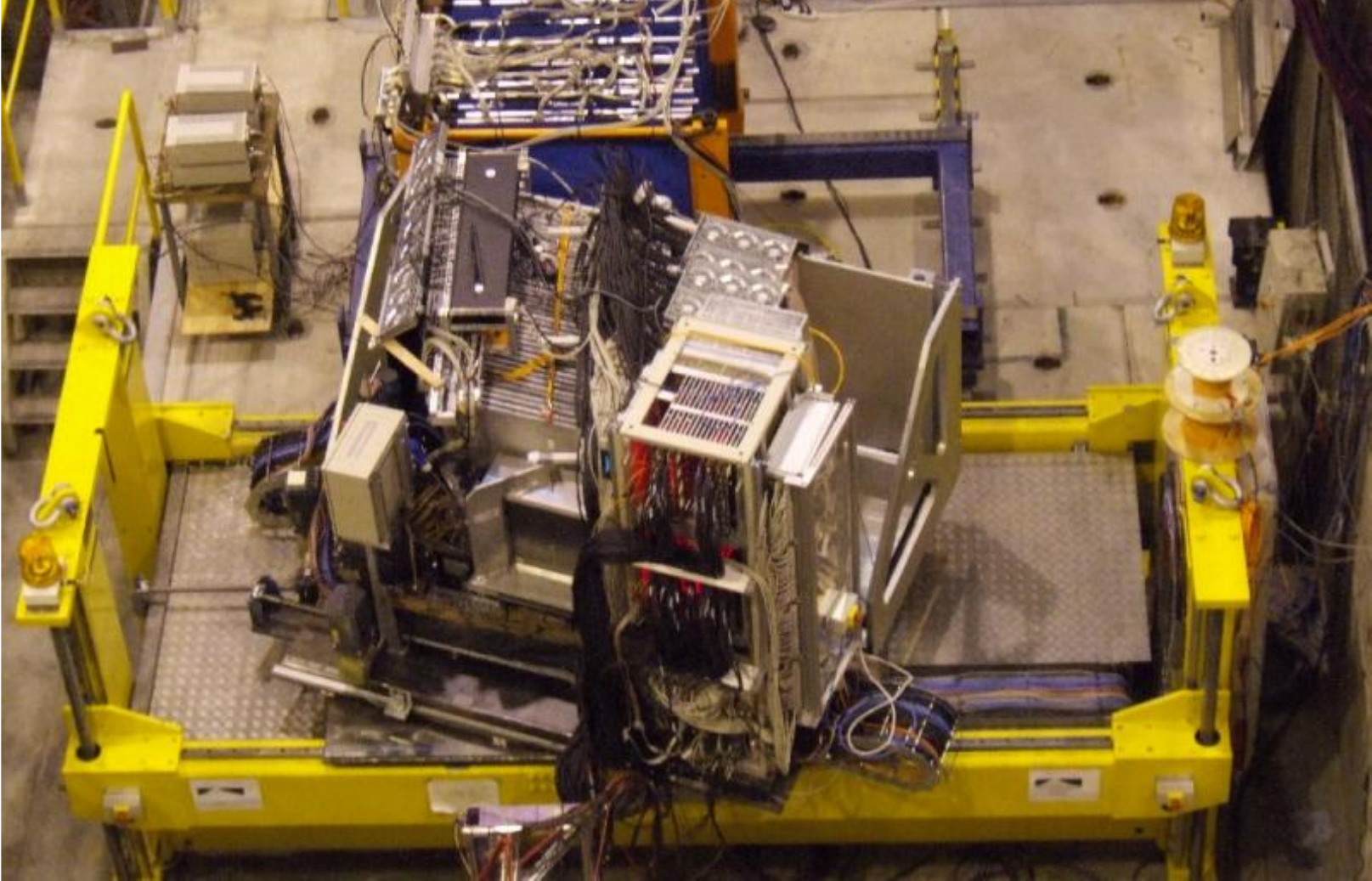}
\caption{\label{fig:tech:testbeam} CALICE test beam set-up with ECAL, HCAL and TCMT at the CERN SPS.}
\end{figure}

Following commissioning runs and initial tests with electrons at DESY, the major CALICE test beam campaign started with a combination of silicon ECAL and scintillator AHCAL, plus TCMT,  at the CERN SPS in 2006-2007, shown in Fig.~\ref{fig:tech:testbeam}.
Until 2012, all major ECAL and HCAL readout-out technologies and HCAL absorber materials were tested in different combinations at Fermilab and CERN, see Table~\ref{tab:Tech:testbeamperidos}.
At both sites, data were also taken with the HCAL in stand-alone mode, without ECAL in front, to validate the performance with electrons and low energy hadrons.
%
\begin{table*}
\caption{\label{tab:Tech:testbeamperidos} Summary of CALICE test beam periods.}
\begin{ruledtabular}
\begin{tabular}{llllll}
Year &	Beam	& W ECAL read-out & HCAL read-out & HCAL absorber & Fe Tail catcher read-out \\ 
 \\ \hline
2006-07 &	CERN SPS	& Silicon pads 	& Scint.\ analogue & Fe	& Scint.\  \\ 
2008	&	FNAL FTBF	& Silicon pads 	& Scint.\ analogue& Fe 	& Scint\    \\
2008-09 & FNAL FTBF	& Scint.\	strips	& Scint.\ analogue& Fe 	& Scint.\   \\
2010   &	FNAL FTBF	& 		--			& RPC digital & Fe & Scint.\    \\
 		& 	CERN PS 	& 		--				& Scint.\ analogue & W 	&  -- \\
2011	&	FNAL FTBF	&Silicon pads 	& RPC digital	& Fe	& RPC   \\
2011	& 	CERN SPS	&		-- & Scint.\ analogue & W 	& Scint.\  \\
2011-12 &	CERN PS, SPS	&		--		& RPC semi-digital& Fe & -- \\
2012	& 	CERN PS, SPS & 	--		& RPC	 digital & W	& RPC  \\
\end{tabular}
\end{ruledtabular}
\end{table*}

Altogether, more than half a billion physics events have been collected, not including calibration data recorded with muons or optical signals. Data management and processing has been based on tools developed for the LDC grid. For this the virtual organisation {\em calice} has been created that is hosted by DESY.

	\section{Calorimeter performance in test beams%
\label{sec:Perf}}

The technologies proposed for the realisation of highly granular calorimeters had not been used before in calorimeters. 
SiPMs were completely novel, silicon pad diodes had been used in smaller devices only, RPCs, GEMs and  micromegas had  typically been combined with strip read-out only, or did not cover large areas.
So the first question to ask was whether these devices function at system level, can be operated reliably, and deliver the expected calorimetric performance in test beams. 
Second, in order to validate the simulations of hadronic showers and arrive at conclusions on the adequacy of the underlying physics models, the new detectors must be understood in terms of simulations. 
Since the evolution of electromagnetic particles can be predicted with much higher precision, the response of the prototypes to electrons or positrons is first used to quantitatively validate the detector modelling of the new prototypes. 
A special issue common to all technologies is the robustness of the embedded read-out chips with respect to possible shower induced malfunctions, which was the subject of dedicated studies. 

\subsection{CALICE silicon tungsten ECAL}

\begin{figure}[h]
\includegraphics[width=0.39\textwidth]{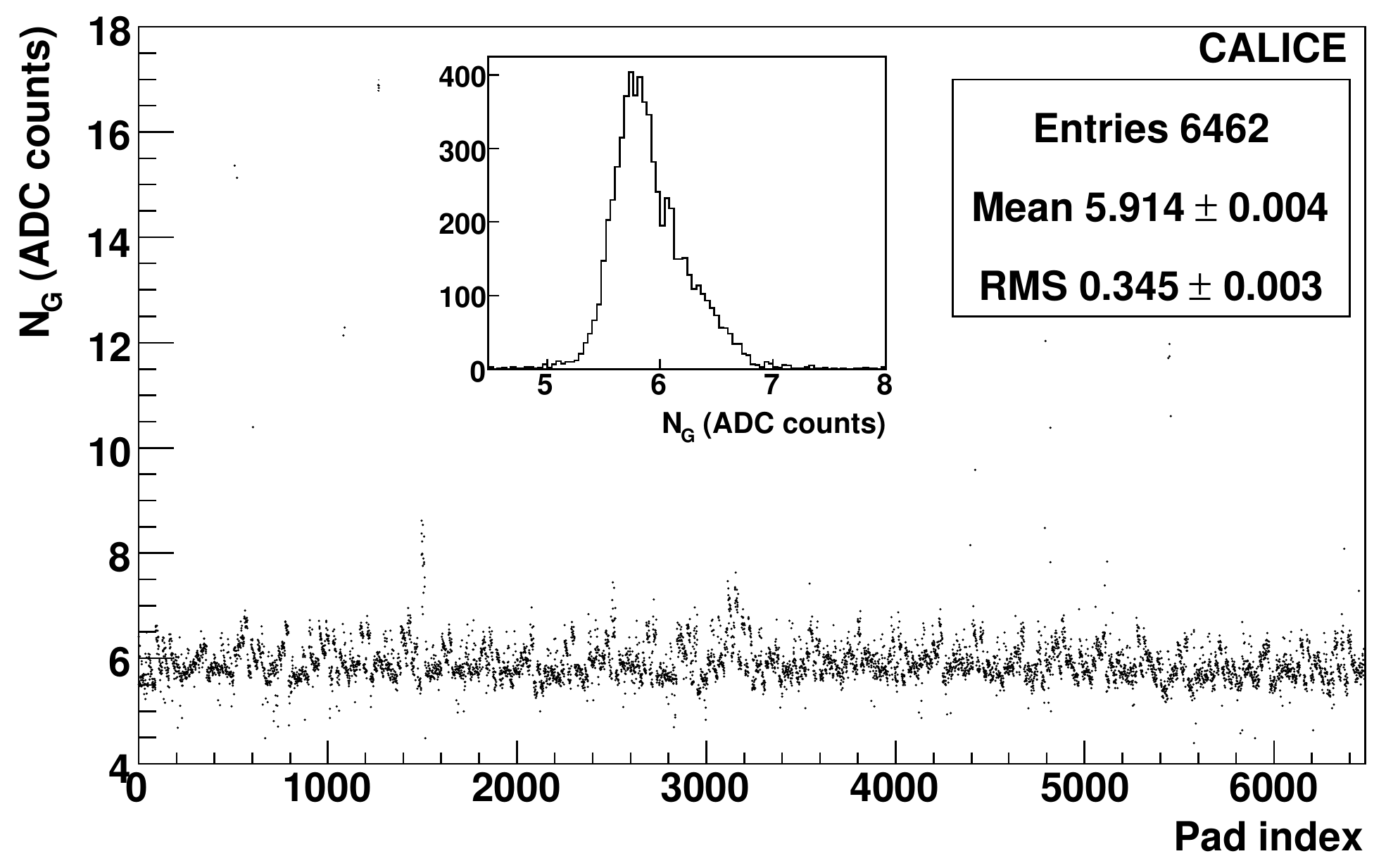}
\hfill %
\includegraphics[width=0.39\textwidth]{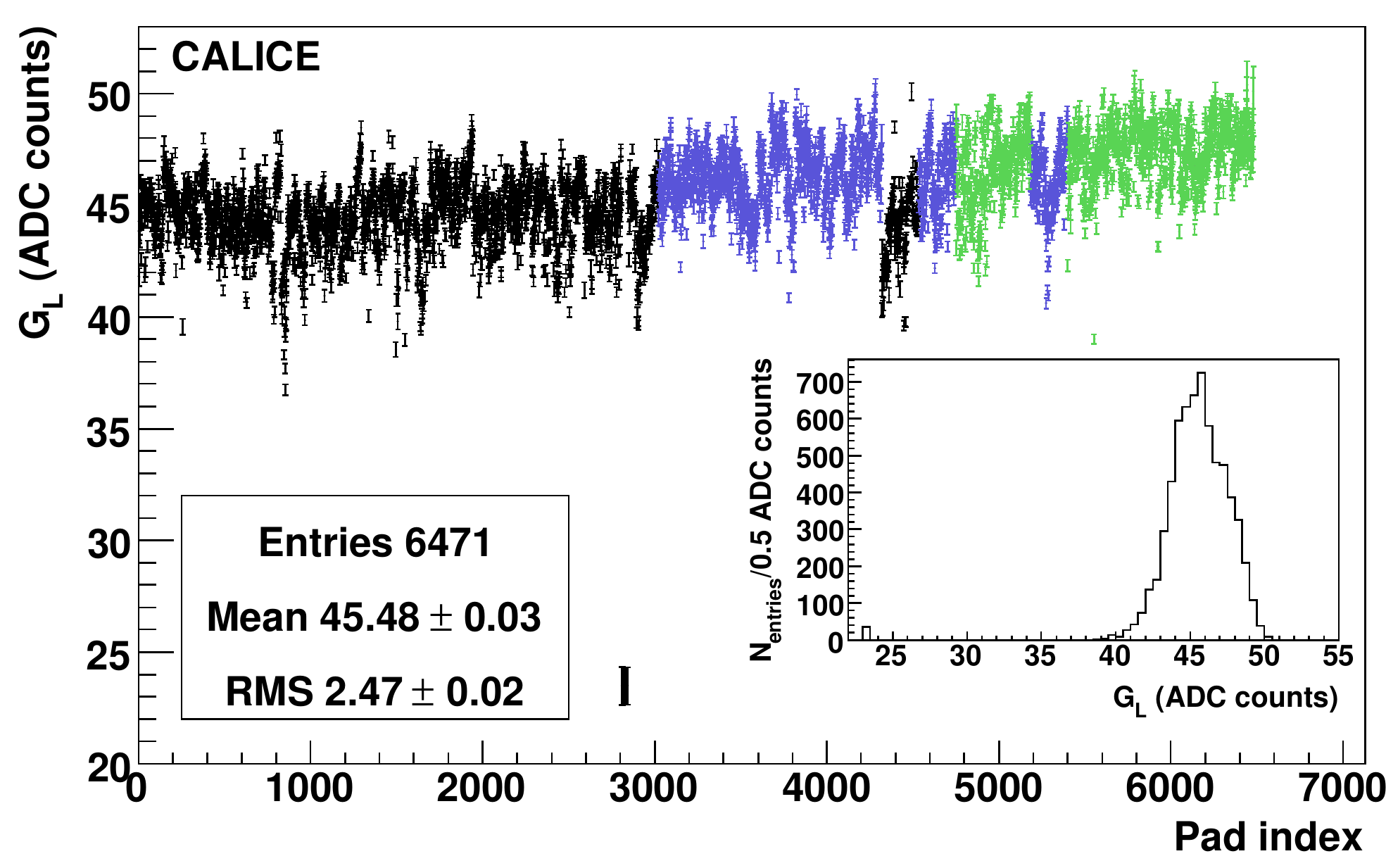}
\hfill %
\caption{Top:  Detector noise of the CALICE SiW ECAL physics prototype as measured in the 2006 beam test campaign at CERN as a function of the
cell number. 
Bottom: The most probable value of the response to minimal ionising particles as a function of cell number. The different colours distinguish wafers from different vendors. The inset histograms display the projections on the y-axis. 
Both from~\cite{Anduze:2008hq}.}
\label{fig:ecalmipnoise}
\end{figure}

Results presented in this review are based on large statistics samples of electrons, pions and muons recorded at CERN and FNAL between 2006 and 2011. 
A detailed overview of the beam test performance of the physics prototype is given in Refs.~\cite{Anduze:2008hq} and~\cite{Adloff:2009zz}. The detector noise is reported to be 13\% of a MIP.  This small value enables single MIP detection in reconstruction algorithms for Particle Flow. The noise and the response to minimum ionising particles are uniform throughout the detector as demonstrated by Fig.~\ref{fig:ecalmipnoise}. 
In~\cite{Anduze:2008hq} issues with coherent noise are reported. 
During the beam test operation, in particular for highly energetic electrons, a square pattern on the wafer periphery has been observed. The reason for that is a capacitive coupling of the guard ring that surrounds the silicon wafer with the actual silicon pads. 
The observation of these 'square events' triggered R\&D
on the guard ring design. The studies conclude that the frequency of square events can be largely reduced by a segmentation of the guard rings~\cite{Cornat:2009zz}.

\begin{figure}[htb]
\centering
       \includegraphics[width=0.8\hsize]{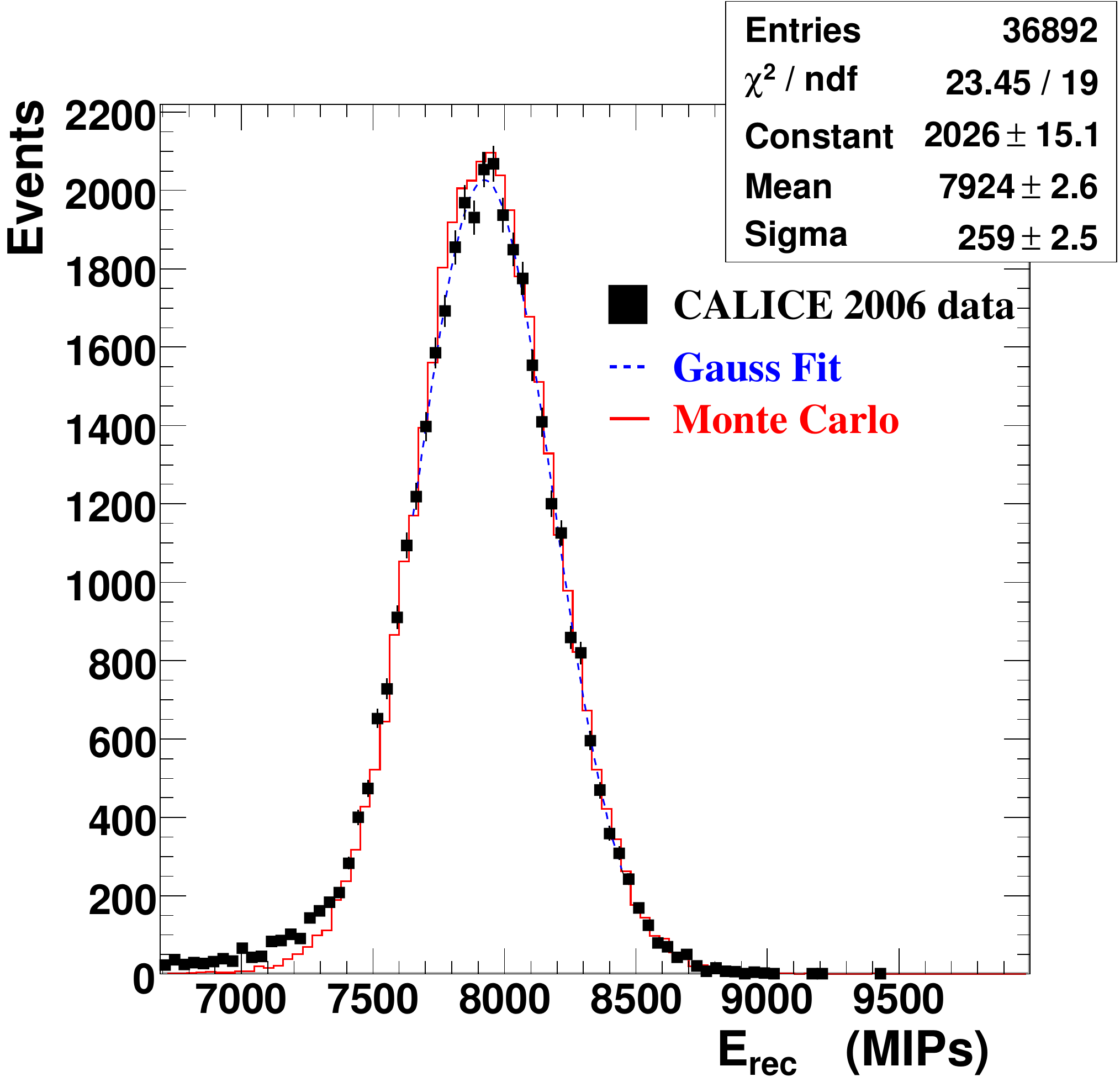}
     
\caption{  \label{fig:ecalresp} Response of the CALICE SiW ECAL to a electron beam with an energy of 30\,GeV.
From~\cite{Adloff:2009zz}.}
\end{figure}

\begin{figure}[h]
\includegraphics[width=0.39\textwidth]{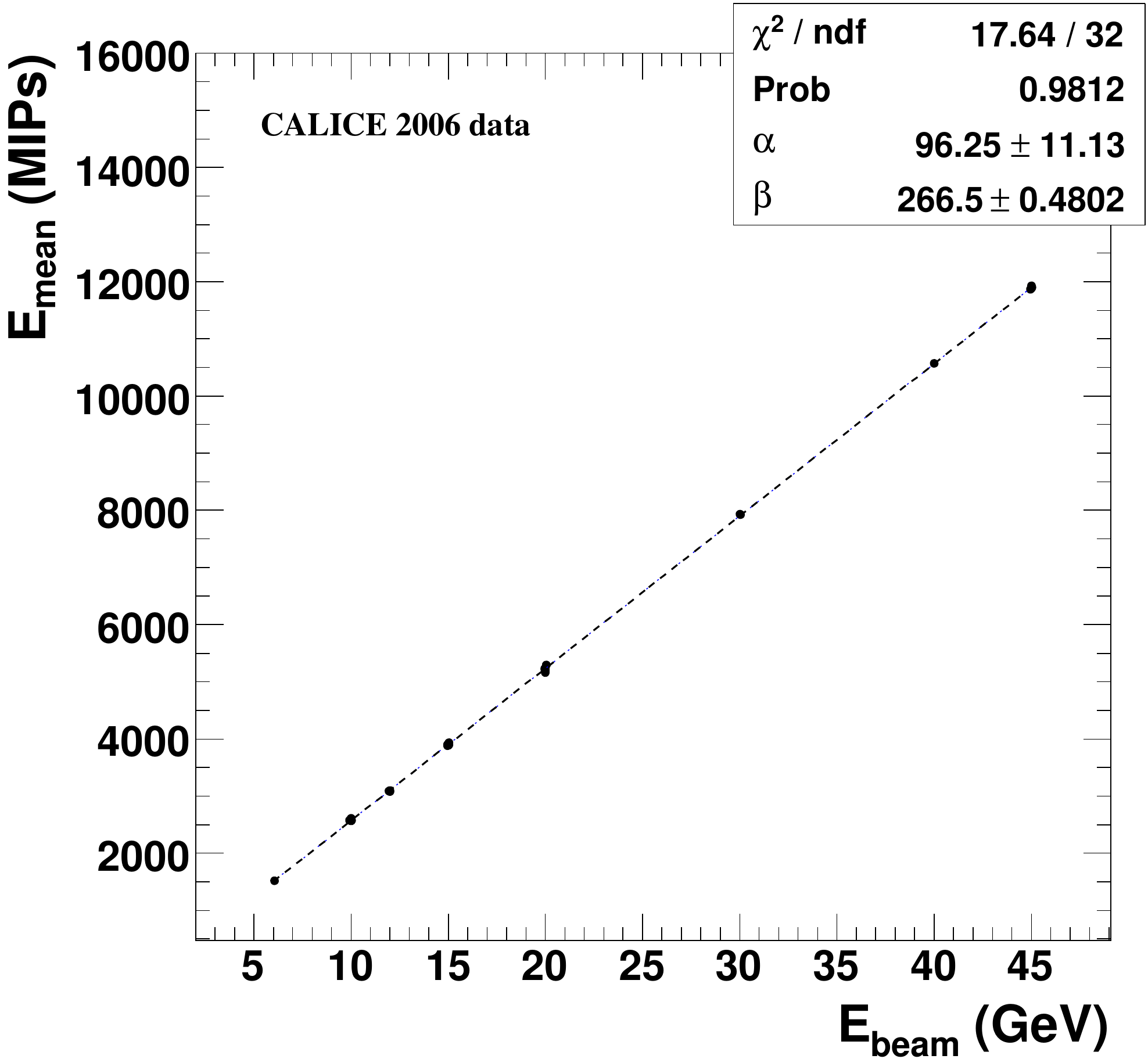}
\hfill %
\includegraphics[width=0.39\textwidth]{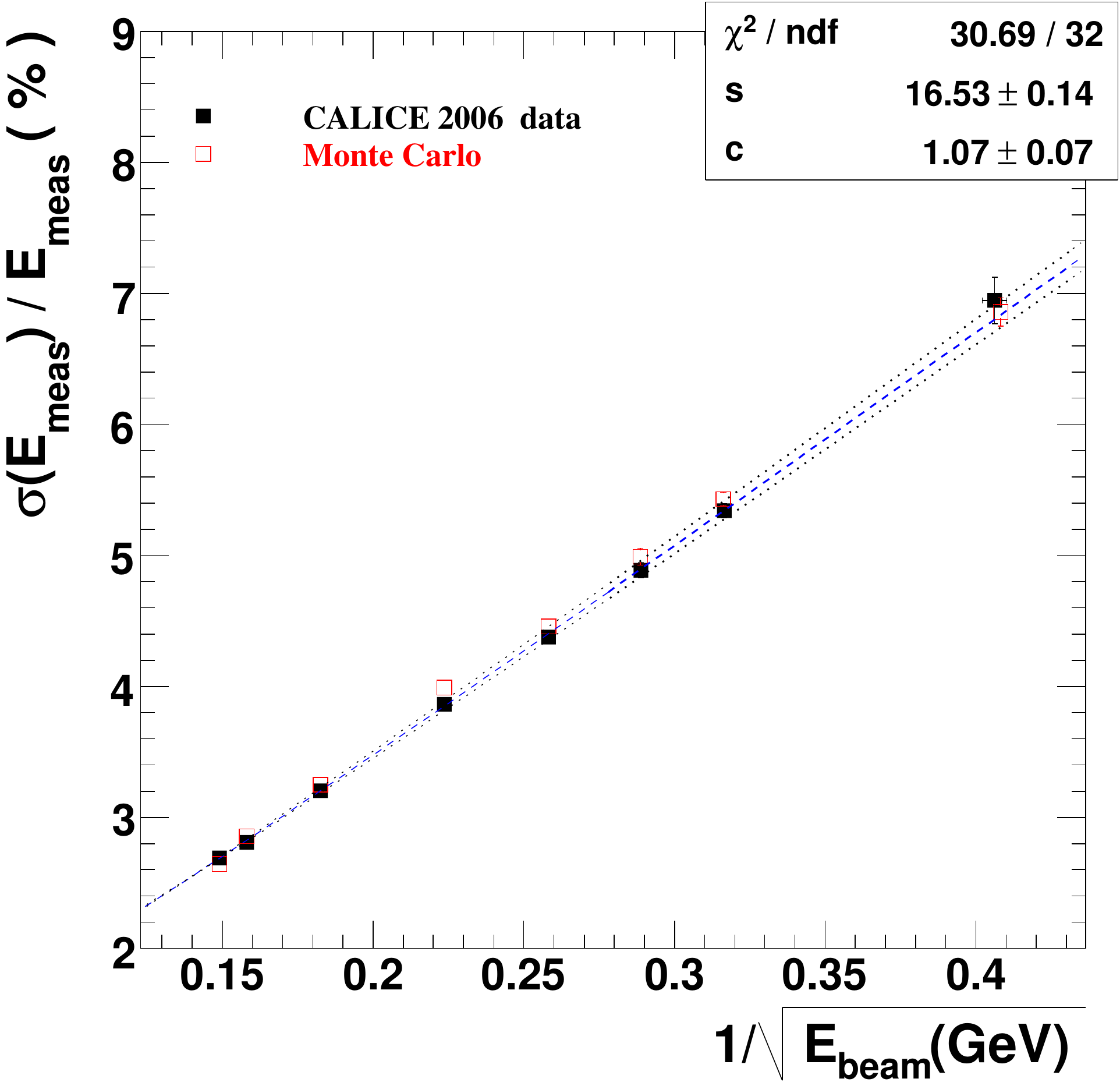}
\hfill %
\caption{Top:  Linearity of the energy response of the CALICE SiW ECAL as a function of the nominal energy. 
The line represents a fit of the form $\beta E_{beam} - \alpha$ to the measured energy $E_{mean}$. Parameters $\alpha$ and $\beta$ are explained in the text.
Bottom: Relative energy resolution ($\sigma(E_{\mathrm{meas})}/E_{\mathrm{meas}}$) as a function of $1/\sqrt{E_{\mathrm{beam}}}$ (solid squares), and its usual parameterisation as $s/\sqrt{E} \oplus c$. 
The values expected from simulation are shown (open squares). The dashed line gives the fitted resolution for data (Equation~\ref{eq:datares}), and the dotted lines correspond to its variation when the beam energy scale is shifted by $\pm$~300~MeV.
Both from~\cite{Adloff:2009zz}.}
\label{fig:ecalcalib}
\end{figure}

A detailed analysis of the response of the physics prototype to electrons is published in~\cite{Adloff:2009zz}. The analysis selects events impinging on the detector reasonably far from the boundaries of the silicon wafers. In addition a correction procedure has been developed for residual losses in the gaps between the wafers. Finally, the total energy deposit is calculated according to
\begin{equation}
 E_{\mathrm{rec}}(\mbox{MIPs})=\sum_i w_i E_i  
\end{equation}
The sum runs over the thirty layers of the prototope and $E_i$ is the energy deposition in one layer.
The weighting factors are given by $w_i = K_i +\eta_i$. The value $K_i=1,2,3$ reflects the varying sampling fraction due to the increase of the absorber material in the three modules, see page~\pageref{siw:fig:physProto}. The value $\eta_i$ corrects for effects of the internal structure of the SiW ECAL. The correction is maximally 7\%.
Based on this weighting, Fig.~\ref{fig:ecalresp} shows the response of the prototype to an electron beam with an energy of 30\,GeV for data and Monte Carlo simulation. Both agree  well 
apart from the tail on the left hand side of the peak. The disagreement is attributed to residual radiative effects (i.e. bremsstrahlung in the beam line material), residual pion background and maybe also energy losses in the space between the wafers.  
The resulting distribution can be well fitted by a Gaussian in the range $[-1\sigma,2\sigma]$ around the mean value $E_{mean}$.
The linearity and energy resolution of the detector is given in Fig.~\ref{fig:ecalcalib}.

The linearity is approximated by a fit of the form $E_{mean}=\beta E_{beam}-\alpha$. The parameter $\beta$ is a simple conversion factor from the beam energy in GeV to the 
scale of the detector response in units of MIP. The parameter $\alpha$ parameterises an offset that according to Ref.~\cite{Adloff:2009zz} is attributed to losses of information, i.e. energy depositions, that are discarded due to the noise cut in the analysis of 0.6 MIP.  
It is found that between 6 and 45\,GeV  the detector response is linear within 1\%. In this energy range the energy resolution is determined to be
\begin{eqnarray}
\label{eq:datares}
 \frac{\sigma (E_{\mathrm{meas}})}{E_{\mathrm{meas}}}  & = &
 \frac{(16.53\pm0.14(\mathrm{stat})\pm0.4(\mathrm{syst}) )\%}{\sqrt{E(\mathrm{GeV})}}  \nonumber \\ 
& \oplus &  \left(1.07\pm0.07(\mathrm{stat})\pm0.1(\mathrm{syst})\right)\%,
\end{eqnarray}

The resulting energy resolution is clearly inferior to that obtained for other calorimeter technologies, e.g.\ for crystals. It should however be stressed at this
point that the detector design emphasises granularity, and thus fine sampling, over the pure calorimetric response. 
 
The high granularity of the SiW ECAL leads naturally to a rich amount of information which can be exploited using advanced, e.g. imaging
processing techniques. One of these techniques is the {\em Hough transformation}~\cite{bib:hough}. The Hough transformation provides a mapping
from a $n$-dimensional feature space onto a $m$-dimensional parameter space, also called Hough space. Briefly, points which
are on a straight trajectory as e.g. generated by a MIP, will all result in the same parameter set in the Hough space. Thus,
by subjecting the calorimeter cells which carry energy at or above one MIP to a Hough transformation, a MIP trajectory
will lead to an accumulation at a given set of parameters. 

\begin{figure}[htb]
\begin{center}
\includegraphics[width=0.8\hsize]{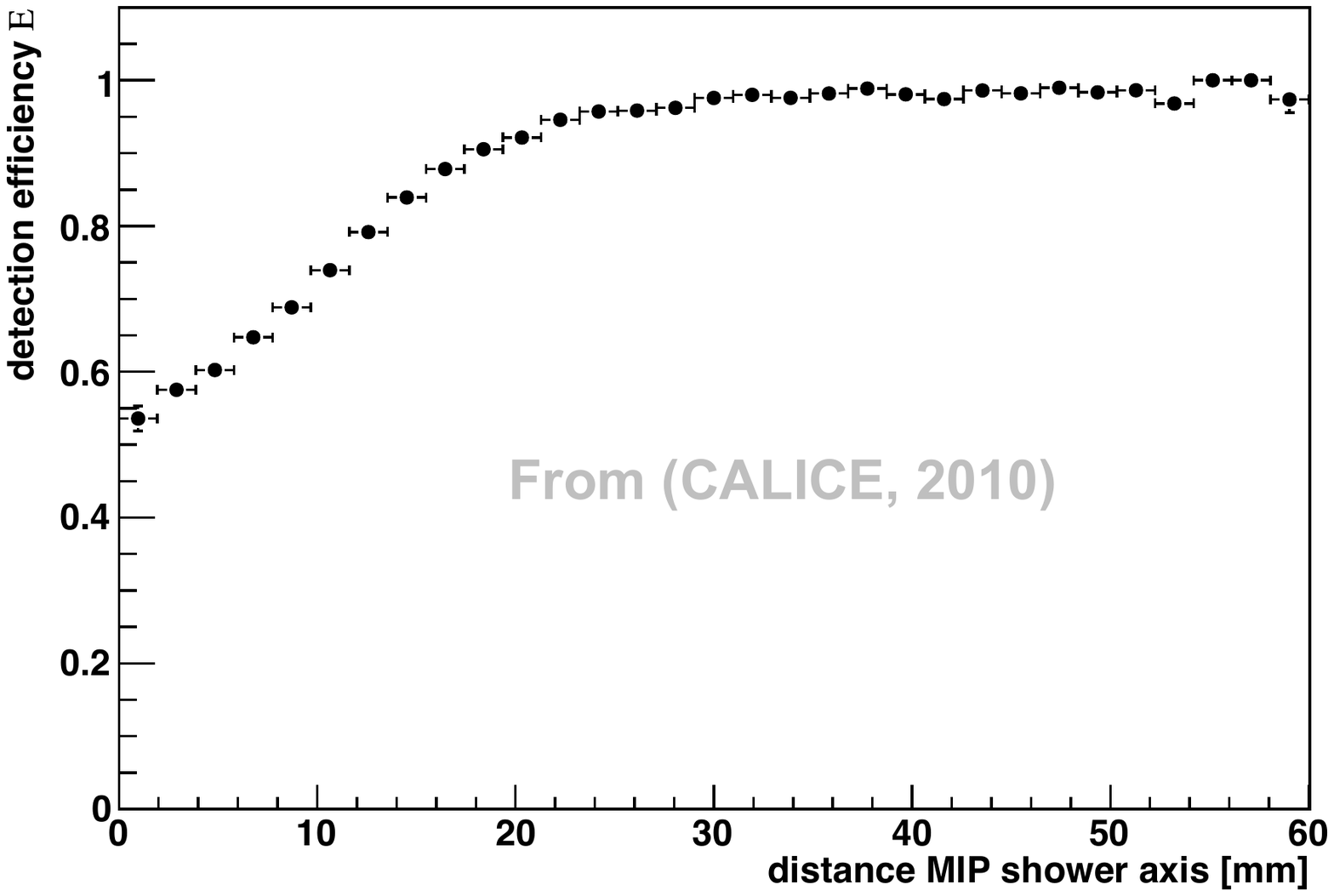}
\end{center}
\caption{Simulated MIP detection efficiency in the CALICE SiW ECAL as a function of the distance between the shower axes of an electron and of a muon. From ~\cite{bib:fehr10,Poschl:2011zz}.}
\label{fig:effdistkappa}
\end{figure}

A result of a simulation study for muons overlaid on 30~GeV electrons carried out in~\cite{bib:fehr10} is shown in Fig.~\ref{fig:effdistkappa}.
It demonstrates that a full separation of close-by particles can be achieved for distances down to 2.5\,cm.

The feasibility of having embedded readout electronics for a calorimeter proposed for a future lepton collider has been studied in~\cite{CALICE:2011aa}. A detailed analysis of noise spectra of the ASICs exposed to high-energy electron beams has revealed no evidence that the noise pattern is altered under the influence of the electromagnetic showers. 
The probability to have fake signals above the MIP level is estimated to be smaller than $6.7\cdot 10^{-7}$ per shower. The probability for a fake signal 
is less than $10^{-5}$ for a threshold of 2/3 of a MIP. For an event of the type $e^+ e^- \rightarrow t \bar{t}$ at $\sqrt{s}=500$\,GeV at a lepton collider about
 2500 cells of dimension $1\times1\,\mathrm{cm^2}$ are expected to carry a signal above noise level which is typically defined to be (60-70)\% of a MIP. The results thus revealed no problems for the design of embedded readout electronics for a detector for a lepton collider.

\subsection{SiD silicon tungsten ECAL}

\begin{figure}[htb]
\centering
       \includegraphics[width=0.8\hsize]{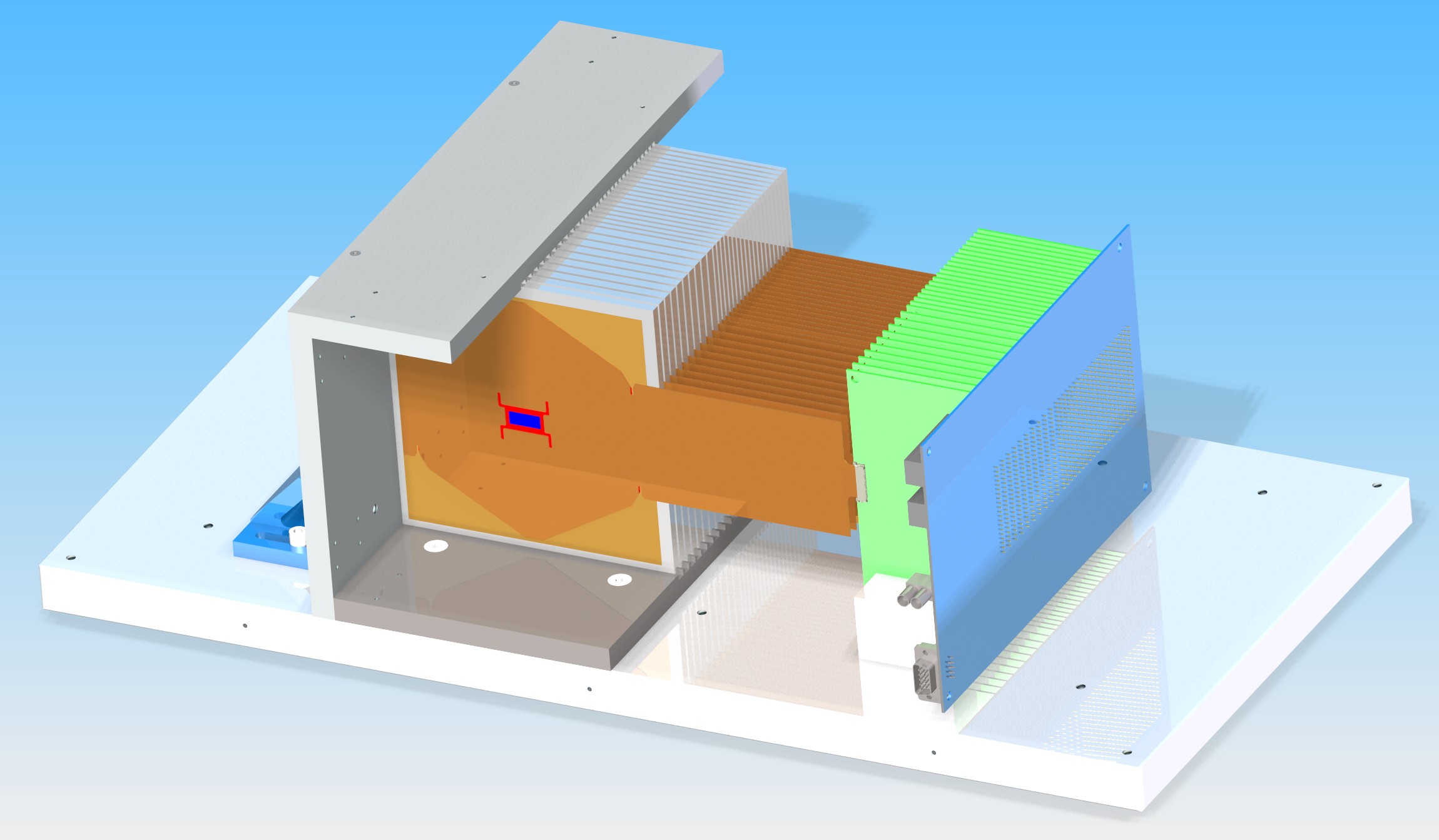}
     
\caption{  \label{fig:SiD_ECAL_test_beam} Configuration of 15~cm size hexagonal sensor planes, each read by one KPix chip placed at the centre and interleaved between tungsten plates for the beam test of the SiD ECAL prototype.}
\end{figure}

In order to demonstrate the feasibility of assembling a highly compact
electromagnetic calorimeter, with printed circuit boards and with direct
bonding of chips to wafers, a first prototype stack for an SiD ECAL has been constructed.
A section of this SiD ECAL with KPiX readout has been exposed to
a 12.1 GeV electron beam at the SLAC ESTB facility. A schematic of the 
test setup is shown in Fig.~\ref{fig:SiD_ECAL_test_beam}.
This agressive design has an active gap between absorber plates of 1.25~mm,
a cell size of 13~mm$^2$, and an effective Moliere radius of 14~mm.

In the first tests, 
a stack of
nine silicon sensor planes and eight tungsten plates (corresponding to
six radiation lengths) was exposed to beam.
The full stack will ultimately consist of 30 layers as in the SiD ECAL design.

\begin{figure}[htb]
\centering
       \includegraphics[width=0.8\hsize]{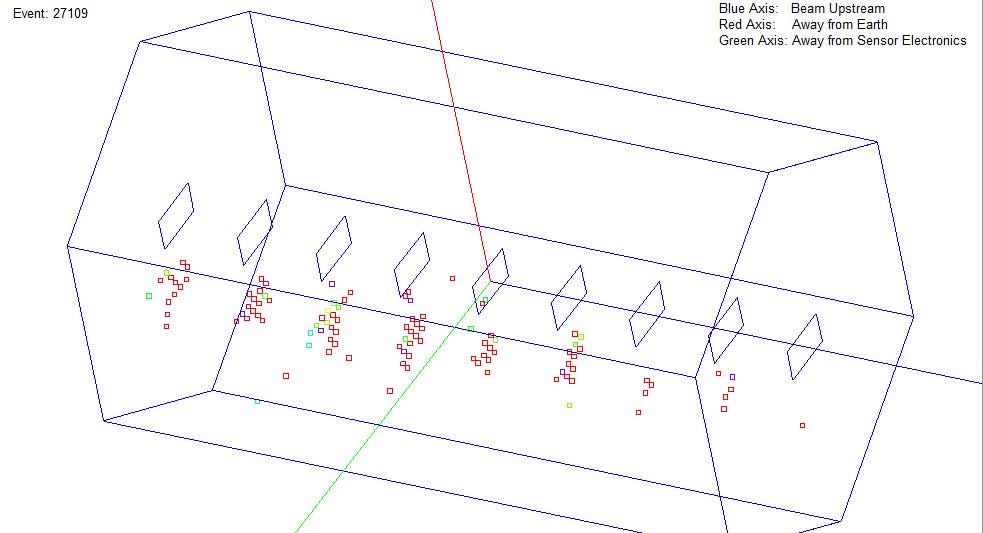}
     
\caption{  \label{fig:SiD_ECAL_shower} A single electron shower event in the SiD ECAL
prototype stack.}
\end{figure}

Data was taken over a four-day period with a beam rate between 0.5 and 5 
electrons per pulse. Fig.~\ref{fig:SiD_ECAL_shower} shows an example of
a single electron shower in the test stack. The development of the
shower, and the transverse distribution of digital hits at each sensor plane,
are clearly visible. The longitudinal profile of electron showers has been studied
and is seen to follow the usual distribution. However, significant crosstalk problems 
were discovered and are being investigated.

\subsection{Scintillator tungsten ECAL}


The first, small ScECAL prototype with about 500 channels was tested in 
2007 using positron beams with energies from 1 to 6~GeV provided by the DESY-II electron synchrotron~\cite{Francis:2013uua}.
The aim was to demonstrate the feasibility of a scintillator strip ECAL with MPPC
readout.
Large-scale tests of Hamamatsu MPPCs in a real detector had not yet been performed.
The test served to establish calibration and correction procedures for this novel type of photo-sensor, which were used in later ECAL prototypes, too. 
They follow the principles developed for the SiPMs of the AHCAL, which are described in the next section. 

%
Calibration runs were performed several times during the beam time using a 3 GeV positron beam,
where the absorber plates were removed from the detector.
Events consistent with a single, non-showering particle passing through the prototype ScECAL were selected and used for the calibration.
Each strip was then individually calibrated with data 
for which the reconstruction of a track recorded in a drift chamber installed upstream showed that a particle passed through the strip.

The MPPC signal is intrinsically non-linear due to its finite number of pixels (1600), which leads to a saturation of its response at high light intensities.
If an input light pulse is shorter than the MPPC recovery time ($\sim 4$~ns for MPPCs used in the prototype), the MPPC response can be parameterised by
\begin{equation}
\label{eq:sipmsaturation}
N_{\mathrm{fired}} (N_{\mathrm{p.e.}}) = N_{\mathrm{pix}}(1 - e^{-N_{\mathrm{p.e.}}/N_{\mathrm{pix}}}),
\end{equation}
where $N_{\mathrm{fired}}$ denotes the number of fired pixels, $N_{\mathrm{pix}}$ the total number of MPPC pixels, and $N_{\mathrm{p.e.}}$ the number of photoelectrons created.
If the input light pulse is longer than the MPPC recovery time, the effective dynamic range is increased due to the possibility
of a single pixel firing several times within the same light pulse.
This effect occurs particularly for the strips using WLS fibres of decay time $\sim$8~ns.
In contrast, directly read-out strips are not expected to have such enhancement because of the shorter decay time of the scintillator itself ($\sim$2~ns),
The single pixel signal of each MPPC was obtained from the spectrum measured in LED calibration runs, so that $N_{\mathrm{fired}}$ could be calculated from the measured ADC counts.
The response of the two types of scintillator strip-MPPC systems was measured using a dedicated apparatus using an ultraviolet LED to inject light into the strip and a photomultiplier to monitor the light yield,
and the measured response functions were used to correct the saturation effect.
To reduce these effects, 
MPPCs with more pixels ($\sim 10000$~pixels in an effective area of 1~mm$^2$) have been developed in the meantime. 

Since the MPPC gain $G$ depends  on the temperature $T$
($C_T =\delta G / \delta T \sim 2\%/$K at 20$^{\circ}$C), correction of this effect is essential.
The dependence of each channel's response 
on temperature $A(T)$ was fitted with a linear function
\begin{equation} 
\label{eq:Tcor}
A(T) = A(T_0) \cdot ( 1 + C_T \cdot (T - T_0))
\end{equation}
where the reference temperature $T_0$ was chosen to be 20$^{\circ}$C.
In the analysis of the positron events
the response of each strip was calibrated using the temperature-dependent calibration function
determined by these fitted functions.

Although plastic film was inserted into the pair of grooves between strips, adjacent strips in the same mega-strip were not perfectly optically isolated.
The cross-talk between neighbouring strips was typically around 10\%, with a relative variation of around 15\% (RMS).
Since the MIP calibration was defined without accounting for cross-talk,
a simple sum over measured strip energies would give an overestimate of the deposited energy in terms of MIPs.
A  correction procedure was applied to estimate the cross-talk contribution from the amplitude recorded in adjacent strips and subtract it from the signal in each strip.

\begin{figure}[htb]
\includegraphics[width=0.8\hsize]{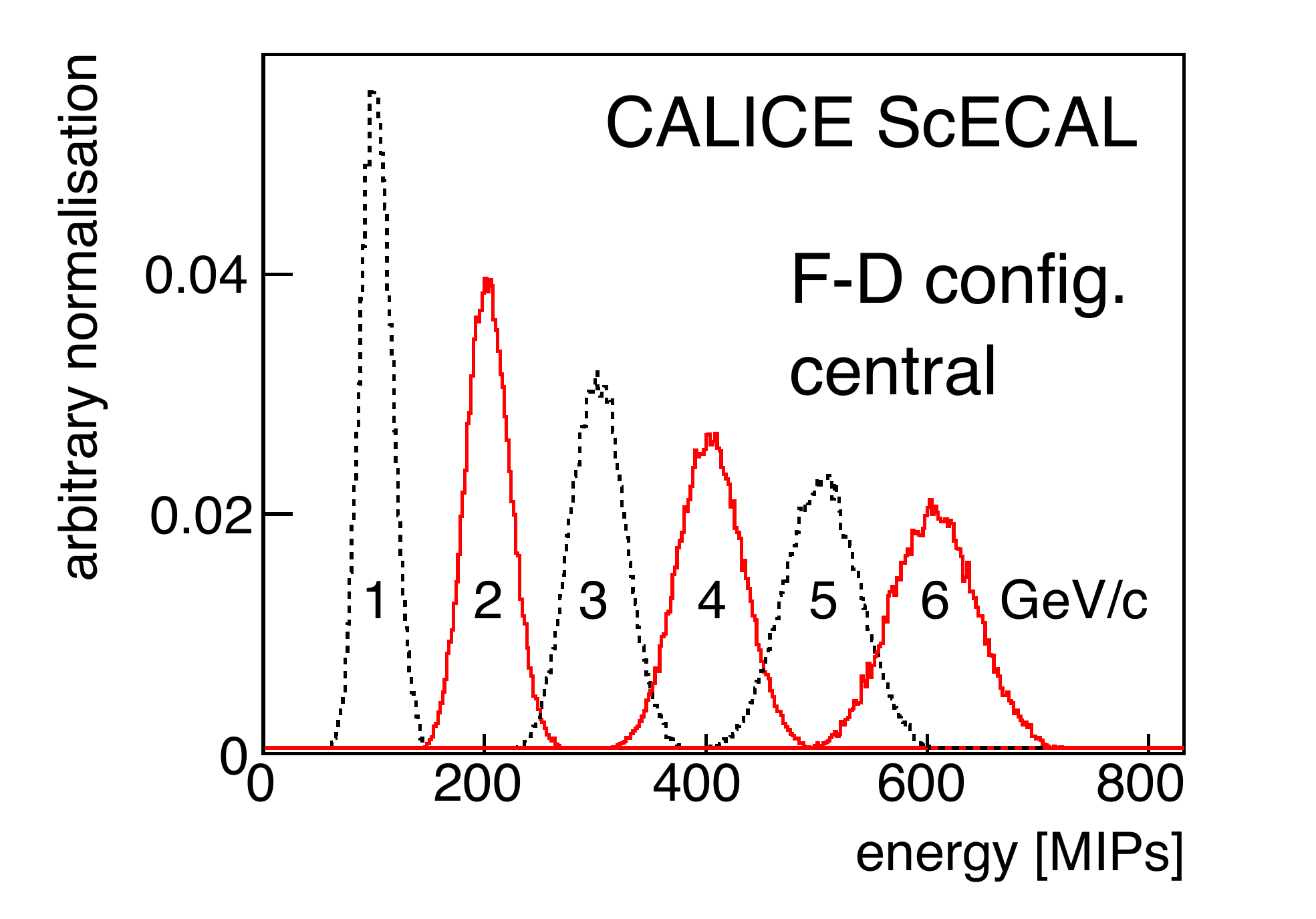}
\caption{\label{fig:scecaldesyspectra}The measured energy spectra of 1-6 GeV positron events collected in the central region
for a detector configuration (F-D configuration)
in which the module with WLS fibres was directly upstream of the module without WLS fibres. From~\cite{Francis:2013uua}.
}
\end{figure}

Fig.~\ref{fig:scecaldesyspectra} 
shows the measured energy spectra of 1 - 6 GeV positron events collected in the central region of the prototype for a detector configuration (F-D configuration), where the type-F module with WLS fibres was directly upstream of the type-D module without WLS fibres.

The successful operation of several hundred MPPCs demonstrates that such a technique is feasible and represents an important milestone in the development of a prototype scintillator strip-based ECAL.
The applied temperature-based corrections to the MPPC response successfully stabilised the prototype's response.
The energy response of this calorimeter prototype was measured to be linear to within 1\% in the energy range between 1 and 6 GeV.
The stochastic terms in the various configurations and regions were measured
to be between 13 and 14\%, while the measured constant terms are between 3 and 4.5\%.
Depending on the true beam energy spread, the intrinsic calorimeter performance may be better than this.
The measured constant term is rather large and a simulation study shows that it has contributions from non-uniformity of the strip response and shower energy leakage due to the limited prototype size, as well as insensitive volume due to the MPPC package. 
It is expected to be reduced in a larger detector with less leakage, and strips with better uniformity

The second physics prototype~\cite{CALICE:2012st} was built using individual small scintillator strips.
The prototype was transversely twice as large as the first prototype, 
and the number of layers was increased from 26 to 30.
The physics prototype was explored with various types of beams:
electrons up to 32~GeV to study  the response to electromagnetic events,
32~GeV muons for the calibration, and  
charged pions of up to 32~GeV to study the hadron response in the combination
with 
the analog hadron calorimeter (AHCAL) and the tail catcher muon tracker (TCMT).
These beams were provided at the Fermilab test beam facility  in 2008 and  2009.
 
\begin{figure}[htb]
\includegraphics[width=0.8\hsize]{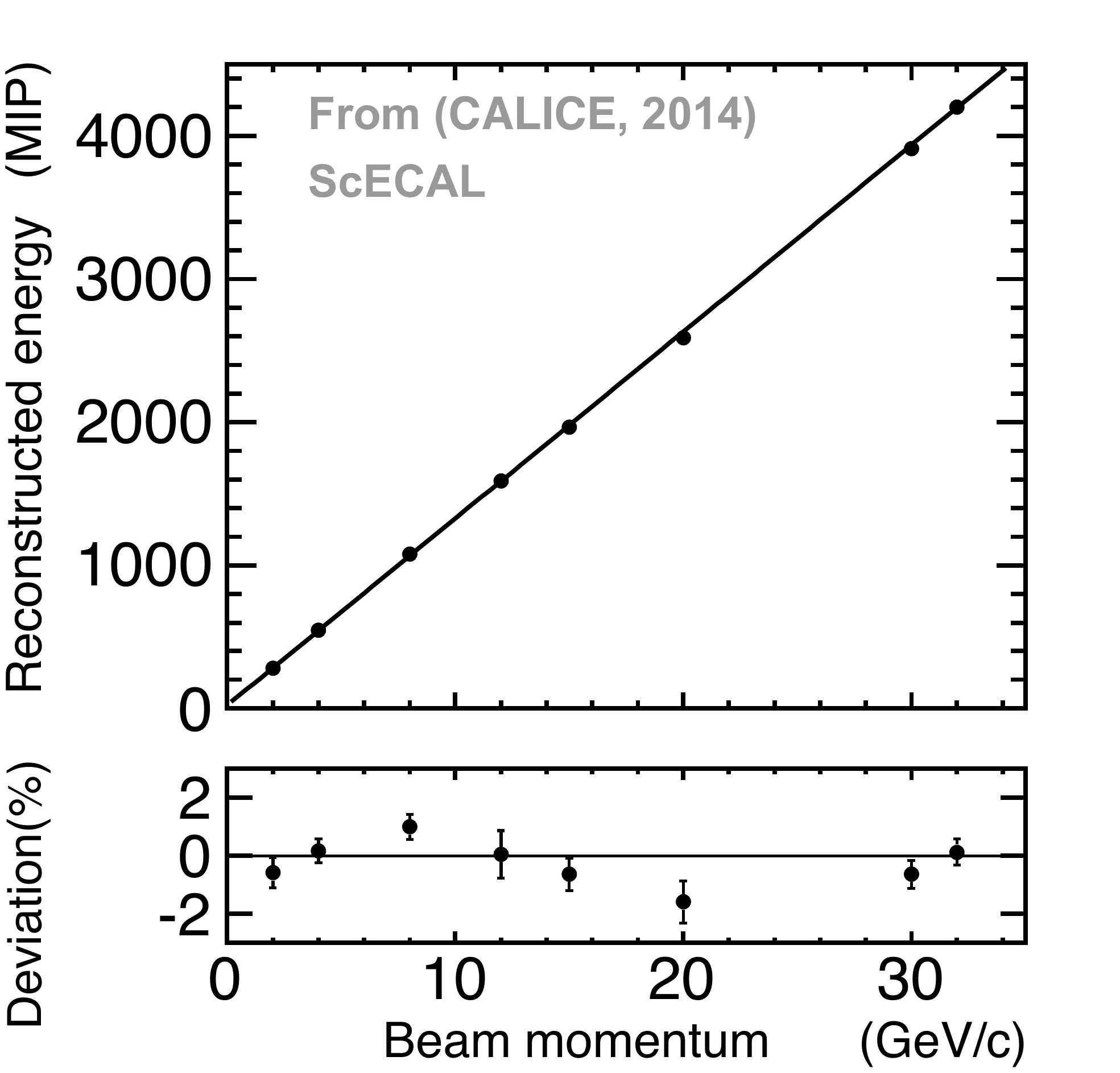}
\caption{\label{fig:ScecalFermilabResults1} Response of the ScECAL physics prototype
to 2 - 32~GeV electrons (top), and
the deviation from the result of a linear fit (bottom). 
From~\cite{CALICE:2012stc,Uozumi:2014}.
}
\end{figure}

\begin{figure}[htb]
\includegraphics[width=0.8\hsize]{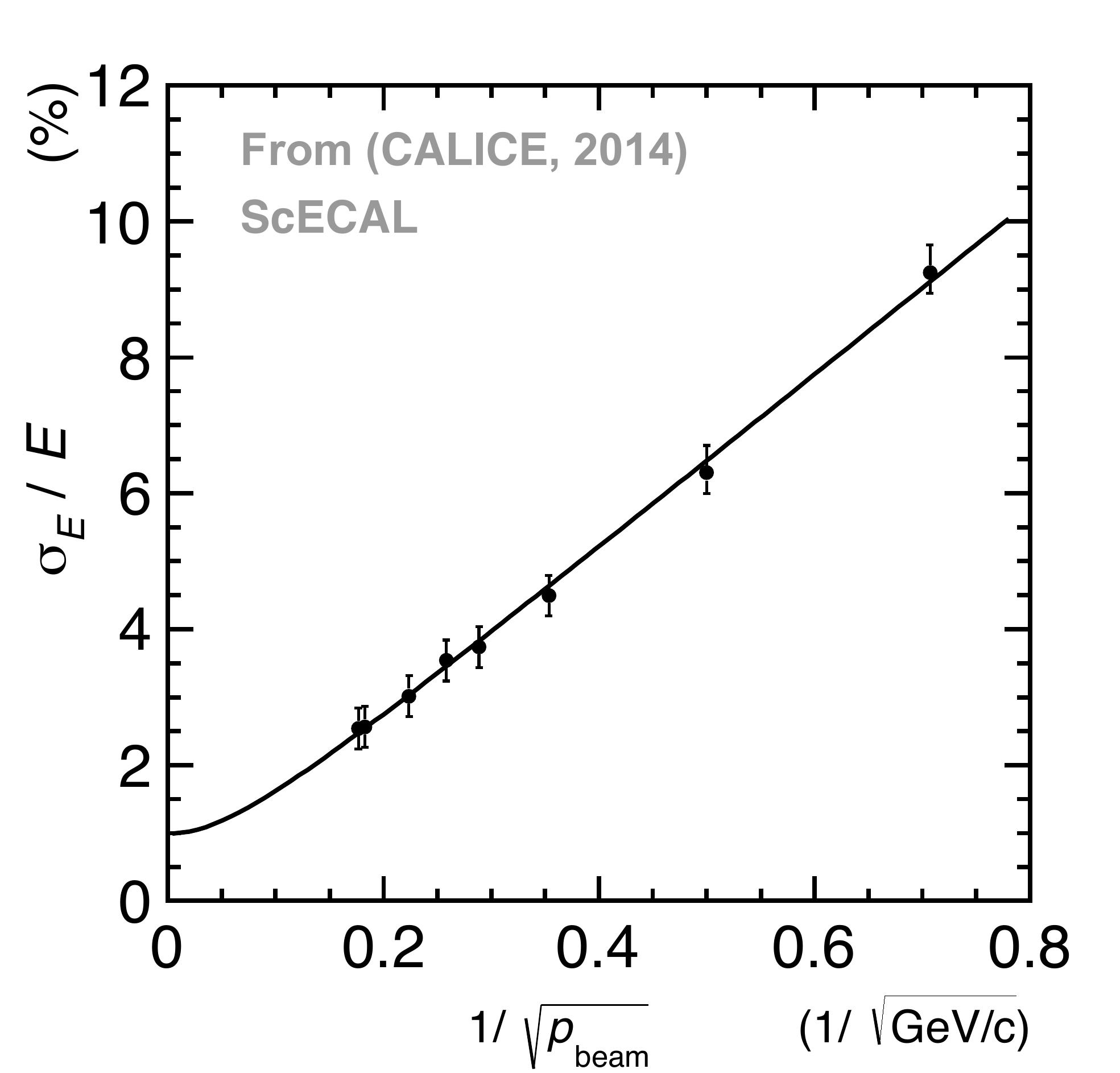}
\caption{\label{fig:ScecalFermilabResults2} Energy resolution of the ScECAL physics prototype
to 2 - 32~GeV electrons as a function of inverse of square root of the beam momentum. From~\cite{CALICE:2012stc,Uozumi:2014}.
}
\end{figure}

Electron events collected in the central region (8cm x 8cm) of the prototype were used to evaluate the linearity and resolution of the measured energy. 
Fig.~\ref{fig:ScecalFermilabResults1}  shows the deposited energy as a function of the momentum of incident beams.
The solid line is the result of a linear fit to the data.
The maximum deviation from linearity is $1.6$\% at 20~GeV.
Fig.~\ref{fig:ScecalFermilabResults2} shows the energy resolution as a function of the inverse of the square root
of the incident beam momentum.
The intrinsic beam momentum fluctuation is estimated to be 
$2.7 \pm 0.3$\% for 2--4~GeV and $2.3 \pm 0.3$\% for 8--32~GeV, respectively, and 
is quadratically subtracted from the resolution. The curve shows the result of a fit to the data
with a quadratic parametrisation of the resolution.
The intrinsic calorimeter resolution for electron beams, after subtraction of the beam momentum spread,
was determined to be
\begin{eqnarray}
\sigma / E  & = & \frac{12.8 \pm 0.1 (stat.) \pm 0.4 (syst.)}{\sqrt{E} } \nonumber  \\
& \oplus &  1.0 \pm 0.1 (stat.) ^{+0.5}_{-1.0}(syst.)~\% \nonumber
\end{eqnarray}
The test beam results show that the physics prototype satisfies the requirements on the linearity and resolution of its electromagnetic response.


\subsection{Scintillator steel or tungsten AHCAL}
\label{sec:Perf:AHCAL}


Following successful operation of a small tile calorimeter with SiPMs~\cite{Andreev:2004uy},
the AHCAL prototype was the first device to use SiPMs on a large scale, with 7608 channels in total. 
Over seven years of test beam operation, and numerous transports between DESY, CERN and Fermilab, the robustness of the technology was convincingly demonstrated.
Since then, SiPMs have been adopted by several high energy physics experiments, e.g.\ Belle~II~\cite{Abe:2010gxa} and CMS~\cite{Lutz:2012yoa} for the read-out of scintillators, and moreover they have conquered a broad range of applications, where they replace classical vacuum photo-multiplier tubes,  e.g.\ in medical imaging.

Using noise data recorded over several years, the long-term stability of the novel photo-sensors was studied in great detail~\cite{collaboration:2010hb},
and no sign of ageing has been found. 
There was a fraction of 2\% of dead channels, due to initial bad soldering, 
which increased by 0.5\% after the transport from CERN to Fermilab. 

\begin{figure}[hbt]
\includegraphics[width=0.7\hsize]{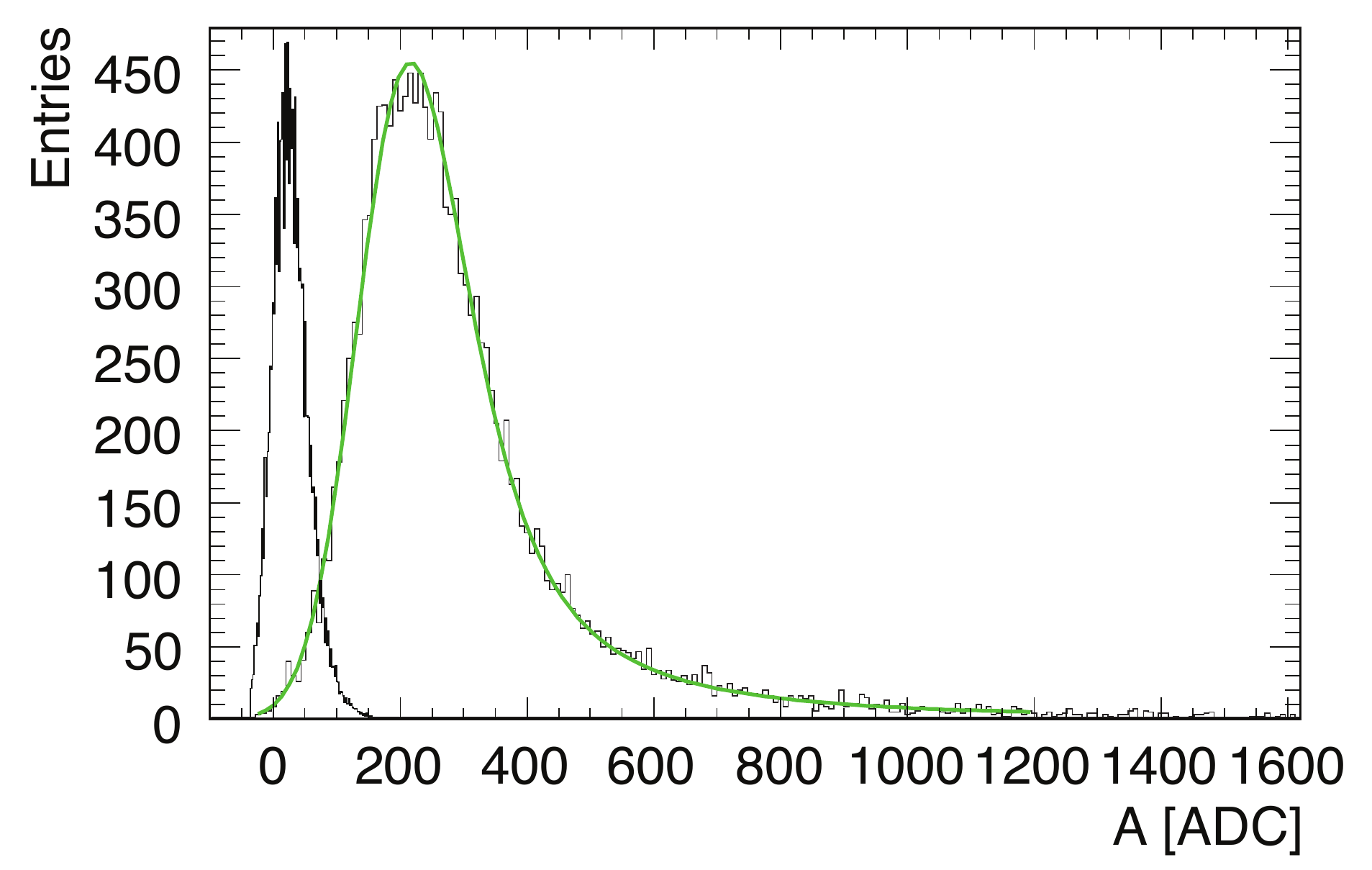}
\includegraphics[width=0.7\hsize]{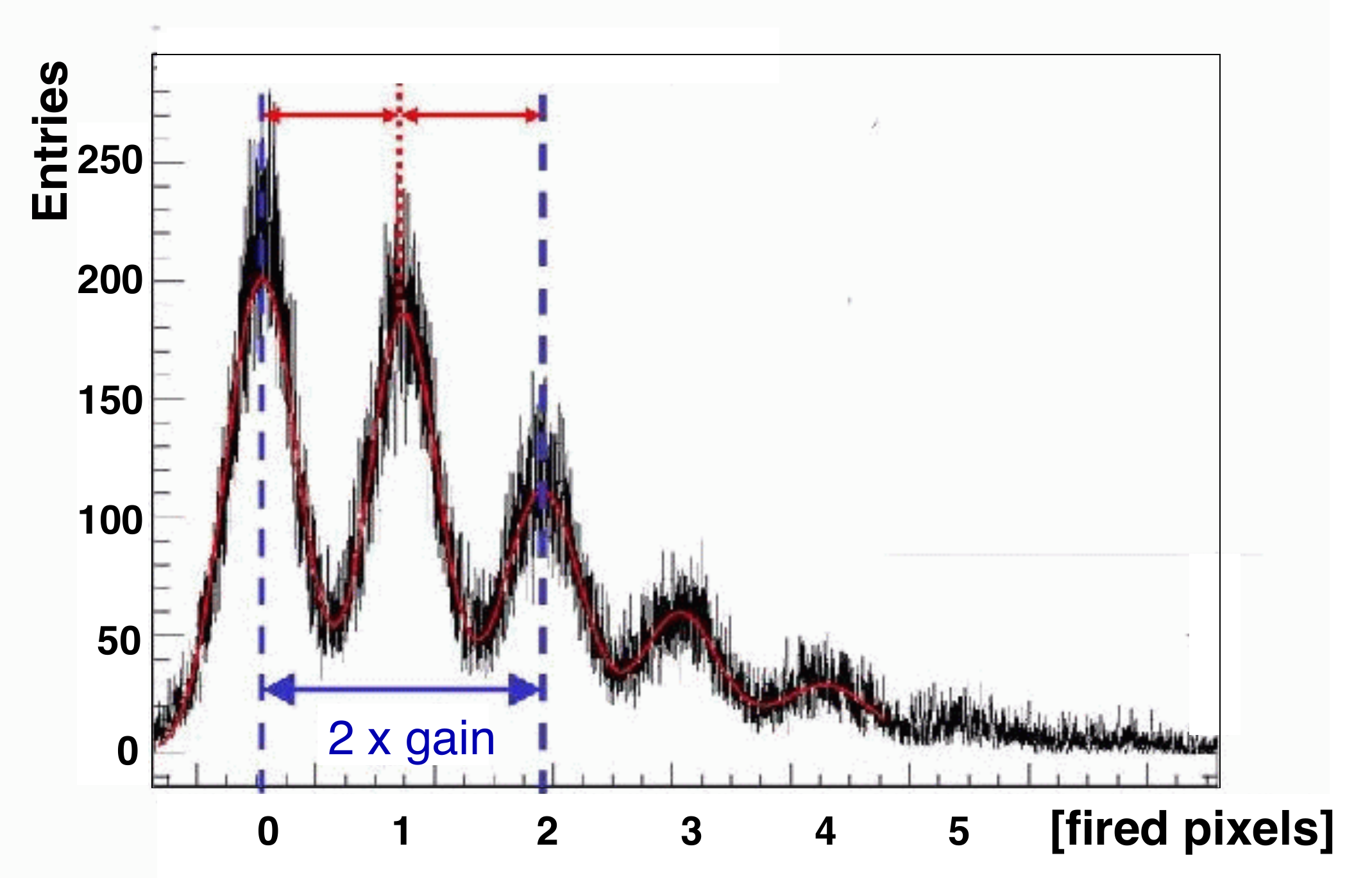}
\caption{\label{fig:Perf:ahcal:calib} MIP distribution of muons with super-imposed fit and noise spectrum for reference (top), distribution of single pixel peaks measured in gain calibration (bottom). Both from~\cite{collaboration:2010hb}.} 
\end{figure}

The response of each detector cell is calibrated~\cite{collaboration:2010rq} using the signal of MIPs, 
see Fig.~\ref{fig:Perf:ahcal:calib}, 
\begin{equation}
E [{\rm MIP}] = A / A_{\rm MIP} \cdot f ( A / A _{\rm pixel} )
\end{equation}
where $E$ denotes the visible energy in units of MIPs, and $A$ the detector response measured in ADC counts. 
The MIP scale $A_{\rm MIP}$ is set by the most probable value of the pulse height spectrum for MIPs.
The correction function $f$ accounts for the exponential saturation of the SiPM response (Eq.~\ref{eq:sipmsaturation}), which is due to the finite number of pixels and finite sensor recovery time, and linearises the response at cell level. 
The correction factor is equal to one for small amplitudes and depends only on the fraction of the sensor pixels fired.
Its argument is the amplitude normalised to the pixel scale $A _{\rm pixel}$, or gain, which is extracted from the separation of single photo-electron peaks in the pulse height spectrum for small, LED induced, amplitudes; see Fig.~\ref{fig:Perf:ahcal:calib}.

\begin{figure}[htb]
\includegraphics[width=0.8\hsize]{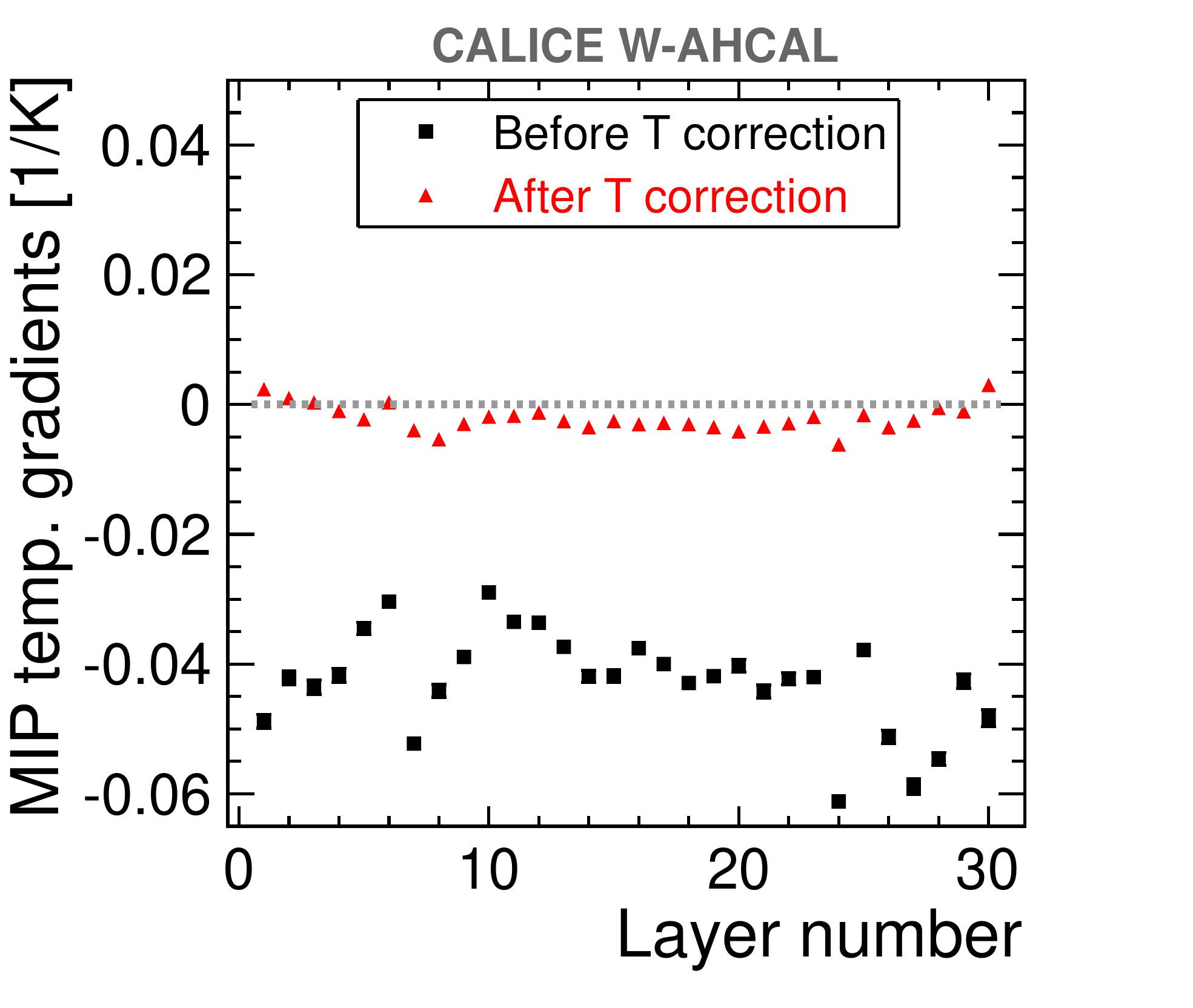}
\caption{\label{fig:Perf:ahcal:tempcor} Temperature gradient of MIP signals before and after correction, for each AHCAL 
layer. From~\cite{Adloff:2013jqa}.} 
\end{figure}

The function $f$ had been determined on the test bench for each SiPM individually, before mounting it on the tile.
Measurements performed {\it in situ} with large LED signals indicated that the saturation level was 20\% lower than on the test bench, which was confirmed by laboratory studies with assembled tile SiPM systems and traced to the fact that the SiPM surface was  only partially illuminated by the fibre. 
The effect was corrected using an average scale factor on the SiPM response curves, and must be accounted for in future test bench procedures. 
SiPM parameters, such as  gain or efficiency of the Geiger discharge, depend on the over-voltage, and the excess of bias over break-down voltage, $\Delta V(T) = V_{\rm bias} - V_{\rm breakdown} (T)$.
The latter increases with temperature $T$, for the SiPMs of the AHCAL prototype by about 50~mV/K, and 
$\Delta V$ is about 2.5~V. 
Therefore both $A_{\rm pixel}$  and $A _{\rm MIP}$ depend on temperature, with coefficients of about $-2\%/{\rm K}$ and 
$-4\%/{\rm K}$, respectively, which is corrected for using the form of Eq.~\ref{eq:Tcor} and the temperature recorded in each active layer during data taking~\cite{Adloff:2013jqa}. 
After correction, the residual dependence is at the level of a few per-mil, 
see Fig.~\ref{fig:Perf:ahcal:tempcor}.
In principle, it is possible to actively compensate this dependence by adjusting $V_{\rm bias}$ according to $T$.

\begin{figure}[htb]
\includegraphics[width=0.9\hsize]{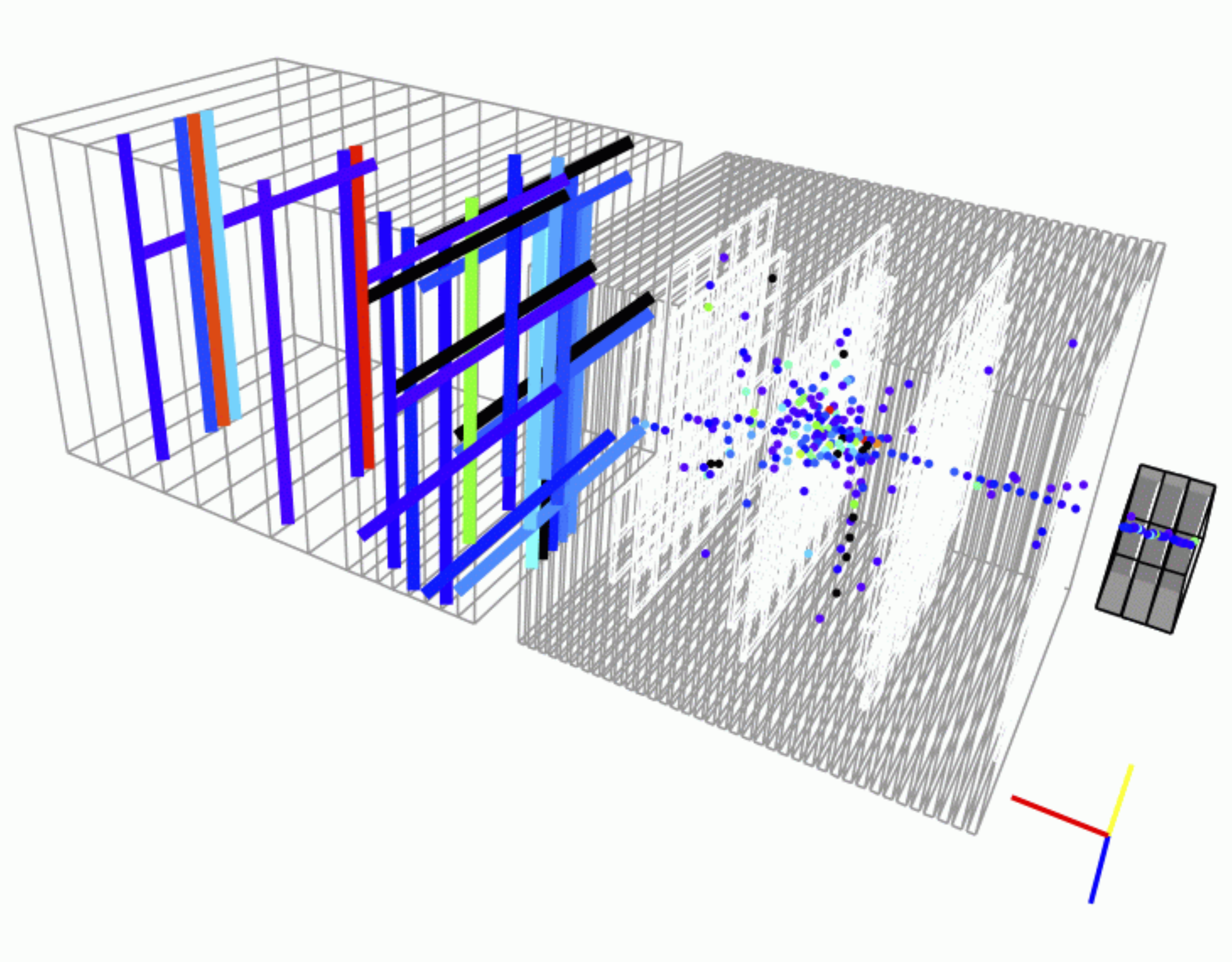}
\caption{\label{fig:Perf:ahcal:evdisplay} Event display of a 20~GeV pion taken from the online monitor. 
The beam enters from the right, the particle traverses the ECAL, interacts in the HCAL and produces signals in the TCMT 
behind. From~\cite{collaboration:2010hb}.} 
\end{figure}

The light yield of a scintillator tile traversed by a MIP corresponds to 13 fired pixels, on average, in the SiPM. 
In the test beam prototype, events were triggered by an external scintillator, and each channel was read out. 
In subsequent reconstruction, only signals with an amplitude above a threshold corresponding to 
$0.5\cdot A_{\rm MIP}$ were retained. 
The efficiency of this noise cut for MIPs is 90-95\%. 
Above this threshold,  the noise hit occupancy was about $2\cdot 10^{-3}$, and the summed noise 
amplitude corresponded to a few hundred MeV. 
These values, and the fine granularity, result in excellent imaging capabilities, which reveal the sub-structure of hadronic showers, see Fig.~\ref{fig:Perf:ahcal:evdisplay}. For example, tracks of charged particles are clearly visible.These are subject to quantitative study later in this article.
With more recent SiPMs, the noise occupancy is expected to decrease by one to two orders of magnitude, due to lower dark rates and steeper decrease of rate with threshold, thanks to suppression of inter-pixel cross-talk.
In general, noise was not subtracted from data. 
Instead, for comparison with simulations, noise events recorded with random triggers were superimposed on the simulated events, before applying the threshold.  

The simulation of the detector response, so-called digitisation, takes the cell-to-cell variation of the calibration, the photo-electron statistics, SiPM saturation, electronic noise and optical crosstalk between tiles into account. 
The effects of non-uniformities in the light response over the area of a tile have been implemented for a dedicated 
study~\cite{Sefkow:2010rt}.  
There is a 100~$\mu m$ wide zone along the edges of the tile where no light is produced, and a reduction of light yield at the positions of fibre and SiPM, due to reduced scintillator material thickness there. 
The effects on the response to electrons and pions were found to be negligible,
which was confirmed for hadrons with test beam data~\cite{CALICE:2013un}.
Thus, to save computing time, they are not simulated by default. 
The simulated events are  processed through the same calibration and reconstruction chain as real data.  
The scale of the simulated response is adjusted by one global parameter to the MIP scale;  no further tuning to data is applied. 

\begin{figure}[htb]
\includegraphics[width=0.8\hsize]{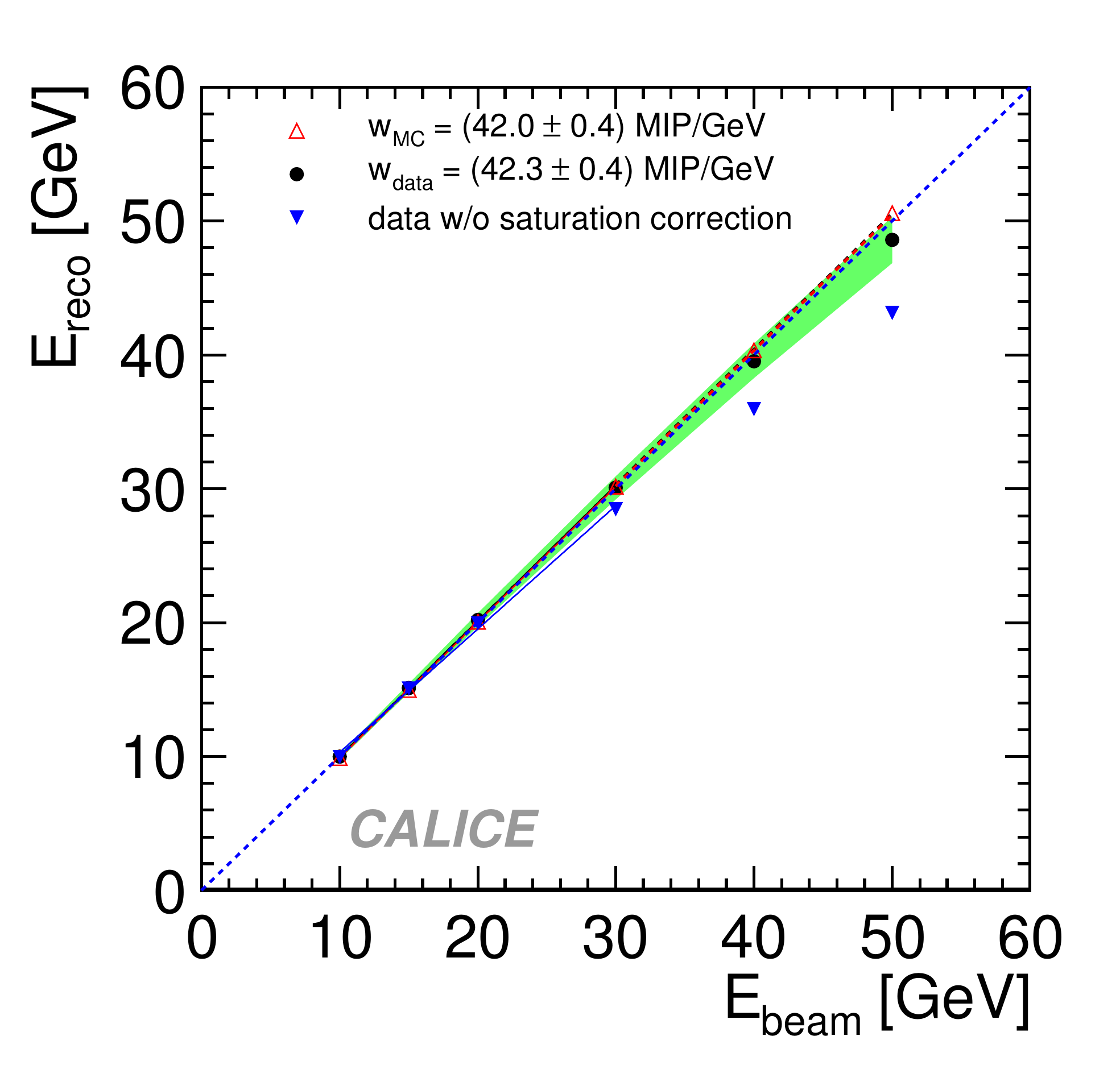}
\caption{\label{fig:Perf:ahcal:emlin} Linearity of the AHCAL response to positrons. Black dots represent data, blue triangles data before saturation correction, and open triangles show simulation results. The dashed line shows exact linearity for reference. The green band represents the systematic uncertainty. From~\cite{collaboration:2010rq}.} 
\end{figure}

For the validation of the detector calibration and simulation, positron induced showers recorded at the CERN SPS test beam have been analysed, in an energy range from 10 to 50~GeV~\cite{collaboration:2010rq}.
%
The sampling structure of the AHCAL corresponds to 1.24~radiation lengths $X_0$ per layer and has an effective Moli\`ere radius $R_M$ of 2.47~cm. 
In order to minimise the noise contribution, the shower energy was summed up over cells within a cylinder of 5~$R_M$ around the extrapolated track and over a length of 20 layers. 
The distribution, for a given beam momentum, is fitted by a Gaussian, with the peak position taken as the mean response, and the width as resolution. 
An electromagnetic energy scale factor $42.3\pm 0.4$~MIP/GeV is extracted from a linear fit from 0 to 50~GeV to the distribution of reconstructed {\it versus} beam energy. 
The reconstructed energy on the electromagnetic scale as a function of beam energy is plotted in 
Fig.~\ref{fig:Perf:ahcal:emlin}. 
Systematic errors, indicated by the green band, are dominated by MIP calibration uncertainties, and, at the highest energies, by those on the saturation correction. 
The deviation from linearity is less than 1\% in the range 10 to 30~GeV, and 3\% at 50~GeV. 
This indicates imperfections in the saturation correction, which in the future will be remedied with SiPMs with a larger dynamic range, and with a more precise test bench characterisation. 
For the study of hadron shower development, this is sufficient, since for a given beam energy the single hit energy spectrum is much softer for hadrons than for electrons. 

\begin{figure}[htb]
\includegraphics[width=0.8\hsize]{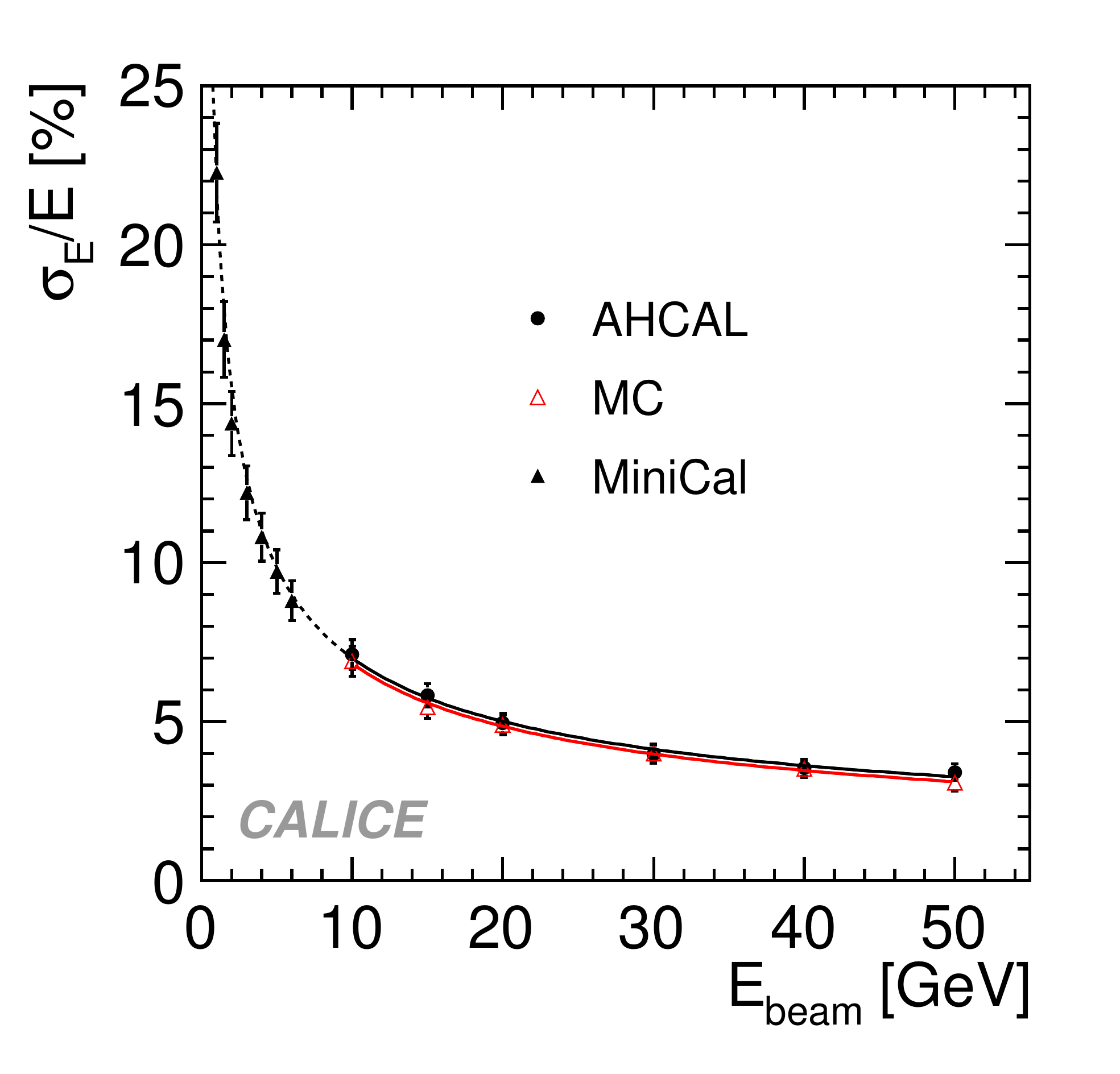}
\caption{\label{fig:Perf:ahcal:emres} Energy resolution of the AHCAL for positrons in data and simulation. The black triangles show data measured with an earlier prototype with the same sampling structure. Error bars represent the quadratic sum of statistical and systematic uncertainties. From~\cite{collaboration:2010rq}.} 
\end{figure}

The resolution for electromagnetic showers is shown in Fig.~ \ref{fig:Perf:ahcal:emres}. 
It agrees with that of a previous prototype~\cite{Andreev:2004uy}, and the data together with the simulation, over an energy range from 1 to 50~GeV. 
The AHCAL resolution data is fitted to the function 
\begin{equation}
\label{eq:Perf:resolfit}
\frac{\sigma_E}{E} = \frac{a}{\sqrt{E}} \oplus b \oplus \frac{c}{E} \; .
\end{equation}
The last term, representing the noise contribution, is fixed using random trigger events,  to 58~MeV. 
The first, originating from stochastic fluctuations in the shower evolution, sampling and signal generation,
is found to be $a=(21.9\pm 1.4)\%$, in excellent agreement with simulation. 
The constant term $b$ is a measure of calibration uncertainties, non-uniformities and instabilities, and is found to be $b=(1.0\pm 1.0)\%$. 
Such effects are not included in the simulation. 
The result thus demonstrates that they are indeed very small and can be neglected for studies with hadrons. 

For the determination of the hadronic response, clean pion samples selected from data recorded at the same CERN SPS beam line (H6) 
have been analysed in the range 
from 10 to 80~GeV~\cite{Adloff:2012gv}. 
The primary inelastic interaction was required to be detected in the first layers of the AHCAL,
by observing a significant increase of recorded energy and hit multiplicity over several consecutive layers. 

The reconstructed energy is obtained from the sum of the energies in the three calorimeters, 
\begin{equation}
\label{eq:Perf:ehad}
E_{had} = E_{\rm ECAL}^{\rm track} + \frac{e}{\pi} (E_{\rm HCAL} + E_{\rm TCMT})\; ,
\end{equation}
where $E_{\rm ECAL}^{\rm track}$ is the measured energy deposited by the particle track in the SiW ECAL, using a conversion factor from simulations, validated by muon data. 
The HCAL energy on the electromagnetic scale, $E_{\rm HCAL}$, is obtained in exactly the same way as for electromagnetic showers, as described above, except that the sum over the hit energies extends  over the entire calorimeter volume. 
The constant factor $e/\pi = 1.19$ accounts for the fact that the HCAL is non-compensating and corresponds to the ratio of electron and pion response measured in the same analysis and averaged over the energy range considered here. 
The first 9 layers of the TCMT have the same sampling structure as the HCAL, 
so its energy $E_{\rm TCMT}$ is reconstructed in the same way. 

\begin{figure}[htb]
\includegraphics[width=0.8\hsize]{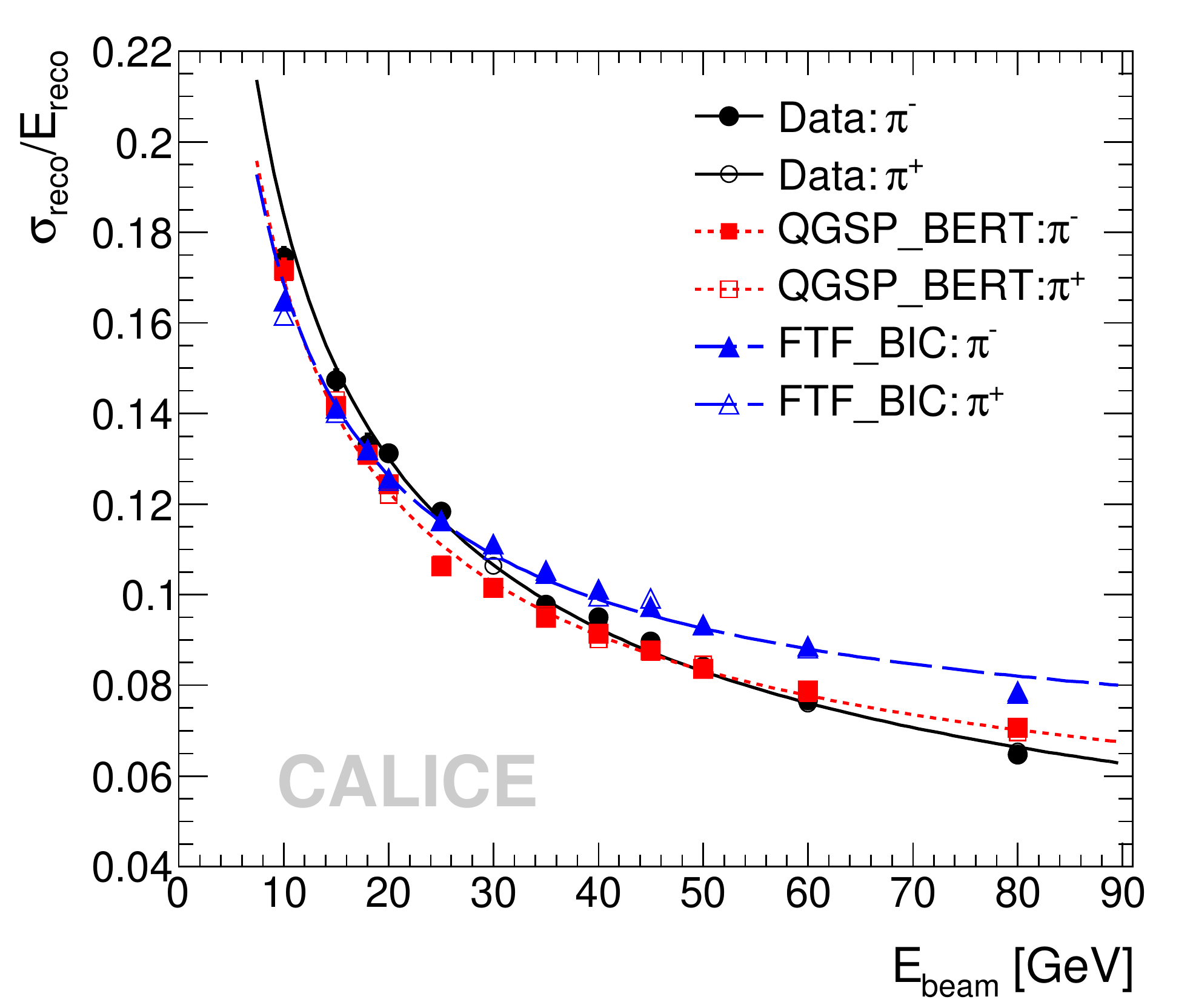}
\caption{\label{fig:Perf:ahcal:hadres} Uncorrected hadronic energy resolution for data as well as simulations, using two physics lists. The curves show fits using Eq.\ref{eq:Perf:resolfit}. From~\cite{Adloff:2012gv}.} 
\end{figure}

The energy distributions for a given beam energy are fit with a Gaussian in the interval $\pm 2\sigma$, and mean and width of the fit are taken as mean reconstructed energy and resolution. 
The linearity of the response to hadrons, selected to start showering early in the HCAL, is within 2\% in the range 10 to 80~GeV. 
The resolution is shown in Fig.~\ref{fig:Perf:ahcal:hadres} as a function of beam energy. 
The data are compared to simulations using two recent physics lists in Geant4 version 9.4, which will be described in more detail in Section~\ref{sec:G4valid}.
Each set of resolution measurements is fit with a function according to Eq.~\ref{eq:Perf:resolfit}. 
The noise term is fixed to 0.18~GeV as measured with random trigger events and is dominated by the TCMT; the noise in the HCAL alone corresponds to 0.06~GeV. 
For data, the stochastic term is $(57.6\pm 0.4)\%$, and the constant term is $(1.6\pm 0.3)\%$.
The simulations predict a somewhat smaller resolution at low energies, and in the case 
of ${\rm FTFP}\_{\rm BERT}$ a worse resolution at high energies. 
In both cases, this leads to smaller stochastic and larger constant terms, 49-52\% and 4-6\%, respectively.  
Overall, the measured resolution falls into the range of expectations based on recent shower models. 

The performance of the AHCAL with tungsten absorber has been evaluated at the CERN PS and SPS test beams, respectively, in the energy ranges 1 to 10~GeV~\cite{Adloff:2013jqa} 
and 10-100~GeV~\cite{CALICE:2013ff}. 
Also here, the event selection ensured a start of the shower evolution in the first few detector layers. 
\u{C}erenkov counter information and topological cuts were used to obtain pure electron, pion and proton samples.

The sampling structure of the tungsten absorber has been optimised for high energy jets and a given total thickness of the detector, to provide good shower containment and to fit inside the magnetic coil~\cite{Linssen:2012hp}. 
It is similar to that of the steel prototype in terms of nuclear interaction lengths, but much coarser in terms of electromagnetic radiation lengths: one layer corresponds to 
0.13~$\lambda_I$ and 2.8~$X_0$. 
Therefore it is expected that the electromagnetic energy resolution is poorer than for steel, and that the intrinsic resolution for hadrons is somewhat worse, too. 
Moreover, since electromagnetic showers are more compact and sampled in fewer layers, the impact of single cell calibration uncertainties is larger, which is reflected in larger systematic uncertainties, in particular for electrons. 
Given the smaller Moli\`ere radius, the transverse granularity needs to be re-optimised. 
For electron data in the lower energy range, a stochastic resolution term of $(29.6\pm 0.5)\%$ is measured, in excellent agreement with simulations, which predict  $(29.2\pm 0.4)\%$, 
as is the constant term of $(0.0\pm 2.1)\%$,.
The hadron resolution is also found to be as expected, with a stochastic term of 63\%, while the constant term is not well constrained at these low energies. 

\begin{figure}[htb]
\includegraphics[width=0.8\hsize]{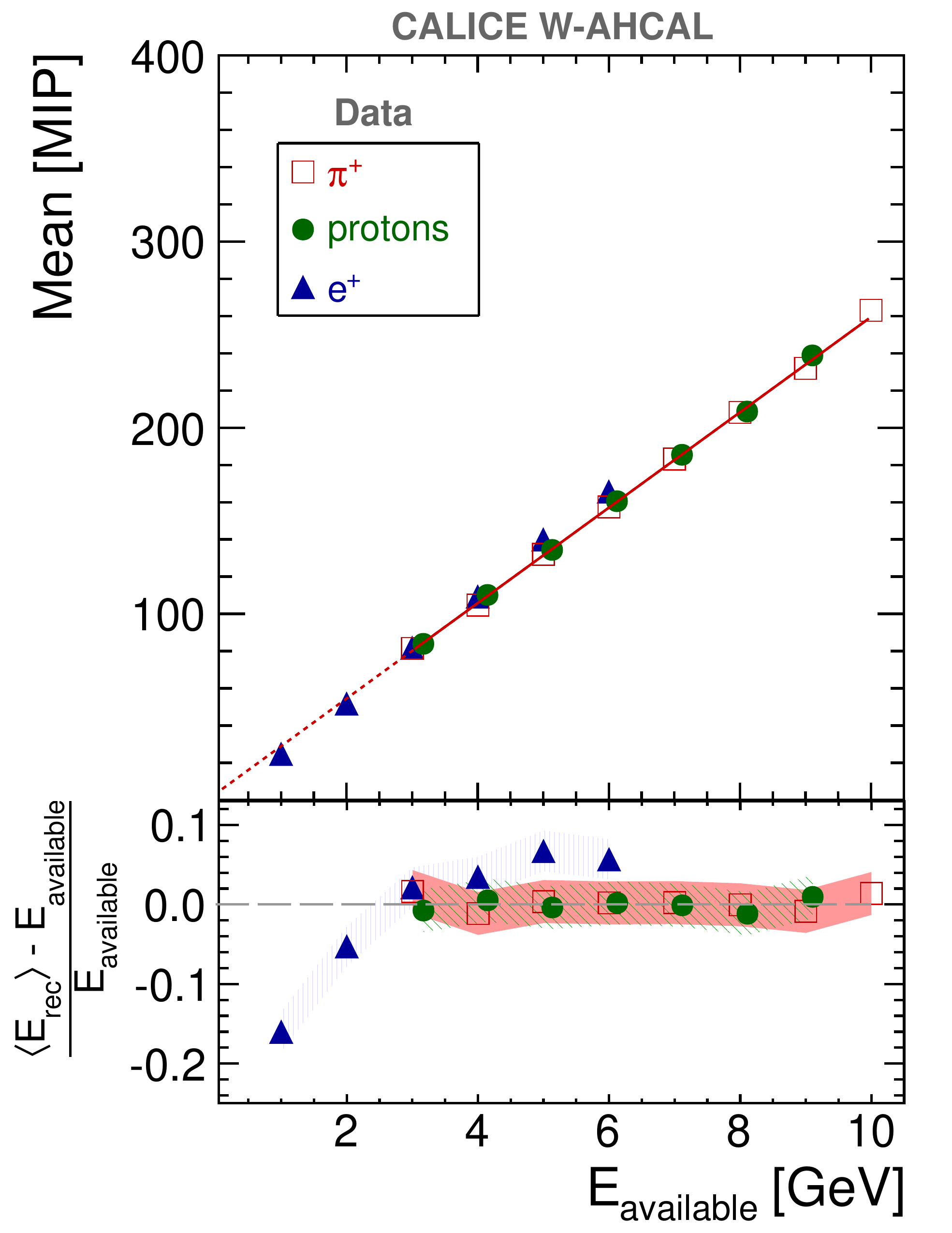}
\caption{\label{fig:Perf:ahcal:wcomp} Dependence of the mean visible energy on the available energy for $e^+$,
$\pi^+$ and protons. The line indicates a linear fit to the $\pi^+$ data, and the bottom part shows the residuals 
relative to this fit. The error bands represent the total uncertainty. From~\cite{Adloff:2013jqa}.} 
\end{figure}

It is noteworthy that the tungsten scintillator combination is nearly compensating and gives very similar response for electrons, pions and protons above 3~GeV.
This response as a function of available energy
is shown in Fig.~\ref{fig:Perf:ahcal:wcomp} for data.
Here, the available energy represents the energy which can be measured in the calorimeter; for pions and electrons it corresponds to the particle energy, for protons it is given by the kinetic energy.
The measurements are very well reproduced by simulations, including the deviation of the electron data from the pion extrapolation, 
which are much smaller than in the case of a steel absorber with $\frac{e}{\pi} = 1.19$.
Preliminary results from the higher energy range confirm the same behaviour up to energies of 100~GeV, and an excellent linearity, see Fig.~56 of~\cite{CALICE:2013ff}.

\subsection{RPC steel or tungsten DHCAL}


The test beam activities of the DHCAL started in 2010 at Fermilab with steel absorber and were completed by 2012 at CERN with a tungsten absorber structure. 
%
At the start of the test beam campaigns, the tail catcher was equipped with the same scintillator strips which had already been used in tests with the AHCAL. As the tests progressed, they were gradually replaced with RPC layers, 
of the same design as in the main stack.
Part of the data was taken with the CALICE silicon-tungsten ECAL placed in front of the DHCAL main stack.  

\begin{figure}[htb]
\includegraphics[width=0.8\hsize]{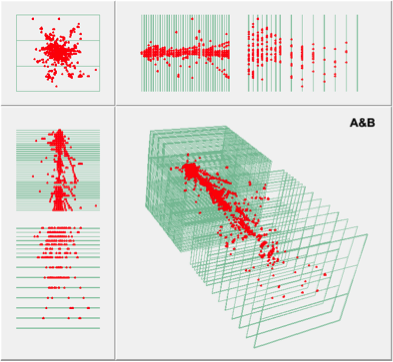}
\caption{\label{fig:DHCAL_event} Event displays of a 120~GeV proton interacting in the DHCAL.}
\end{figure}

A dedicated run was performed without absorber plates. In this case, the 2 mm steel and 2 mm copper cover plates of the detector cassettes together with the glass and readout boards of the RPCs served as the only absorber material. Thus, each layer corresponded to a thickness of only 0.4 radiation lengths or 0.04 interaction lengths. The minimal amount of absorber material provided the most detailed event pictures recorded with the DHCAL, see Fig.~\ref{fig:DHCAL_event}.

\begin{figure}[htb]
\includegraphics[width=0.7\hsize]{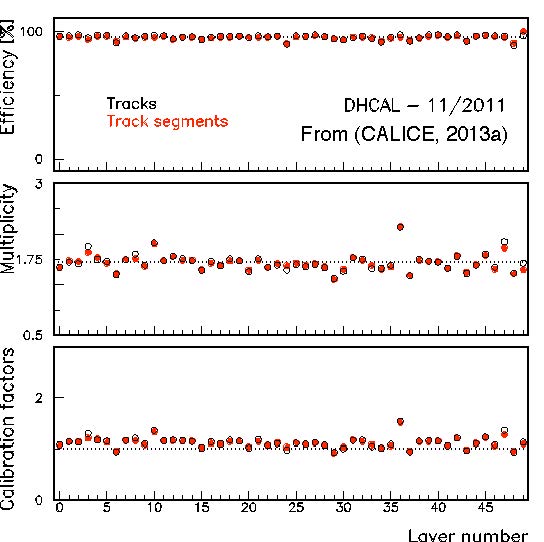}
\includegraphics[width=0.7\hsize]{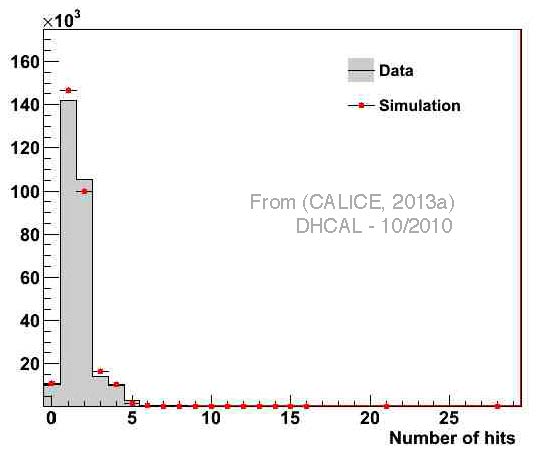}
\caption{\label{fig:DHCAL_muon} The performance parameters of the DHCAL as function of layer number (top), hit multiplicity distribution (bottom).
From~\cite{CALICE:2013th,Repond:2013}.}
\end{figure}

Muon events were collected with the 32 GeV secondary beam and a 3 m long beam blocker placed into the beam. Muons provided an excellent tool to monitor the performance of the detector elements, i.e. the MIP detection efficiency and the average pad multiplicity of the RPCs. To measure these, either tracks in the entire DHCAL or track segments spanning only five layers were reconstructed~\cite{CALICE:2013th}. To avoid a bias of the measurements, in either case, the layer for which the performance parameters were assessed, was not utilised in the track reconstruction. 
Fig.~\ref{fig:DHCAL_muon} shows the efficiency, average pad multiplicity, and the product of the two, and the calibration factors (after normalisation to the average value over the entire detector) as function of layer number. This very uniform performance was obtained in the run without absorber plates, where the cooling of the cassettes was made easier due to the large gap between detector elements. When placed in the absorber structure the uniformity of the response was not quite as good.

For purposes of tuning the simulation of the RPC response it is useful to measure the average muon response per layer. Fig.~\ref{fig:DHCAL_muon} shows the response per layer averaged over all layers in the DHCAL. The data are compared to Geant4-based simulations using a two-exponential model to describe the induced electrical charge in the pad plane. A satisfactory agreement between data and simulation is observed.

The muon response is utilised to equalise the response of the individual RPCs in the DHCAL~\cite{CALICE:2013ft}. Different schemes are being investigated: a) full calibration (application of calibration factors obtained from the efficiency and average pad multiplicity), b) density weighted calibration (application of the same calibration factors as in the full calibration, but to each hit individually and weighted by a function dependent on the density of hits surrounding a given hit), and c) a hybrid of the two approaches. In all three cases, the calibration procedure results in a reduced spread of the mean response in runs of the same beam energy and in distinct improvements to the linearity of the response as function of beam energy.

\begin{figure}[htb]
\includegraphics[width=0.7\hsize]{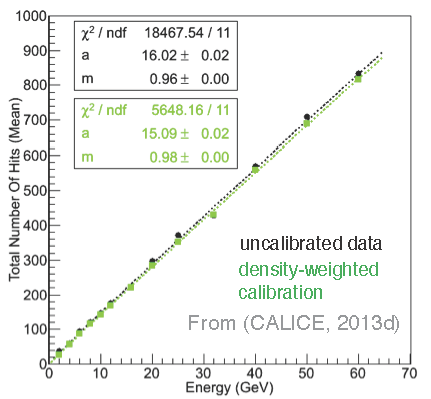}
\includegraphics[width=0.7\hsize]{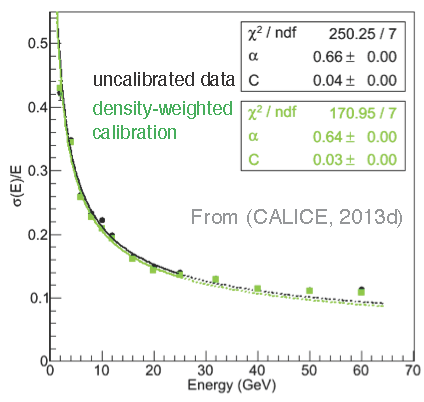}
\caption{\label{fig:DHCAL_Fe} Response and resolution for pions as measured at Fermilab with steel absorber plates. From~\cite{CALICE:2013ft,Bilki:2013}.}
\label{fig:dhcal-linres-f}
\end{figure}

Using the density-weighted calibration Fig.~\ref{fig:DHCAL_Fe} shows the average number of hits as function of the pion beam energy, as measured with the steel absorber plates in the Fermilab test beam~\cite{CALICE:2013ft}. The response is fitted with a power law, $N = a\cdot E^m$. The exponent of 0.974 indicates a slight saturation, compared to an exponent of unity corresponding to a perfectly linear response. The resolution is also shown in Fig.~\ref{fig:DHCAL_Fe}. As expected, due to the finite size of the readout pads and the ensuing saturation at higher energies, at energies above about 35 GeV the resolution remains constant and fails to improve further with higher energies. Below 30 GeV a fit to $C \oplus \alpha / \sqrt{E}$ results in a stochastic term of 64\%, in good agreement with expectations based on simulation. 

\begin{figure}[htb]
\includegraphics[width=0.7\hsize]{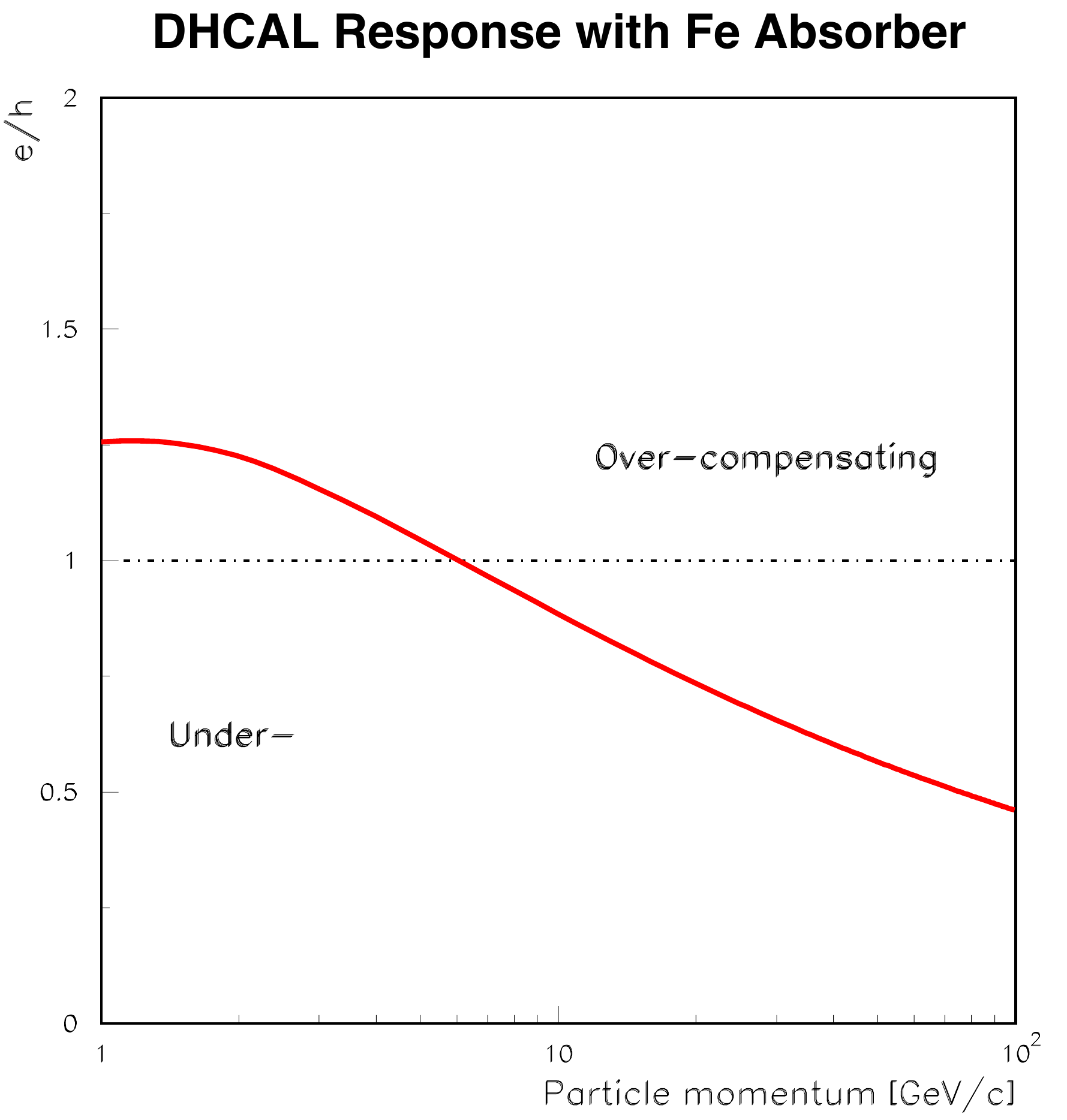}
\caption{\label{fig:DHCAL_comp} The ratio of the electromagnetic to hadronic response in the simulated Fe-DHCAL as function of particle momentum.}
\end{figure}

The responses to positrons and pions of the same momentum are different in the DHCAL, see Fig.\ref{fig:DHCAL_comp}. At momenta below about 4 GeV/c the positrons provide a higher response (larger number of hits), which is a typical feature of non-compensating calorimeters.  On the other hand, at momenta above 8 GeV the response to positrons is suppressed due to the finite size of the readout pads, the high density of electromagnetic showers, and the ensuing saturation of the response. At these momenta the number of hits for pions exceeds the corresponding number for positrons, a phenomenon called over-compensation. In the intermediate momentum range the response of the DHCAL is approximately compensating.  

Even though the density-weighted calibration treats positrons and pions differently and takes into account the local density of hits when determining the calibration constants, the procedure is not to be confused with software compensation, which attempts to correct for differences in the response to electromagnetic or hadronic sub-showers. The calibration procedure only corrects for differences in the overall performance of individual chambers, such that after the correction all chambers show the same average response. Thus the calibration factor for a specific hit in an individual RPC depends on a) the difference in performance of the RPC to the average performance of the stack b) the local density of hits surrounding the specific hit and c) the particle type. Therefore, unlike for software compensation, if a given chamber's performance is exactly average, no corrections are applied. Obviously, if the procedure were to be applied to digitised Monte Carlo data, also no corrections would be applied.

\begin{figure}[htb]
\includegraphics[width=0.7\hsize]{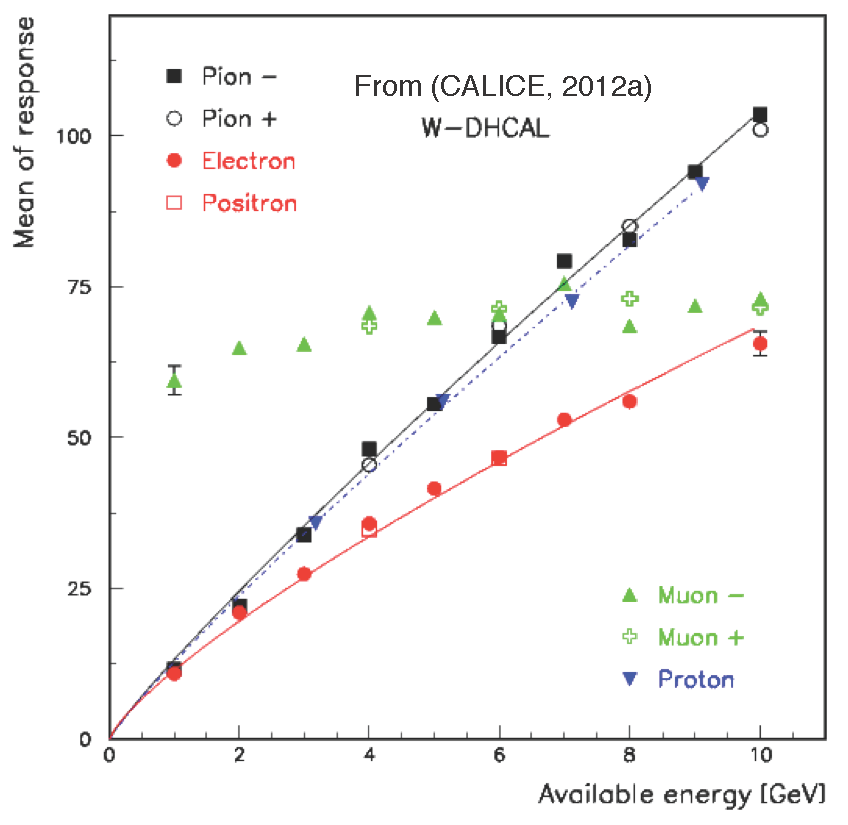}
\includegraphics[width=0.7\hsize]{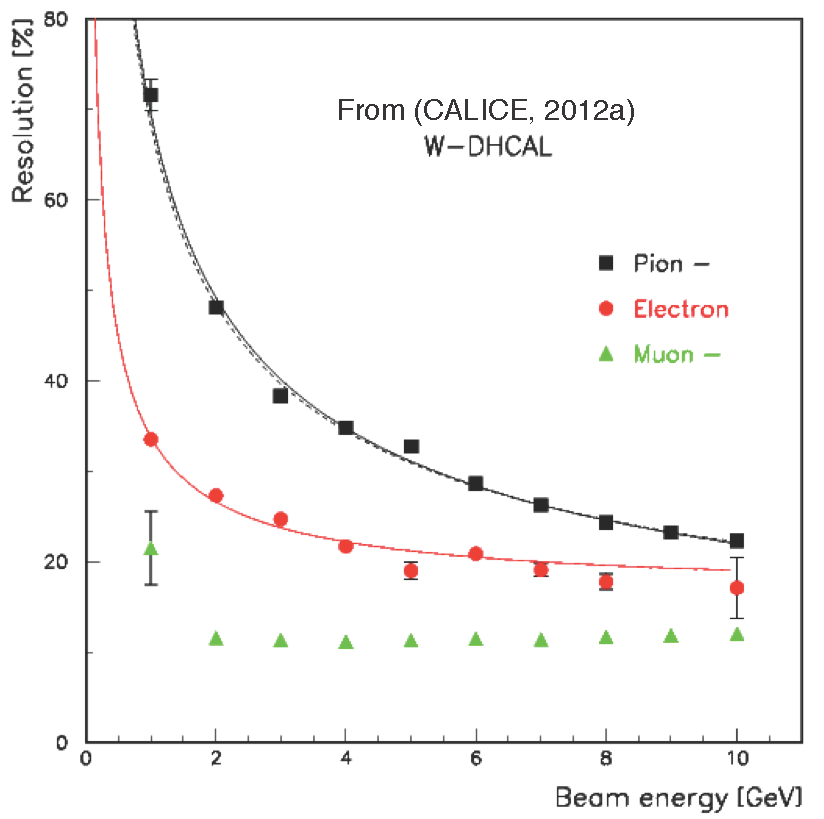}
\caption{\label{fig:DHCAL_W} Response (number of hits) and resolution for electrons, muons and pions as measured in the PS with tungsten absorber plates. 
From~\cite{CALICE:2012tn,Repond:2012}. }
\end{figure}

Due to the higher elevation of the CERN site, the RPCs showed a higher gain at CERN than at Fermilab (see also~\cite{Bilki:2009gc} for details on the environmental dependence of the RPC response). To obtain a similar performance, the default high voltage was decreased from 6.3 to 6.1 kV~\cite{CALICE:2012tn}. The increased thickness in radiation lengths of the tungsten absorber plates, compared to the steel plates used at Fermilab, resulted in a highly suppressed electromagnetic response. 
Fig.~\ref{fig:DHCAL_W}  shows the response as function of energy as measured at the CERN PS. The electron response is seen to be significantly smaller than the hadronic response. In other words, the W-DHCAL is seen to be overcompensating in the entire energy range. Furthermore, the hadronic response is approximately 30\% smaller than the corresponding response measured at Fermilab with steel absorber plates. Over the entire energy range the hadronic response is seen to saturate, with an exponent of m = 0.90. Fig.~\ref{fig:DHCAL_W} shows the resolution for electrons, muons and pions as function of beam momentum. Due to the smaller hit count, the pion resolution is somewhat degraded compared to the case with steel absorber plates. However, in the context of a PFA, it should be recalled that only the neutral hadron component of a jet, on the average 10\%, is measured directly in the calorimeter.

In general, the DHCAL performed very well in the various test beams. Nevertheless, there were a small number of design and operational issues. 
\begin{enumerate}
\item[a)] Lower efficiency at the edges of the chamber. 
\item[b)] Loss of efficiency over time for some chambers, possibly due to the high-voltage contact. 
\item[c)] Environmental dependence of the response, mainly affecting the pad multiplicity~\cite{Bilki:2009gc}. 
\end{enumerate}


The Fermilab and CERN test beam set-ups have been simulated with a Monte Carlo program based on the Geant4 package and a standalone program RPCSIM for the simulation of the response of RPCs. 
The spatial coordinates of any energy deposition in the gas gap of an RPC was recorded for further analysis. In the following these energy deposition locations are named {\it points}.

For each generated {\it point}, the RPCSIM program generates a signal charge $Q$, distributes this charge over the pads, sums up all charges on a given pad, and applies a threshold {\it T} to identify the pads with hits. 

The signal charges are generated according to the measured~\cite{Drake:2007zz} spectrum of avalanche charges, as was  obtained with cosmic rays. The induced charge in the plane of the readout pads is assumed to decrease as a function of lateral distance R from a given {\it point}. This decrease is parameterized as the sum of two exponentials.
 	
The emulation of the RPC response depends on six parameters, of which five were tuned to reproduce the distribution of hits per layer as measured with muon tracks. The sixth parameter, related to the suppression of avalanches close-by to other avalanches, was tuned such as to reproduce the response to positrons.

With all parameters of the RPCSIM program tuned, the program is now ready to predict the response to hadron beams, without having lost its predictive power.  The comparison of the measured and simulated response to pions is ongoing and will be subject of future publications. However, in a first iteration, the RPCSIM parameters were tuned to the data collected with a small-scale DHCAL prototype, the so-called Vertical Slice Test (VST)\cite{Bilki:2008df,Bilki:2009wp,Bilki:2009gc,Bilki:2009xs,Bilki:2009ym}. 
Due to the limited depth of the VST only the forward part of hadronic showers could be measured, while most of the energy leaked out the back of the stack. Nevertheless, the distribution of the number of hits could be well reproduced with Geant4 and the tuned RPCSIM program, giving confidence in the overall approach to simulate the DHCAL response.

\subsection{RPC steel SDHCAL}
\begin{figure}[htp]
\begin{center}
\includegraphics[width=0.4\textwidth]{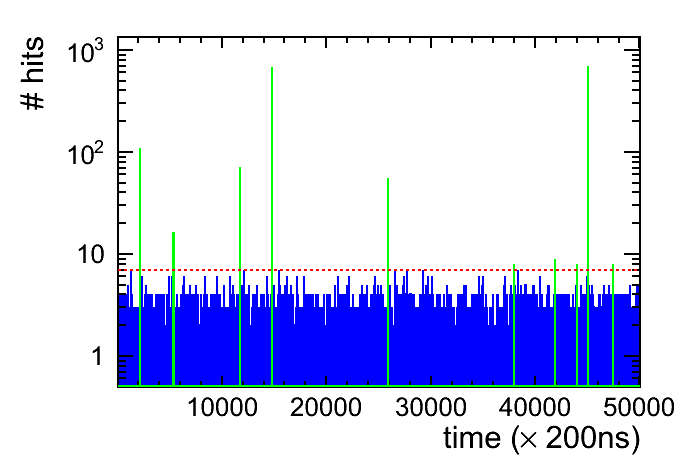}
\includegraphics[width=0.4\textwidth]{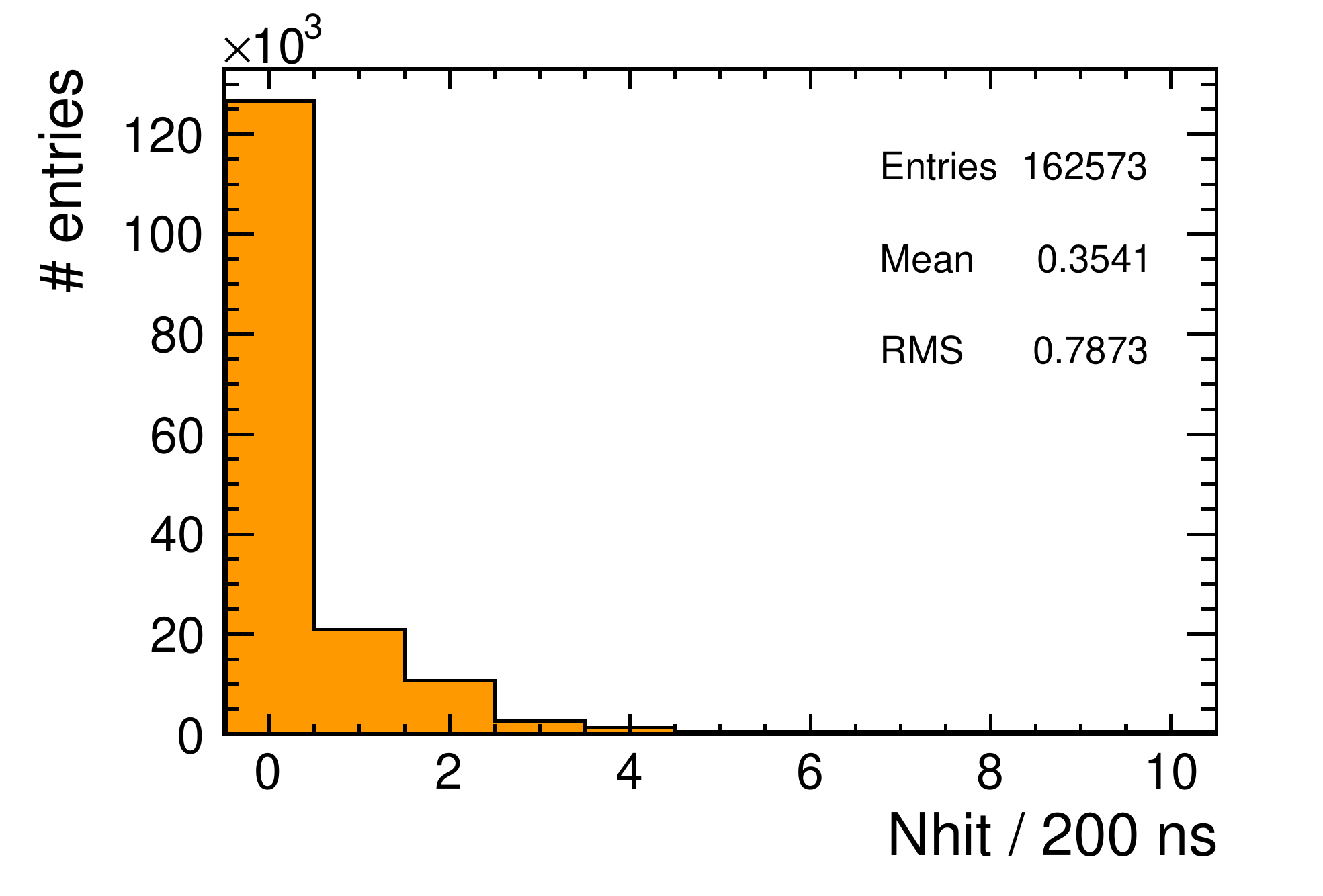}
\caption{
Top: Number of hits in physics events (green) and number of noise hits in the CALICE SDHCAL as function of time in units of clock cycles of the HARDROC ASIC. Bottom:  Total number of noise hits within 200\,ns. From~\cite{bib:can-037,Boudry:2012}.
}
\label{fig:sdhcal-noise}
\end{center}
\end{figure}

The SDHCAL, featuring three readout thresholds encoded in 2 bits per cell instead of only one, has been subject to a large scale test beam campaign in 2012. The current main results of the performance are summarised in~\cite{bib:can-037}. No notable problems are reported from this running period. The minor shortcomings concern coherent noise and a small number of dead ASICs.
At first sight, the SDHCAL is quite similar to the DHCAL. However, the detector readout is significantly different. The HARDROC ASIC used in the SDHCAL is already adapted to the
expected mode of operation at the future ILC. Operated in self-triggering mode it stores up to 128 events in an internal buffer. The entire detector has to be read out if the buffer of any of the ASICs is full. This requires an excellent control of the noise level of the detector. Fig.~\ref{fig:sdhcal-noise} shows in its top part the hit spectrum  as a function of elapsed time. The number of hits attributed to physics events is significantly different from the number for noise events. The subdivision of the $x$ axis into 200\,ns bins is motivated by the clock frequency of the HARDROC ASIC. The bottom part shows the number of noise hits within 200\,ns. This spectrum has its maximum number of entries at zero hits and the resulting average is 0.35 hits/200\,ns. 

\begin{figure}[htp]
\begin{center}
\includegraphics[width=.3\textwidth]{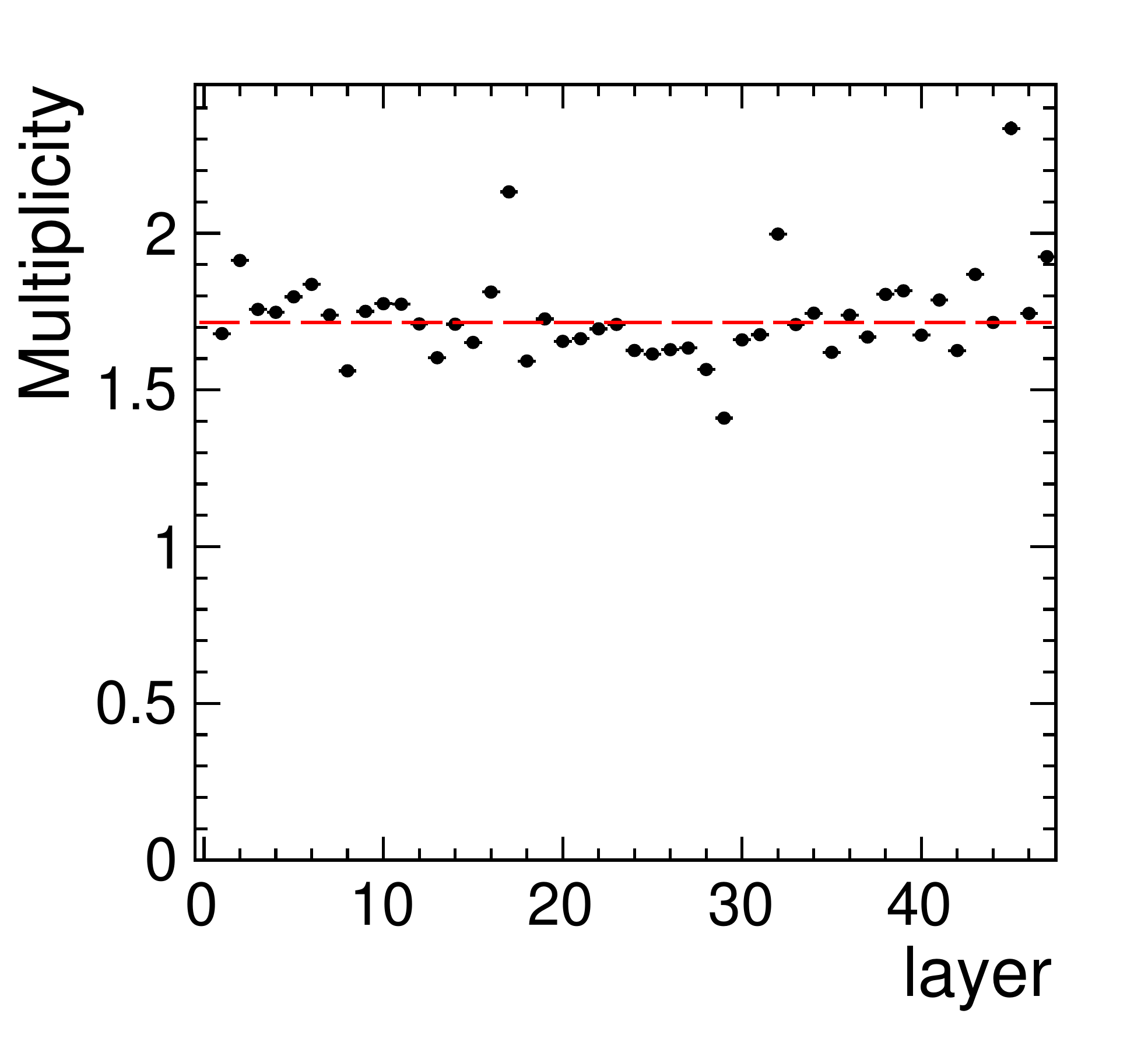}
\includegraphics[width=.3\textwidth]{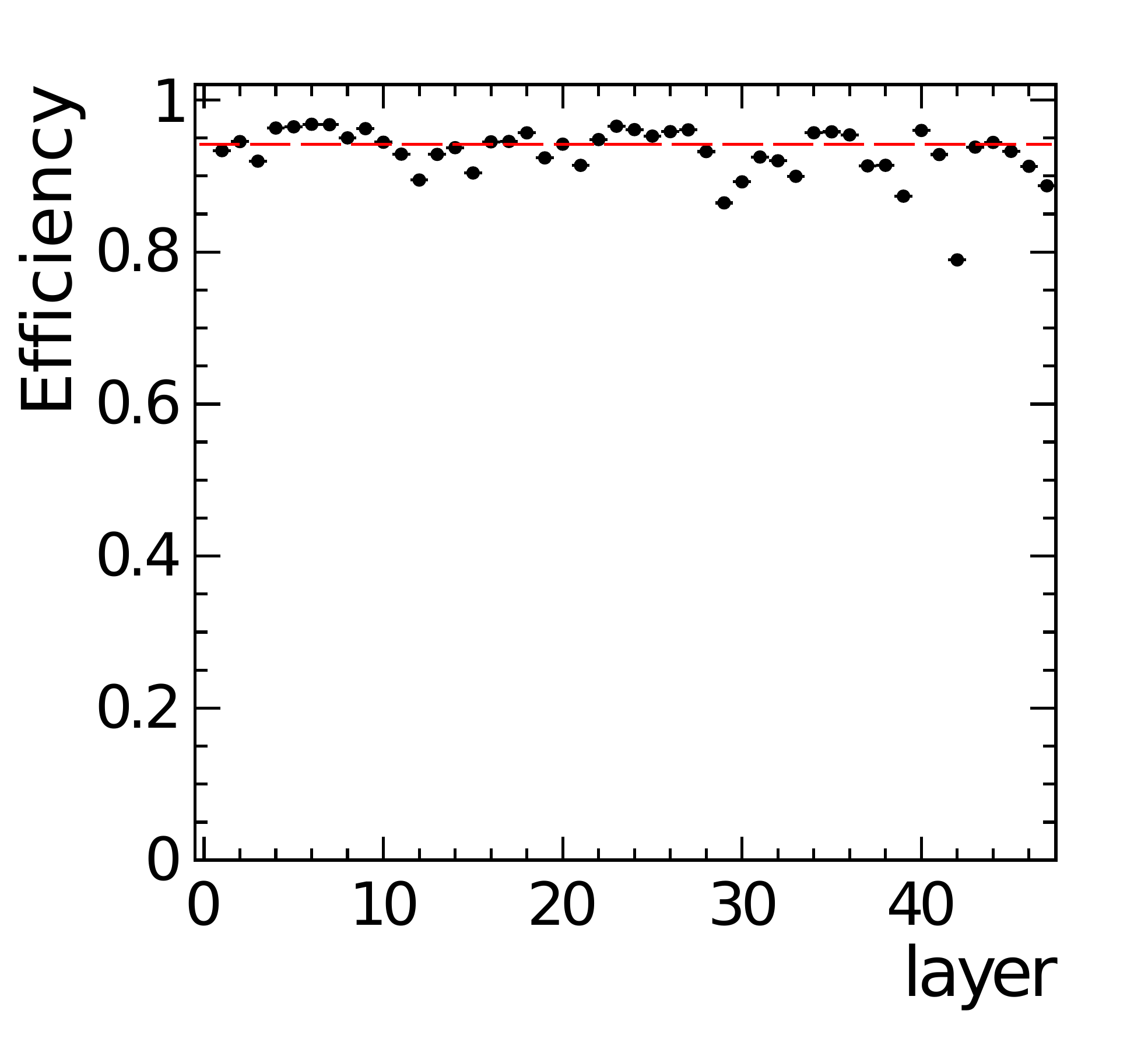}
\caption{Top: Average hit multiplicity in the CALICE SDHCAL chambers as a function of the layer. Bottom: Average efficiency of the CALICE SDHCAL chambers as a function of the layer. From~\cite{bib:can-037,Boudry:2012}.}
\label{fig:eff-multi-per-layer}
\end{center}
\end{figure}

The hit efficiency and the hit multiplicity are important benchmarks for the performance of a gaseous calorimeter.  As shown in Fig.~\ref{fig:eff-multi-per-layer} for the CALICE SDHCAL an efficiency of about 95\% and a hit multiplicity of about 1.7 hits have been measured with muons crossing the detector. These numbers are independent of the layer. It is interesting to note that these numbers have been confirmed with track segments within hadronic showers.  
They depend on environmental conditions in a similar way as for the DHCAL. 

\begin{figure}[htp]
\begin{center}
\includegraphics[width=.35\textwidth]{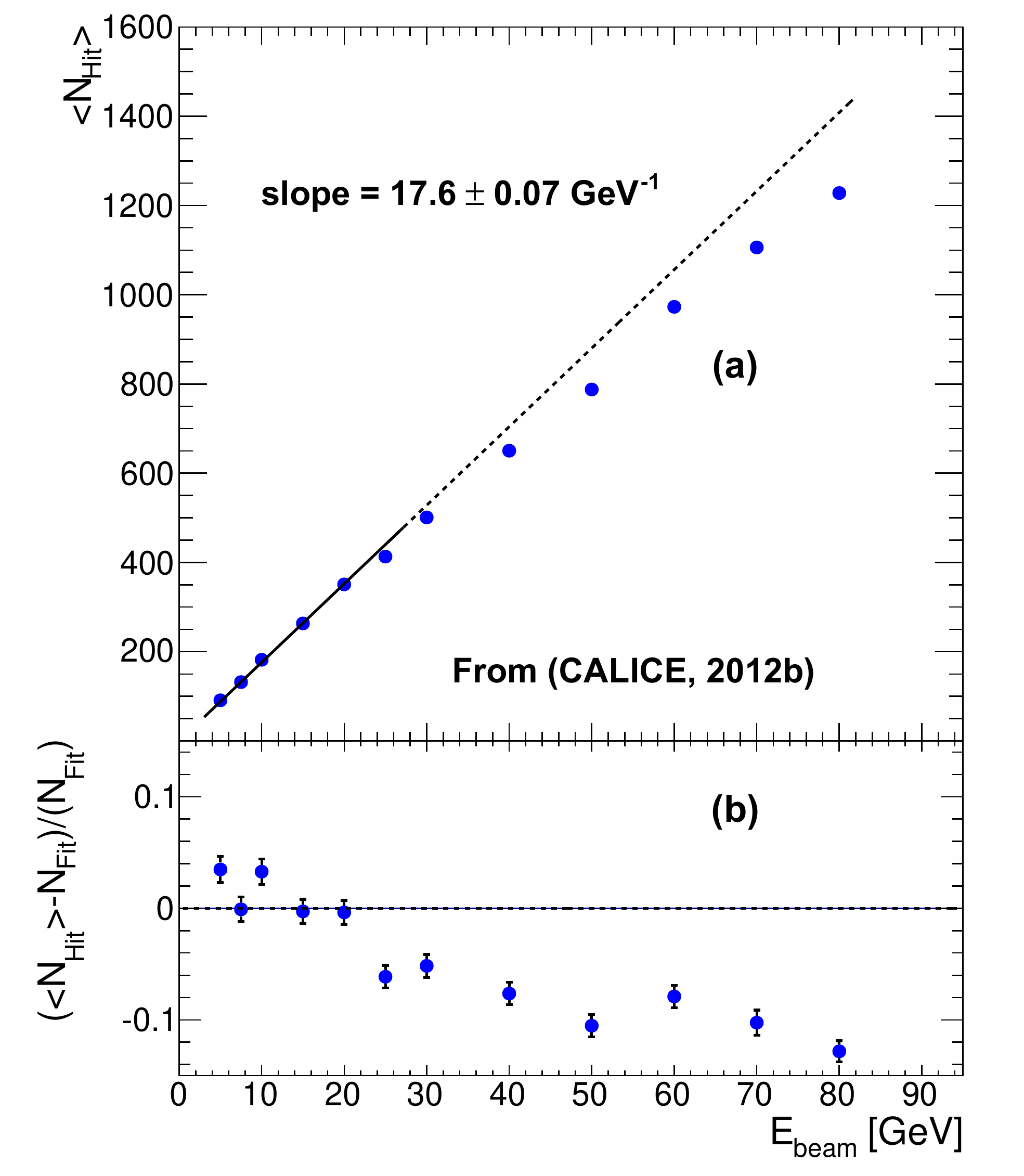}
\includegraphics[width=.35\textwidth]{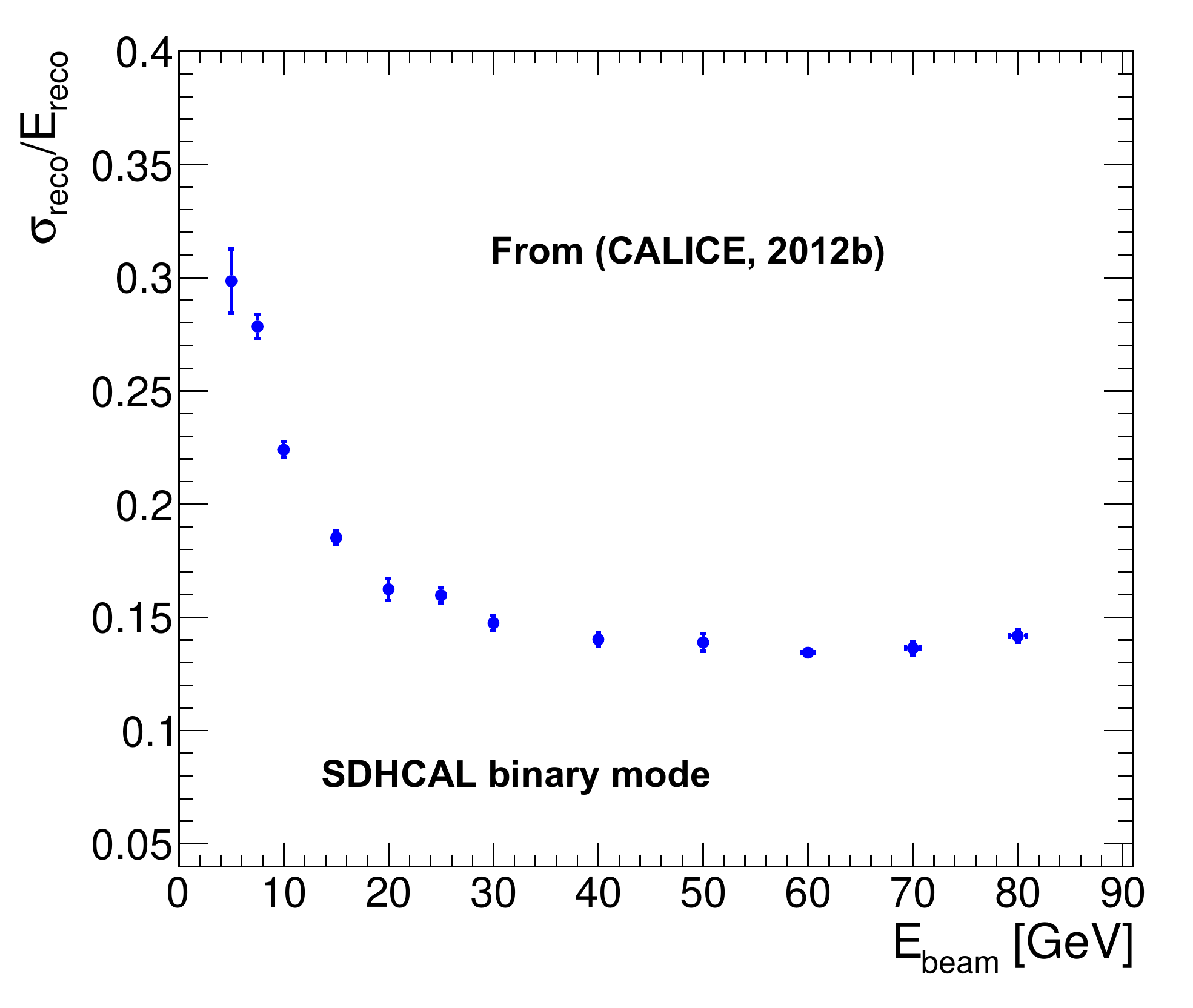}
\caption{Top: Raw response of the CALICE SDHCAL as a function of the energy of primary pions. Bottom: Energy resolution obtained after correction for the non linearity as explained in the text. From~\cite{bib:can-037,Boudry:2012}.}
\label{fig:sdhcal-linres-bin}
\end{center}
\end{figure}

An important cross check of the performance of the CALICE SDHCAL are the results obtained in pure binary mode. This will allow for judging the benefit of having additional thresholds and for comparisons with the CALICE DHCAL. 
The raw response of the calorimeter is given in the top part of  Fig.~\ref{fig:sdhcal-linres-bin}. Saturation effects set in for an energy of the primary pion of about 30\,GeV. To account for this non-linearity the reconstructed energy is calculated as a function of the number of hits. This function takes the form $E=(C+D\cdot N_{hit}) \cdot N_{hit} $, with $C=0.0543$ and $D=9\cdot 10^{-6}$. By construction the reconstructed energy is linear as a function of the energy of the primary particle. The resulting energy resolution is shown in the bottom part of Fig.~\ref{fig:sdhcal-linres-bin}. The shape of the resolution curve agrees with the qualitative expectation. It flattens out at energies above 30\,GeV and approaches a value of about 15\%. 
Both observations are broadly compatible with the results obtained for the DHCAL according to Fig.~\ref{fig:DHCAL_Fe}. Understanding the different contributions to the resolution in terms of simulations, and of the differences between the prototypes, is still ongoing.  
%

\begin{figure}[htp]
\begin{center}
\includegraphics[width=0.5\textwidth]{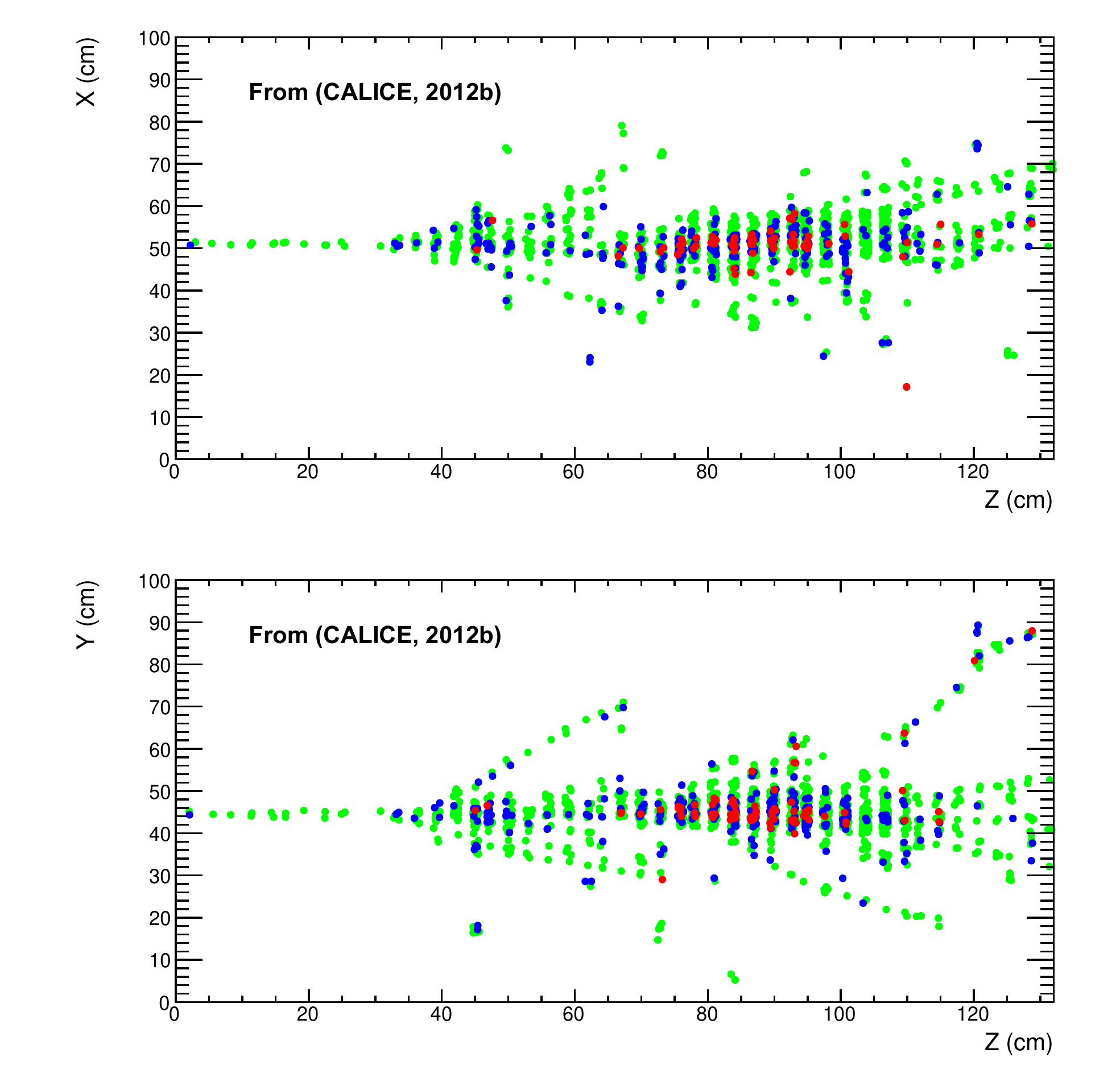}
\caption{Interaction of a pion with an energy of 80~GeV in the CALICE SDHCAL. The top part shows the $xz$ projection of the hit positions and the lower part the $yz$ projection, respectively. Red coloured hits indicate hits in which a charge above the highest of three thresholds has been measured.  Blue indicates hits with a charge deposition between the second and the third thresholds and green those between the first two thresholds. From~\cite{bib:can-037,Boudry:2012}.}
\label{fig:event_display_thresholds_pion}
\end{center}
\end{figure}

By virtue of the HARDROC ASIC the SDHCAL implements three readout thresholds in order to moderate saturation effects in dense shower regions. A natural question is whether
these thresholds allow for a true (even still rough) measurement of the different energy depositions. This question is even more justified since the RPC chambers are operated in avalanche mode, and the measured charge is only weakly correlated with the primary deposited energy in a given cell. 
Fig.~\ref{fig:event_display_thresholds_pion} shows an 80\,GeV pion that interacts in the SDHCAL. The incoming particle acts MIP like in early stages of the shower before the interaction. The core of the interaction is populated with red coloured entries that represent high energy deposition. These regions of high energy depositions are surrounded by hits associated with smaller energy depositions. Thus, the expected rough features of highly energetic hadronic showers are reflected by the multi-threshold read-out. Quantitative results using the three thresholds will be presented in Sec.~\ref{sec:sd-rec}. 

\subsection{Tests with alternative technologies}

Additional alternative technologies have been suggested for implementing
particle flow calorimetry. They have not been developed to the extent of
those already described, but are included here for completeness as they
may be used in actual future detector systems.



For the ECAL, a highly granular solution has been proposed that uses Monolithic Active Pixel Sensor
technology. Three first-generation sensors for digital electromagnetic calorimetry 
consist of 168 x 168 pixels with the required size of 50 microns by 50 microns. 
These sensors were tested~\cite{Stanitzki:2011zz}
using sources and lasers, and 
a beam at CERN. 
The addition of a deep p-well feature and a high resistivity epitaxial layer made the sensor close to 100\%
efficient for MIPs.

\begin{figure}[htb]
\centering
       \includegraphics[width=0.7\hsize]{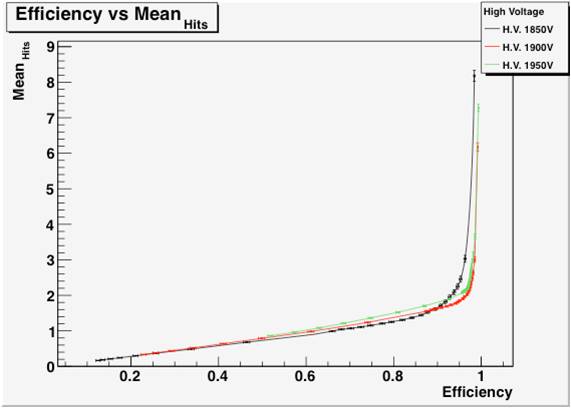}
	   
\caption{ \label{fig:GEM_hit_eff} The number of anode pads above threshold versus GEM chamber efficiency. 
From~\cite{White:2012kpa}.}
\end{figure}

For the GEM-based DHCAL approach, beam tests have been made with several 30 x 30 cm$^2$ double-GEM foil
chambers, with 1 cm$^2$ anode pads at Fermilab. Both the KPiX and Argonne DCAL readout systems
were tested. 
The important results~\cite{White:2012kpa} of these tests were that the
chambers operated stably in both beam types, and a hit multiplicity vs. efficiency curve that shows that
this technology is a very good candidate for use in a DHCAL - see Fig.~\ref{fig:GEM_hit_eff}.

\begin{figure}[htb]
\centering
       \includegraphics[width=0.7\hsize]{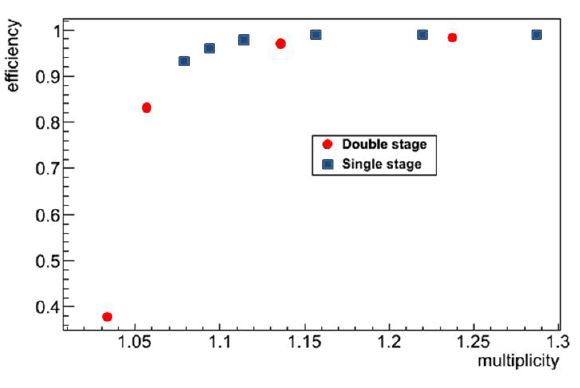}
	   
\caption{ \label{fig:THGEM_hit_eff} Chamber efficiency vs. hit multiplicity for thick-GEM
prototypes. From~\cite{Arazi:2013ldf}.}
\end{figure}

For the thick-GEM option, single and double resistive well chambers, 10 x 10 cm$^2$ were
exposed to a muon beam at CERN. The resulting~\cite{Arazi:2013ldf} efficiency vs. hit multiplicity curves
are shown in  Fig.~\ref{fig:THGEM_hit_eff}, showing that low multiplicity at high efficiency 
was obtained - again indicating the suitability of this approach for use in a DHCAL.

\begin{figure}[htb]
\centering
       \includegraphics[width=0.7\hsize]{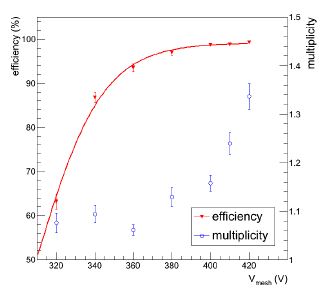}
	   
\caption{ \label{fig:Micromegas_hit_eff} The number of anode pads above threshold versus micromegas chamber efficiency.}
\end{figure}

Tests have been made with 1 m$^2$ micromegas chambers in beam at CERN, with a goal of using
this technology for a semi-digital HCAL.
The limited spatial dimensions of the signal from the
avalanche in the micromegas results in a low hit multiplicity - about 1.15, indicating the
suitability of this technology also for DHCAL applications. Results on the hit multiplicity 
vs. efficiency~\cite{Chefdeville:2012} are shown in Fig.~\ref{fig:Micromegas_hit_eff}.

In later tests, four 1 m$^2$ micromegas planes were installed in a RPC SDHCAL stack and
data recorded. The distribution of numbers of hits in the planes vs. the position of the
planes in the stack was found to be in good agreement with expectations.
\section{Weighting, software compensation and combined performance%
\label{sec:Weighting}}


The detector performance for single hadrons and for jets depends on the combined resolution -- for energy and topology -- of ECAL and HCAL together, possibly complemented by a tail catcher / muon tracker (TCMT). 
It is a particular strength of the CALICE validation approach to take an integral view of the calorimeter system and conceive the beam tests as combined set-ups from the beginning. 
The combination of energy measurements is discussed here, while the next section is devoted to the combined particle flow performance. 

The imaging power of the calorimeters offers additional means to improve the particle energy measurements. 
The fine segmentation provides ideal conditions for the application of software compensation methods, based on local or global energy or hit density.
The energy reconstruction of a semi-digital calorimeter makes also use of weighted combinations of the hit multiplicities for each of the multiple thresholds.
Finally, leakage can be estimated on the basis of topological observables such as the shower start point or activity in the rear part of the HCAL. 

\subsection{Software compensation in the AHCAL}

For most calorimeters, the response, i.e.\ the ratio of the signal generated  in the active part to the total deposited energy, is different for electromagnetic and for hadronic showers, as was discussed in Section~\ref{sec:Intro}. 
Therefore event-to-event-fluctuations in the electromagnetic content of the shower contribute to the hadronic energy resolution, unless the difference is compensated. 
This is done either by design, by using active and passive materials in an optimised sampling ratio, such that the electromagnetic response is suppressed and the hadronic is enhanced, or by so-called {\it software compensation}. 
The latter method exploits the fact that for the heavy absorber materials of hadron calorimeters, the radiation length is much smaller than the nuclear interaction length, such that the electromagnetic parts of the shower are much more compact and have a higher density. 
Therefore compensation can be achieved at the reconstruction stage if energy deposits in regions of high energy density are given a lower weight than those in less dense regions. 
This requires three-dimensional segmentation of the read-out.
	One of the first applications was in the CDHS experiment~\cite{Abramowicz:1980iv}, 
it was further developed and applied for the H1~\cite{Andrieu:1993tz,Issever:2004qh} and the
ATLAS calorimeter~\cite{Cojocaru:2004jk}.

Due to the fine 3D segmentation,  particle flow calorimeters offer excellent conditions for the successful applications of such methods. 
A local (LC) and a global (GC) compensation method have been studied~\cite{Adloff:2012gv} using data with steel absorber and the same event selection as described in Section~\ref{sec:Perf:AHCAL}.
The first method uses the local energy density in the shower to determine weights for each hit, 
while the second applies a global correction factor to the whole reconstructed energy, 
which is based on the hit energy distribution in the shower. 
The weights and parameters for these methods have been obtained from data and then been used for independent data samples as well as for simulations. 

\begin{figure}[htb]
  \includegraphics[width=0.7\hsize]{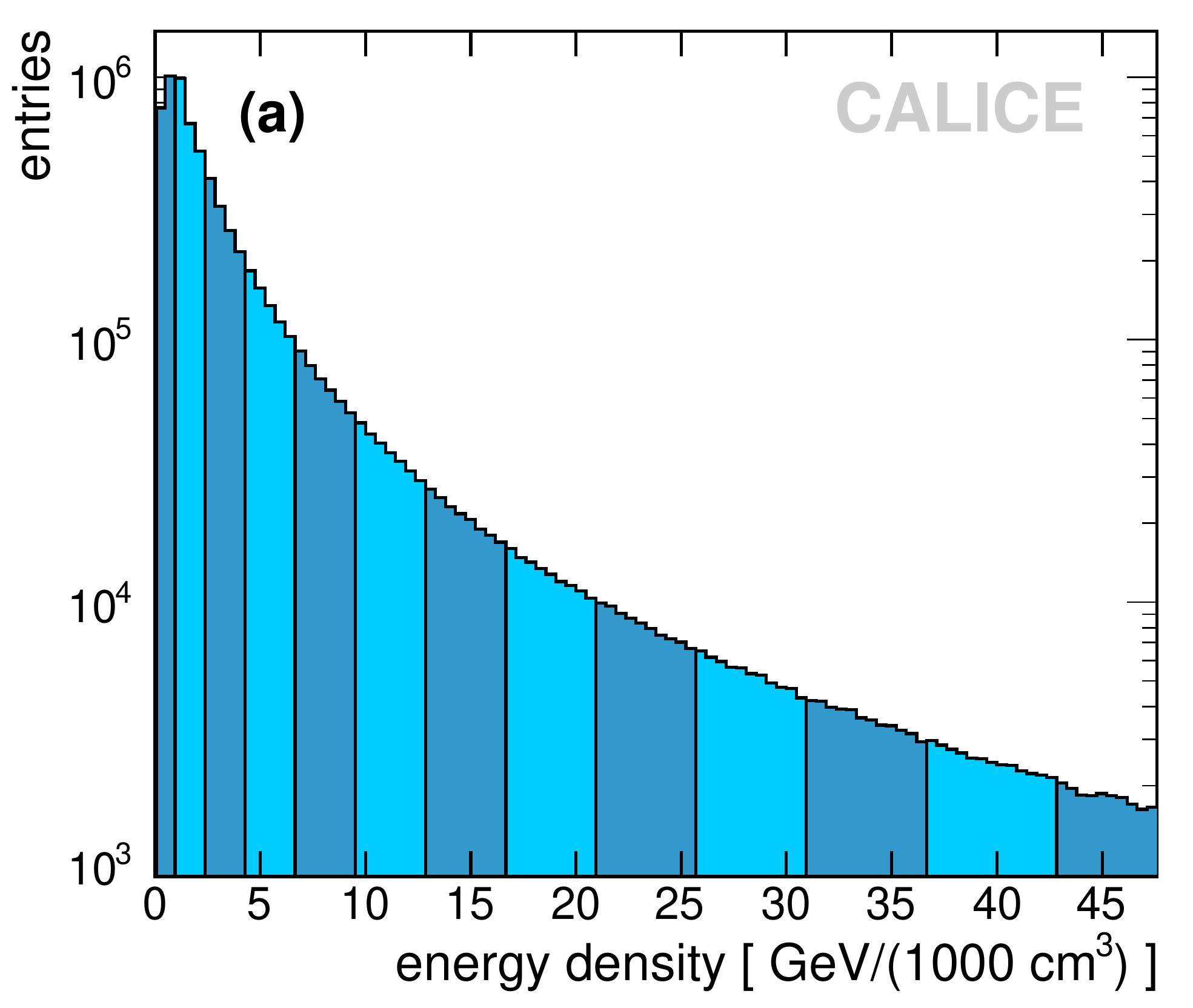}
  \includegraphics[width=0.7\hsize]{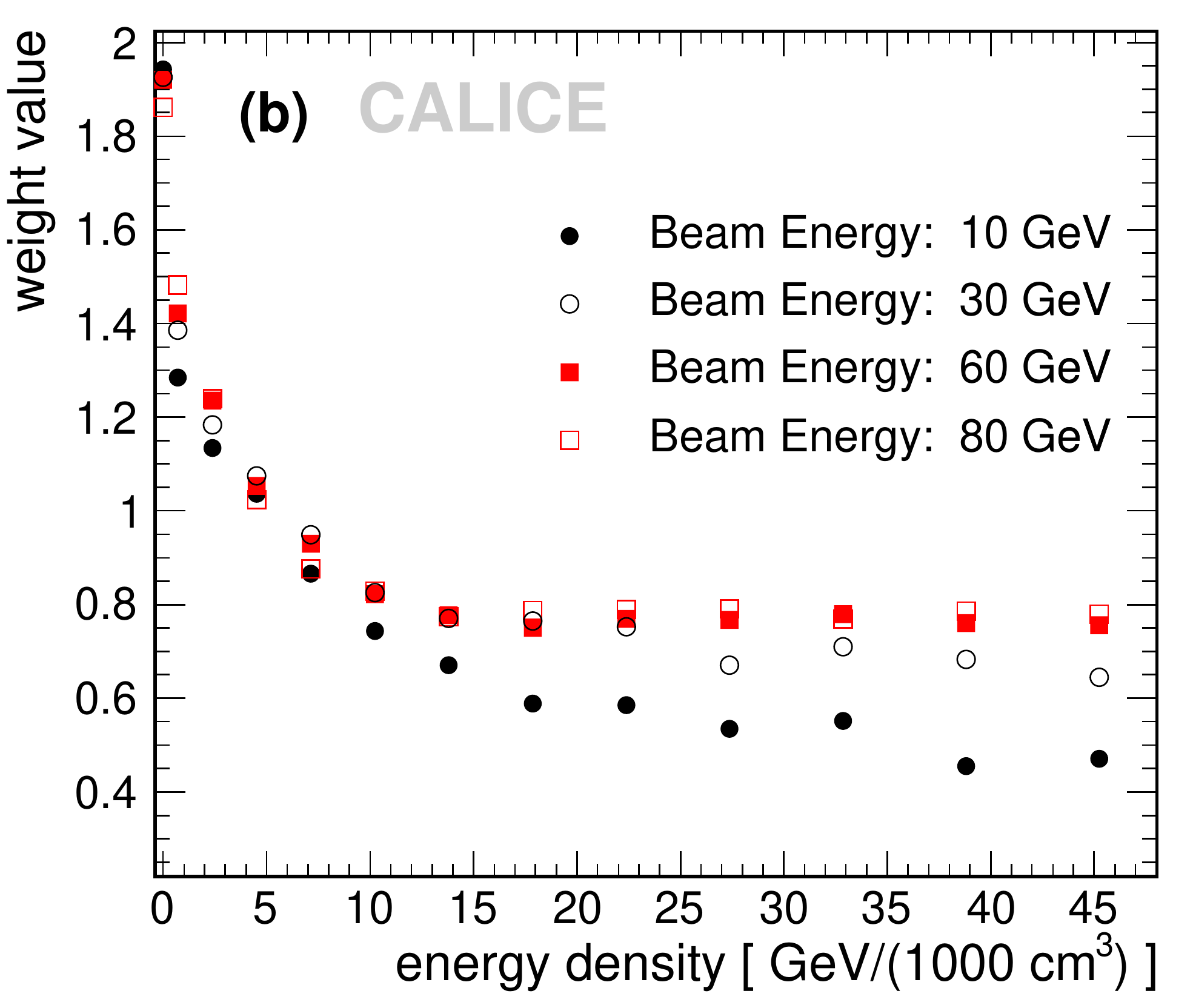}
\caption{\label{fig:Weight:localcomp} Hit energy distribution in the AHCAL for 40~GeV pion shower test beam data (top), optimal weights derived from data, as a function of energy density and particle energy (bottom). From~\cite{Adloff:2012gv}.}
\end{figure}

For the LC method, the weighted energy is reconstructed according to 
\begin{equation}
 \label{eq:weight:local-e-sum}
  E_{\mathrm{LC}} = E_{\mathrm{ECAL}}^{\mathrm{track}} + \frac{e}{\pi} \cdot  \left( \sum_{i} \left( E_{\mathrm{HCAL},i} \cdot \omega_{i} \right) + E_{\mathrm{TCMT}}\right)
\end{equation}
-- see also Equation~\ref{eq:Perf:ehad}. Here in addition $\omega_{i}$ is introduced, the energy density dependent weight applied to the cell energy $E_{\mathrm{HCAL},i}$.
The density is here simply the cell energy divided by the cell size. 
Fig.~\ref{fig:Weight:localcomp} shows the energy density distribution for 40~GeV pion showers. 
The distribution is binned as indicated, and in each bin the weights are determined such that the energy resolution is optimised, by minimising  $\chi^{2} = \sum_i (E_{\mathrm{LC},i} - E_{\mathrm{beam}})^2$. 
The results of this procedure are also shown in Fig.~\ref{fig:Weight:localcomp} for data.
The weights depend not only on the hit energy, but also on the total particle energy. 
For interpolation between the bins, this energy dependence is parameterised. 
Since the true energy is not known {\it a priori} for data,  the unweighted energy is used instead for the determination of weights. 
It was found that no further iteration is necessary. 

In the global method, a correction factor $C_{\mathrm global}$ is calculated for each shower from the ratio of the number of hits below a given threshold, and the number of hits below the mean hit energy for this shower. 
This factor is smaller for events with a larger electromagnetic component and thus fewer low energy hits. 
In practice, the correction proceeds in two steps, where first the global correction is applied to improve the estimate of the energy, 
 $E_{\mathrm{shower}} = C_{\mathrm{global}} \cdot \left( E_{\mathrm{HCAL}} + E_{\mathrm{TCMT}} \right)$,
and second an energy-dependent correction function $P_{\mathrm{global}}$ restores linearity:
\begin{equation}
 \label{eq:weight:global-e-sum}
E_{\mathrm{GC}} = E_{\mathrm{ECAL}}^{\mathrm{track}} + E_{\mathrm{shower}} \cdot P_{\mathrm{global}}(E_{\mathrm{shower}})\; .
\end{equation}

\begin{figure}[htb]
 \includegraphics[width=0.7\hsize]{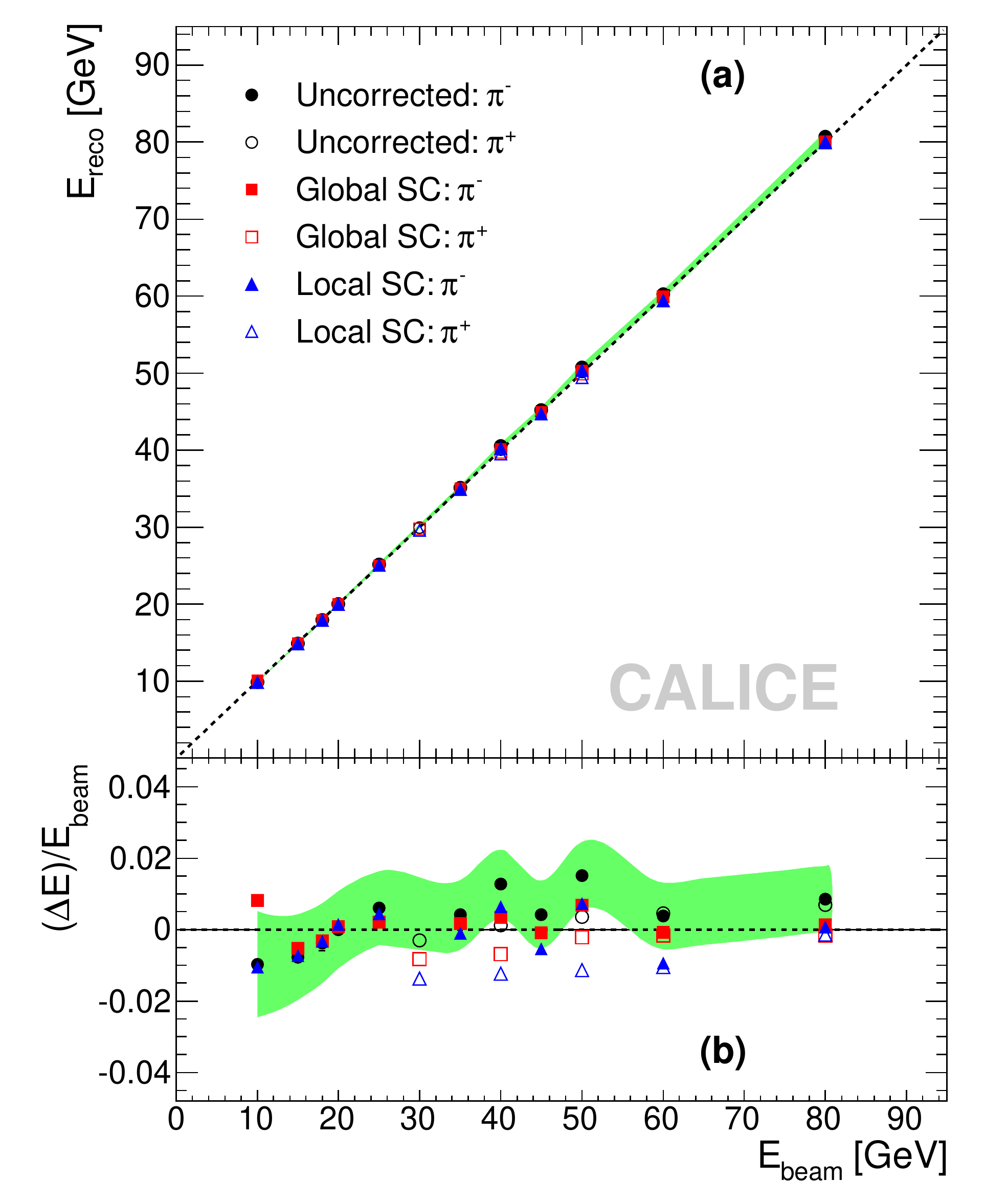}
\includegraphics[width=0.7\hsize]{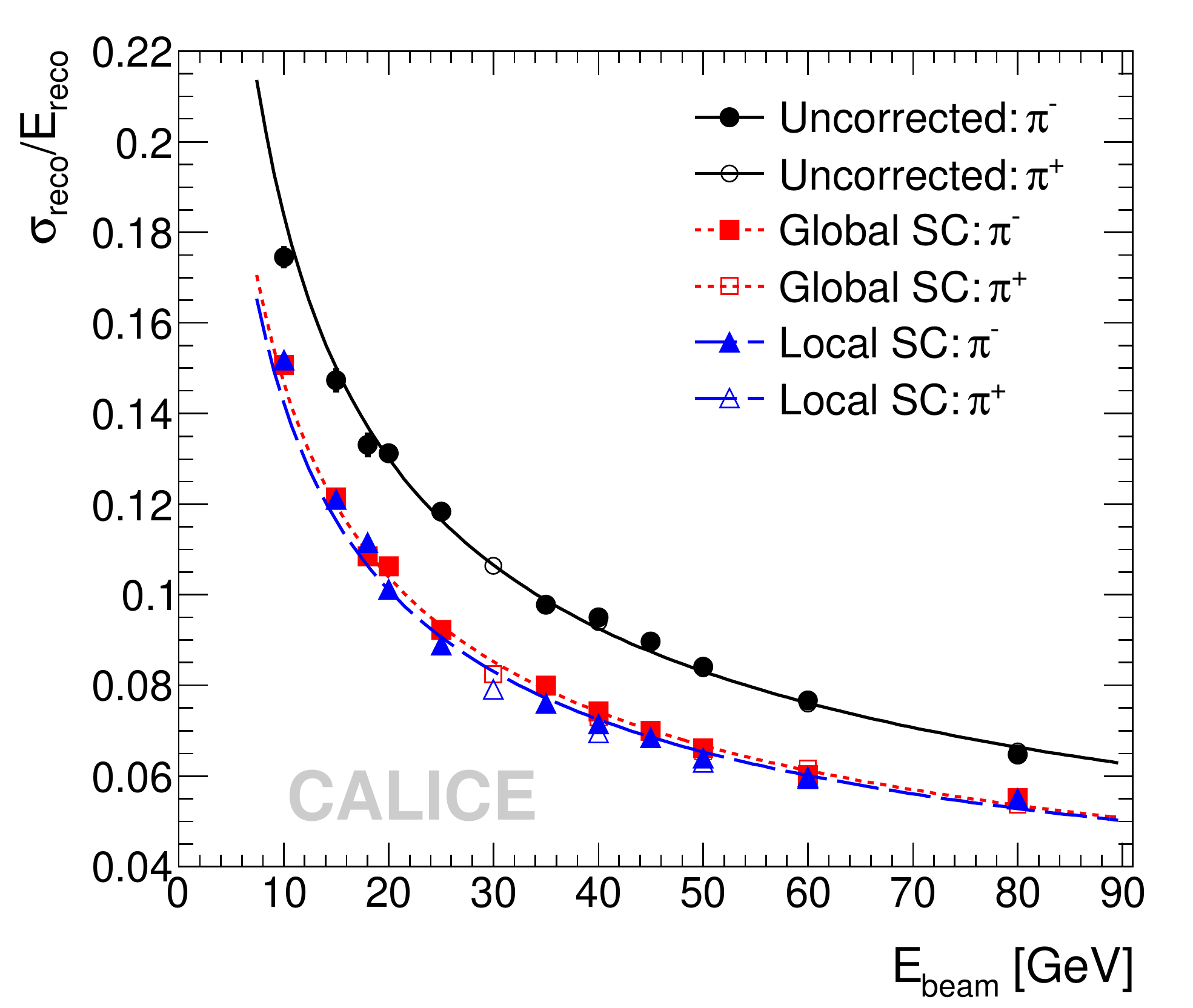}
\caption{\label{fig:Weight:wresol} Linearity and resolution for unweighted AHCAL data, and for the two software compensation methods. From~\cite{Adloff:2012gv}.}
\end{figure}

The results of applying these procedures, after training them on data, to independent data sets, are shown 
in Fig.~\ref{fig:Weight:wresol}.
For both compensation methods, the linearity of the AHCAL is preserved or improved, and better than $\pm 1.5\%$. 
The resolution is improved by $12-15\%$, depending on energy. 
Numerical results of fits using Eq.~\ref{eq:Perf:resolfit} are shown in Table~\ref{tab:Weight:compresol}.
\begin{table}[htb]
\begin{ruledtabular}
\begin{tabular}{lccc}
& a [\%] & b [\%] & c [GeV]   \\
\colrule
uncorrected (Eq.~\ref{eq:Perf:ehad})& 57.6$\pm$0.4 & 1.6$\pm$0.3 & 0.18  \\
local compensation (Eq.~\ref{eq:weight:local-e-sum}) &  44.3$\pm$0.3  & 1.8$\pm$0.2 & 0.18 \\
global compensation (Eq.~\ref{eq:weight:global-e-sum}) &  45.8$\pm$0.3  & 1.6$\pm$0.2 & 0.18  \\
 \end{tabular}
\end{ruledtabular}
\caption{\label{tab:Weight:compresol} Stochastic, constant and noise term contributions to the resolution of the CALICE AHCAL determined with a fit of Equation~\ref{eq:Perf:resolfit} to data. From~\cite{Adloff:2012gv}.}
\end{table}
A stochastic term around 45\% and a constant term below 2\% are achieved, which is also reasonably well reproduced in simulations. 
The results show that a particle flow calorimeter with high granularity can be realised which still has a very good purely calorimetric performance. 
Software compensation is still to be integrated into particle flow algorithm, where some improvement for the jet energy performance can be expected, in particular at the lower end of the linear collider jet energy range, where the intrinsic HCAL resolution dominates. 
In addition, it can help to sharpen the calorimeter cluster association to charged particle tracks; here one can use the measured momentum for the proper choice of weights. 
%

\subsection{Semi-digital reconstruction} \label{sec:sd-rec}
As was shown in Section~\ref{sec:Perf} the spatial distribution of hits above the three thresholds in the RPC-SDHCAL meets the intuitive expectation. 
The reconstruction now seeks to exploit the additional information in order to improve linearity and resolution of the energy response. In~\cite{bib:can-037,bib:can-037a} 
the reconstructed energy $E_{rec}$ is expressed as a function of the number of hits $N_i,\,i=1..3$ above the three thresholds. In this approach $E_{rec}$ is given by
\begin{equation}
E_{rec} = \alpha N_1 + \beta N_2 + \gamma N_3
\end{equation}

\begin{figure}[!h]
\begin{center}
\includegraphics[width=0.45\textwidth]{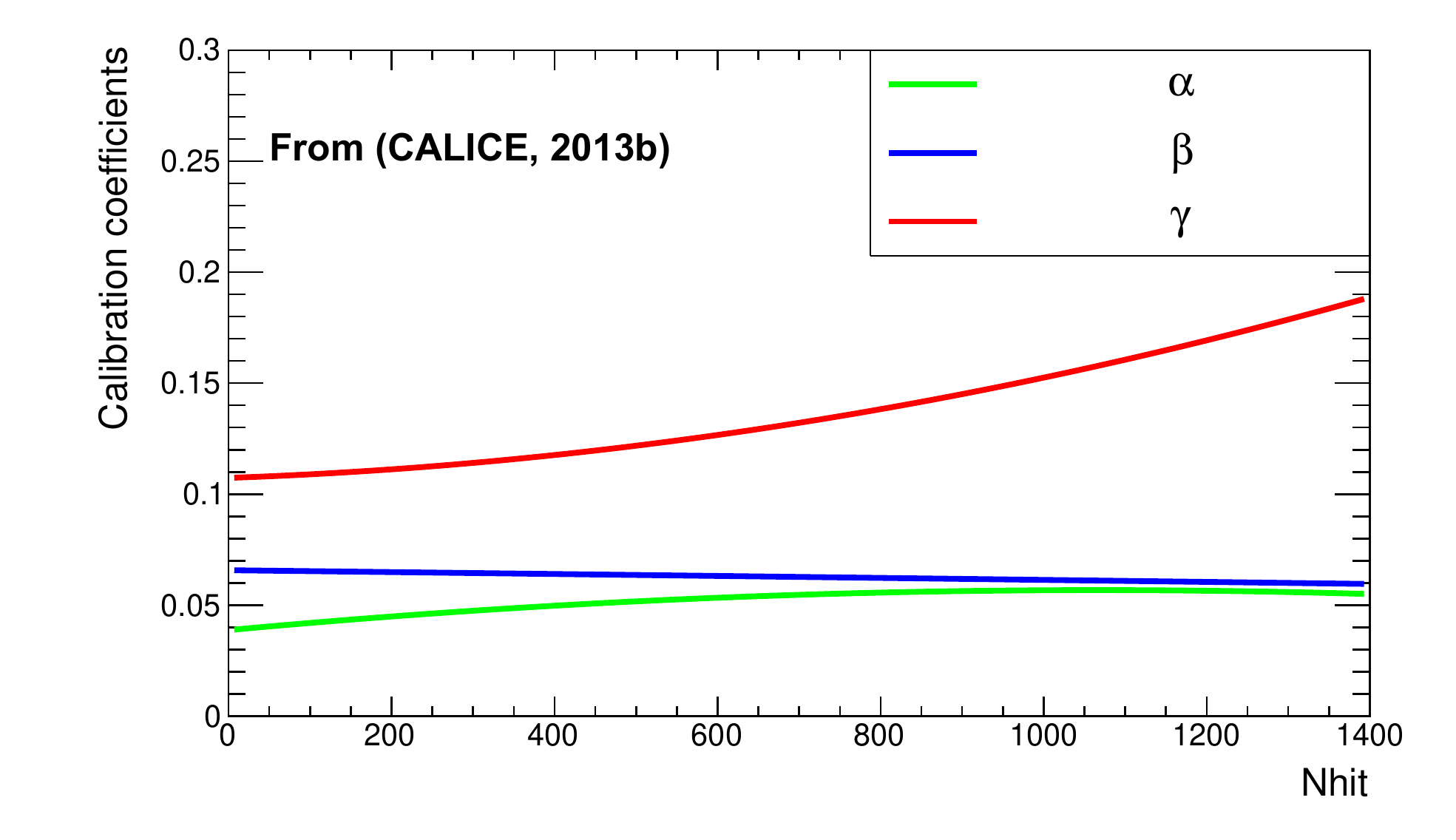}
\caption{Evolution of the coefficient $\alpha$ (green), $\beta$ (blue) and $\gamma$ (red) as a function of the total number of hits. 
From~\cite{bib:can-037a,Steen:2014ufa}.}
\label{fig:evolution}
\end{center}
\end{figure}

The coefficients $\alpha, \beta, \gamma$ are a function of the total number of hits $N_{hit}=N_1 + N_2 + N_3$. The functional dependence is derived from the minimisation of a $\chi^2$ test variable that uses a subset of the data at each of the energy points of $10, 20, 30, 40, 50, 60$\,GeV. According to~\cite{bib:can-037} the best results are obtained by a 2nd order polynomial. The coefficients as a function $N_{hit}$ are shown in Fig.~\ref{fig:evolution}. The parabolic dependency is most clearly visible for the coefficient $\gamma$ associated with the highest readout threshold. 
Optimising the weight for each threshold without strict relation to the cell energies, and letting the weights vary with total number of hits -- or energy -- contains some elements of software compensation, e.g.\  as applied in the local method for the AHCAL. 
In this context it is interesting to note that due to the parabolic dependency the relative importance of the highest threshold increases with an increasing number of hits that is in first order proportional to the deposited energy.

\begin{figure}[!ht]
\begin{center}
\includegraphics[width=0.23\textwidth]{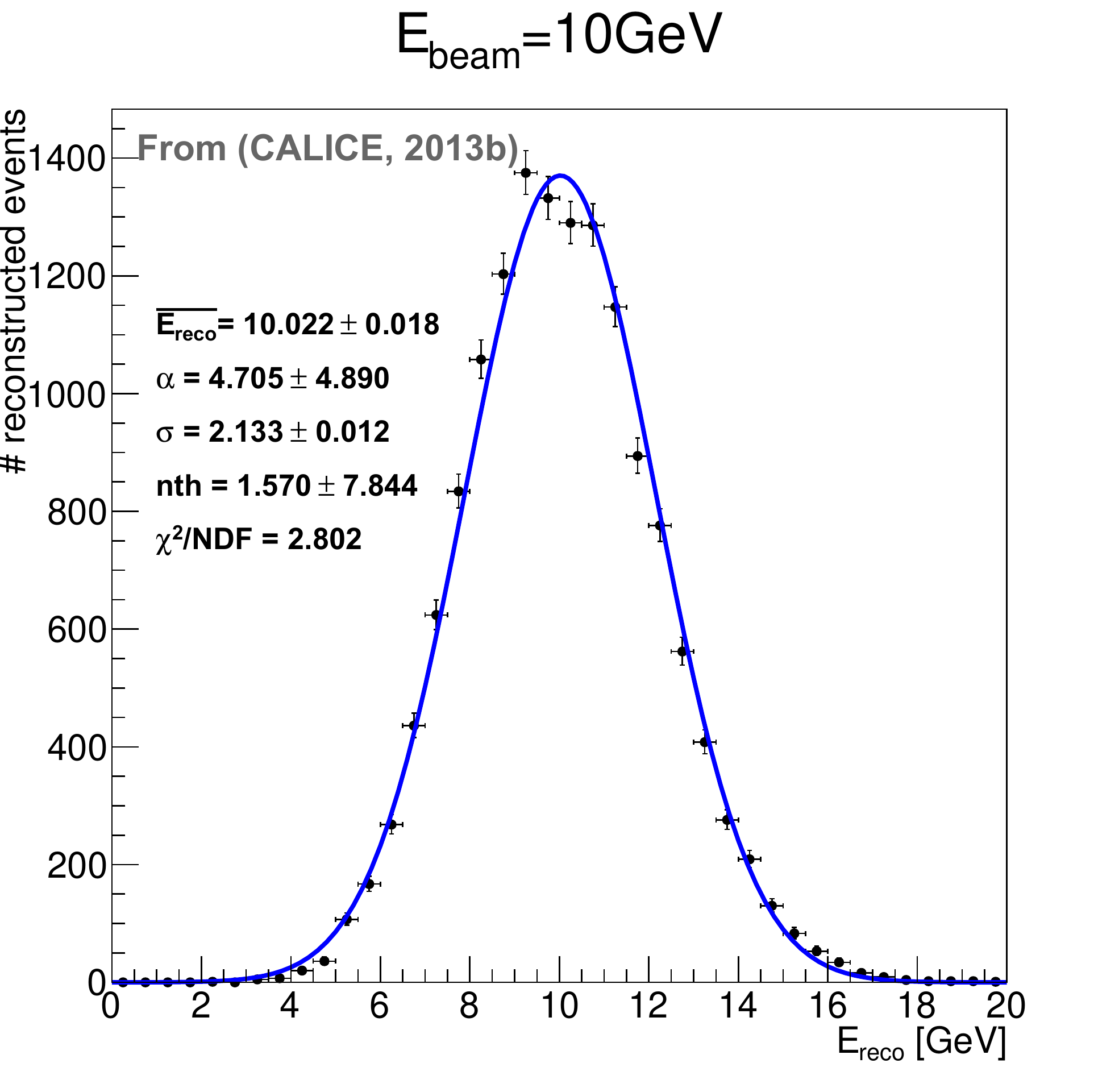}
\includegraphics[width=0.23\textwidth]{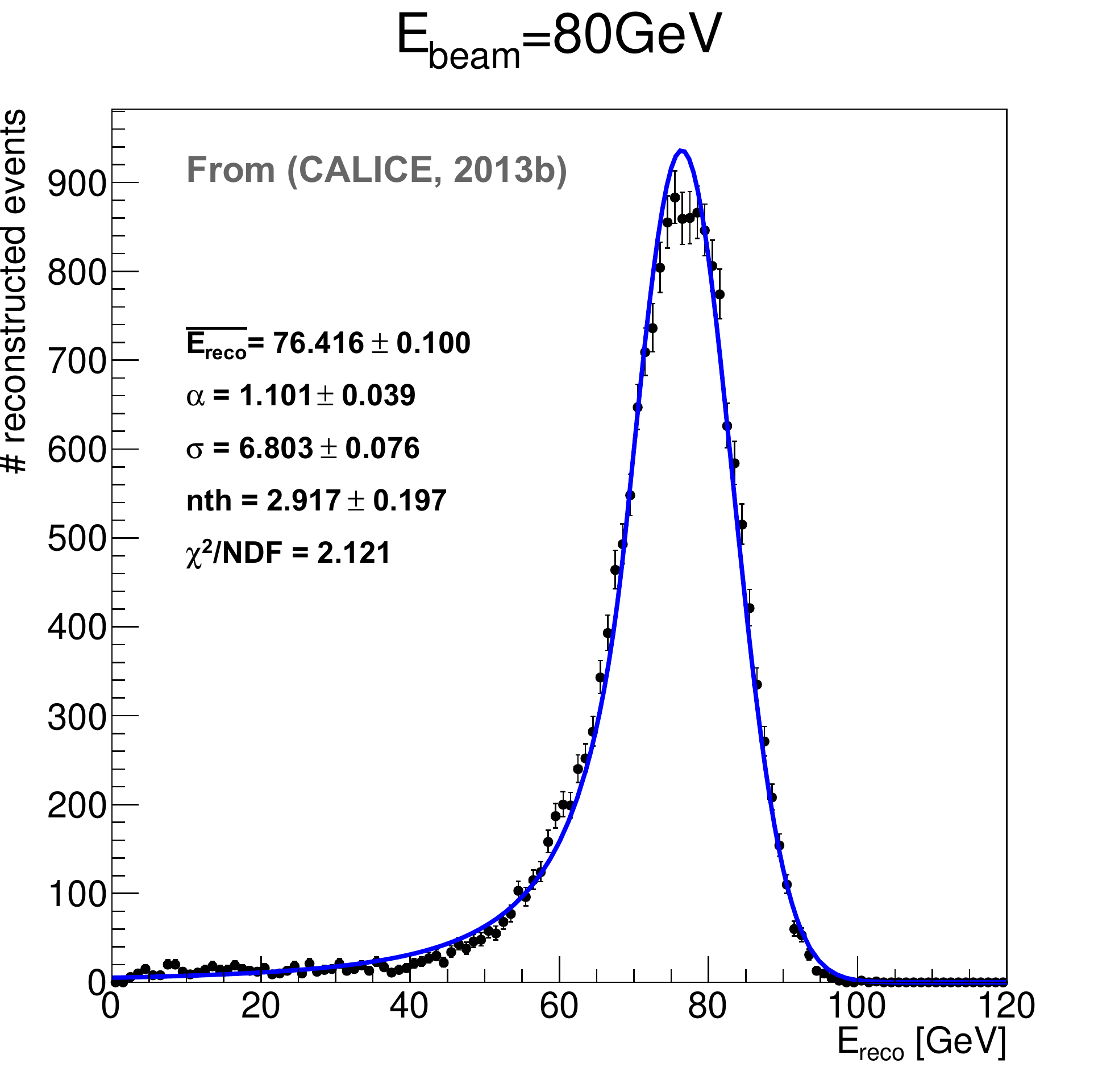}
\\
\caption{Energy reconstruction with the RPC-SDHCAL for pion showers of 10 and 80 GeV (right) using the threshold information. The Crystal Ball function is fitted to the distributions.
From~\cite{bib:can-037a,Steen:2014ufa}.}
\label{fig:enereco}
\end{center}
\end{figure}

Using these coefficients the response of the detector for the example of pions with energies of 10\,GeV and 80\,GeV is demonstrated in Fig.~\ref{fig:enereco}. At small energies the response is Gaussian like. At higher energies a low energy tail reflects the loss of information due to saturation effects. The distributions are fitted with the Crystal Ball function~\cite{Gaiser:1982yw} that can take this information loss adequately into account. The mean of this Crystal Ball function is close to the nominal beam energy. 

\begin{figure}[!ht]
\begin{center}
\includegraphics[width=0.4\textwidth]{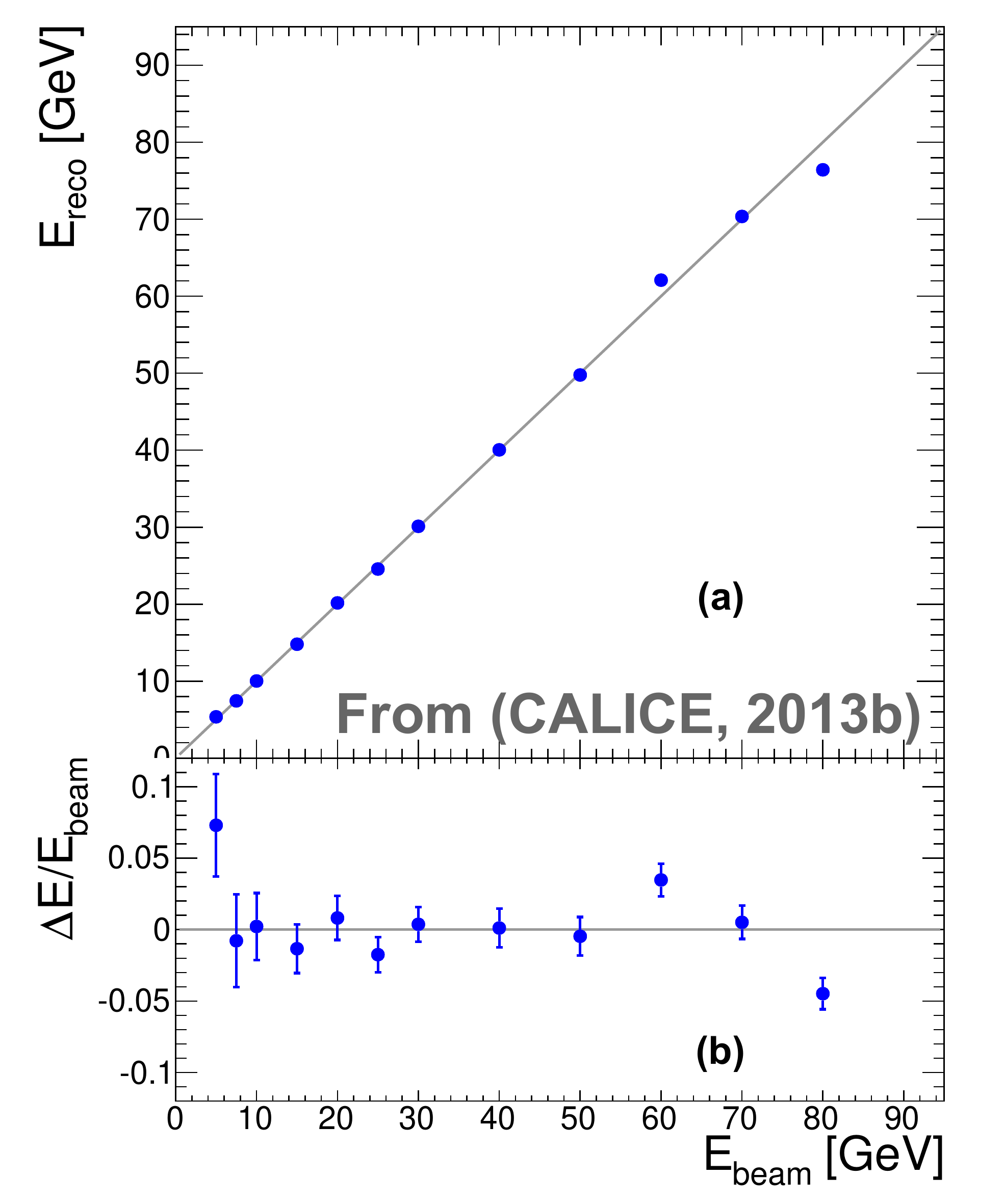}
\caption{(a): Mean reconstructed energy $E_{reco}$ in the RPC-SDHCAL for pion showers and (b): relative deviation $\Delta E/E_{beam}=(E_{reco}-E_{beam})/E_{beam}$ of the pion mean reconstructed energy from the beam energy $E_{beam}$.
From~\cite{bib:can-037a,Steen:2014ufa}.}
\label{fig:linearity}
\end{center}
\end{figure}

As shown in Fig.~\ref{fig:linearity} the detector response is linear within 5\% over an energy range between 10 and 80\,GeV. 

\begin{figure}[!ht]
\begin{center}
\includegraphics[width=0.4\textwidth]{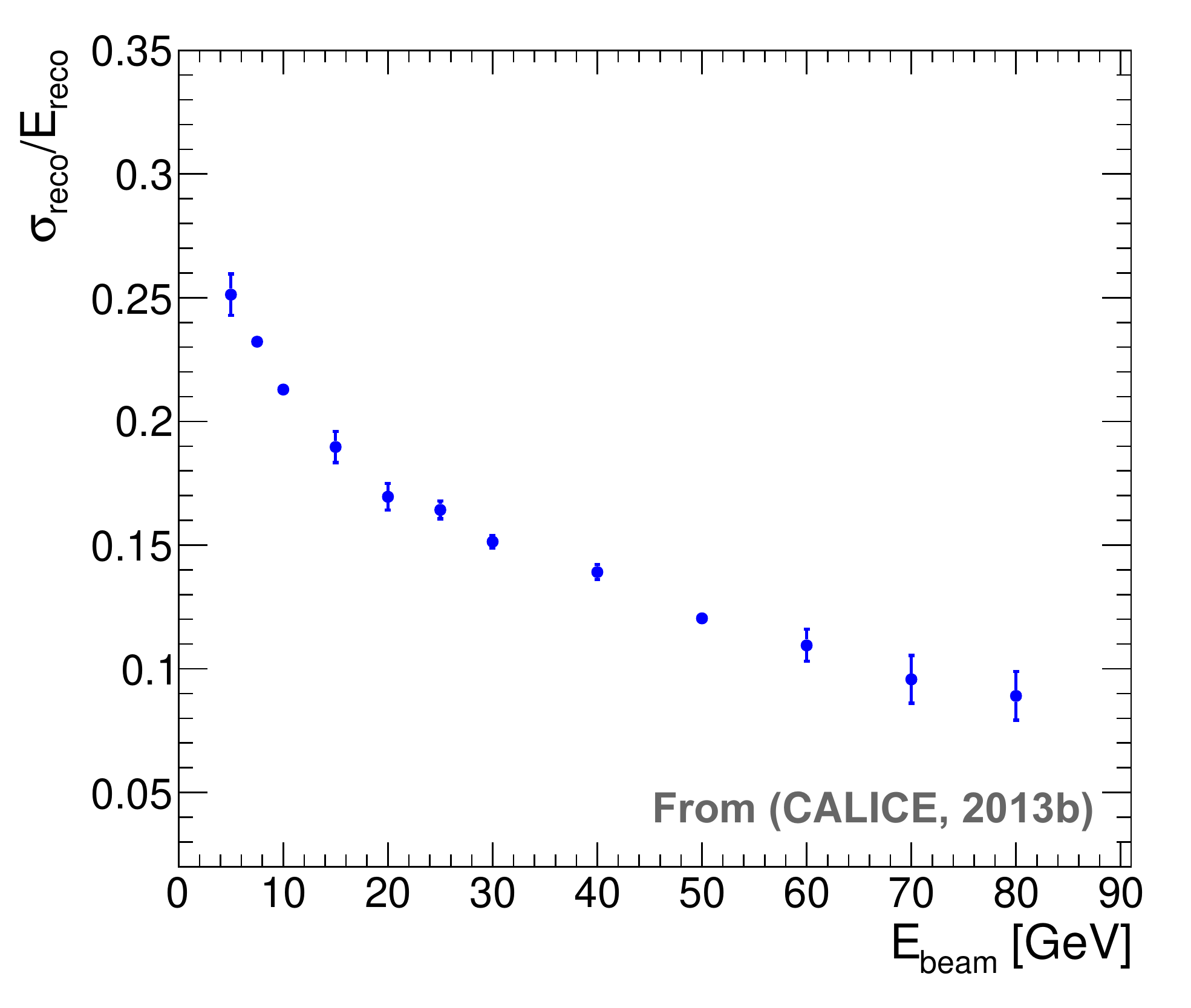}
\caption{The energy resolution $\frac{\sigma_{reco}}{E_{reco}}$ of the RPC-SDHCAL of the reconstructed pion energy $E_{reco}$ as a function of the beam energy $E_{beam}$. From~\cite{bib:can-037a,Steen:2014ufa}.}
\label{fig:resolution}
\end{center}
\end{figure}

The energy resolution over this energy is given in Fig.~\ref{fig:resolution}. It seems that the semi-digital approach allows for an improved control of the saturation effect that is expected to set in towards higher energies. Therefore semi-digital calorimetry might become a way to soften limits of a pure digital approach.  

For the micromegas option, a large scale prototype does not yet exist. The study of the benefit of additional thresholds beyond the pure digital approach is therefore so far based on Monte Carlo
studies only~\cite{bib:mmegas-chef13}. There it is assumed that a second threshold is available for the energy information. In terms of the hits $N_0, N_1$ above thresholds $0$ and $1$, the reconstructed energy can be written as 
\begin{equation}
E_{rec}(N_0,N_1)=C(N_0 + D N_1)  
\label{eq:erec-mm}
\end{equation}

\begin{figure}[!ht]
\begin{center}
\includegraphics[width=0.35\textwidth]{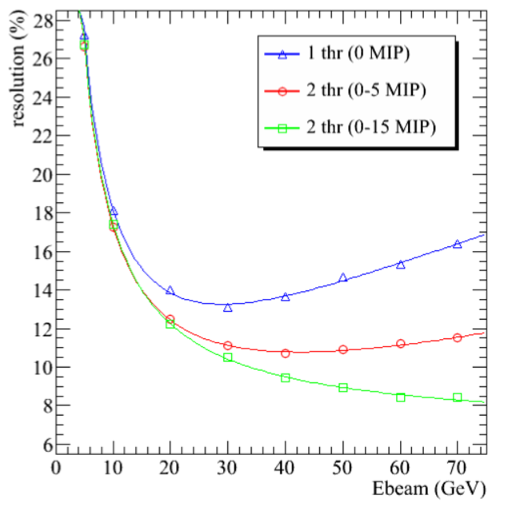}
\caption{
Energy resolution of a micromegas-SDHCAL for a pure digital approach and two settings of the second threshold. 
The results are taken from a Monte Carlo study of a large micromegas-SDHCAL.
From~\cite{bib:mmegas-chef13}.}
\label{fig:mmegas-resolution}
\end{center}
\end{figure}

The coefficient $C$ is a simple conversion factor from hits to energy derived from the linear part of the spectrum, i.e. where $E_{rec} \propto N_0$. It is thus the coefficient $D$ that quantifies the contribution of the number of hits above the second threshold. It is determined analytically from Eq.~\ref{eq:erec-mm} by requiring that the reconstructed energy $E_{rec}$ equals the beam energy $E_{beam}$, i.e.
\begin{equation}
D = 1/N_1 (C^{-1} E_{beam} - N_0)
\end{equation}
The relative importance of the hits above the second threshold increases with energy.
The benefit of an additional threshold depends considerably on the actual value of the second threshold. This is demonstrated in 
Fig.~\ref{fig:mmegas-resolution}. The 'unnatural' increase of the energy resolution towards high energies is inverted for a high second threshold, here set to 15 MIP.

Digital and semi-digital reconstruction schemes have not yet been implemented into the Pandora particle flow reconstruction algorithm. As for the case of software compensation, the use of total energy or total number of hits per shower for the calculation of weights is non-trivial in the case of dense jet environments. 

\subsection{Combining ECAL, AHCAL and TCMT}


In a collider detector, there is generally an electromagnetic section in front of the hadron calorimeter, and ECAL and HCAL, and possibly TCMT,  measurements must be combined in order to reconstruct hadron or jet energies. 
It is their combined performance that counts for physics, and it can be worse than that of the HCAL alone, for example, if the ECAL is highly non-compensating.  
In any case, weighting factors are needed to account for the different response in each detector section. 

The response is expected to approximately scale with the sampling fraction, the ratio of visible to totally deposited energy, and for the same type of active detector and passive absorber, with active and passive material thicknesses. 
In addition, there are surface effects, due to the short range of soft photons, which introduce a dependence on sampling {\em frequency}, too. 
Moreover, the variation of the response with particle type results in a dependence on shower `age', since the particle composition in the early stages of shower evolution is different than in later stages;
therefore optimal weighting factors may deviate from the first approximation according to sampling. 

In the CALICE prototypes, the response is measured in units of MIPs, and is then converted to the electromagnetic scale in GeV, which gives the best estimate for electrons and photons. 
For hadrons, there is an additional correction factor to account for the average $e/\pi$ ratio, 1.19 in the AHCAL case.
To give a numerical example, in the AHCAL with steel absorber, the conversion factor is 
$R_{\rm HCAL}=(42~{\rm MIPs}/{\rm GeV})^{-1} = 0.024~{\rm GeV}/{\rm MIP}$, and for the first section of the silicon ECAL it is 
$R_{\rm ECAL}=(266~{\rm MIPs}/{\rm GeV})^{-1}= 0.0034~{\rm GeV}/{\rm MIP}$.
In simulations, the most probable value of MIP energy deposition in the scintillator is found to be 816~keV, so the sampling fraction of the AHCAL is 3.4\%. 
The absorber thicknesses of the first, second and third ECAL sections are 1.4, 2.8 and 4.2~mm, respectively.
The TCMT has the same absorber thickness as the HCAL in its first section, and 5 times more in the second. 
Therefore, a first approximation for the total hadronic energy would be
\begin{eqnarray}
E_{\rm had}  & =  &  (E_{\rm ECAL 1} + 2 E_{\rm ECAL 2} + 3 E_{\rm ECAL 3}) \cdot (R \cdot e/\pi)_{\rm ECAL} \nonumber \\
& + & (E_{\rm HCAL }+ E_{\rm TCMT 1} + 5 E_{\rm TCMT 2}) \cdot (R \cdot e/\pi)_{\rm HCAL} \nonumber \\
& = &  a_1 E_{\rm ECAL 1} + a_2 E_{\rm ECAL 2} + a_3 E_{\rm ECAL 3} \nonumber \\ 
& + & a_4 E_{\rm HCAL } + a_5 E_{\rm TCMT 1} + a_6 E_{\rm TCMT 2}
\end{eqnarray}
Note that in the case of the tungsten HCAL absorber the individual WHCAL response and $e/\pi$ has to be inserted. 
 
CALICE has collected data with the following combinations: silicon ECAL + AHCAL, scintillator ECAL + AHCAL, silicon ECAL + DHCAL. 
Results are given here for the first~\cite{CALICE:2009ft,Simon:2009an},
while analysis for the others is still on-going. 
The data set is the same as the one studied in Section~\ref{sec:Perf}, however, in the event selection, requirements on shower containment in the HCAL have been made. 
For the reconstruction of the energy, the six weighting factors $a_i$ -- three for the ECAL, one for the HCAL and two for the TCMT -- have been determined in an optimisation procedure,  minimising 
$\chi^2 = (E_{\rm beam} -\sum_{i=1}^6 a_i E_i)$ for all runs in the energy range from 10 to 80~GeV. 
From the result, three internal inter-calibration factors were extracted, 1 : 1.124 : 1.629 for the ECAL and 
1 : 4.55 for the TCMT. 
These were fixed and the external inter-calibration factors determined with a second $\chi^2$ minimisation.  The result is given in Table~\ref{tab:Weight:singlweight}. 
\begin{table}[htb]
\begin{ruledtabular}
\begin{tabular}{lc}
Detector & conversion factor [GeV/MIP]   \\
\colrule
ECAL & 0.00827 \\
HCAL & 0.0293 \\
TCMT & 0.0337 \\
 \end{tabular}
\end{ruledtabular}
\caption{\label{tab:Weight:singlweight} Conversion factors $R \cdot e/\pi$ from MIP to the hadronic energy scale for each detector in the complete CALICE test beam set-up, as determined in a $\chi^2$ minimisation, see text.}
\end{table}
In particular for the ECAL, both the internal and external inter-calibration factors deviate significantly from the expectation based on the simple scaling considerations outlined above. 
This indicates that correlations in the shower and the shower age effects mentioned earlier do play a role. 
In principle one could use weights depending on the reconstructed shower start point, but this has not yet been tried. 

\begin{figure}[htb]
 \includegraphics[width=0.9\hsize]{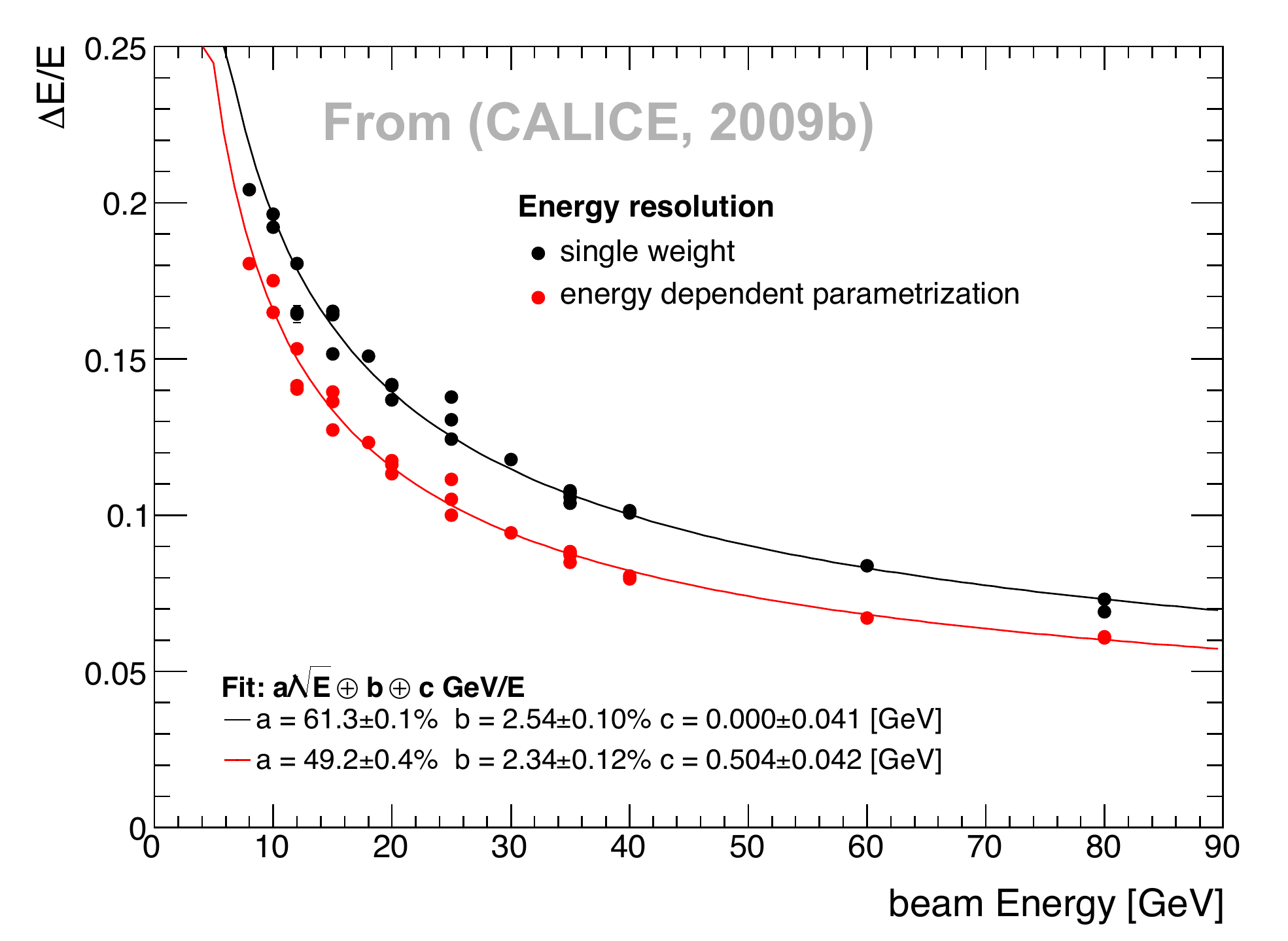}
\caption{\label{fig:Weight:combresol} Hadron energy resolution for the combined ECAL + AHCAL + TCMT set-up, using single weights for each detector, or energy density dependent weights.(From ~\cite{CALICE:2009ft,Simon:2009an})}
\end{figure}

In addition to the method using single weights, the local software (LC) compensation method described above was extended to incorporate the ECAL and the TCMT. 
Weights depending on energy and energy density have been optimised using the same $\chi^2$ minimisation, but keeping the internal inter-calibration constants from the single weight method. 
Both methods result in a very good linearity, with less than 5\% excursion over the full range. 
The resolution is shown in Fig.~\ref{fig:Weight:combresol} for the two cases. 
For the combined set-up, it is almost as good as for the AHCAL alone, and the improvement using the software compensation method is also very similar. 
The constant terms are slightly larger; however, for this preliminary study no temperature correction had been applied yet. 

\subsection{Impact of the tail catcher}

Particle flow methods require the ECAL and the HCAL to be located inside the solenoidal coil, such that the total depth, and thus containment, of the calorimeter system is constrained by cost considerations and technical limitations of the coil size. 
Even small shower leakage contributes significantly to the energy resolution, in particular at higher energies, since its event-to-event fluctuations are larger than its  mean value. 
Two means are at hand to correct each particle measurement: the energy seen in the TCMT, and, thanks to high granularity, shower topology based leakage estimates. 

The impact of the TCMT on the hadronic energy resolution has been investigated in a dedicated study~\cite{CALICE:2012aa}.
First, the amount of leakage has been directly measured, by comparing the energies reconstructed in the the ECAL and HCAL alone with that measured in the complete set-up, including the TCMT. 
The data were taken at the CERN SPS in 2006 where only 30 absorber layers of the AHCAL had been installed, so the first, finely segmented section of the TCMT was treated as part of the HCAL. 
This emulates a structure with a depth of 5.9 nuclear interaction lengths $\lambda_I$ for ECAL and HCAL, which is typical for ILC detector designs.  
In the energy range from 10 to 80~GeV, the fractional leakage increases from 3\% to 8\%, with RMS values from 8\% to 12\%. 
The distribution of leakage energy exhibits long tails, such that the fraction of events with more than 10\% leakage ranges from 3 to 6\%.  

\begin{figure}[htb]
 \includegraphics[width=0.9\hsize]{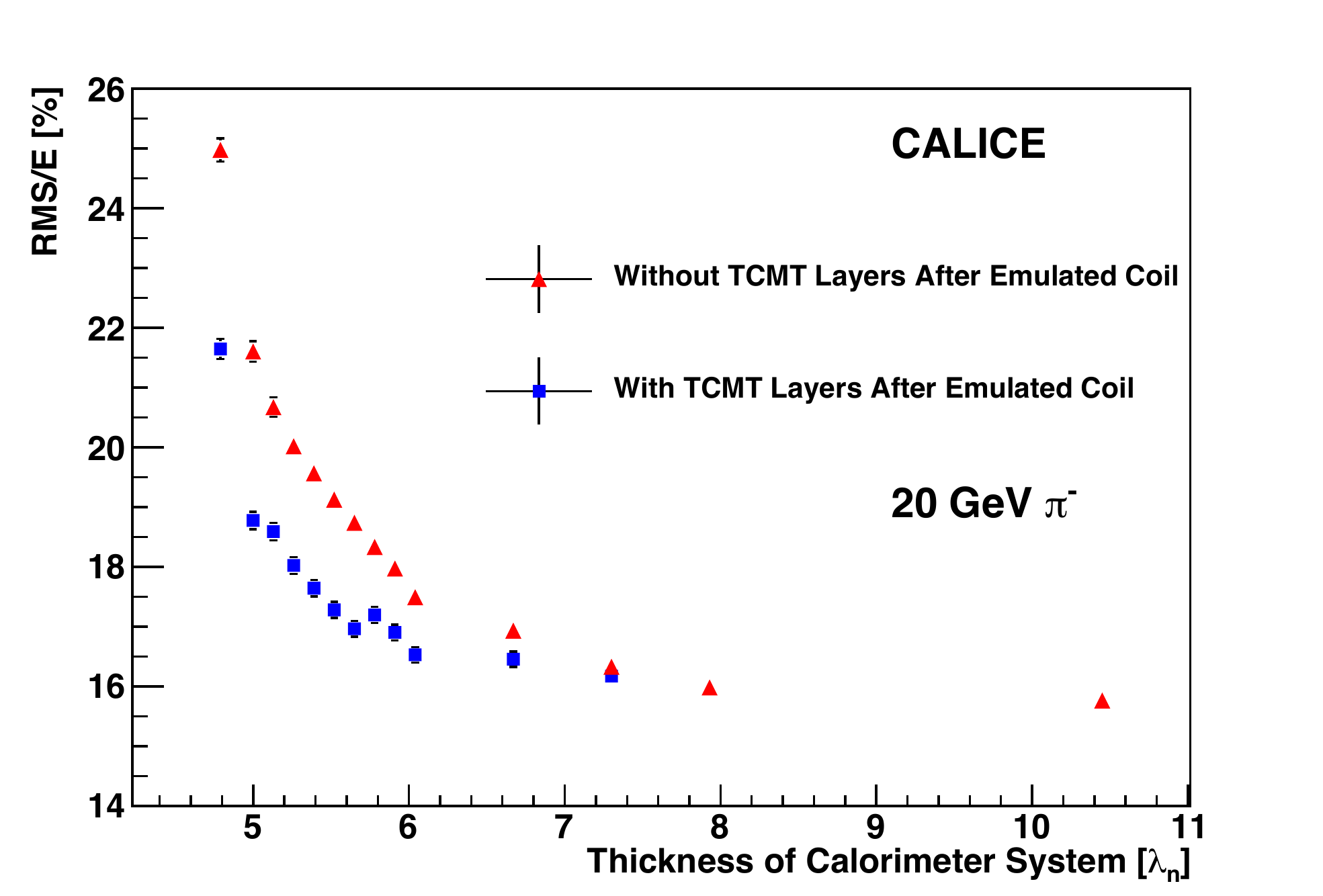}
\caption{\label{fig:Weight:tcmt} Hadron energy resolution (RMS) with and without the TCMT after an emulated coil, as a function of ECAL plus HCAL absorber thickness.
From~\cite{CALICE:2012aa}.}
\end{figure}

The degraded resolution can be recovered by adding the TCMT energies; however, this would not be realistic, since in a real detector the coil between HCAL and TCMT represents un-instrumented material, typically corresponding to 1.5 to 2$\lambda_I$. 
In order to obtain a realistic estimate of resolution improvement due to the TCMT, the effect of the coil material was emulated by excluding a corresponding number of TCMT layers from the energy sum. 
The resolution is measured as RMS, to maintain the sensitivity to the leakage-induced tails.  
Results for 20~GeV pions are shown in Fig.~\ref{fig:Weight:tcmt} for cases with and without sampling behind the emulated coil, as a function of total ECAL plus HCAL absorber thickness. 
As expected, the effect is reduced as the calorimeter thickness increases, but for a typical depth of about 5.5$\lambda_I$ the improvement is still 6 to 8\%.
With rising energy, the improvement gets more significant; at 80~GeV, it corresponds to 16\%.

The impact of a tail catcher on the jet energy resolution depends on the abundance of high-energy particles with sizeable leakage, and therefore also increases with jet energy. 
The TCMT has been included in simulation and reconstruction for the optimisation of the HCAL thickness of the ILD detector using the Pandora algorithm.
In comparison to classical calorimetry, the particle flow approach drastically reduces the sensitivity to leakage, since charged hadrons are measured with the tracker and only neutrals are affected. 
Nevertheless, for jet energies above 100~GeV and calorimeter thickness below 6$\lambda_I$ (for the HCAL alone) the benefits of the TCMT for the jet resolution are clearly visible~\cite{Thomson:2009rp}.

\subsection{Leakage estimation using the shower topology}

The detailed reconstruction of each shower shape provides additional means to derive event-by-event corrections for leakage. 
However, one has to keep the large fluctuations in the evolution of hadronic showers in mind; a simple extrapolation of the visible part is not sufficient.
Showers which apparently finished their development may "re-appear" in a deeper detector section, since a high momentum neutral hadron carried leading energy away from a hard interaction to deposit it there, without leaving a trace in several interaction lengths of traversed material in-between.
Still, the fraction of energy deposited in the rearmost part of the HCAL is statistically correlated with the amount of leakage,
and the variation of the depth of the first interaction
represents the largest single source of leakage fluctuations.  

\begin{figure}[htb]
 \includegraphics[width=0.8\hsize]{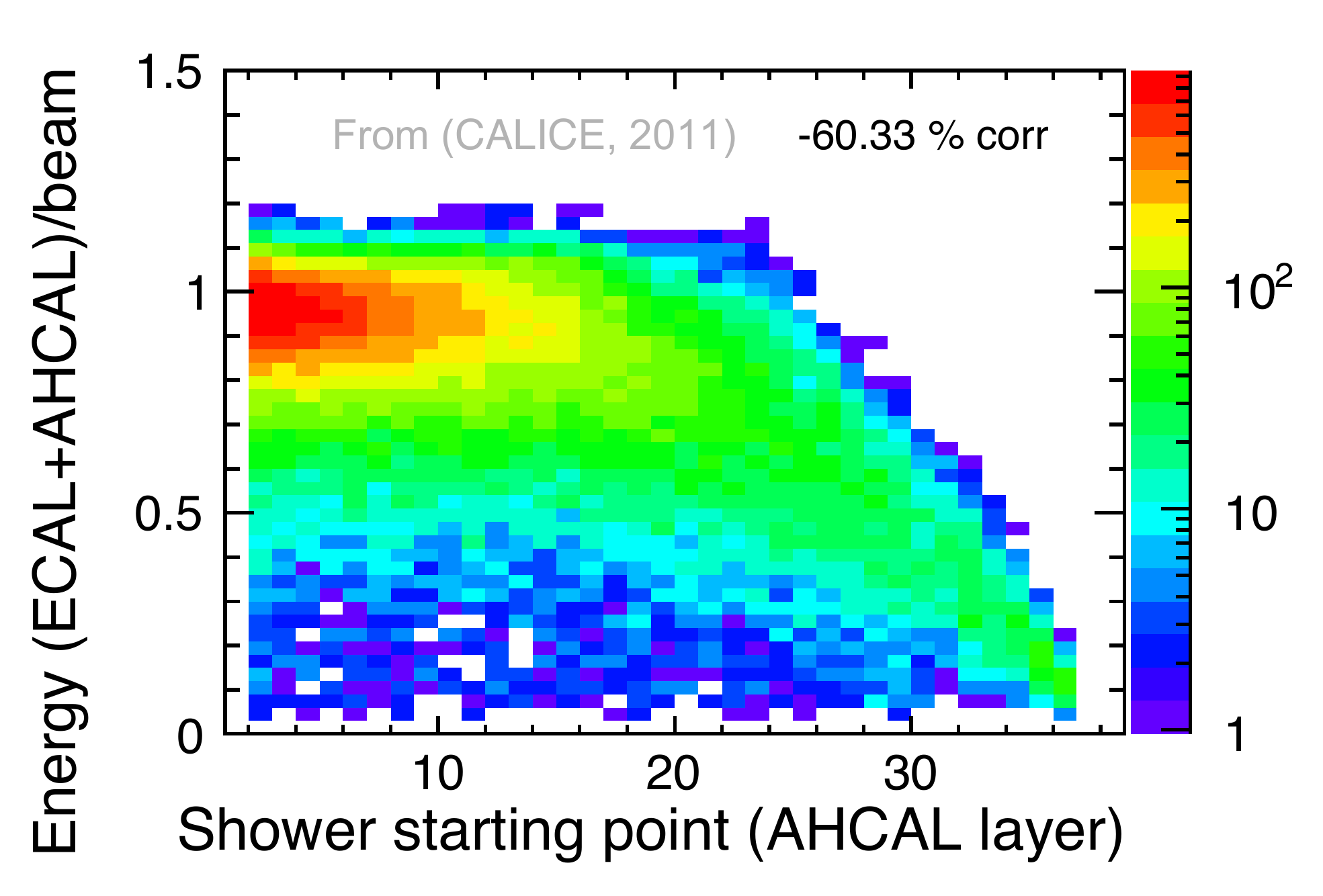}
 \includegraphics[width=0.8\hsize]{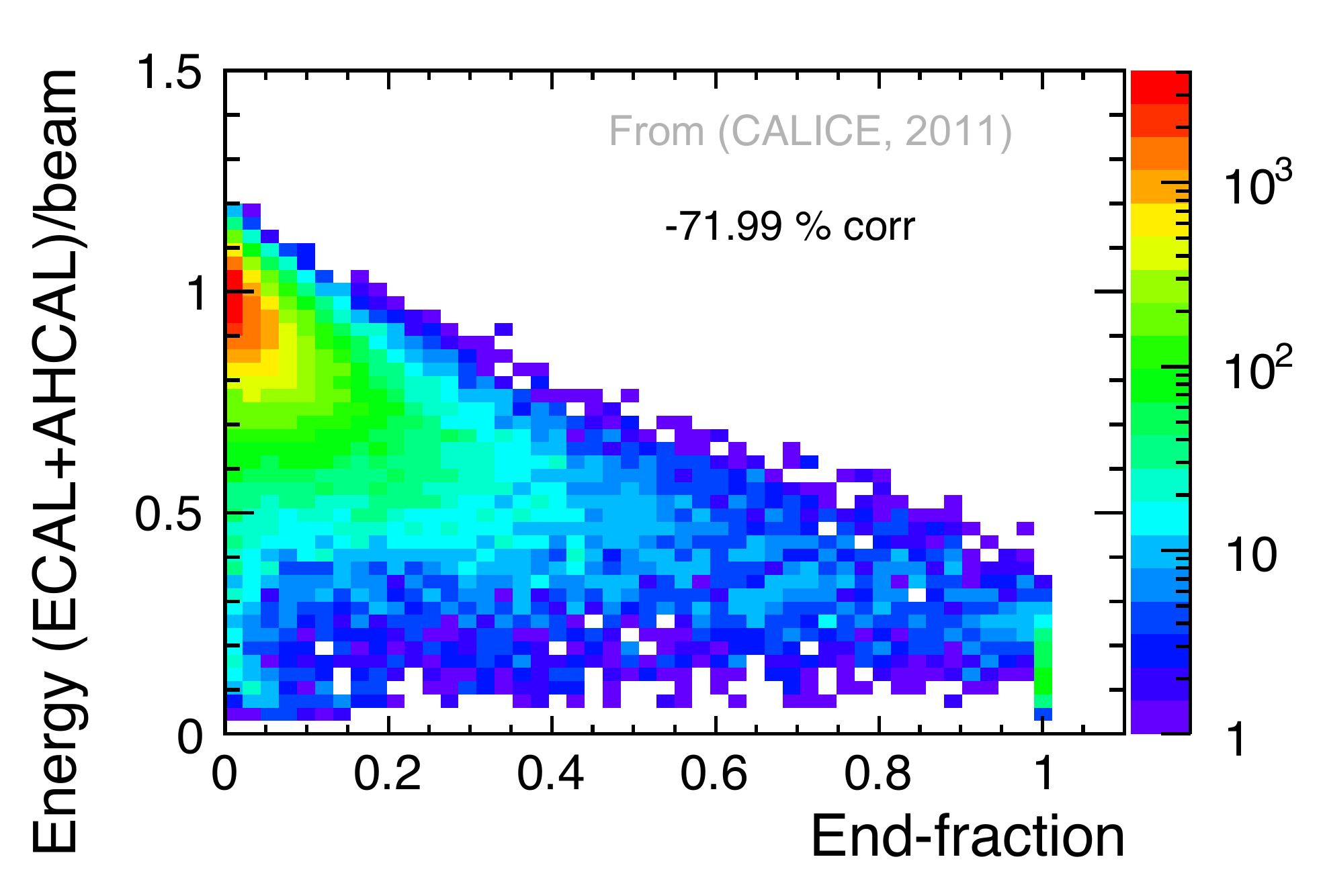}
\caption{\label{fig:Weight:leakcorrel} Correlation between the energy, reconstructed in ECAL plus AHCAL, normalised to the beam energy, and the shower starting point ({\it top}), and end fraction ({\it bottom}), for 80~GeV pions.
From~\cite{CALICE:2011tn,Zutshi:2012}.}
\end{figure}

These two observables have been used in a correction procedure developed 
in~\cite{CALICE:2011tn} and applied to data taken at energies ranging from 8 to 100~GeV with the ECAL, AHCAL and TCMT combined set-up at the CERN SPS in 2007. 
The starting point has been reconstructed by detecting the first layers with energy and multiplicity above thresholds, and the end fraction is the energy in the rearmost four AHCAL layers, normalised to the sum of ECAL and AHCAL deposits. 
Their correlation with the energy reconstructed in ECAL and AHCAL alone is shown in 
Fig.~\ref{fig:Weight:leakcorrel}.
In the case of 80~GeV pions, the correlation is found to be 60 to 70\%, respectively. 
The figures clearly demonstrate the large fluctuations: events with an early start or a small end fraction still do exhibit large leakage in some cases; therefore a statistical combination of several observables is promising. 

\begin{figure}[htb]
 \includegraphics[width=0.6\hsize]{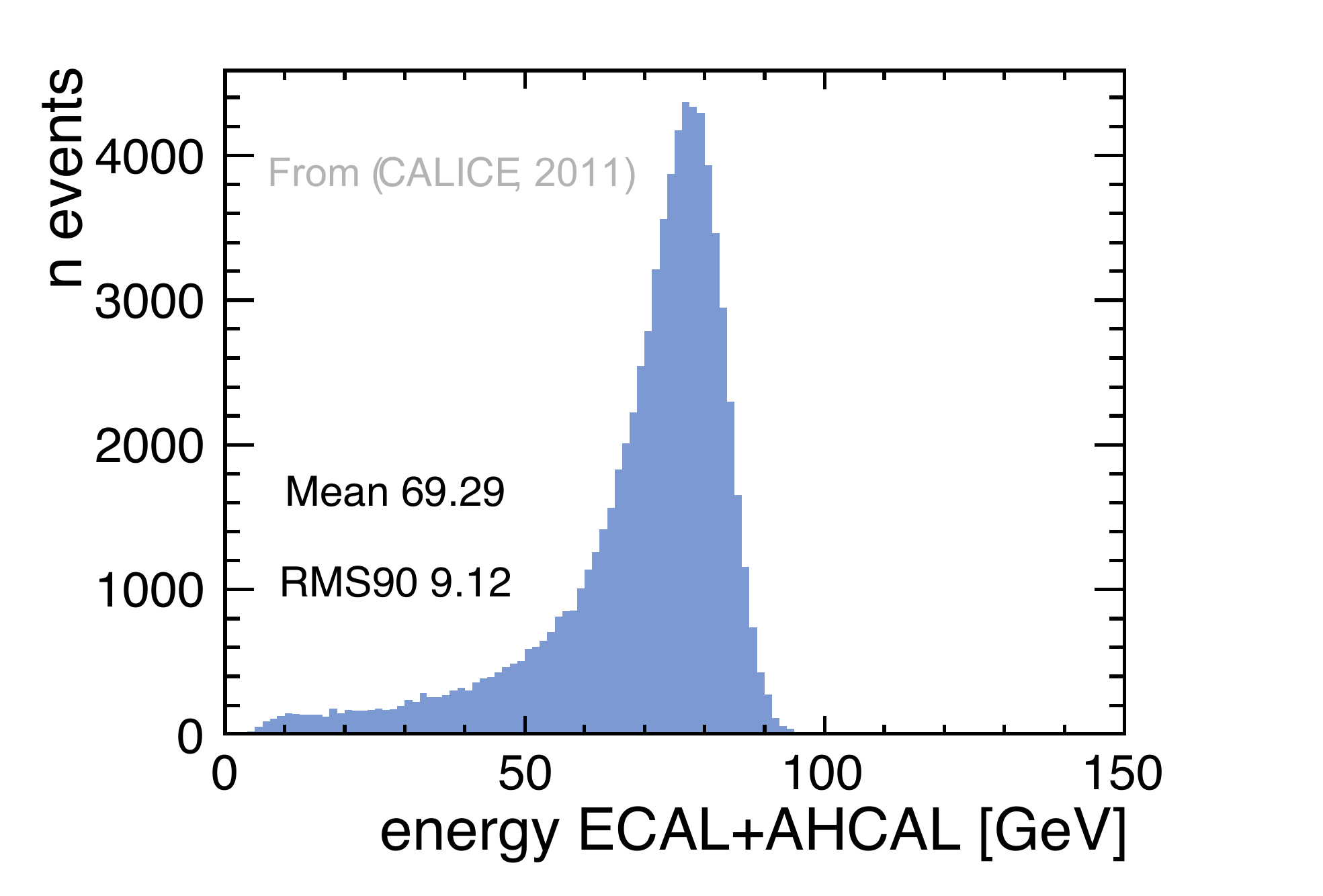}
 \includegraphics[width=0.6\hsize]{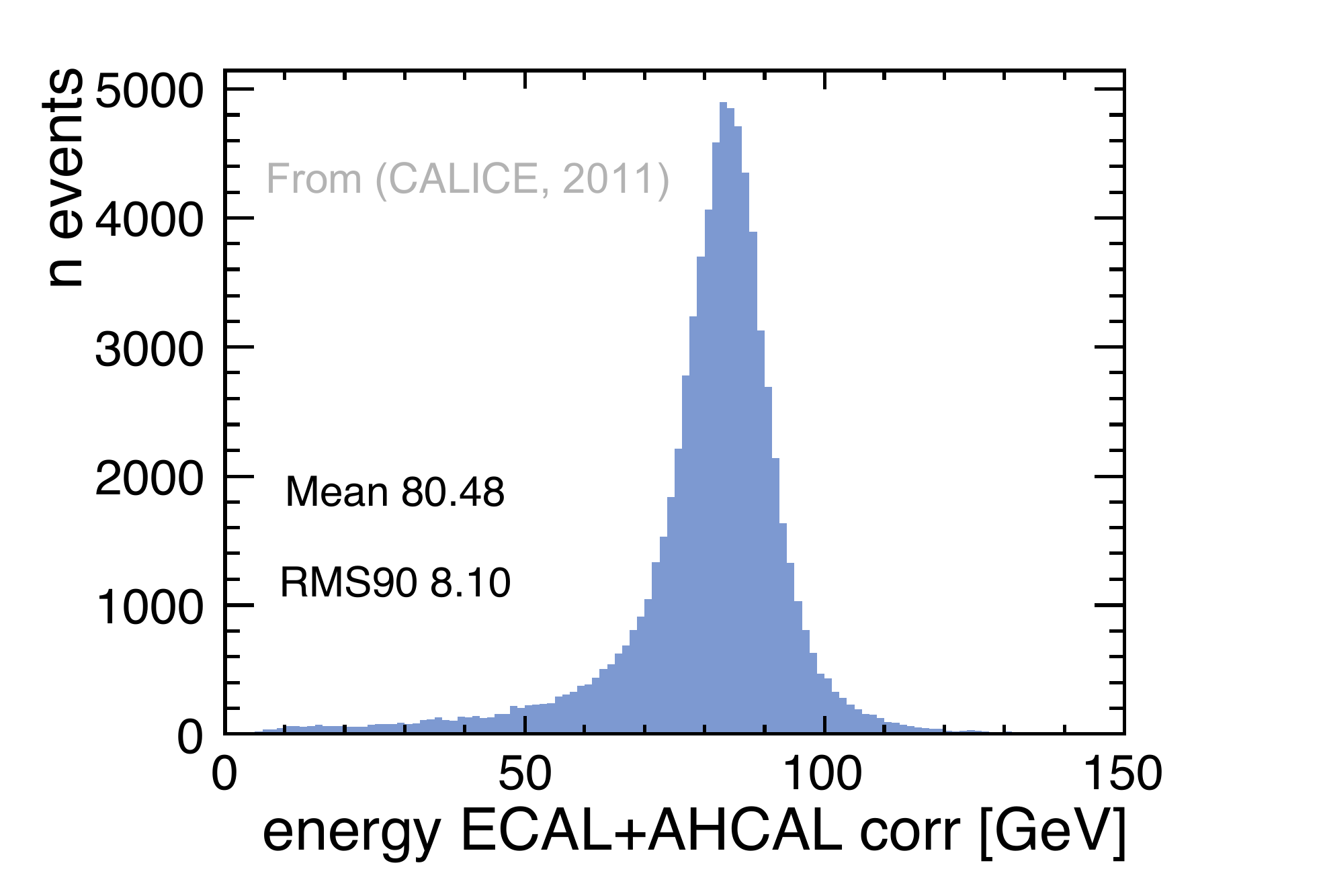}
\caption{\label{fig:Weight:leakcorrect} Reconstructed energy in ECAL plus AHCAL, before  ({\it top}) and after ({\it bottom}) applying the correction based on shower start and end fraction, for 80~GeV pions.
From~\cite{CALICE:2011tn,Zutshi:2012}.}

\end{figure}

The correlations are different for different energies.
To obtain an unbiased procedure, an energy-dependent correction is derived as follows: the data from all energies are combined and binned in two dimensions in the two observables starting point and end fraction. 
In each bin 
a correction function depending on the reconstructed energy is extracted by
parameterising the correlation with the beam energy with a polynomial fit. 
The function is flat and has values near one for early starts and small leakage, but exhibits a stronger energy dependence for late starts and large end fractions, where correction factors increase from 1 to 4 with energy in the studied range.  
Fig.~\ref{fig:Weight:leakcorrect} shows the distribution of reconstructed energy in ECAL plus AHCAL before and after applying the correction to an independent sample of 80~GeV test beam data. 
At this energy, the improvement in relative RMS resolution is 25\%. At lower energies the degradation and thus the improvement is less pronounced. 
Similar results were obtained for Monte Carlo simulations. 

A possible future development would be to implement such corrections into the particle flow reconstruction algorithm, for example for a re-assessment of the required absorber depth. 
One should expect a somewhat reduced performance of the correction, since in a dense jet environment both the determination of the end fraction and the reconstruction of the shower start for neutral hadrons will be disturbed.   
Yet, some net improvements should remain, in particular for energetic jets, which would be complementary to the information provided by the TCMT. 
It is unlikely that the TCMT can be substituted completely by topological leakage estimation, because the latter become ineffective for very late showers or complete "sail-through" of primary hadrons as MIPs. 
These rare cases may hardly affect the average jet performance, but they may become important for missing energy signatures of rare processes. 

These considerations of topological leakage estimation, and the relation of the fine hadronic calorimetry and the tail-catcher, are input to the process of ILC detector design optimisation, which is ongoing.

	\section{Calibration issues%
\label{sec:Calib}}


The highly granular particle flow calorimeters have channel counts, which exceed those of existing detectors, at the LHC for example, by several orders of magnitude. 
The question thus arises how these large numbers can be handled with respect to calibration and monitoring, what precision is actually required, and how it can be obtained and maintained. 
This section discusses general calibration issues for particle flow calorimeters.
Aspects specific to each technology are summarised in Section~\ref{sec:Perf} on performance. 
The discussion is limited to the single particle energy calibration; corrections at jet level are part of the particle flow reconstruction and need to take tracking information on an event-by-event basis into account. 

\subsection{Calibration scheme}

Calibration as a general term is used for several aspects of the calorimeter reconstruction. 
For the channel-to-channel normalisation we use the term equalisation, to be distinguished from the corrections of time-dependent effects, induced for example by temperature or pressure variations. 
Tracing such variations is called monitoring. 
Establishing an absolute scale in units of GeV is again a separate task, and different scales, electromagnetic, hadronic or weighted scales need to be distinguished. 
If applied at the particle level, they may depend on the clustering definition. 
Other corrections, such as for dead materials, may be applied at the particle level, too.  

The calibration of the electromagnetic and hadronic response of the calorimeter proceeds in the 
following general steps: 
\begin{enumerate}
\item Test bench characterization of sensor parameters at cell level
\item Inter-calibration of the electronic response of all individual cells using 
muon test beams, and conversion to the MIP scale
\item Verification of the electromagnetic scale and linearity using electron 
beams impinging directly on the detector modules  
\item Determination of the hadronic response using
hadron test beams
\item Determination of combined ECAL and HCAL hadronic response, 
including weighting procedures
\item Verification of dead material corrections at inter-module connections 
using hadron test beams
\item In-situ validation and monitoring using kinematic constraints, tracker information and track segments in hadronic showers 
\end{enumerate}
To a large extent these steps have been carried out for the test beam data, providing a basis for extrapolation to a full collider detector.  

\subsection{Channel equalisation and energy scales}

For silicon and scintillator, the muon response defines the MIP scale. 
For gaseous detectors, the response is proportional to the product of
efficiency and pad multiplicity which are also determined with muon
beams. 

The inter-calibration with muon beams must be done for all cells and all detector layers. 
Thanks to the modular design, this can be done  with the assembled 
modules, but also with the bare active layers before insertion into the absorber.
For example, in the CERN test beam 12 hours were needed for the AHCAL to acquire sufficient 
statistics on a stack with a square meter front face and 38 layers. 
This would translate into about two months for an entire ILC detector, or less, if more layers 
are aligned after each other in the beam. 
The analysis of the calibration data can be massively parallelised. 

The response to electromagnetic showers on the MIP scale can be uniquely
predicted by simulations and verified in test beams with known energy. 
In practice, the electromagnetic scale is useful only for the linear range for electrons and hadrons,
i.e\ for the silicon and scintillator detectors. 
For these also a hadronic scale can be defined, since the deviations from 
linearity are small (less than a few per-cent up to 80 GeV).
For the gaseous calorimeters, the hit multiplicity noticeably deviates from a linear behaviour for energies above 30~GeV,
and additional procedures or weighting techniques need to be applied to obtain a linear scale.

The weighted energy scale  depends on the applied algorithm. 
Software compensation methods and semi-digital reconstruction have been discussed in Section~\ref{sec:Weighting}. 
Additional corrections will be necessary to account for un-instrumented 
regions or additional material from support structures, electronics and 
service lines, at the ECAL-HCAL transition and at inter-module boundaries. 
This must be extracted from simulations which need to be benchmarked in 
test beams with realistic ECAL and HCAL prototypes  combined. 
Apart from the inter-calibration, which must be done for every individual active detector element, 
it is assumed that all studies addressing the absolute electromagnetic  and weighted scales can be done 
with single representative sample structures.  

\subsection{Monitoring and {\it in situ} techniques}

The above calibration scheme needs to be complemented by monitoring
techniques in order to take time-dependent variations into account.
The general approach is that if the MIP scale -- or the MIP hit multiplicity -- is 
maintained and under control, all derived scales are stabilised as well. 

\begin{figure}
\begin{center}
\includegraphics[width=0.45\textwidth]{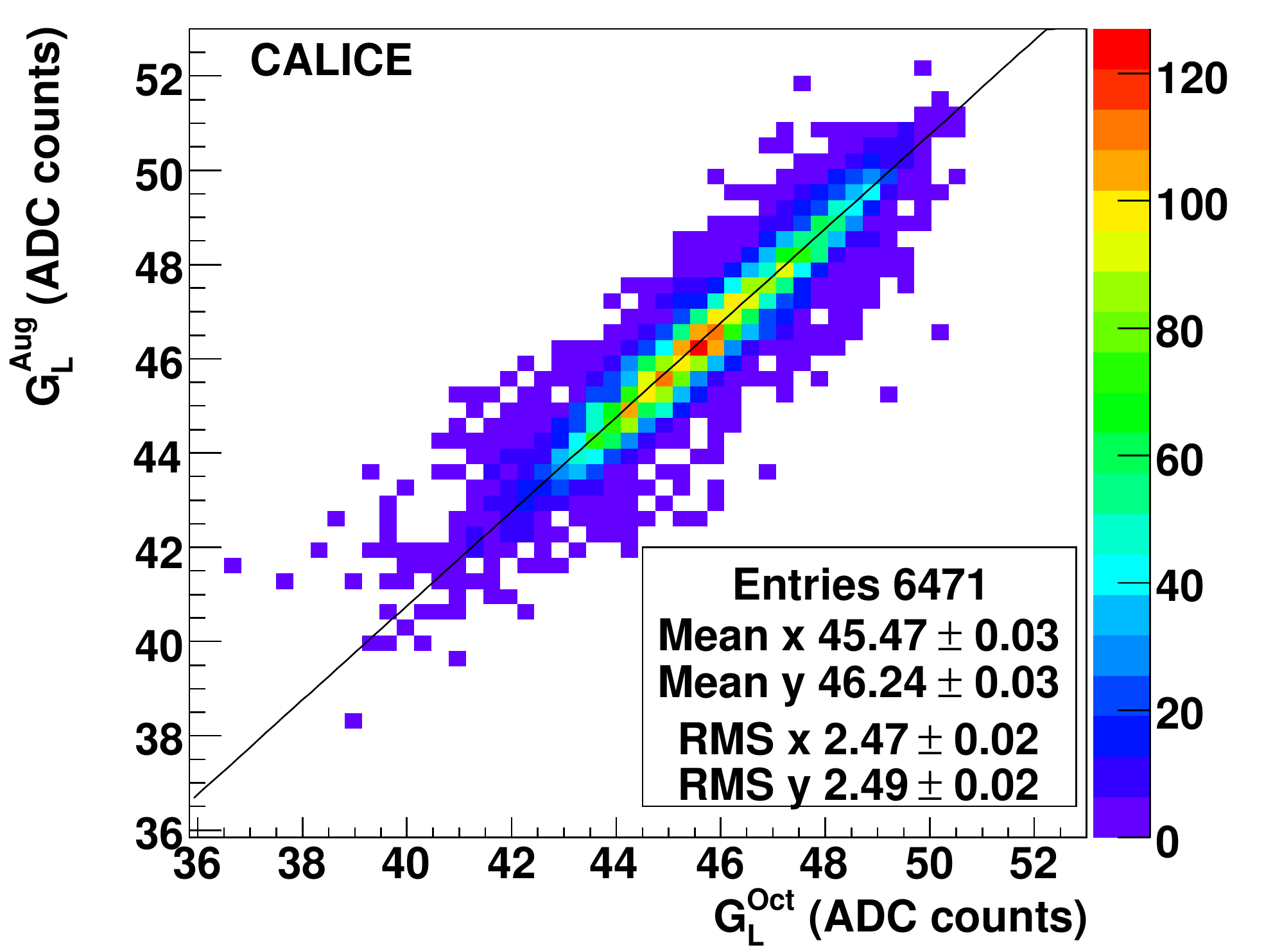}
\end{center}
\caption{%
Comparison between calibration constants obtained in two different data taking periods in 2006. Results are taken from~\cite{Anduze:2008hq}.}
\label{siw:fig:corr_calib}
\end{figure}

Test beam experience has demonstrated that the MIP scale of the silicon-based ECAL is intrinsically stable
over long time periods to the per-cent level, where the variations are mainly due to different experimental conditions (e.g. cable length) at the beam test sites. 
The calibration constants showed no influence from external factors like temperature.
The correlation of calibration constants obtained for different periods of data taking in 2006~\cite{Anduze:2008hq} are 
shown in Fig.~\ref{siw:fig:corr_calib}. 
It was  shown~\cite{bib:phd_hengne,Rouene:2014pva} that the correlation coefficient  is 83.8\% between the calibration constants obtained at FNAL in 2008 and at CERN in 2006. 
Considering that many operations like mounting, un-mounting, and shipping occurred between 2006 and 2008, this high correlation coefficient demonstrates the stability with time of the SiECAL prototype. The same level of correlation exists between calibration constants derived for the beam tests in 2008 and 2011 at FNAL. 
The absolute calibration of the ECAL can be verified and adjusted  by comparison with the
tracker or using electrons and photons kinematically constrained such as in Bhabha events or Z$\rightarrow e^+e^-$ decays.

Variations of the MIP scale of scintillation detectors are mainly due to changes 
of the electronic response of the photo-sensor,  induced by changed thermal conditions,
which have successfully been corrected using the known temperature dependence; see Section~\ref{sec:Perf}.
The photo-sensor gain stability can also be monitored by measuring the spacing between 
peaks in the pulse-height spectrum attributed to small, discrete numbers 
of registered photo-electrons, which does not require LED light stability. 
In principle it is also possible to adjust the voltage in order to 
compensate the temperature variation, and use the gain 
to watch the stability. 
The hit multiplicity of gaseous detectors mainly varies as a consequence of variations of 
temperature and pressure which are continuously monitored.  
The use of radioactive sources is not necessary according to present understanding. 
Changes in the amplification of the read-out chain were checked 
independently and found to be much smaller than those of the sensors. 
The pedestals of the read-out electronics are regularly monitored using 
random trigger events; this also detects and monitors dead or noisy channels. 

Due to the underground location of collider detectors, the orientation of the detector layers, 
the power pulsing, and due to the high granularity, cosmic rays might not be 
sufficient for monitoring the MIP scale {\it in situ}.   
However, thanks to the excellent imaging capabilities of the calorimeters, MIP-like 
track segments can be identified in hadronic showers and used for calibration
purposes, see Section~\ref{sec:G4v-tracks}.
The potential for in-situ calibration of the AHCAL in the ILD detector was studied in
simulations - for details see~\cite{CALICE:2009:et}. 
Although typically two tracks are found in each shower which are used for the 
calibration of 20 cells, it is even at the Z resonance not possible to 
obtain a channel-by-channel calibration within realistic running times. 
However, the method is well suited for the determination of average 
corrections for a sub-section of the detector, e.g.\ a layer in a module. 

%
%

\subsection{Required Accuracy}

A common feature of all particle flow calorimeter technologies is their relative insensitivity to any sort of stochastic calibration or alignment uncertainty for individual cells.
The large number of cells required for the topological resolution is an asset rather than a burden, since the precision 
with which cell level information needs to be known scales with $\sqrt{N}$, where $N$ is the number of channels contributing to a shower. 
Even in the case of the scintillator HCAL, which is the coarsest of the detectors considered, this amounts to about 10 cells per GeV. 

In contrast, coherent systematic effects must be corrected with higher precision, depending on the fraction of the detector affected. If it is the entire calorimeter, the required precision is given by the constant term aimed at, about 1\% for the ECAL and 2-3\% for the HCAL. 
The challenge of the high granularity is that time-dependent corrections cannot be applied at cell level, and cell-wise corrections require stability over time to reach statistical precision. 
On the other hand, since every cell is individually read out, one is free to form averages over space or time according to the specific problem, but finding the optimal averaging procedure and identifying the leading effects is often an involved analysis and intimately related to understanding the detector and its systematics. 
The procedures needed in practice can only be developed from real data.
Such studies form an important part of the test beam data analysis, and they are also the reason why each generation of prototypes must undergo beam tests at system level again to obtain realistic performance figures.  

Using fully detailed simulations of the ILD detector and reconstruction 
based on the Pandora particle flow algorithm, different
scenarios of statistically independent as well as coherent mis-calibration 
effects have been modelled, affecting the entire AHCAL or parts (module layers) of it. 
Purely statistical variations, like those arising from calibration errors 
or random ageing effects, hardly affect the energy resolution at all. 
However, they may degrade the in-situ MIP calibration capability. From this, 
a moderate requirement of the inter-calibration stability to be ensured by 
hardware design of $\pm$10\% is derived. 

Coherent effects which could for example arise from uncorrected temperature 
variation induced changes of the response are potentially more harmful,
if they affect the entire detector. 
However, these are easy to detect, and even a 5\% variation only 
mildly propagates into the jet energy resolution. Systematic effects 
shifting sub-sections like layers are unnoticeable unless they exceed about 
15\%, comfortably in range of the {\it in situ} calibration method accuracies. 

\begin{figure}[htb]
\begin{center}
  \includegraphics[width=.35\textwidth,angle=-90]{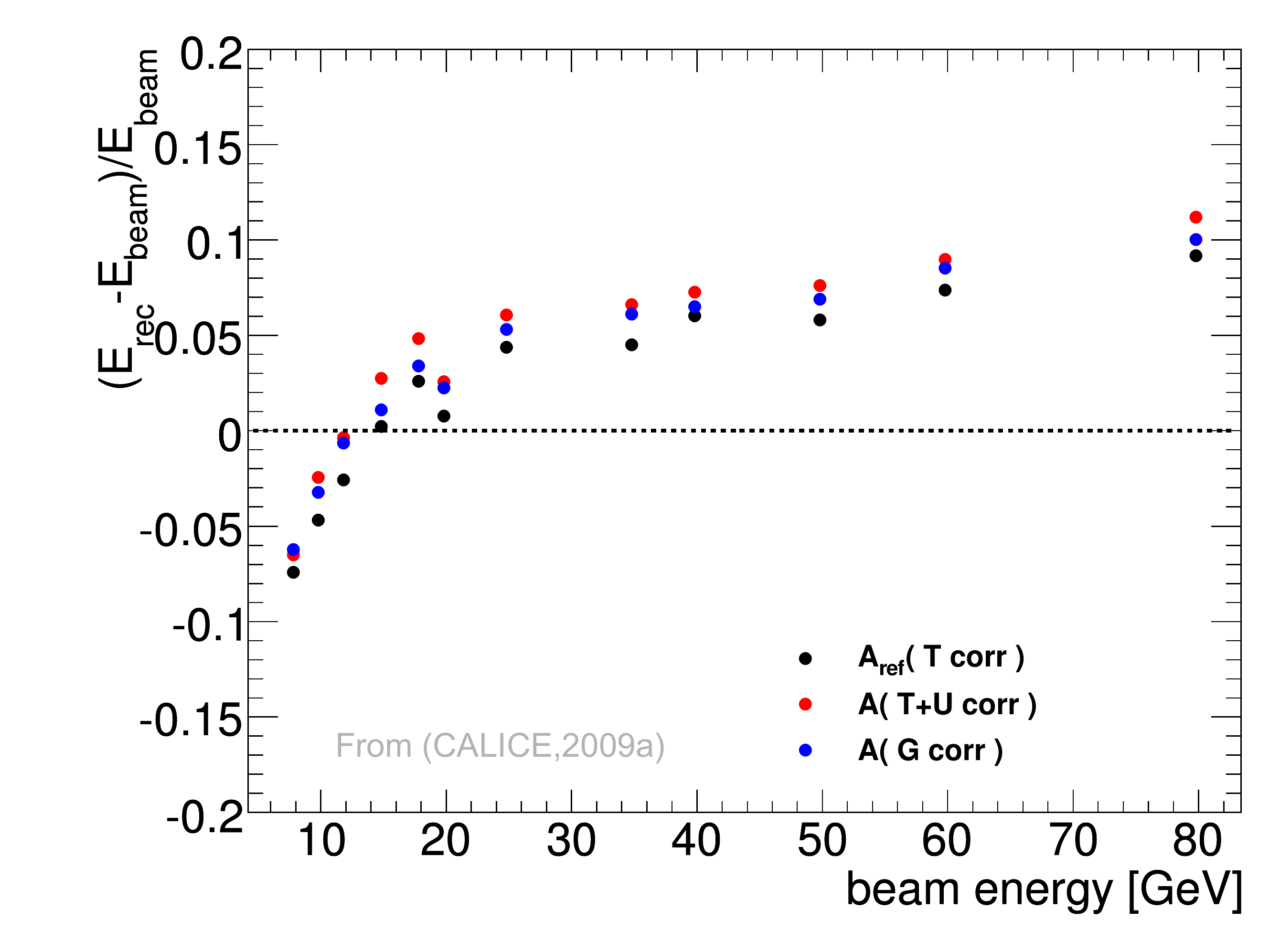}
  \caption{\label{fig:Calib:ahcalscale}
    Residuals from linearity of reconstructed hadron showers in the AHCAL,
    in the range 8 to 80 GeV: using calibration samples from muon runs at FNAL transported to the ``collider'' 
    run conditions and the CERN calibration as reference. Corrections were applied based on either temperature (T) and voltage (U), or on observed photo-sensor gain (G), and additionally on the {\it in-situ} 
    MIP stub correction layer-by-layer. From~\cite{CALICE:2009:et}.
  }
\end{center}
\end{figure}

The validity of these simulation-based estimates has been verified 
by treating the AHCAL test beam experiment at CERN and FNAL 
like a collider detector, using 
cell-by-cell inter-calibrations only from data taking at a different site, 
under different conditions and after having it exposed to disassembly, 
transport and re-assembly influences~\cite{CALICE:2009:et,Abe:2010aa}. 
Applying only in-situ monitoring 
techniques,  the scale was re-established and the resolution reproduced,
see Fig.~\ref{fig:Calib:ahcalscale}. 
Imperfections absent in any simulation showed up, but were successfully
compensated using a combination of techniques.

Overall, the high granularity and channel count provides net
advantages for calibration. On one hand, due to the law-of-large-numbers
suppression of statistical effects, the requirements on individual cell 
precision are very relaxed. Coherent effects, on the other hand, can be 
studied with any desired combination of channels, be it layers, longitudinal
sections, electronics units or according to any other supposed hypothesis 
of systematic effects. The high degree of redundancy and the full 
information for each channel provide maximum flexibility, without having to 
rely on intrinsic homogeneity as in the case of internal
optical or analog summing.

%
	\section{Tests of Geant4 shower simulation models%
\label{sec:G4valid}}


\subsection{Shower simulation models}

The simulation of hadron shower evolution is challenging because there is no single model that describes hadronic interactions over the full  range of energies that play a role in each single shower, from the first hard interaction, possibly several TeV, down to the MeV range of nuclear binding effects.
Therefore different approaches must be combined, and transition regions or even gaps, where none is strictly applicable, must be bridged in order to describe the energy deposition in a calorimeter in its entirety. 
The first simulation codes, e.g.\ in the GEISHA framework~\cite{Fesefeldt:1985yw}, were based on phenomenological parameterisations (LHEP) of the interaction and aimed at reproducing average shower properties in calorimeters. 
However, energy conservation was not ensured event by event, so correlations within the shower were not described. 

\begin{figure*}[htb]
  \includegraphics[width=0.8\hsize]{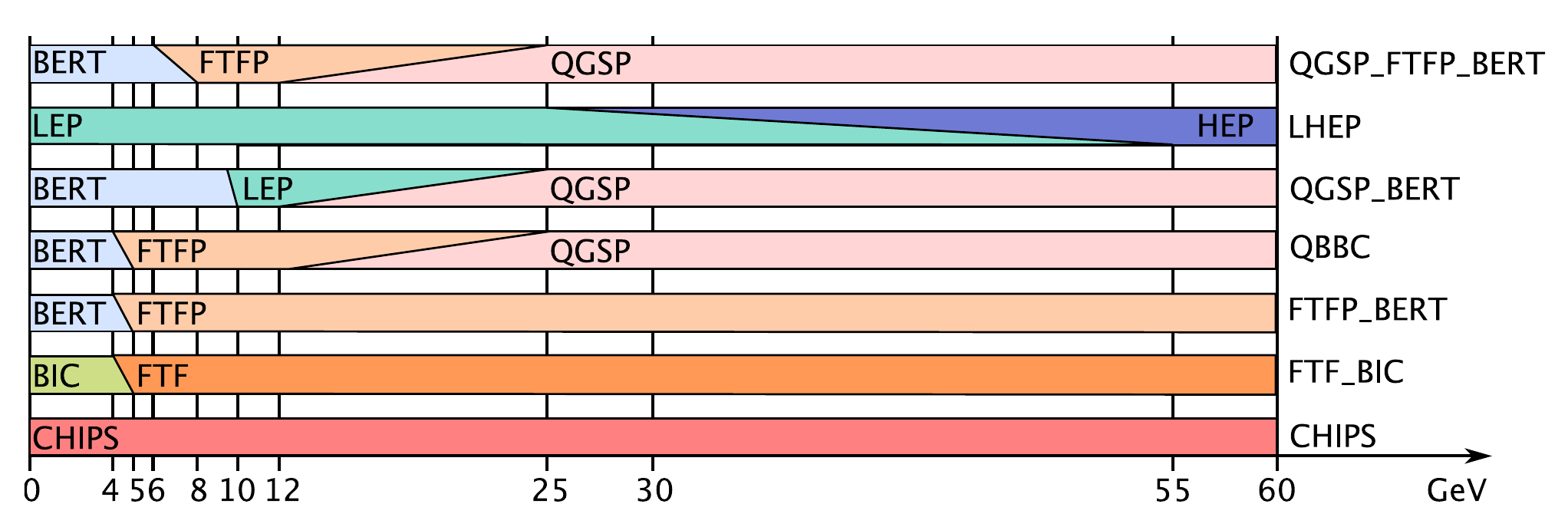}
\caption{\label{fig:G4v:models} Energy ranges of shower simulation models used in Geant4 physics lists.}
\end{figure*}

In the past decade, significant progress has been made in replacing these empirical parameterisations by more fundamental, theory-driven models within the Geant4 framework~\cite{Agostinelli:2002hh,Allison:2006ve}. 
The main models are high energy quark gluon string based models (QGS, FTF) for hard interactions in the range 
above 5 -- 10~GeV, the Bertini (BERT) or binary (BIC)  models for the intra/nuclear cascade in the intermediate region, and for processes below about 200~MeV de-excitation models including a pre-compound (P) nucleus stage, and nucleon evaporation. 
These are combined into so-called physics lists, with the energy ranges depicted in
Fig.~\ref{fig:G4v:models}. 

In the transition regions models are selected randomly. 
The models are not tuned to calorimetric measurements, so-called "thick target" data, which give little hint on which detail needs adjustment, but rely on "thin target" data, differential cross sections for the scattering of hadrons off nuclei, while thick target data represent the benchmark for validation.  
The development was strongly motivated by the needs of the LHC experiments and more recently also by the demands of particle flow calorimetry. 
The focus was originally on describing mainly detector response, and shower extension, albeit with the coarse spatial resolution of LHC detectors. 
It is now moving towards internal structure, particle composition and even time evolution. 

At the time when the CALICE test beam programme was proposed, there were significant uncertainties in the simulation of shower shapes, and the predictions for the radius, for example, varied between models by up to a factor of two~\cite{Mavromanolakis:2004qz}. 
In the meantime, following detailed comparisons with LHC test-beam data, major revisions of all stages of the simulation code have been undertaken~\cite{Apostolakis:2010eu,Dotti:2011zz}
For example, the range of the string-based model has been extended towards lower energy, and in particular the diffraction based Fritiof model does not need any recursion to the old parameterised models to bridge the gap to the low energy cascade models. 
Also an experimental list was implemented, based on the chiral invariant phase space (CHIPS) model, 
which covers the full energy range, but is not finally tuned yet. 
With these improvements, a good description of the LHC test-beam data is achieved using the QGSP-BERT physics list, 
which is the most commonly used at the LHC. 
The detector response agrees within a few percent~\cite{Adragna:2009zz,Abat:2010zz}. However, some issues remain, namely the radial shower extension is underestimated by 
about 15\%~\cite{Adragna:2010zz}.

A good description of global shower properties, except possibly for the radius, can thus be expected for the CALICE data, too. 
However, particle flow reconstruction makes use of the shower structure in much more detail, for example at the re-clustering or fragment removal stage. 
Even though the algorithms were shown~\cite{Thomson:2009rp} to be rather robust against the choices of physics lists, a test of the simulation at the level of detail which the fine segmentation of the CALICE prototypes allow, forms an essential basis of the validation of particle flow calorimetry and supports the simulation of multi-particle states in complex detector configurations. 
As will be shown, new observables like the charged track multiplicity are becoming accessible which are intimately related to the internal shower structure. 

\subsection{Hadrons in the silicon tungsten ECAL}


Applying a simple power law and given the depth of about one interaction length, about 60\% of the hadrons contained 
in a jet will interact in the volume of the SiW ECAL. It is therefore important to understand the interactions of 
hadrons in the SiW ECAL. This has been addressed in Refs.~\cite{Adloff:2010xj}, for energies between 8 and 80\,GeV 
and more recently~\cite{Bilki:2014uep} for energies between 2 and 10\,GeV. The latter test a region where many Geant4 physics lists feature a transition between models for hadronic cascades, see Fig.~\ref{fig:G4v:models}. The earlier results report on comparisons between data and hadronic shower models  as implemented in Geant4 version 9.3. 

\begin{figure}[htb]
\begin{center}
\includegraphics[width=0.3\textwidth]{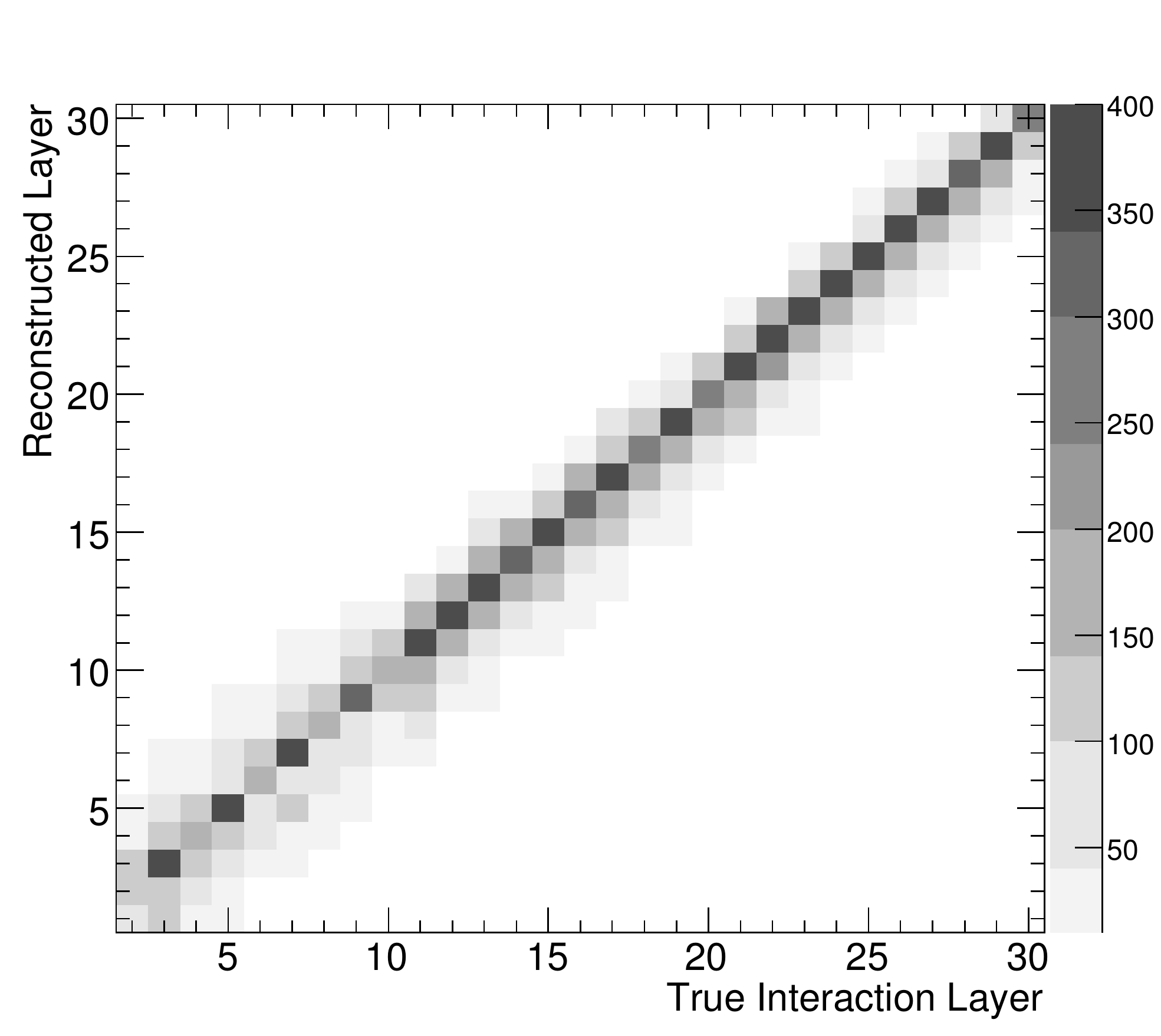}
\includegraphics[width=0.3\textwidth]{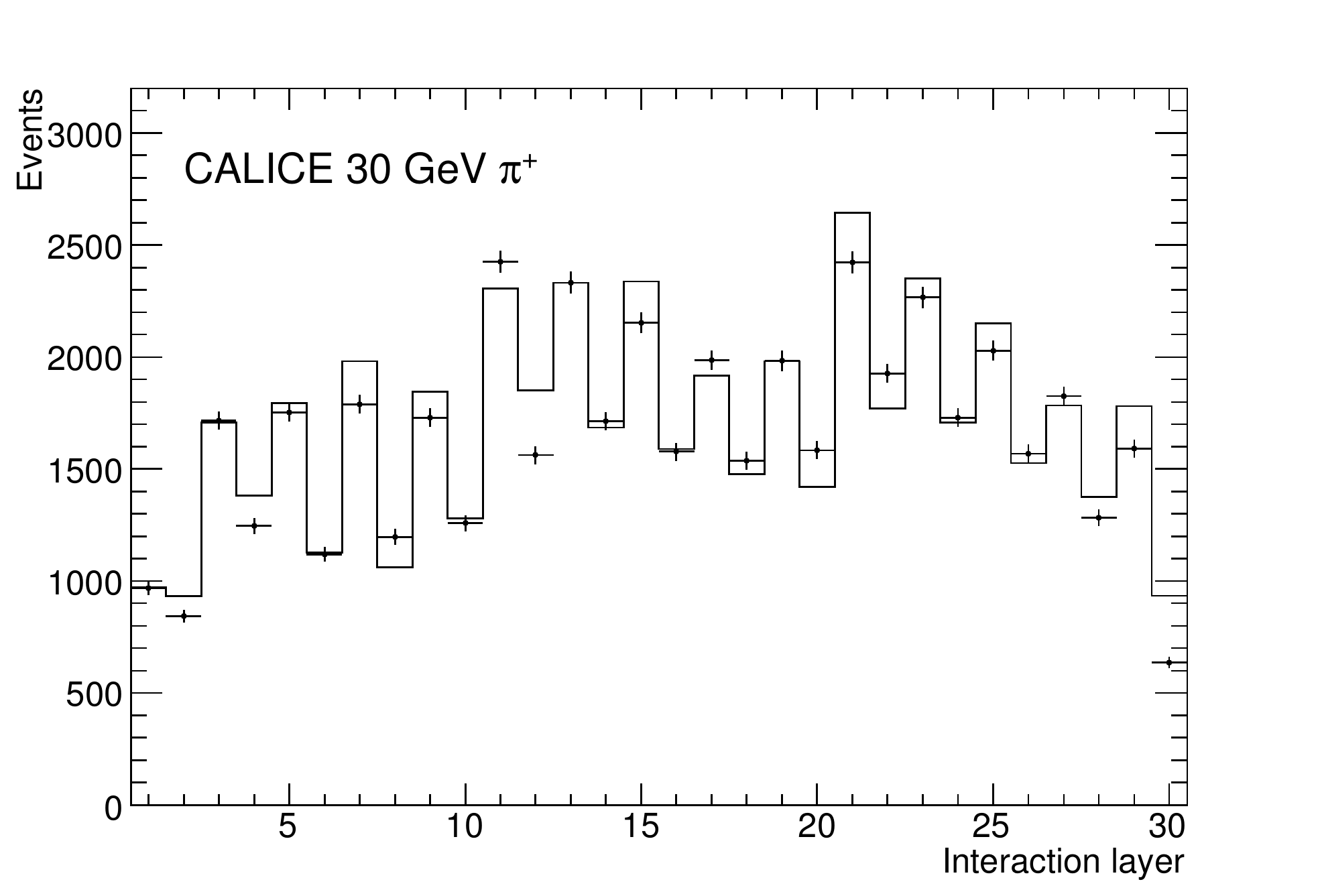}
\end{center}
\caption{
 Top:  Comparison between the interaction layer reconstructed in the SiW ECAL and the interaction layer as extracted from the GEANT4 event record.
This example corresponds to a 20~GeV $\pi^-$ beam simulation, using the QGSP\_BERT physics list.  Bottom:  Distribution of the reconstructed interaction layer in the  ECAL for 30~GeV data (points), compared with Monte Carlo predictions using the QGSP\_BERT physics list (solid histogram). 
From~\cite{Adloff:2010xj}.}
\label{fig:hadreac}
\end{figure}

\begin{figure}[htb]
\includegraphics[width=0.33\textwidth]{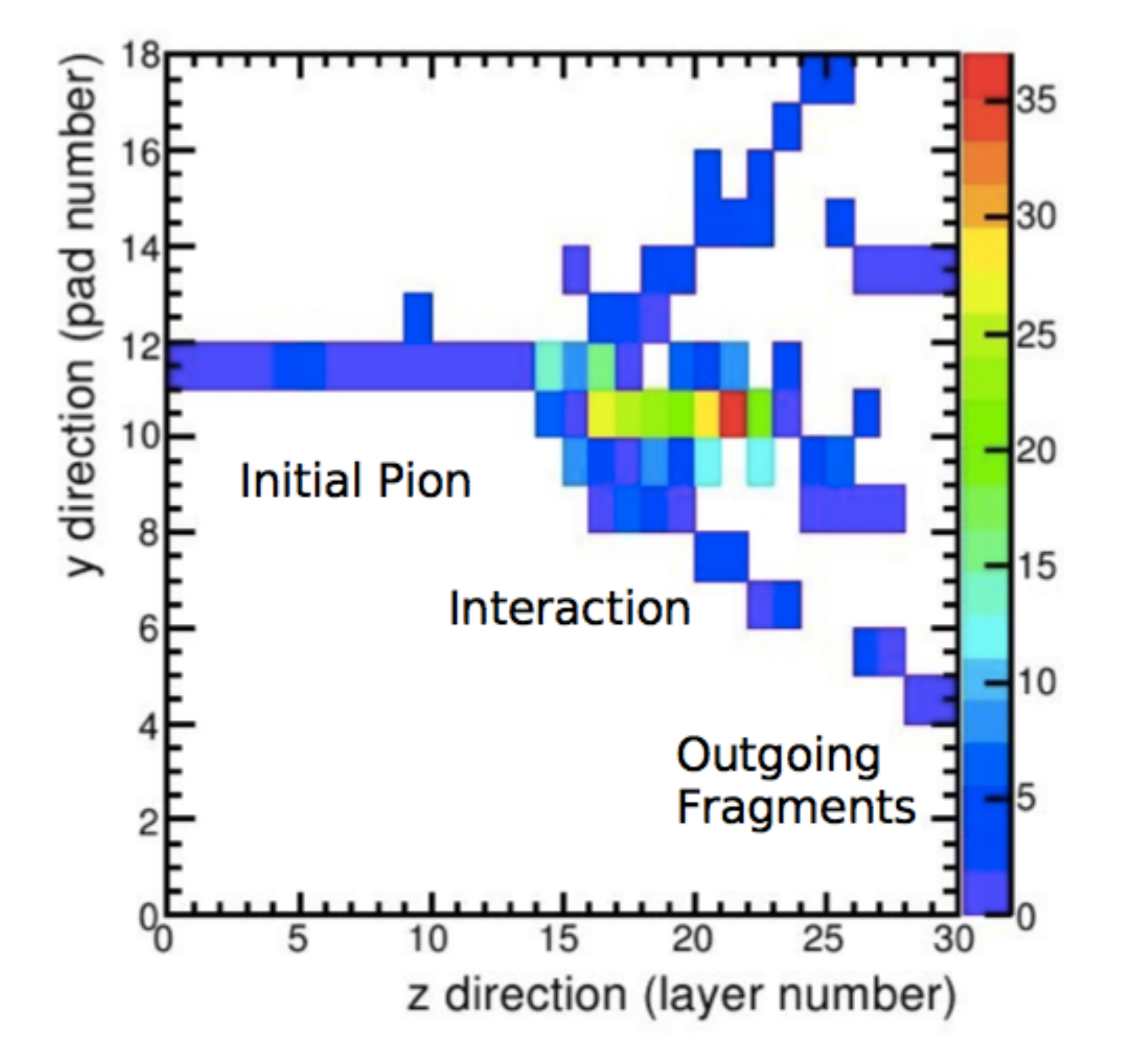}
\includegraphics[width=0.32\textwidth]{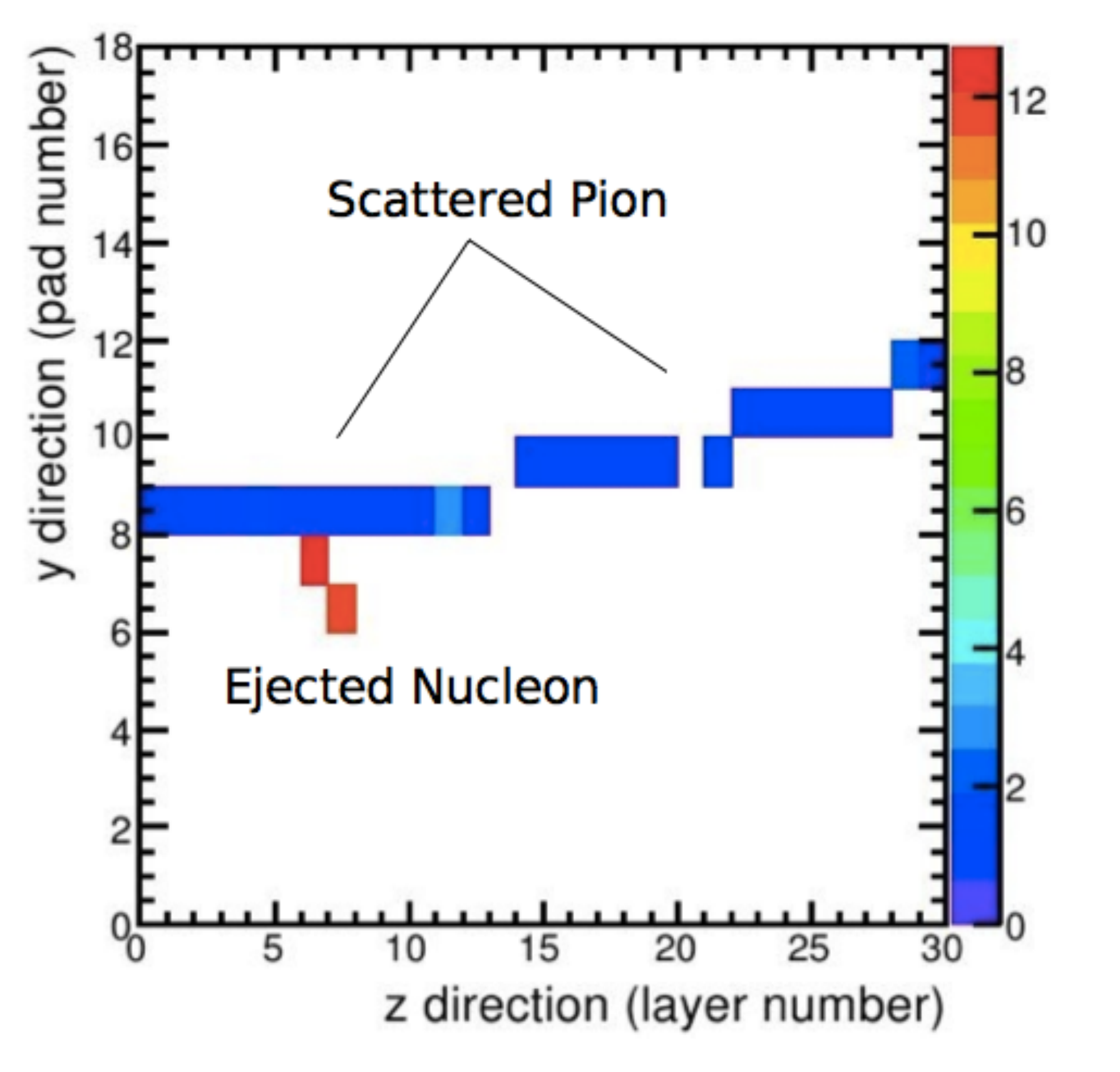}
\caption{Top: Inelastic reaction of a primary hadron in the SiW ECAL with a sizeable  release of energy and secondaries. Bottom: Inelastic reaction leading only to a local sparse deposition. From~\cite{Adloff:2010xj}.}
\label{fig:hadreacex}
\end{figure}

Exploiting the longitudinal granularity of the calorimeter, the interaction point of interactions of primary hadrons can be identified by 
comparing energy depositions and hit densities in subsequent detector layers. The correlation between the reconstructed and
the true interaction point, as available from Monte Carlo simulation, is shown in the top part of Fig.~\ref{fig:hadreac} for pions of an energy of 20\,GeV. A good correlation is seen.
Interaction layers as found in data and in Monte Carlo simulation are compared in the bottom part of Fig.~\ref{fig:hadreac}. Here, the  QGSP\_BERT physics list is used. The good agreement found in this case holds also when testing other Geant4 physics lists. The efficiency to find the interaction point correctly within two layers  is around 80\% for energies above 10\,GeV and still as high as 60\% for energies at 2\,GeV. At high energies the interaction point can be determined from an absolute energy increase after the
interaction of the primary pion with the absorber material. Due to shower fluctuations this absolute increase is less prominent at small energies. However, the high longitudinal granularity allows for measuring in addition the relative increase of the deposited energy and for determining the interaction point by that means. Examples of the two scenarios are shown in Fig.~\ref{fig:hadreacex}.

\begin{figure}[htb]
\centering \includegraphics[width=0.4\textwidth]{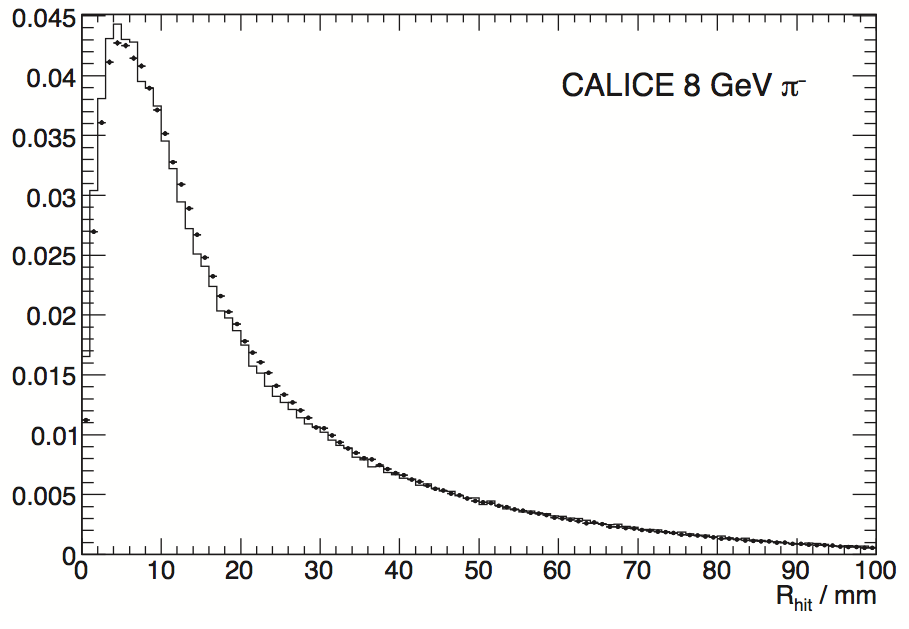}
           \includegraphics[width=0.4\textwidth]{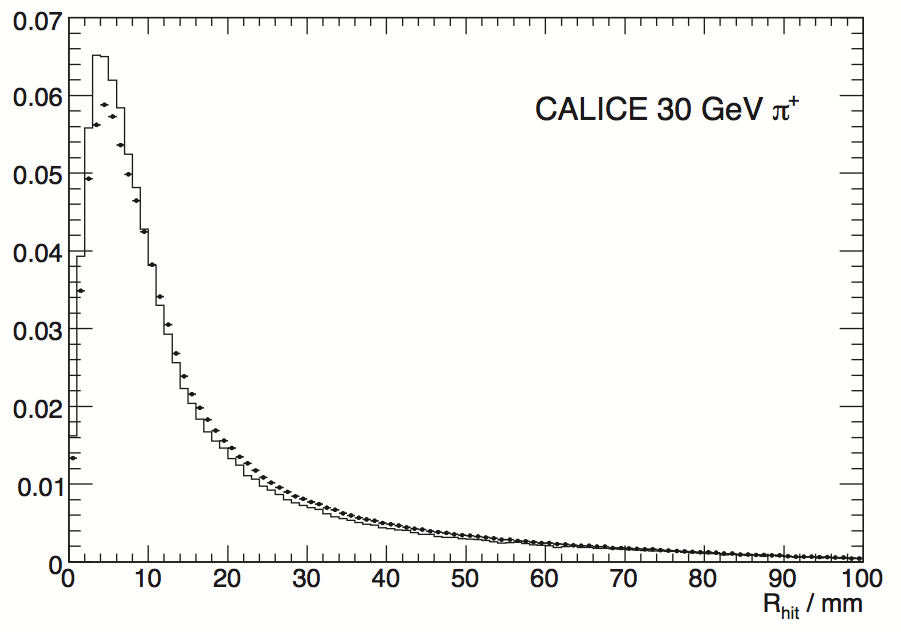}
\caption{\label{fig:Trans_QGSP_BERT} Radial distribution of hits (energy weighted) for SiW ECAL data at
energies of 8 and 30\,GeV (points with errors) compared with Monte Carlo (solid histograms)
 using the QGSP\_BERT physics list. The distributions are normalised to unity. From~\cite{Adloff:2010xj}.}
\end{figure}

The amount of overlap of showers generated by close by particles is governed by the transverse shower radius of the particles. 
In turn it is easy to understand that the amount of overlap influences directly the precision which can be achieved by PFA 
algorithms. It is thus of major importance that the transverse properties of a hadronic shower are well modelled by Monte Carlo 
simulations. Fig.~\ref{fig:Trans_QGSP_BERT} shows the comparison of the transverse shower profile for 8\,GeV and 30\,GeV 
pions incident on the calorimeter surface. Again, the default physics list is chosen to be 
QGSP\_BERT. 

\begin{figure}[htb]
\centering \includegraphics[width=0.45\textwidth]{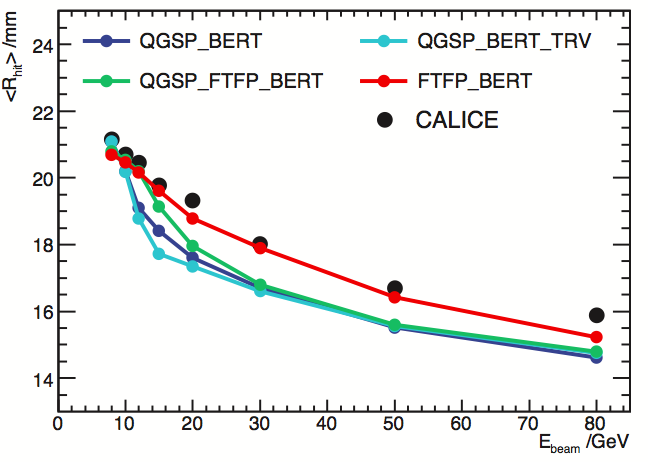}
\caption{\label{fig:TransMean} Mean energy-weighted shower radius in the SiW ECAL as a function of beam energy.
The data are compared with the predictions of simulations using different Geant4 physics lists. 
From~\cite{Adloff:2010xj}.}
\end{figure}

The mean values of the introduced transverse profiles for several physics lists and energies are given in Fig.~\ref{fig:TransMean}. 
From here and from the earlier result it can be concluded that those models which
implement the Bertini cascade~\cite{Guthrie:1968ue} give a good description at low energies. At high energies however, all models predict smaller
shower radii than observed.

\begin{figure}[htb]
\centering \hspace{2mm} \includegraphics[width=0.305\textwidth]{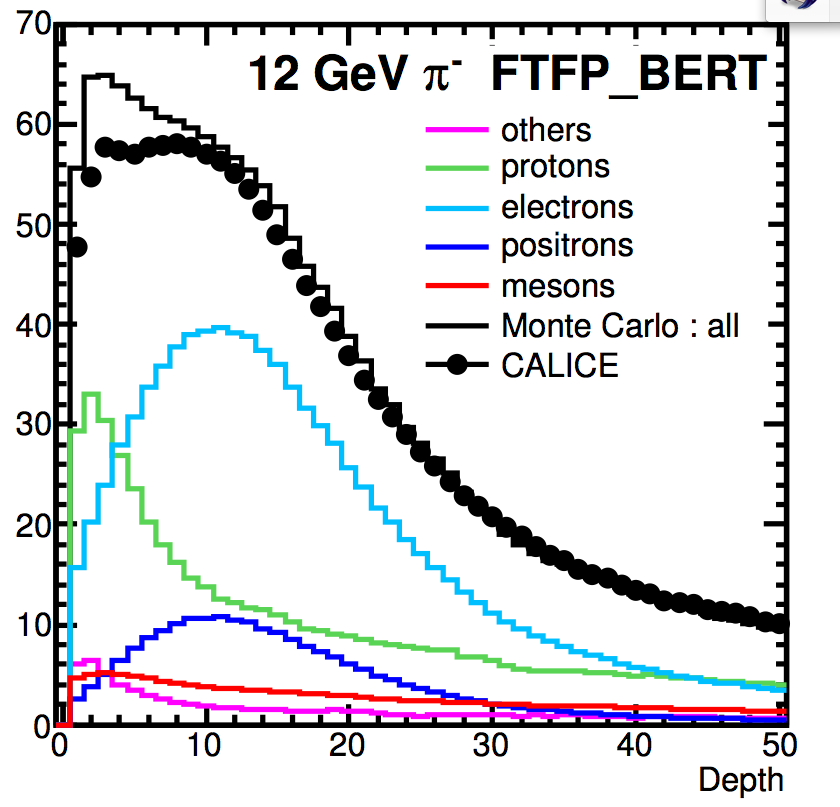}
\centering  \includegraphics[width=0.305\textwidth]{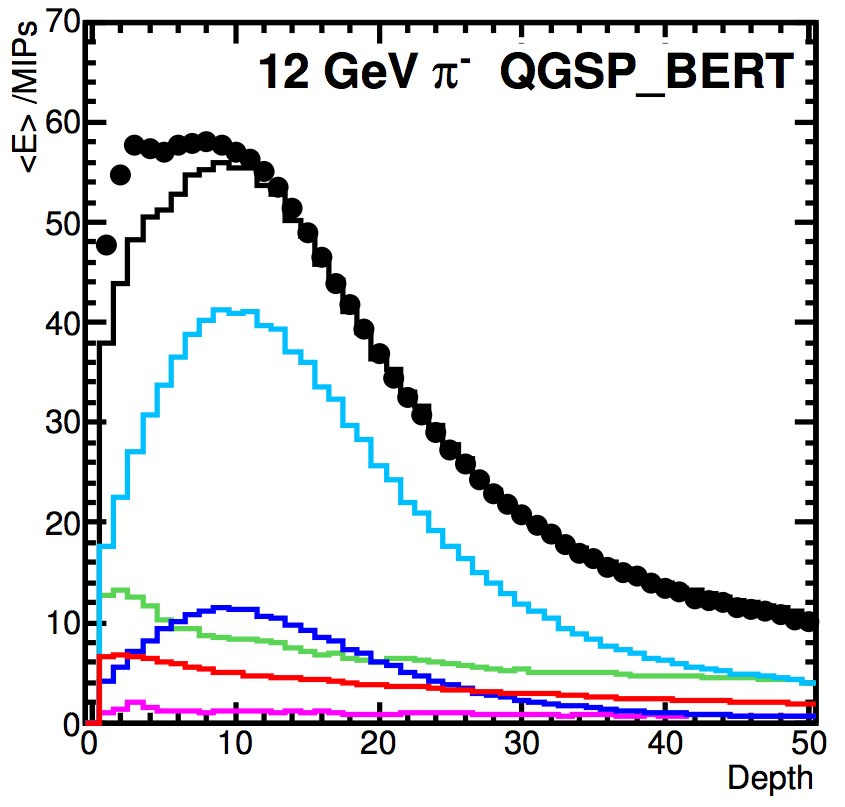}
\caption{\label{fig:Long12} SiW ECAL longitudinal energy profiles for 12~GeV $\pi^-$
data (shown as points), compared with simulations using two physics lists.
The mean energy in MIPs is plotted against the depth after the initial interaction, in units of effective 1.4~mm
tungsten layers. The total depth shown corresponds to $\sim20\;X_0$ or $0.8\;\lambda_{\mathrm{int.}}$.
The breakdown of the Monte Carlo into the energy deposited by different particle categories is also indicated. From~\cite{Adloff:2010xj}.}
\end{figure}

The longitudinal profile is composed of contributions from several components. Fig.~\ref{fig:Long12} shows the longitudinal profile
as a function of the shower depth after the interaction point for pions with an energy of 12\,GeV compared with the 
predictions obtained by the QGSP\_BERT and FTFP\_BERT physics lists. 

\begin{figure}[htb]
\centering \includegraphics[width=0.4\textwidth]{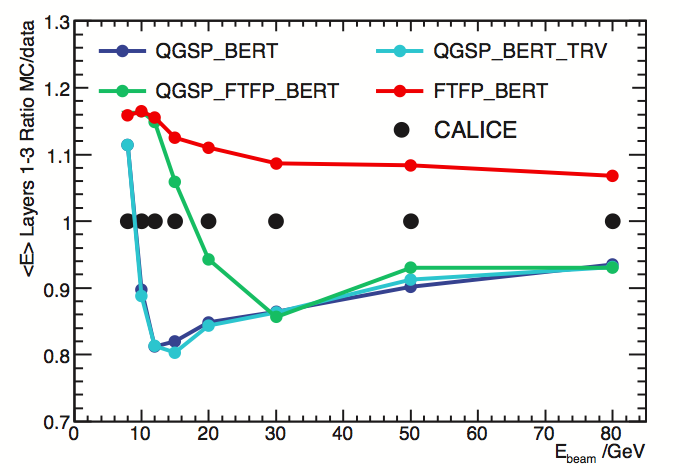}
\includegraphics[width=0.4\textwidth]{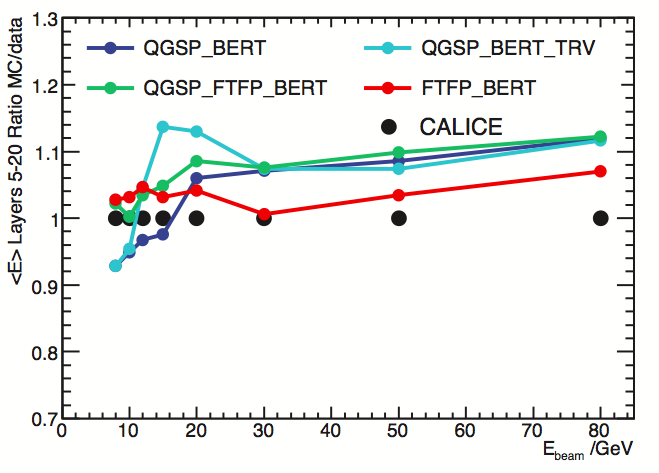}
\includegraphics[width=0.4\textwidth]{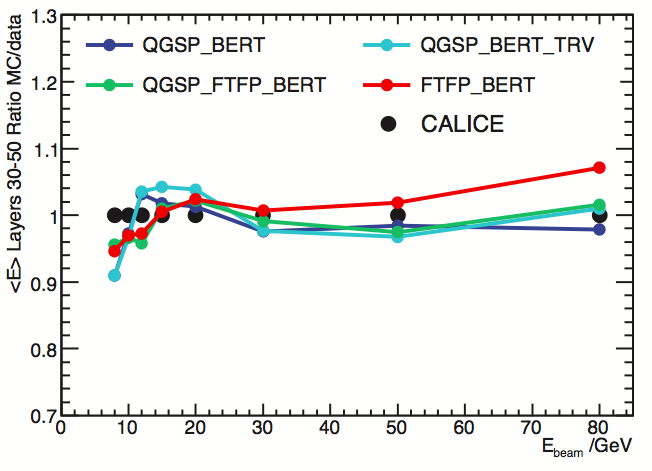}
\caption{\label{fig:LayerRatios} Ratio of simulation to data for
three different regions of the SiW ECAL longitudinal energy profile for pions: layers 1-3,
dominated by nuclear breakup (top); layers 5-20, dominated by electromagnetic showers (centre);
and layers 30-50, dominated by penetrating hadrons (bottom). From~\cite{Adloff:2010xj}.}
\end{figure}

The profile of the Monte Carlo prediction is broken down into its various components. It is clearly visible that there are significant differences between the models and the data and
between the models themselves. This is particularly true for short range components generated by heavily ionising particles
such as protons. This first separation into various domains of the shower is continued in the study presented in Fig.~\ref{fig:LayerRatios}.
Here the energy depositions in different layers after the interaction point are shown in greater detail. Again, the early stages
of the shower created by nuclear breakup are not well described by either of the models with discrepancies between 10\% and 20\%. The list FTFP\_BERT overshoots the data while both lists labeled as QGSP overshoot the data at smaller energies while they undershoot the data at higher energies.
For bigger shower depth FTFP\_BERT agrees with the data within 5\% and is around twice as good as the QGSP based models. 

%
\subsection{Hadron shower shapes in the scintillator AHCAL with steel and tungsten}

A large number of studies have been performed in order to test the simulation of showers in the scintillator AHCAL, in as much detail as possible, using data taken with steel and tungsten absorbers, with pions, protons and kaons. 
Both classical and novel observables were studied, and it is impossible to give here a full account of how well each model describes every given quantity or distribution.  
The purpose of the present discussion is rather to give an overview of which features have been the subject of validation studies, and to provide an overall picture of the level of adequacy of the simulations.  

Events have been selected to provide pure samples of hadronic showers, starting in the AHCAL,\ i.e. leaving only minimum ionising traces in the ECAL, if that was installed upstream. 
Electron, muon and anti-proton contaminations to the selected negative pion sample can be neglected.

The detector simulation had been validated with electron data as described in 
Section~\ref{sec:Perf}. 
For the simulation of hadron showers, additional effects become important. 
Low energy neutrons can travel for long times, up to many microseconds, before they interact and possibly produce a signal in the detector. 
It depends on the read-out technology and timing characteristics of its electronics, whether this actually contributes to the measured energy. 
In the case of the AHCAL, a cut of 150~ns was applied according to the shaping time of the pre-amplifiers. 
In addition, low energy protons, as produced by elastic scattering of slow neutrons in hydrogenous material, can in principle have very high local ionisation energy loss. 
However, in scintillator materials shielding effects lead to a saturation of  light production. 
This is empirically described by Birks' law~\cite{Birks:1964zz}, which is implemented in the Geant4 simulation. 
Both effects have been found to be important for the proper description of the hadronic response of the AHCAL. 
If not taken into account, the signal contribution from neutrons is overestimated, which results in too wide radial profiles and too large delayed signals.  
For silicon or gas, which are much less sensitive to neutrons, this is less critical. 

\begin{figure}[tb]
  \includegraphics[width=0.7\hsize]{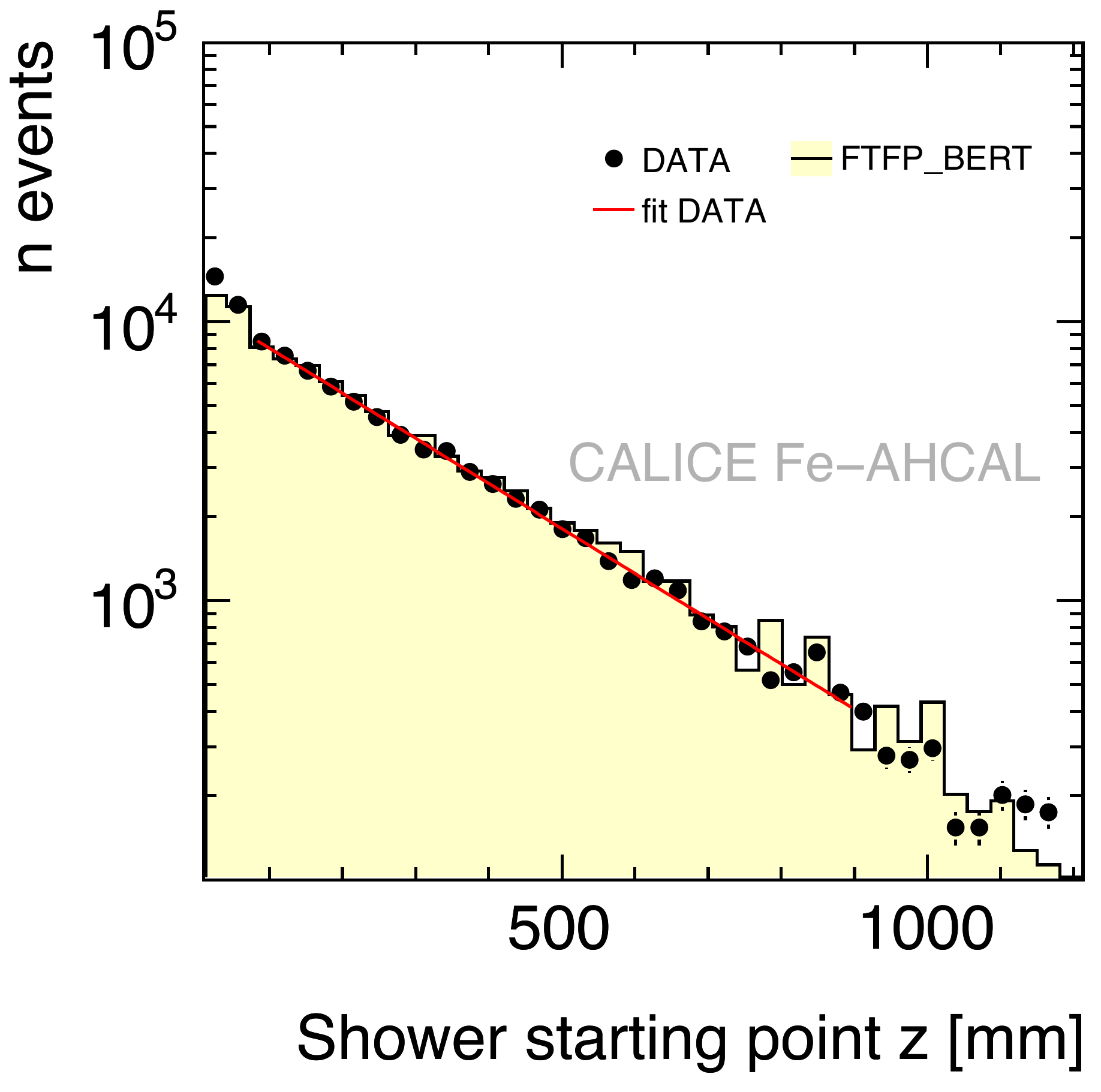}
  \includegraphics[width=0.7\hsize]{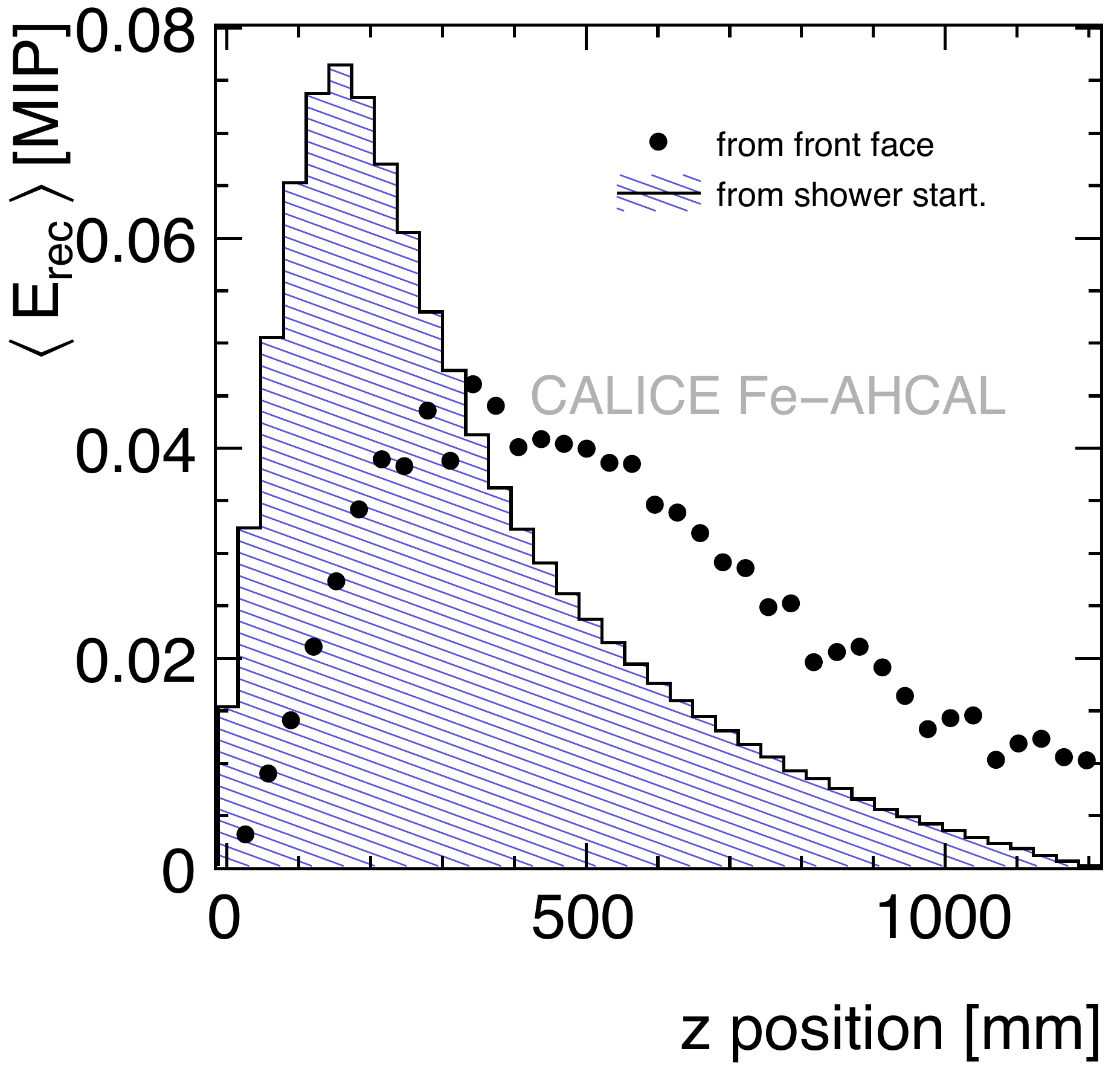}
\caption{\label{fig:G4v:longdeconv} Top: Distribution of the measured layer of the first hard interaction  in data and simulation, both normalised to the same arbitrary number of events. Bottom: Longitudinal profiles as a function of distance from the AHCAL front face and from the reconstructed shower start point, both normalised to unity. The figures are for 45~GeV pions.
From~\cite{Adloff:2013mns}.}
\end{figure}

For the analysis of longitudinal profiles in the steel AHCAL~\cite{Adloff:2013mns}, the capability to reconstruct the shower start from the three-dimensional hit distribution has been used to de-convolute the distribution of average energy per layer into a spectrum of depth of the first hard interaction, and a profile measured from this first interaction point. 
This provides a cross-check of the material composition and nuclear absorption properties in the simulation, and a profile which is more directly sensitive to the physics processes at different stages of the shower evolution, as already seen in the silicon tungsten ECAL discussed above. 
An example is shown in Fig.~\ref{fig:G4v:longdeconv}.

From a fit of an exponential function to the distribution of the starting points, the effective pion interaction length has been extracted. 
As expected, there is no visible dependence on energy in the range from 8 to 80~GeV. 
The average is $\lambda_{\pi} = (26.8 \pm 0.46)$~cm, where the error includes statistical effects as well as systematic variations due to the choice of the fit range. 
The result is in good agreement with expectations based on the material composition of the detector as used in the simulation, which gives 28~cm, and also provides a cross-check of the starting point reconstruction algorithm.
It is also well reproduced by recent physics lists, within 4-6\%, but not by the outdated LHEP parameterisation. 

The comparison of the shower profiles in Fig.~\ref{fig:G4v:longdeconv} shows that the profile measured from the shower start is not only more compact, but also smoother, since layer-to-layer variations in the detector response are averaged out. 
These variations are due to different numbers of channels with bad connections or failed calibrations, which are not corrected for, but included in the simulation. 
The comparison with simulation models is made with the distributions from the shower start, and 
systematic uncertainties due to such detector effects are obtained for each bin by comparing the distributions for different, but fixed, starting points in the detector. 

\begin{figure*}[hbt]
\centerline{
\includegraphics[trim=0 0 10 10, clip, height=5cm]{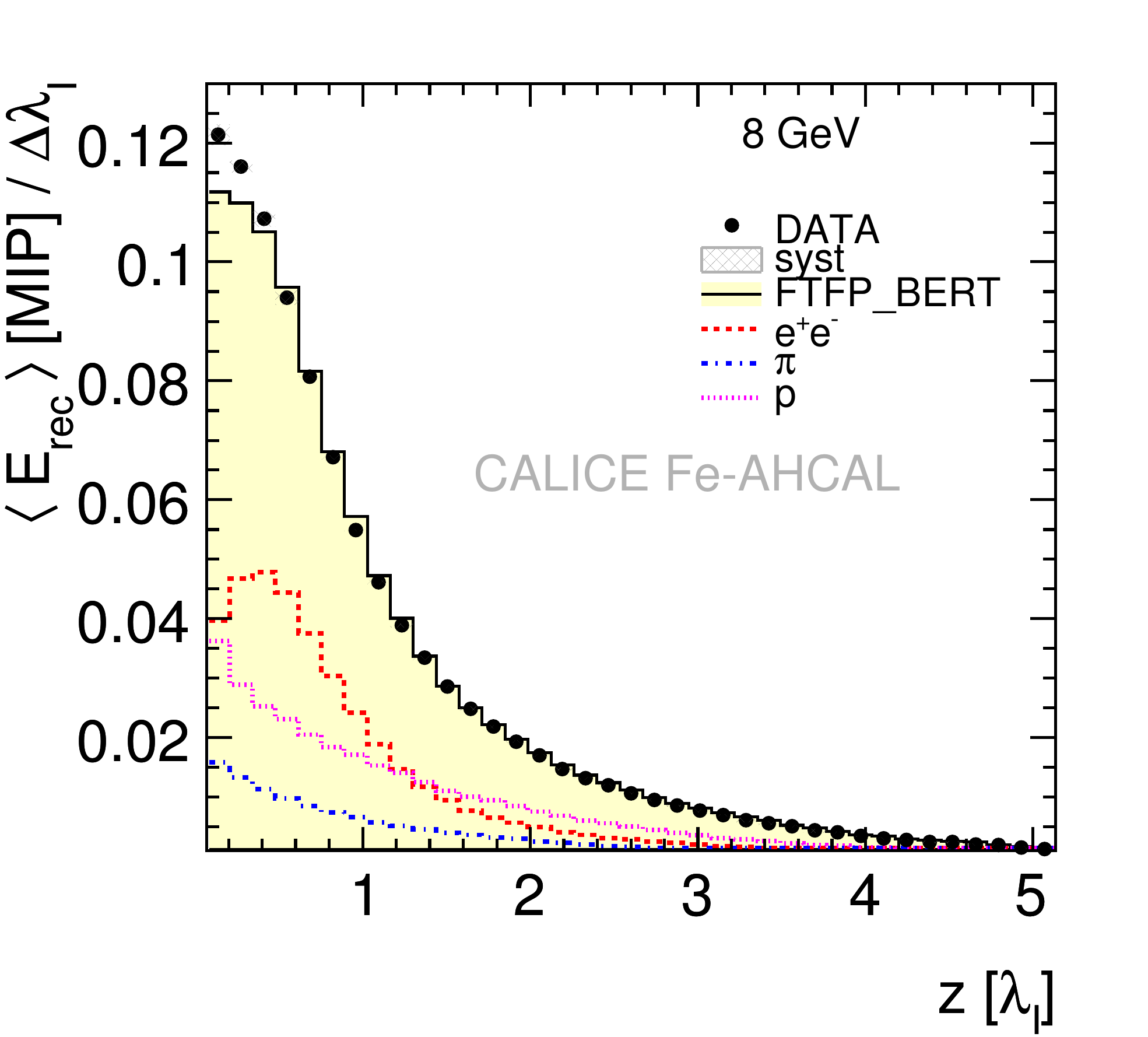}\hspace{-4.mm}
\includegraphics[trim=80 0 10 10, clip, height=5cm]{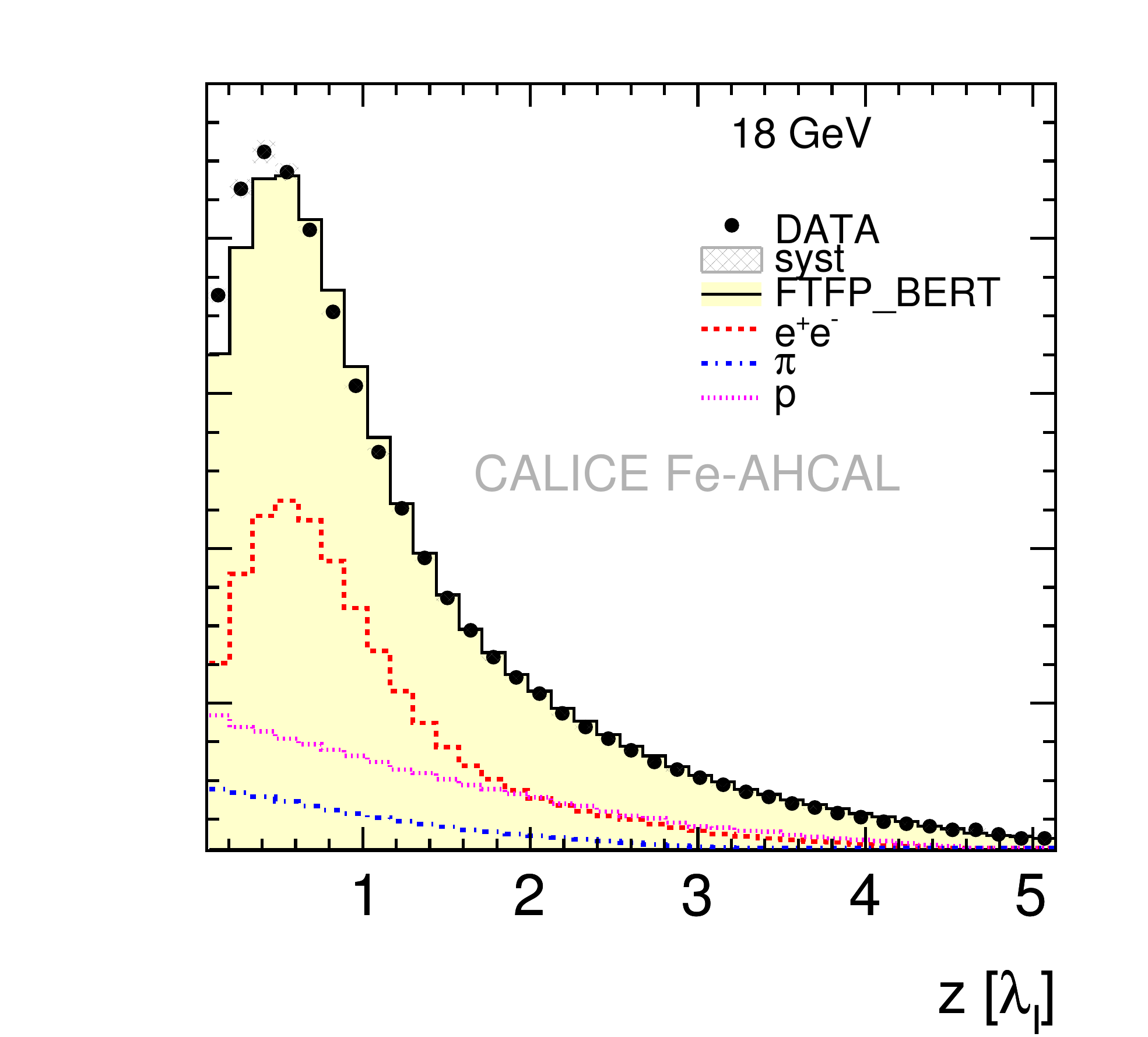}\hspace{-4.mm}
\includegraphics[trim=80 0 10 10, clip, height=5cm]{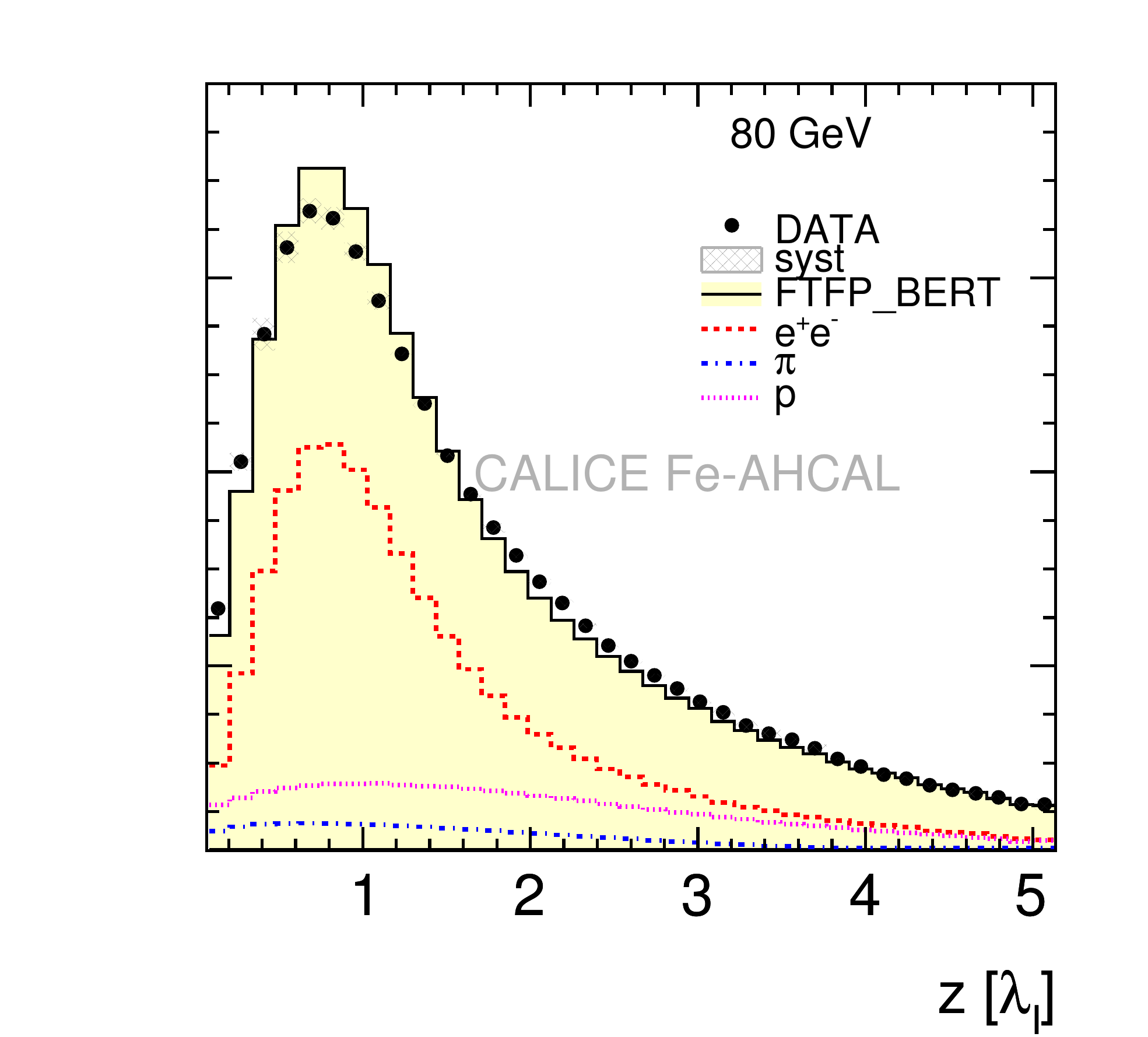}
}
\vspace{0mm}
\centerline{
\hspace{-1.5mm}
\includegraphics[trim=0 0 10 10, clip, height=5cm]{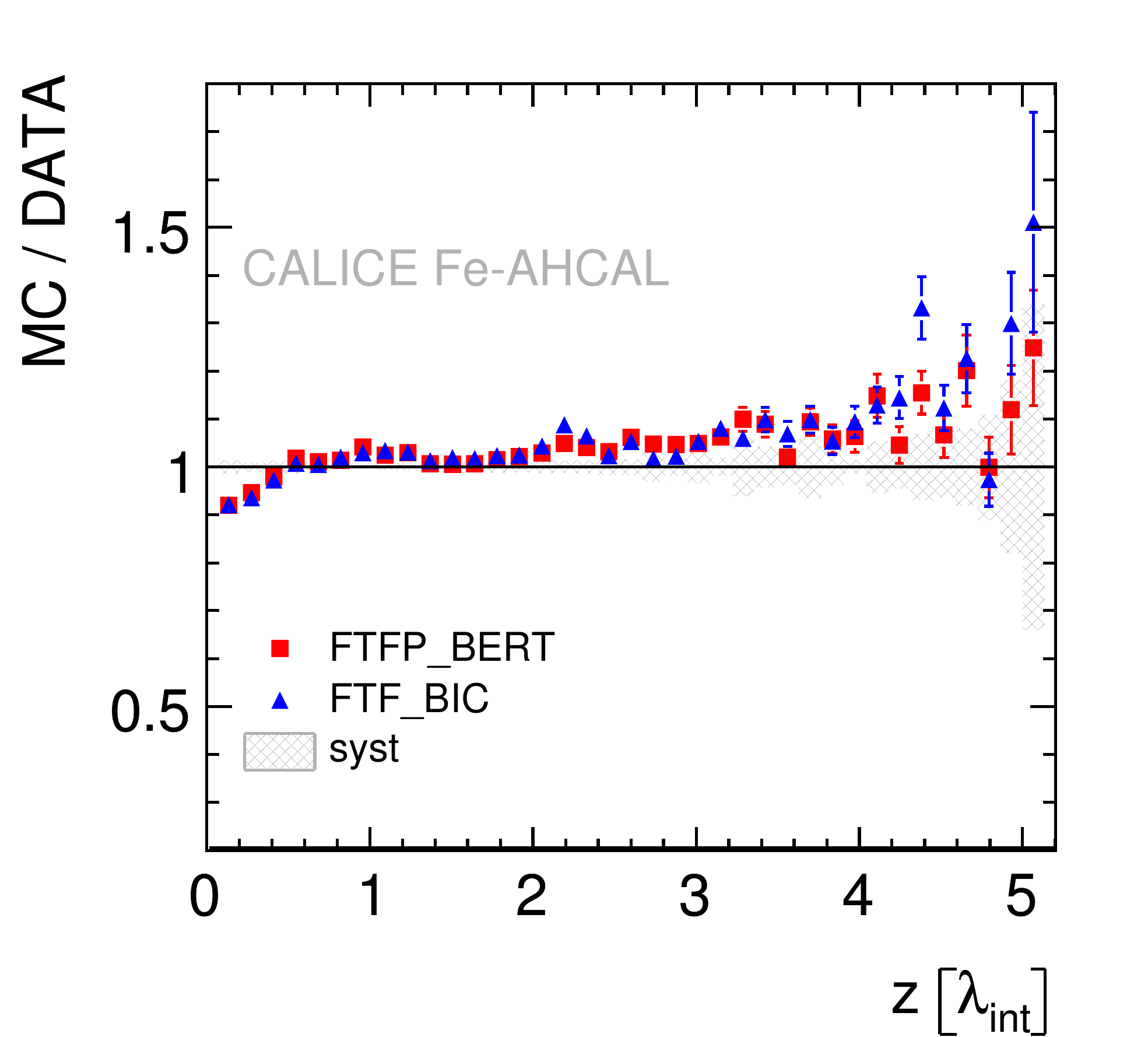}  \hspace{-5.mm}
\includegraphics[trim=80 0 10 10, clip, height=5cm]{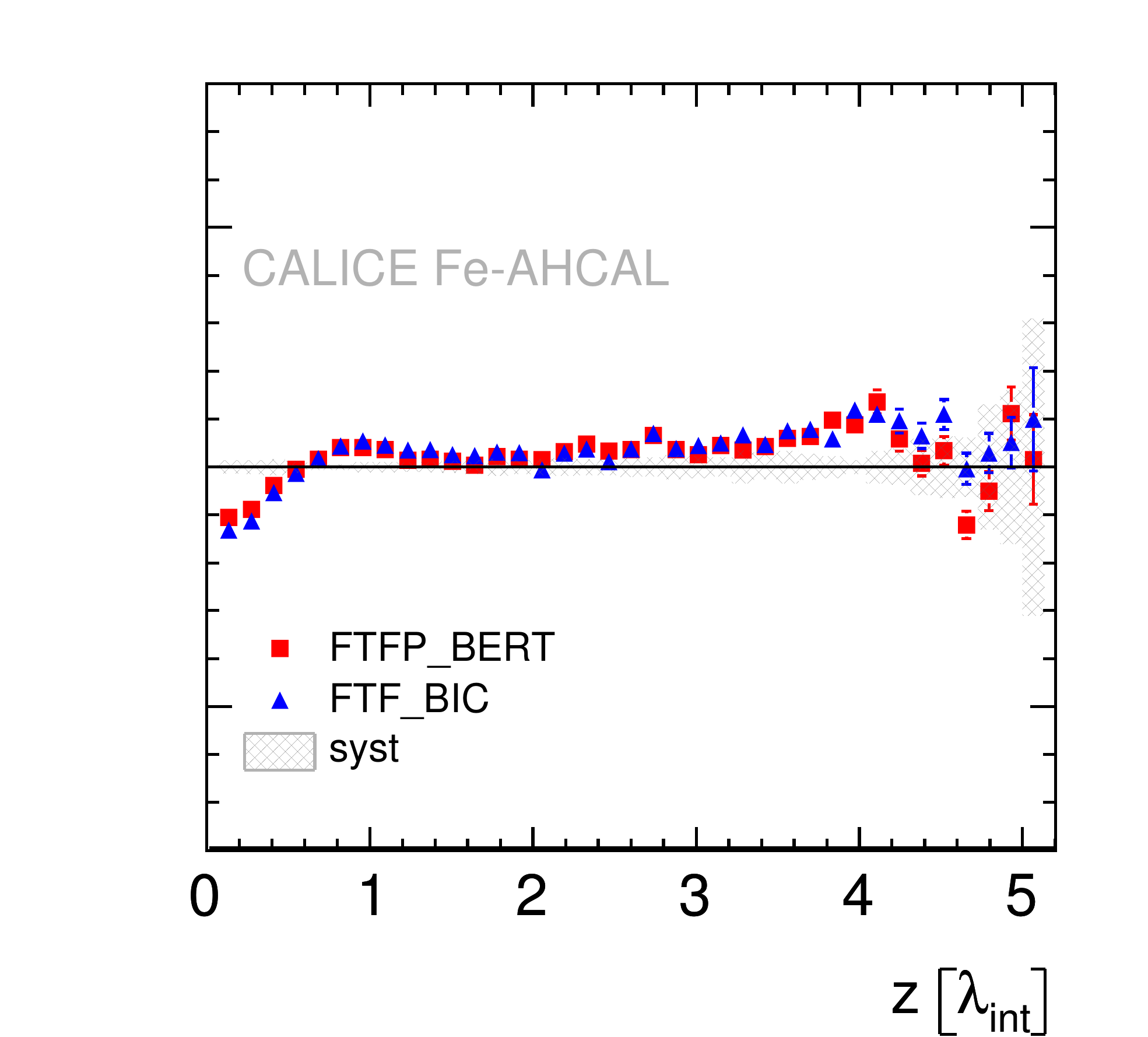}  \hspace{-5.3mm}
\includegraphics[trim=80 0 10 10, clip, height=5cm]{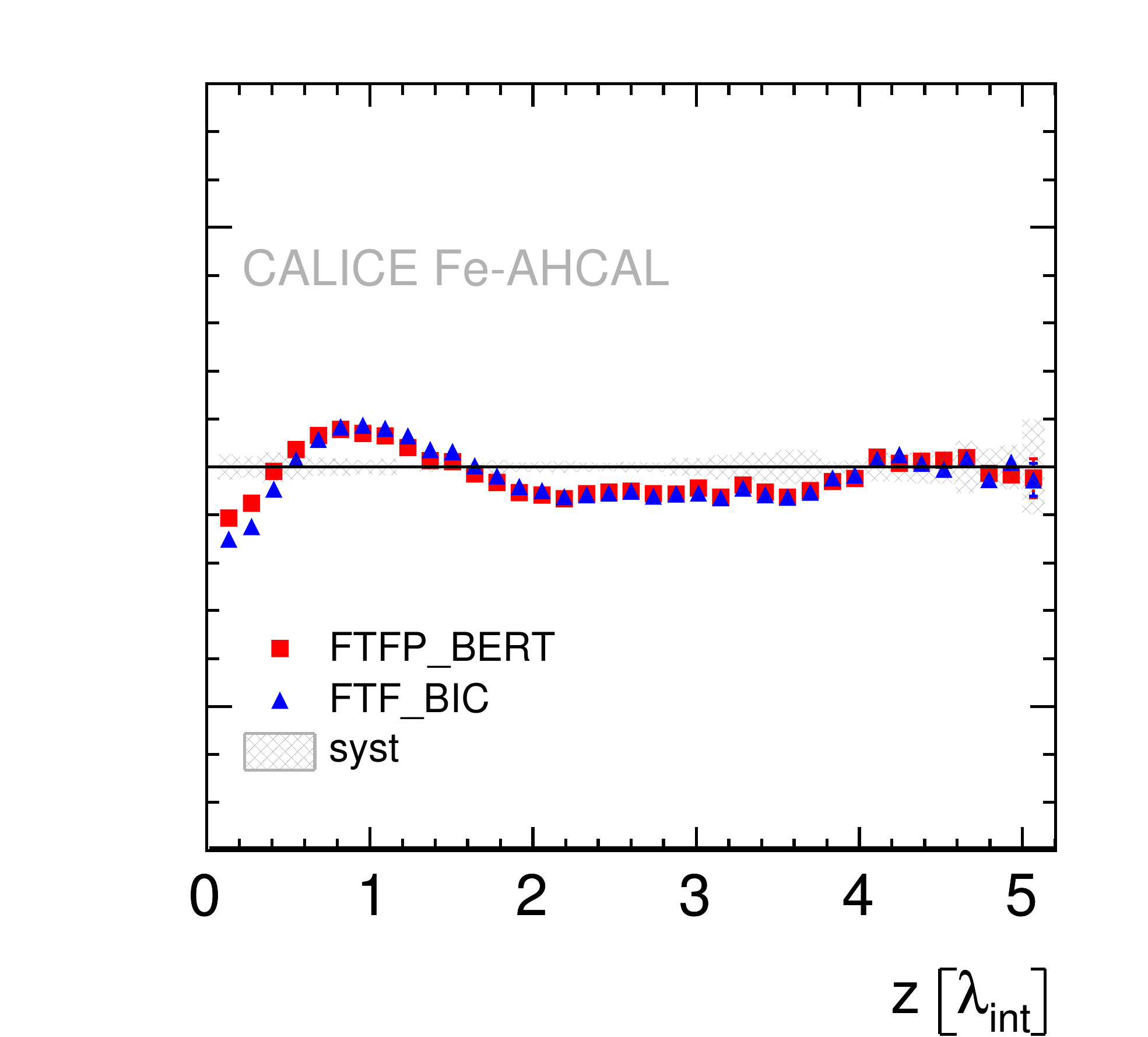}
\vspace{0mm}
}
\caption[Longitudinal shower profiles from shower starting point in data and Monte Carlo]{Mean longitudinal shower profiles from shower starting point for 8\,GeV (left column), 18\,GeV (center column) and 80\,GeV (right column) pions. First row: For data (circles) and for the \texttt{FTFP\_BERT} physics list (histogram). Second row: Ratio between Monte Carlo and data for several physics lists. All profiles are normalized to unity. The grey area indicates the systematic uncertainty on data. $\langle \mathrm{E}_{\mathrm{rec}} \rangle / \Delta \lambda_{\mathrm{I}}$ is the average deposited energy in a $\Delta \lambda_{\mathrm{I}}$ thick transverse section of the calorimeter. $\mathrm{z}$ is the longitudinal coordinate, expressed in units of $\lambda_\mathrm{I}$.
From~\cite{Adloff:2013mns}.}
\label{fig:G4v:longprof}
\end{figure*}

The profiles for three different energies are shown in Fig.~\ref{fig:G4v:longprof} together with simulations and their decomposition. 
%
The method of decomposing the simulated shower energy depositions is described in~\cite{Kaplan:2011uma}. 
The electromagnetic component dominates the shower maximum, while hadronic and electromagnetic depositions are comparable in magnitude in the start and in the tail of the shower. 
At 8 and 18~GeV, the profiles are described within a few per-cent, whereas at 80~GeV, the simulated showers have a somewhat more pronounced maximum. 
This is very similar for the QGSP and QBBC physics lists, too. 

\begin{figure}[tb]
\includegraphics[trim=0 0 45 0, clip, width=0.6\hsize]{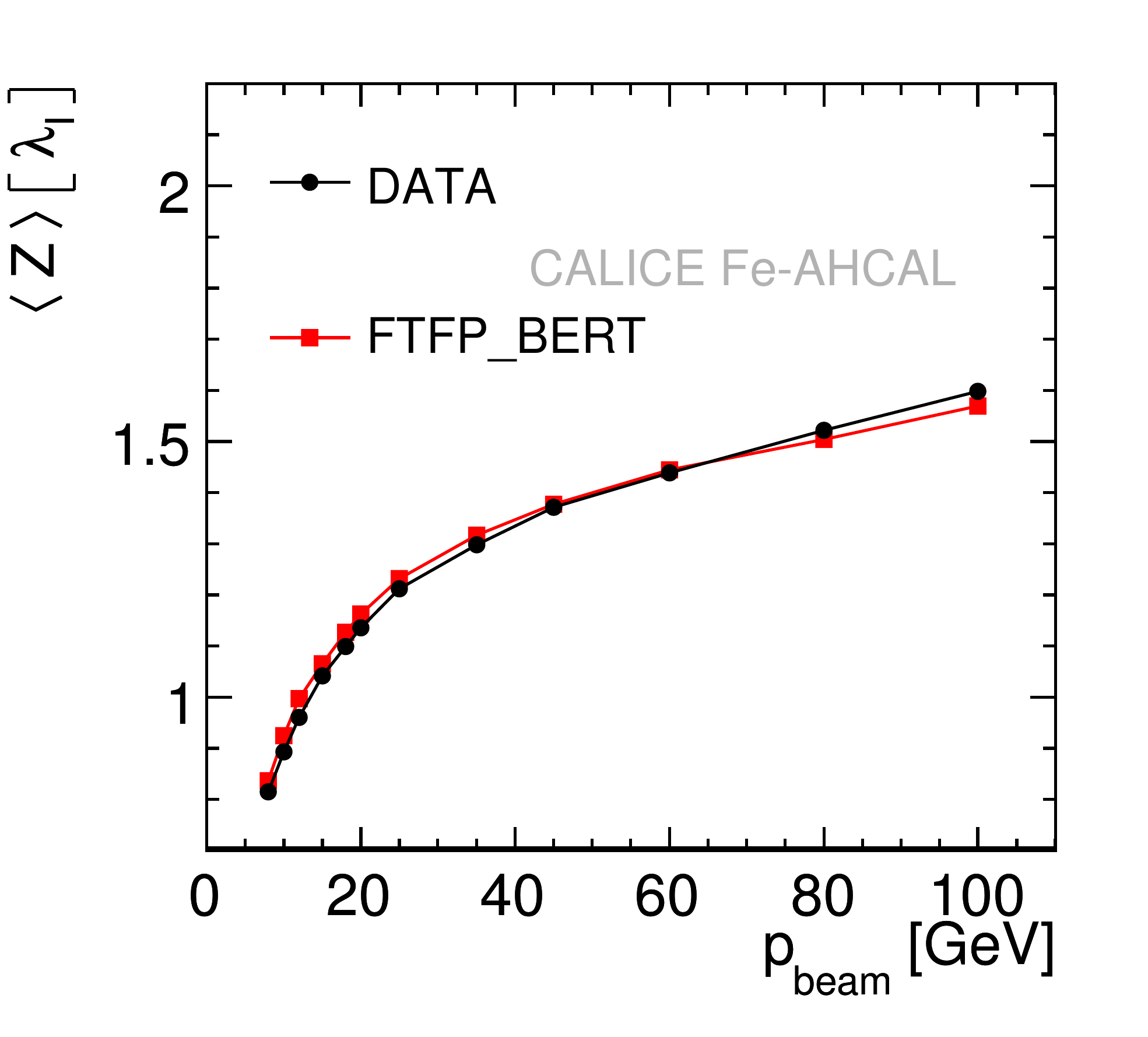} 
\includegraphics[trim=0 0 45 0, clip, width=0.6\hsize]{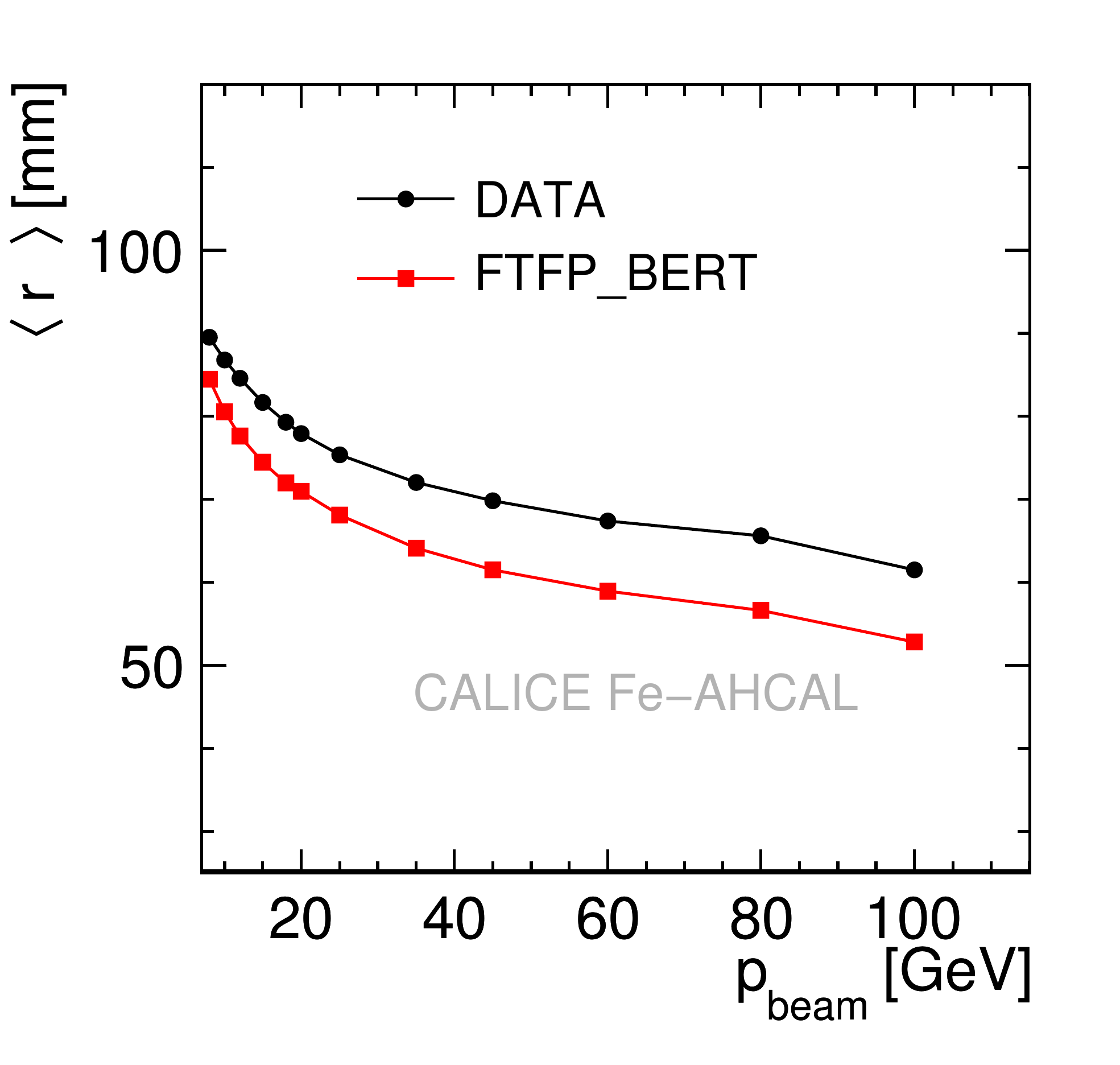}  
\caption{\label{fig:G4v:zr-edep} Energy-weighted mean longitudinal position (top) and radius (bottom) of  pion showers as a function of beam momentum, for data and FTFP-BERT simulations. 
From~\cite{Adloff:2013mns}.}
\end{figure}

The position of the maximum, and the centre-of-gravity (energy-weighted mean longitudinal position of the shower), are well reproduced for all physics lists, as shown for one example in Fig.~\ref{fig:G4v:zr-edep}. 
The energy dependence of the energy-weighted mean shower radius is also shown, together with the FTFP-BERT prediction. 
The radial shower profile is less well reproduced: here the CALICE results confirm the findings of LHC test-beam experiments.   
All physics lists overestimate the core, which is dominated by electromagnetic depositions, and under-estimate the tail of the approximately exponentially falling radial distribution. 
The disagreement increases with energy and exceeds 10\% at the upper edge of the probed range. 
While this leaves room for improvement in the shower modelling, we like to recall that the systematic studies in~\cite{Thomson:2009rp} have shown that the particle flow performance is rather insensitive to the choice of the physics list from a rather broad set. 
%

The studies made with tungsten as AHCAL absorber are essential for the validation of the CLIC detector performance, but they are also relevant for the ILC calorimeters, since all electromagnetic sections are made from tungsten and will influence the development of the majority of hadronic showers, even if the HCAL itself is made from steel. 
However, the ECAL prototypes themselves are not large enough to measure the spatial extent of hadron showers. 
The interactions on the heavy and neutron-rich absorber nuclei produce a significantly larger number of neutrons than in steel, such that the shower composition is different, and the validation of simulations for steel cannot be straightforwardly extrapolated to tungsten, in particular not in the case of hydrogenous scintillator material sensitive to soft neutrons. 

\begin{figure}[tb]
  \includegraphics[width=0.6\hsize]{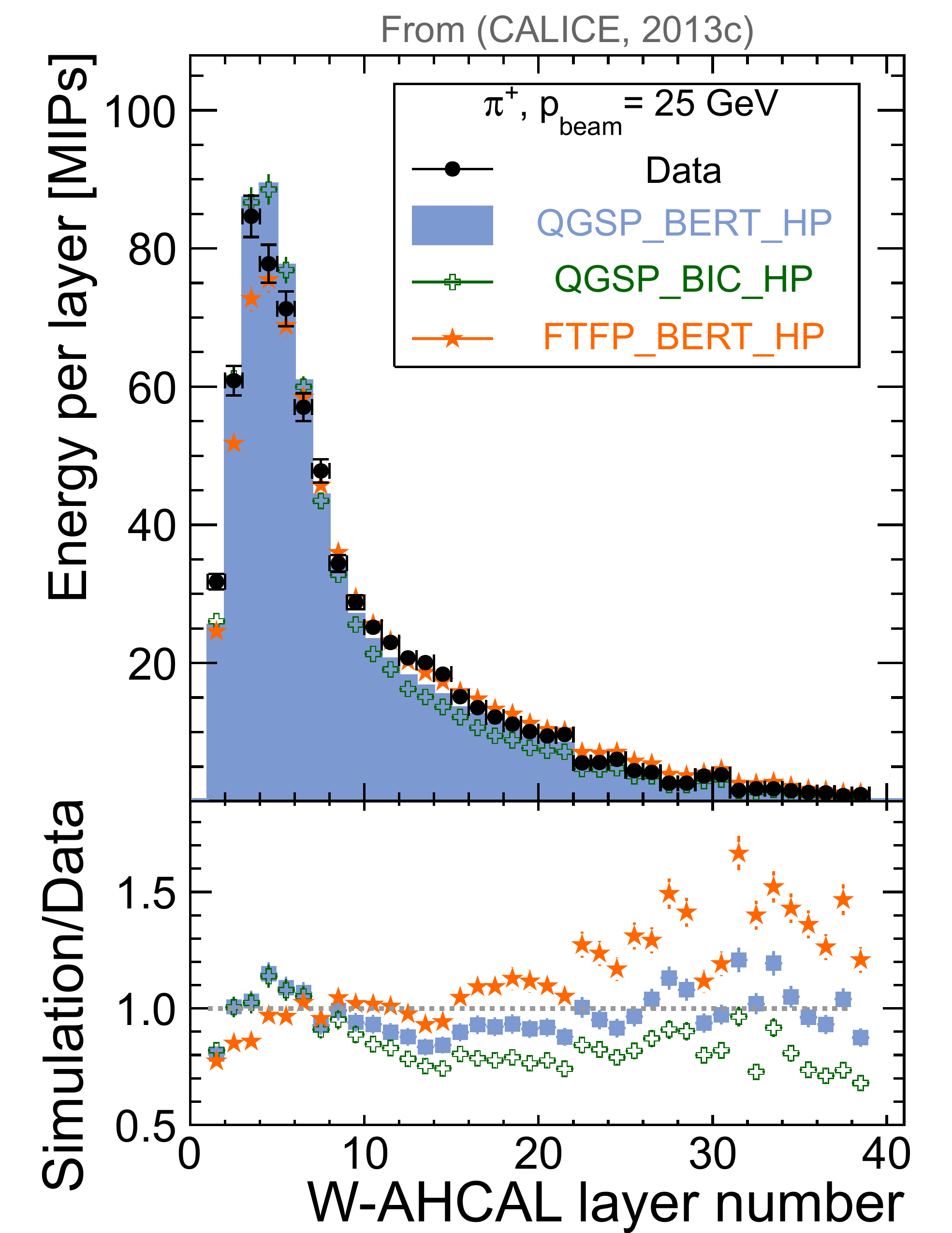}
  \includegraphics[width=0.6\hsize]{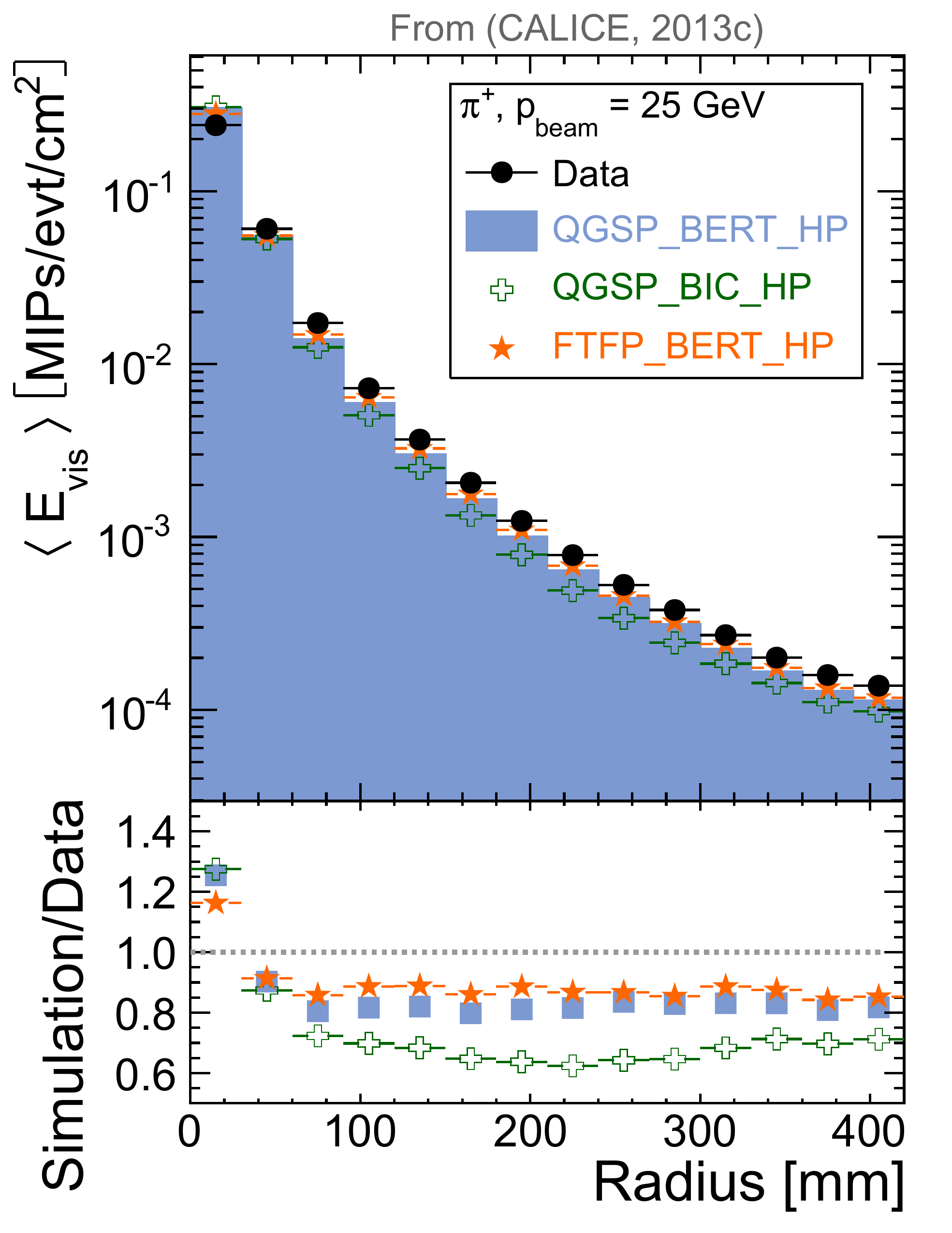}
\caption{\label{fig:G4v:Wprofiles} Energy-weighted longitudinal  (top) and radial (bottom)  pion shower profiles for W-AHCAL data and different Geant4 physics lists. From~\cite{CALICE:2013ff,Sicking:2013}.}
\end{figure}

The tungsten data, recorded below 10~GeV~\cite{Adloff:2013jqa} and between 10 and 100~GeV~\cite{CALICE:2013ff}, present a picture which is very similar to the steel data, in terms of model comparisons.
The simulations give somewhat too pronounced longitudinal and radial shower maxima for pions at high energies, but otherwise there is very good agreement, including the overall response and linearity; see also Section~\ref{sec:Perf}. 
Two example profiles are shown in Fig.~\ref{fig:G4v:Wprofiles}.
The top part can be compared with the 18~GeV data in Fig.~\ref{fig:G4v:longprof} . 
The longitudinal shower shape in tungsten is different from that in steel, even if displayed as a function of depth measured in units of $\lambda_I$, because the ratio of interaction length to radiation length $X_0$ is very different, 
about 25 for tungsten {\it versus} 10 for steel. 
Therefore the electromagnetic part of the shower is more compact than the hadronic, resulting in a more pronounced shower maximum and an overall somewhat shorter shower. 
It is interesting that on the other hand the mean shower radius, measured in mm, is very similar for both materials, which reflects the fact that the transverse shower scale is not governed by the nuclear cross sections but by other effects, which are  less dependent on the mass of the nucleus, like the angular distribution of the scattering products. 

\begin{figure}[tb]
  \includegraphics[width=0.7\hsize]{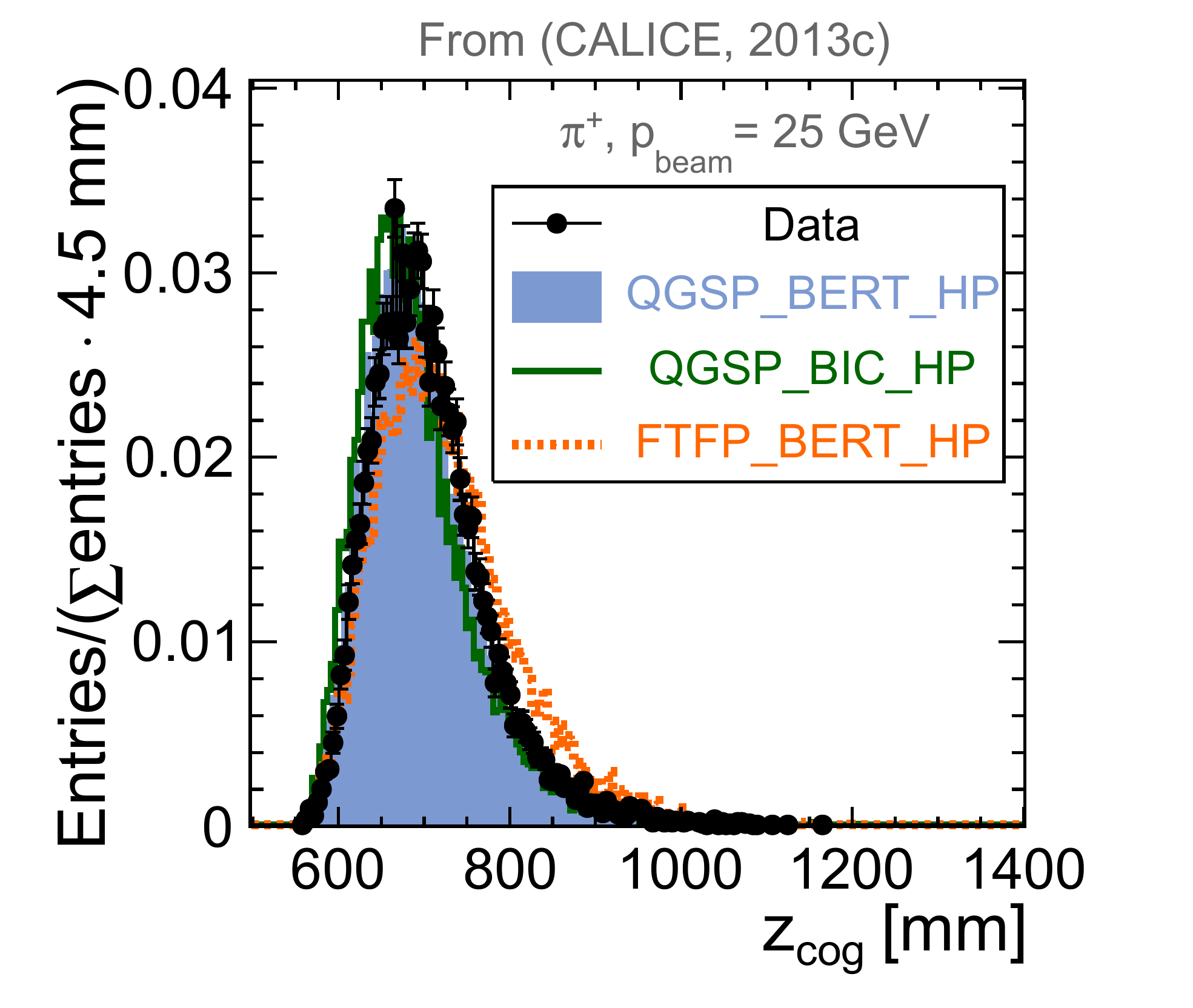}
  \includegraphics[width=0.7\hsize]{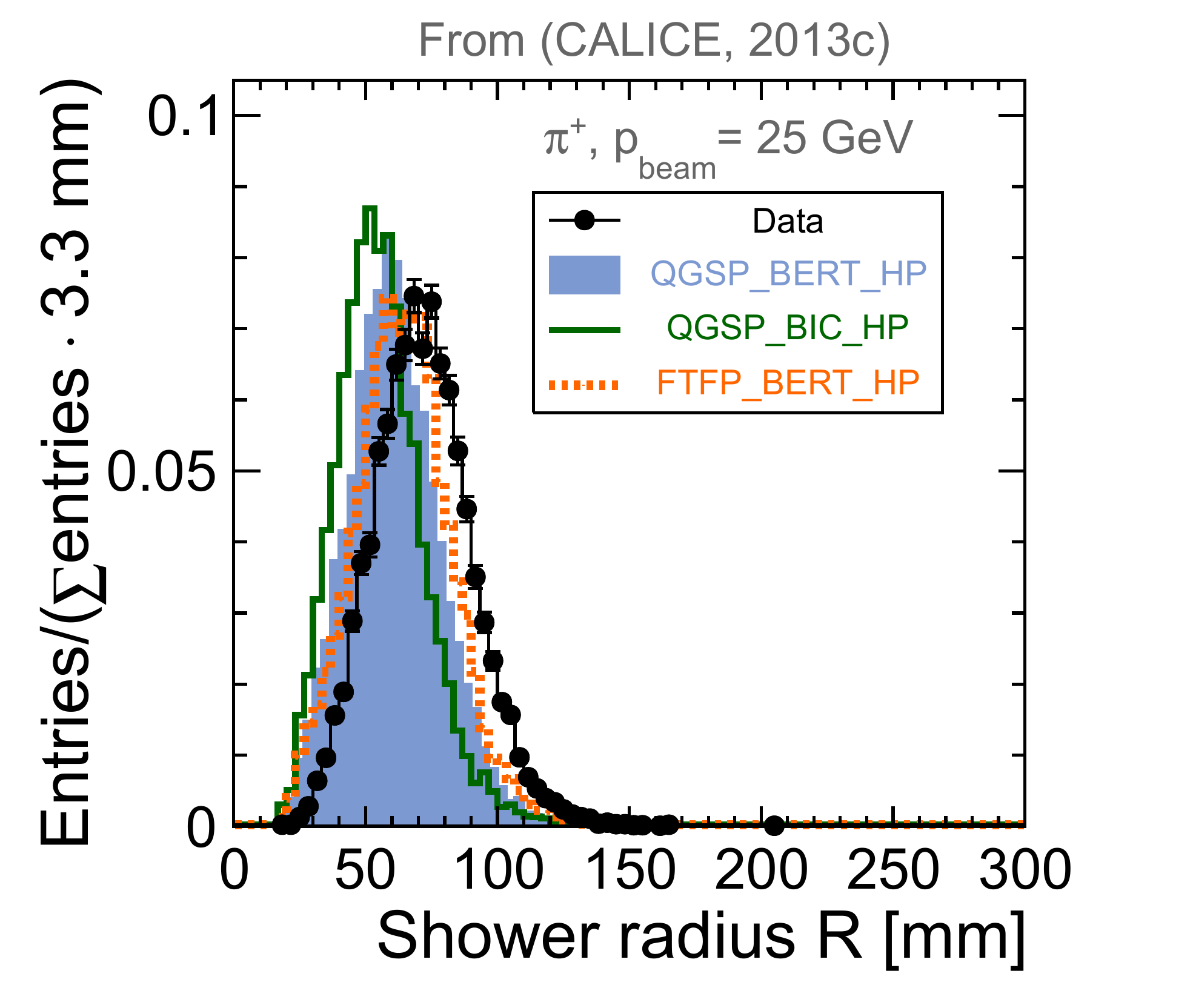}
\caption{\label{fig:G4v:zr-distr} Distribution of energy-weighted mean longitudinal position (top) and radius (bottom) of  pion showers for W-AHCAL data and different Geant4 physics lists.
From~\cite{CALICE:2013ff,Sicking:2013}.}
\end{figure}

For a proper modelling of particle flow performance it is important that not only the mean longitudinal and radial extent of the showers are reasonably well reproduced, but also the event-to-event (or shower-to-shower) fluctuations. 
As an example, Fig.~\ref{fig:G4v:zr-distr} shows the distribution of energy-weighted mean $z$ position and radius. As already stated, the mean radius is underestimated, but the widths of both distributions, characterising the typical event-to-event variation of these quantities, are well reproduced. 

In an ideal particle flow detector, the hadron calorimeter would measure the energies of only neutrons and K$^0_L$ mesons. 
Although in each shower many types of hadrons are present, they inherit some properties from the incident particle. 
For example, neutron -- or proton -- induced showers are expected to have a smaller electromagnetic component,  when compared to pions, since baryon number conversation reduces the phase space for $\pi^0$ production, and strangeness conservation plays a role in kaon induced showers. 
Moreover, in the case of protons only the kinetic energy is available for producing calorimeter signals, and for the same beam momentum this is smaller than the corresponding pion energy. 
Therefore, as neutral beams are not available, test beam data with tagged protons and charged kaons are of particular interest. 

\begin{figure}[tb]
  \includegraphics[width=0.67\hsize]{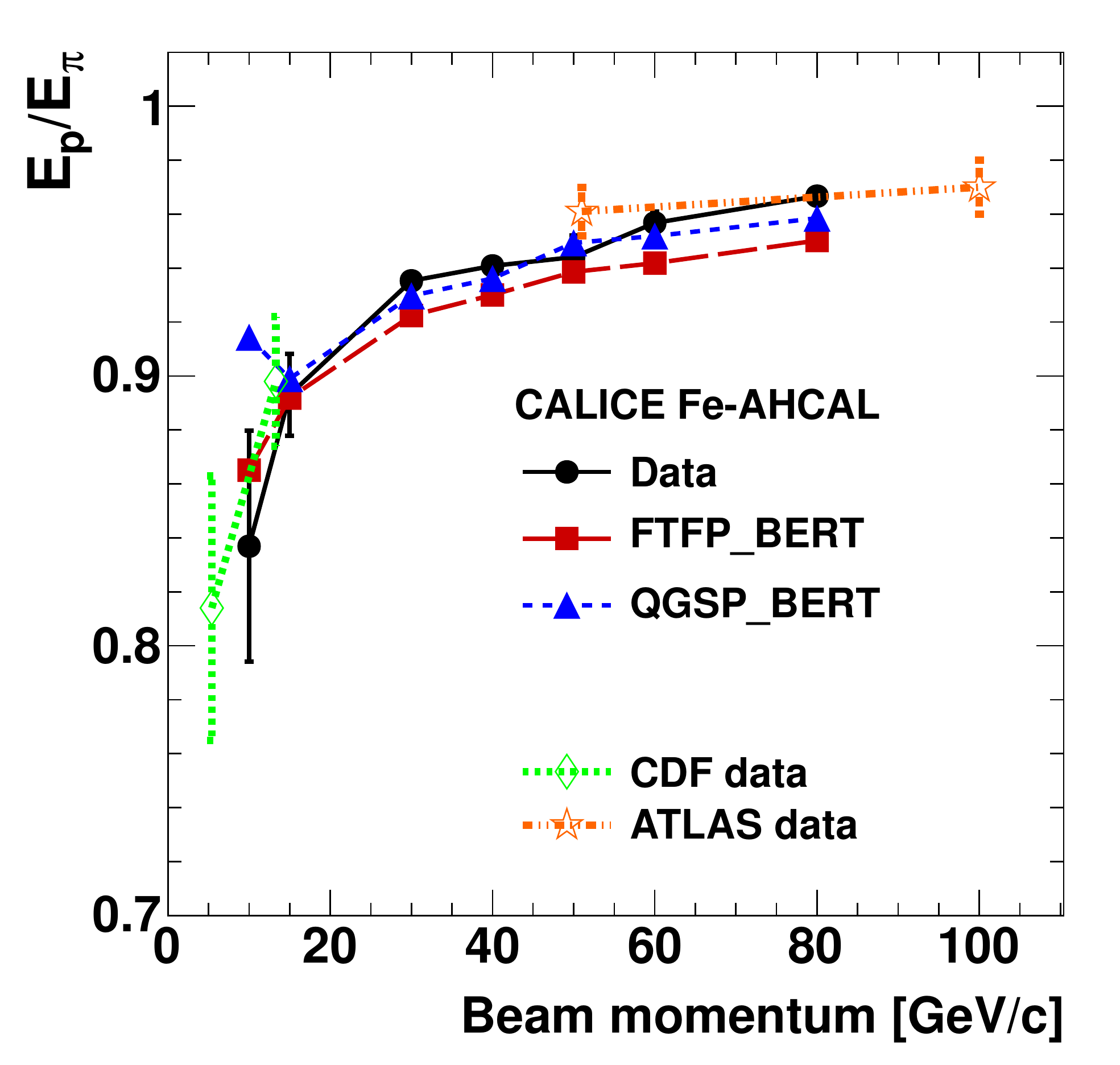}
  \includegraphics[width=0.8\hsize]{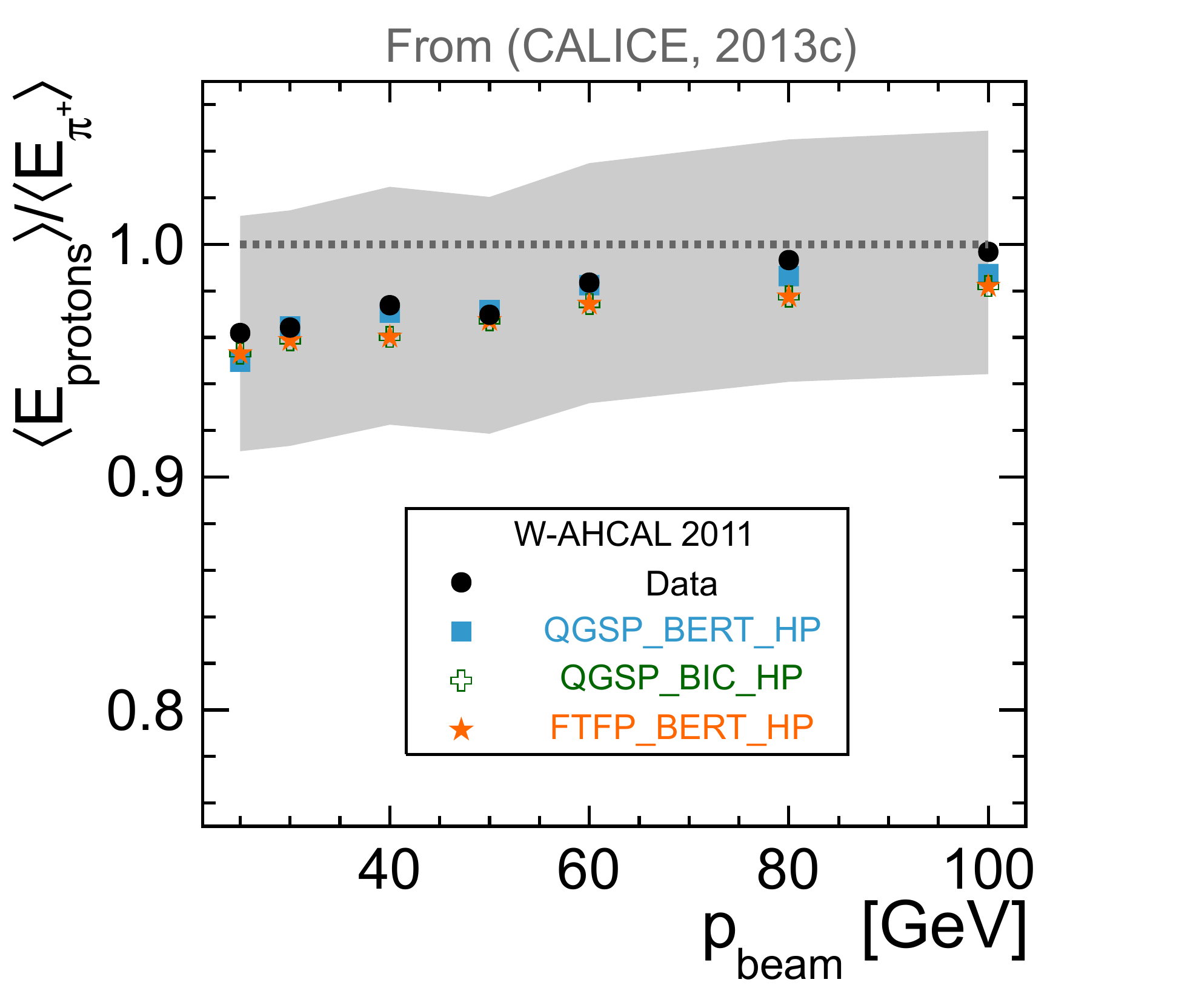}
\caption{\label{fig:G4v:protons} Ratio of AHCAL response to protons relative to pions, as a function of beam momentum, for steel (top) and tungsten (bottom) absorber.
From~\cite{Bilki:2014bga} (top), \cite{CALICE:2013ff,Sicking:2013} (bottom).}
\end{figure}

For the study of proton and kaon showers, single or differential \u{C}erenkov counter information was used.
For tungsten data, the purity  was better than 99\% for proton data at beam energies below 10~GeV, and better than 80\% for protons and kaons at higher energies. 
For the case of steel absorber~\cite{Bilki:2014bga}, the proton purity ranged from 64 to 95\% 
and the pion admixture is corrected for.
The agreement of simulations with data for shower shapes of protons and kaons is equal or better than for pions. 
The ratio of the response to protons relative to that for pions, $p/\pi = E_p/E_{\pi}$ is shown in 
Fig.~\ref{fig:G4v:protons} together with data from two other scintillator steel hadron calorimeters, the CDF end plug calorimeter~\cite{Liu:1997mw}
and the ATLAS Tile HCAL~\cite{Adragna:2010zz}.
It exhibits the expected slight decrease at low energies for steel, attributed to the smaller available energy, reduced electromagnetic content and the non-compensating nature of the calorimeter. 
A similar behaviour was observed in beam tests with a CMS barrel calorimeter prototype~\cite{Abdullin:2009zz}.
In the case of tungsten~\cite{CALICE:2013ff} with its nearly compensating behaviour, this is  less pronounced, 
see Fig.~\ref{fig:G4v:protons}. 
For kaons, no differences were expected within experimental precision, and were not seen in data. 
The resolutions are very similar for all hadron types. 

\subsection{Charged track segments in hadronic showers%
\label{sec:G4v-tracks}}

Highly granular calorimeters offer an excellent opportunity to study the internal structure of a hadronic shower. 
Hadronic showers are not amorphous clouds of spatially varying energy density, but they exhibit a rich sub-structure. 
The cascade evolution leads to a tree-like structure, with partially visible and invisible branches. 
There are centres of dense activity -- due to electromagnetic sub-showers, with size characterised  by $X_0$, and short-ranged nuclear evaporation products -- located around points of hard interactions, and regions with sparse signals in-between, in which leading  interaction products travel distances of the order of a nuclear interaction length, leaving only a MIP-like signature, or none at all. 
Particle flow algorithms make use of this sub-structure, for example by using the pointing information of tracks to or from clusters of energy deposition. 
It is thus highly relevant to ask whether this is adequately modelled in the simulations, and highly granular calorimeters are needed to study this. 

\begin{figure}[tb]
  \includegraphics[width=0.8\hsize]{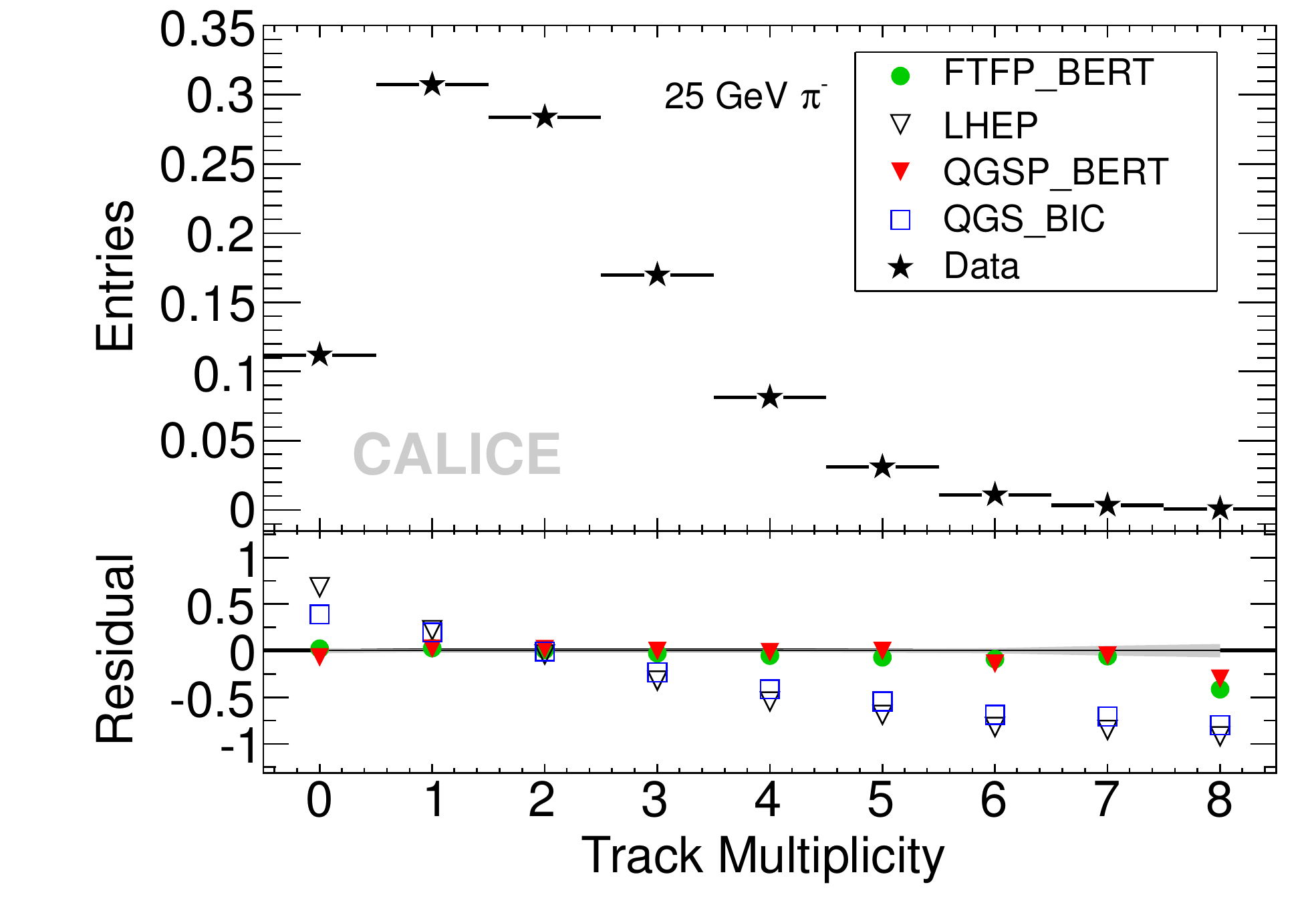}
  \includegraphics[width=0.8\hsize]{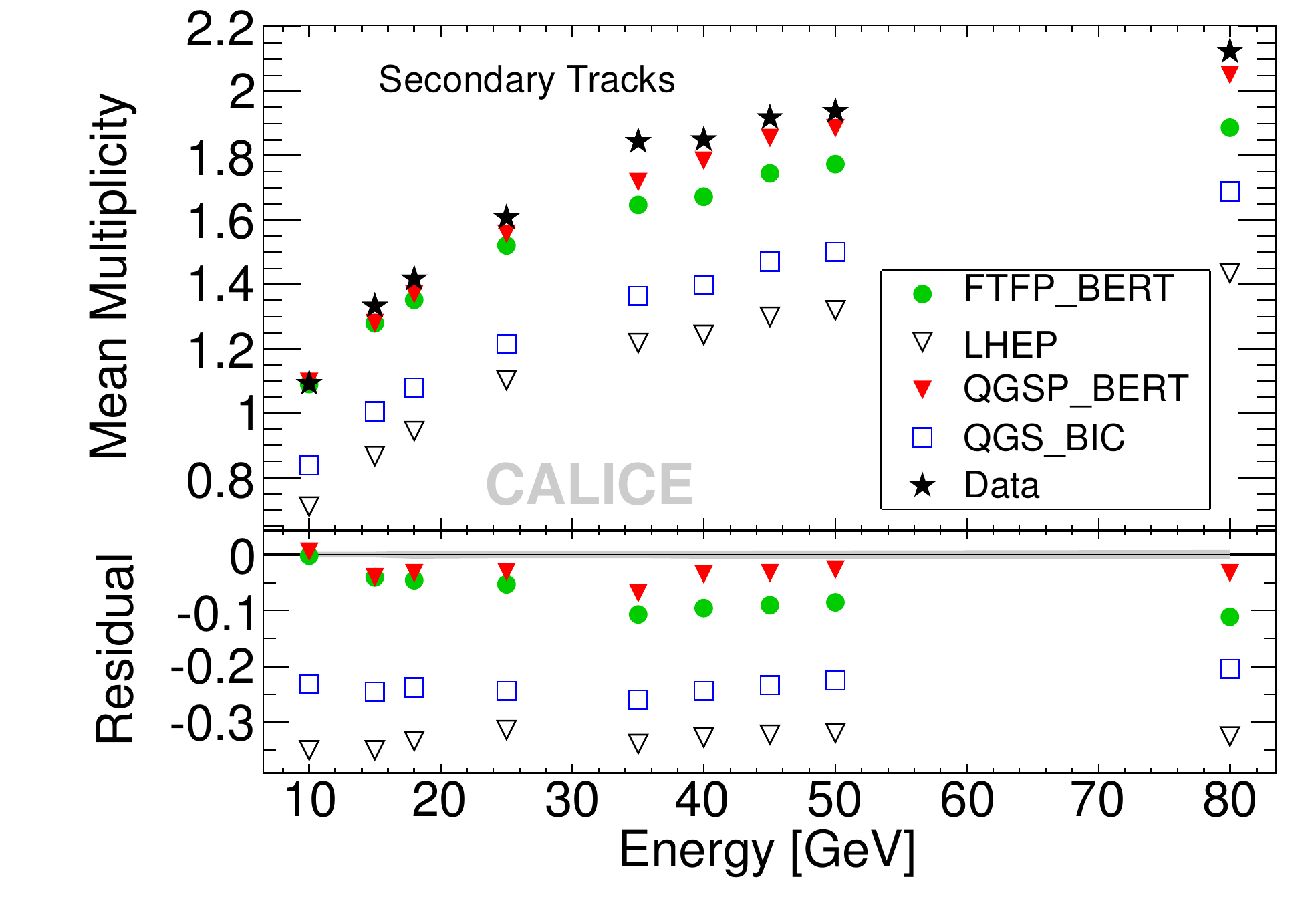}
\caption{\label{fig:G4v:trackseg} Track multiplicity distribution for 25~GeV pion showers (top), and mean secondary track multiplicity in the Fe-AHCAL as a function of beam energy (bottom). The lower panels show the residuals of the comparisons with different physics lists.
Both from~~\cite{Adloff:2013vra}.}
\end{figure}

%
For the study using the AHCAL~\cite{Adloff:2013vra}, tracks are searched for using a nearest-neighbour type algorithm on isolated hits, and Hough transformations in a subsequent filter stage in order to reject unphysical tracks with kinks or jumps. 
It was shown that the number of tracks found with this technique is indeed correlated with the number of hard charged tracks produced in the simulated interactions. 
The distribution of the track multiplicity for 25~GeV pion showers is shown in Fig.~\ref{fig:G4v:trackseg} and compared to different physics lists. 
The bottom figure displays the evolution of the mean multiplicity with shower energy.  
Here, the primary track of the incoming hadron has been excluded, and only secondary tracks starting in the third layer or later are shown. 
The multiplicity is well reproduced by recent physics lists, in particular QGSP-BERT, while the older parameterisations (LHEP, also used in QGSP-BIC) fail to reproduce the showers at this level of detail. 
A similar conclusion is drawn from the distribution of track lengths. 
Although there are still imperfections visible, this agreement is remarkable and a triumph for the modern, theory-driven shower models. 
%

\begin{figure}[!ht]
\includegraphics[width=.4\textwidth]{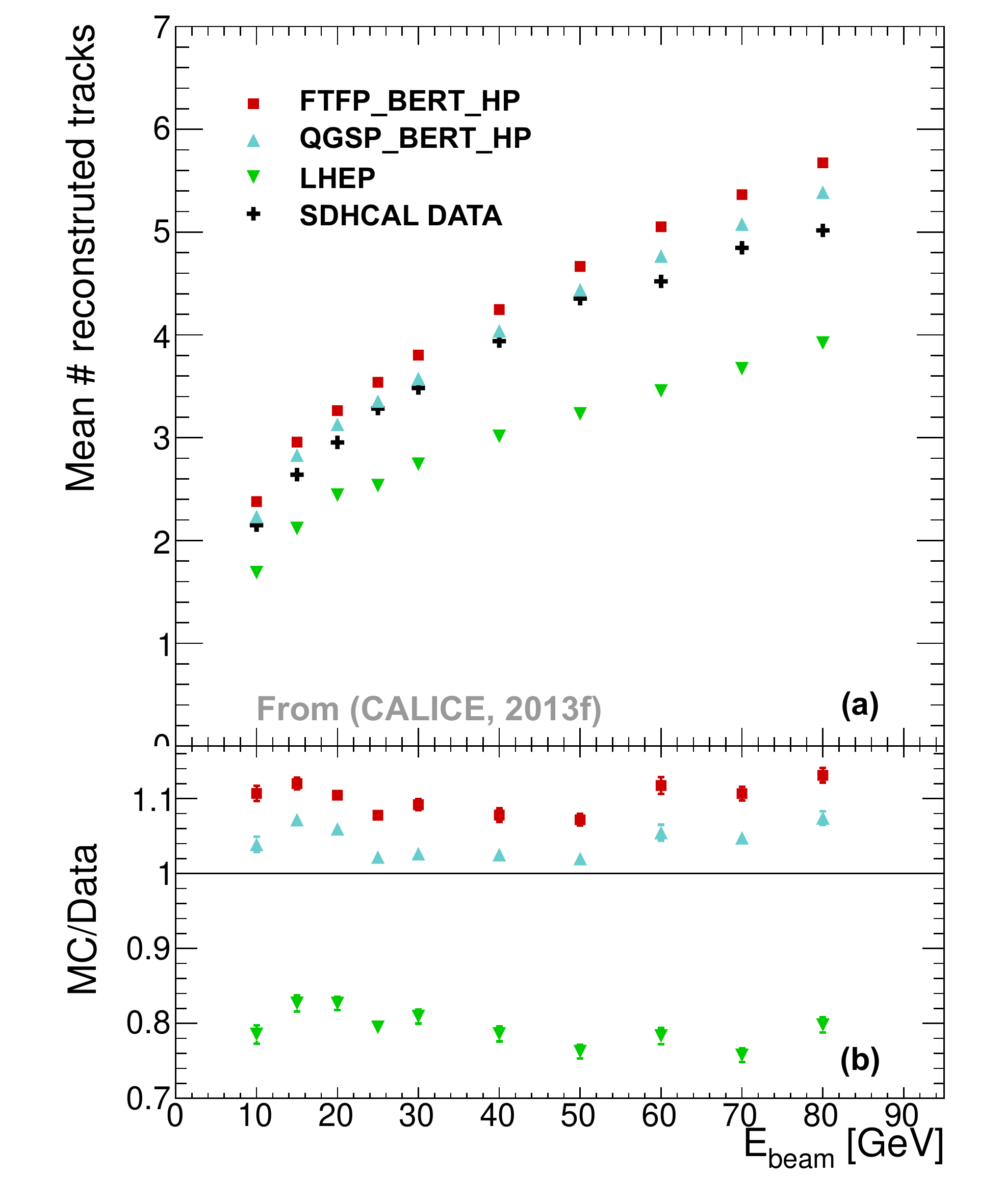}
\caption{Mean number of reconstructed tracks in hadronic showers in the SDHCAL as a function of the beam energy (a) and the ratio between simulation and data (b). 
Black crosses correspond to data. Red squares correspond to FTFP\_BERT\_HP physics list,  blue triangles to QGSP\_BERT\_HP and green triangles to LHEP. The digitizer used in the simulation was tuned using hadron data as explained in the text.
From~\cite{CALICE:2013fs,Laktineh:2013}.}
\label{fig:trackMulti-Length}
\end{figure}

A similar study has been made,  using the RPC-SDHCAL, in which the secondary tracks of the internuclear cascade of the hadronic showers are reconstructed~\cite{CALICE:2013fs}
After separating these charged hadrons from the electromagnetic core or more generally from the interaction region, tracks are reconstructed by means of the Hough Transformation. The mean number of reconstructed tracks is shown in Fig.~\ref{fig:trackMulti-Length} as a function of energy. 

As expected it increases with increasing energies. The logarithmic increase reflects the fact that at higher energies the electromagnetic fraction of the hadronic shower increases which disfavours the production of measurable tracks. The number of tracks is adequately reproduced by the physics lists of type QGSP and FTFP where the data favour slightly the QGSP approach. 

\begin{figure}[!ht]
\includegraphics[width=.32\textwidth]{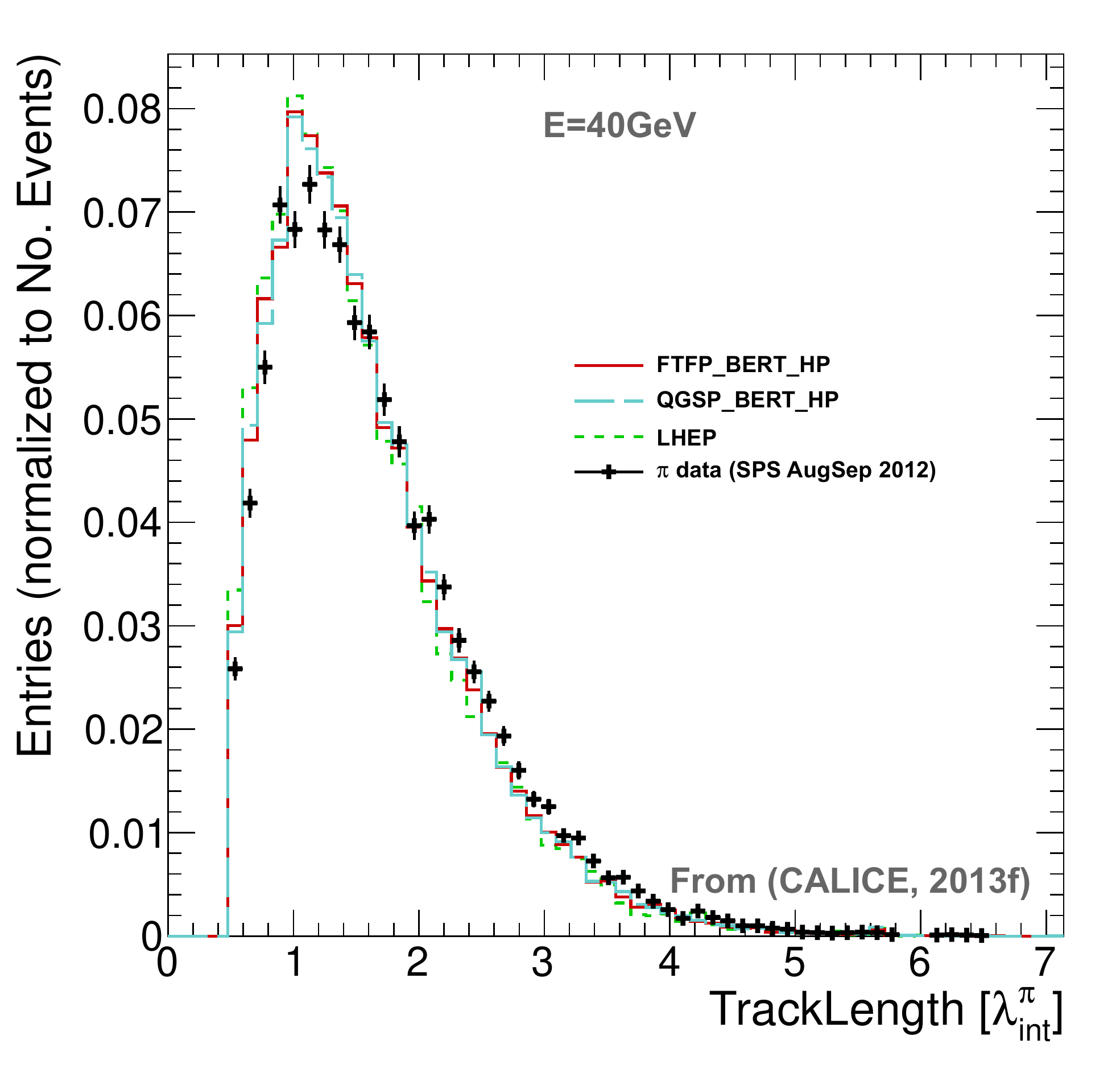}
\caption{The track length in the CALICE SDHCAL for 40~GeV hadron showers compared with predictions from various Geant4 physics lists. From~\cite{CALICE:2013fs,Laktineh:2013}.}
\label{fig:tracklength}
\end{figure}

The depth of the prototype allows for measuring the length of tracks in the hadronic shower. It is expected that a majority of particles travel a distance equivalent to about one pion interaction length $\lambda_{I}^{\pi}$. Fig.~\ref{fig:tracklength} demonstrates that this is indeed the case. The distribution of the track lengths is adequately described by all tested Monte Carlo models over a broad energy range. This confirms that the basic pion-nucleon cross sections are correctly implemented in the Geant4 physics lists.  

\subsection{Shower evolution with time} 

The timing capabilities of calorimeters do not only depend on the signal generation in the active part, and subsequent processing in the read-out electronics, but hadronic showers also exhibit an intrinsic time structure,  as observed e.g.\ in~\cite{Caldwell:1992te}, which is due to delayed nuclear de-excitation processes and the slow propagation of low energy neutrons. 
In experiments at colliders with high interaction or background rates, such as CLIC,  this leads to pile-up. 
It was shown~\cite{Linssen:2012hp} in simulations that even under such harsh conditions particle flow methods can be successfully applied, by combining calorimeter topological and timing information, and exploiting the high granularity of both. 
Moreover, it has been suggested to apply timing cuts in order to suppress neutron-induced hits, which are less well correlated in space with the shower axis, by means of timing cuts and thus ease the particle flow reconstruction~\cite{Abe:2006bz}. 
A full validation of particle flow must thus include a test whether the time evolution of hadronic showers is adequately modelled. 

For this purpose, dedicated add-on tungsten timing test-beam (T3B)~\cite{Simon:2013zya} instrumentation has been constructed and operated together with the W-AHCAL, W-DHCAL and Fe-SDHCAL in the SPS beam at CERN.  
Its active part consisted of 15 read-out elements with 3x3~cm$^2$ active area each, either scintillator tiles with MPPCs, or an RPC with 15 pads. 
They were installed behind the last layer of the absorber structures and aligned radially from the beam axis to the outer part of the HCAL. 
In both cases the read-out was performed by custom pre-amplifiers and fast digitisation electronics with 1.25~GHz sampling frequency and large buffer depth. 
Pulse shape analysis allowed the reconstruction of the time of the first hit with nano-second accuracy. 
Due to the fine segmentation the probability of a second hit is small, so using the first does not introduce a bias. 
In the case of the AHCAL, the T3B read-out was synchronised with that of the main calorimeter, such that using events with the shower start reconstructed in different layers, the timing could be investigated not only as a function of shower radius, but also of shower depth. 

\begin{figure}[tb]
  \includegraphics[width=0.95\hsize]{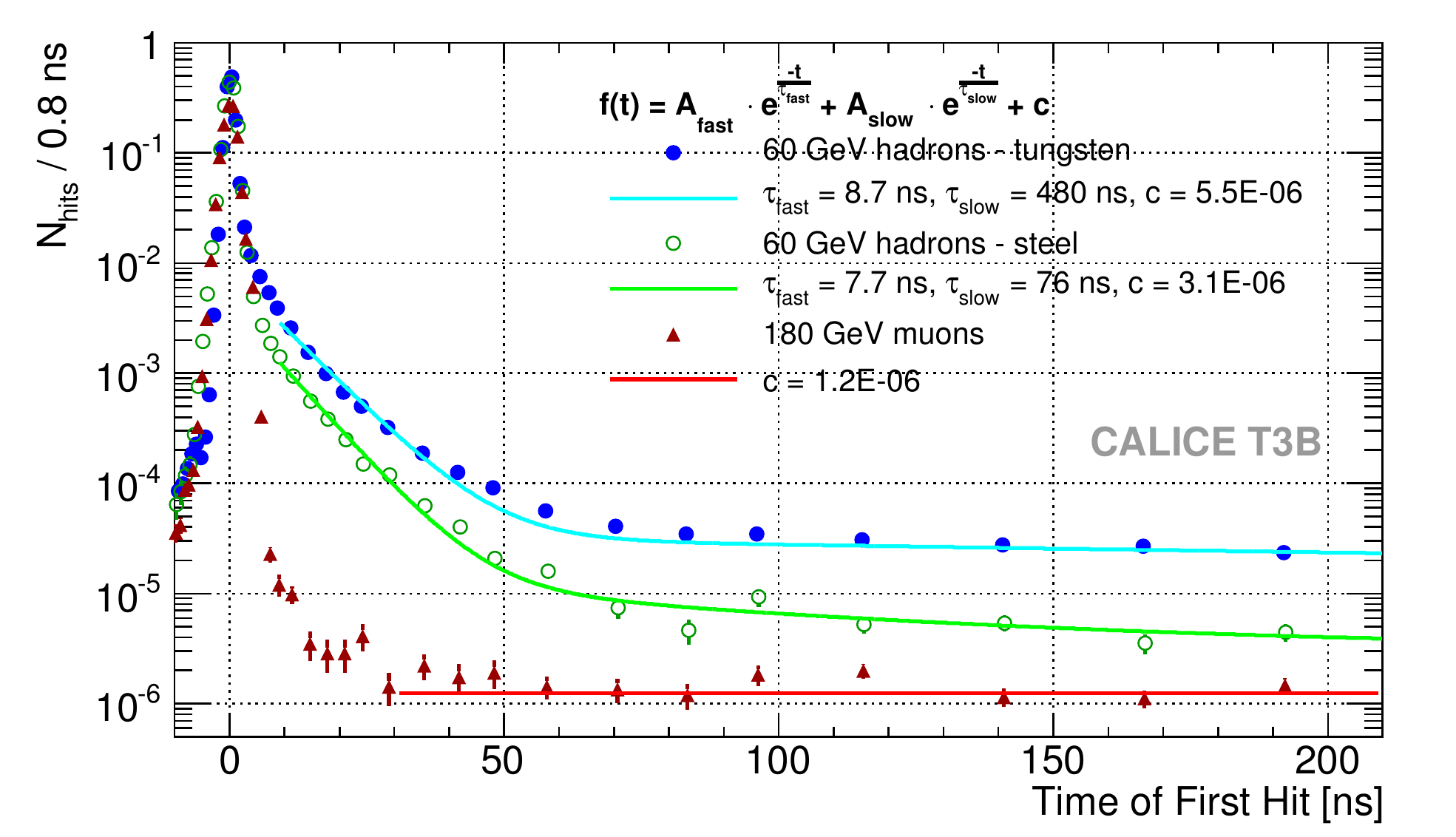}
  \includegraphics[width=0.8\hsize]{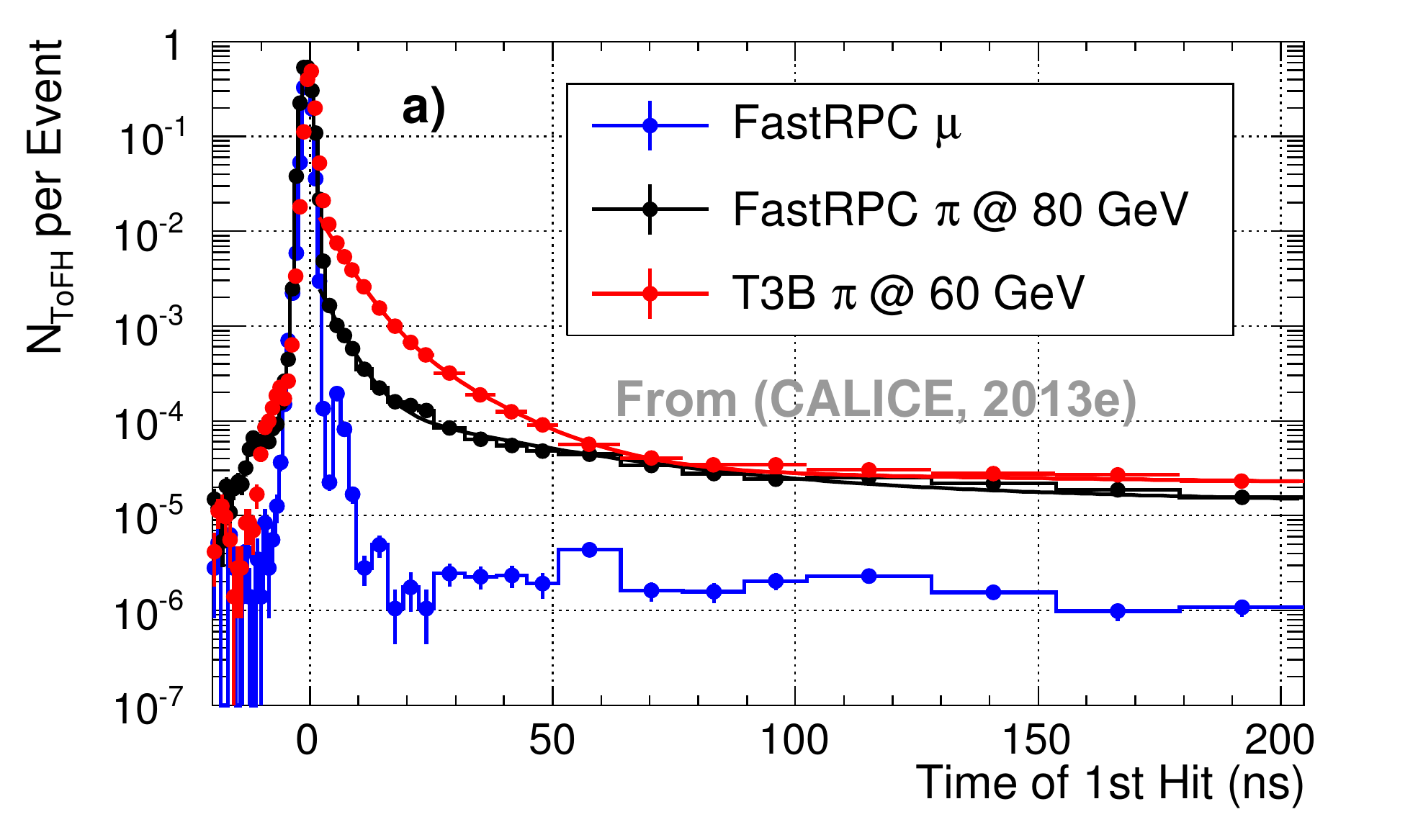}
\caption{\label{fig:G4v:t3btime} Distributions of the time of first hit for hadrons and muons, measured with scintillator in steel and tungsten (top), and comparison between scintillator and gas for tungsten (bottom). 
From~\cite{Adloff:2014rya} (top) and~\cite{CALICE:2013fth,Simon:2013nka} (bottom).}
\end{figure}

Fig.~\ref{fig:G4v:t3btime} shows the time distributions measured with scintillator for hadron showers in tungsten and steel absorber~\cite{Adloff:2014rya}. 
The distribution for muons is also shown, to illustrate the time resolution of the set-up, including effects of trigger, sensor, electronics and reconstruction. 
In contrast to the prompt muon response that for hadrons exhibits sizeable delayed contributions, which are significantly more pronounced for tungsten than for steel. 
This was expected from simulation, due to the higher abundance of neutrons in showers in tungsten. 
The late hits in tungsten are also more energetic (several MIPs) than in steel, where most are close to the threshold of 0.5~MIP. 
The bottom part of the figure shows a comparison of the time spectrum measured with scintillator and with an 
RPC~\cite{CALICE:2013fth}.
In the region between 10 and 50~ns, the delayed component is much smaller for the gaseous detector in the same absorber, which is attributed to the much reduced sensitivity to neutrons, since the hydrogen content and density is very small compared to that of plastic scintillator.

\begin{figure}[tb]
  \includegraphics[width=0.8\hsize]{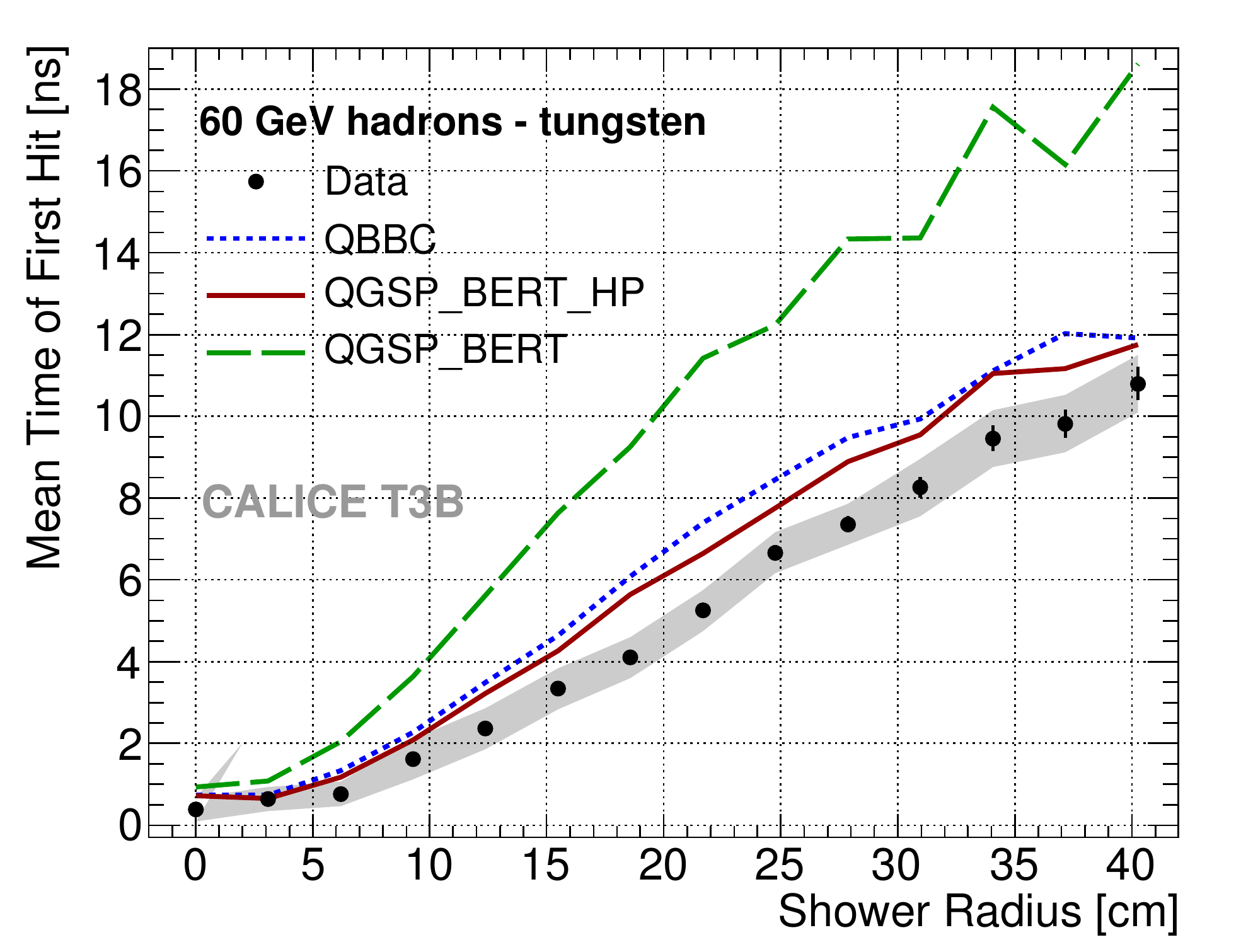}
\caption{\label{fig:G4v:t3bmc} Mean time of first hit as a function of radius, measured with scintillator for hadrons in tungsten, for data and different physics lists. 
From~\cite{Adloff:2014rya}.}
\end{figure}

In Fig.~\ref{fig:G4v:t3bmc} the timing measurements are compared with simulations. 
They show for tungsten the average time of the first hit as a function of radial distance from the beam axis. 
The delayed component becomes more important at large radius and, as could be shown for tungsten, in the rear part of the shower. 
The QGSP-BERT physics list with the designation HP (for high precision), and also the QBBC list, include a specialised treatment of neutron transport and a detailed implementation of energy-dependent absorption cross sections. 
In the case of tungsten, only these lists are able to reproduce the data reasonably, while the QGSP-BERT list without this treatment grossly overestimates the delayed component. 
In the case of steel, all lists give an acceptable description. 
%

%

%
\section{Tests of particle flow algorithms%
\label{sec:PFlow}}




The performance of a particle flow detector depends on its overall design parameters, like radius or magnetic field strength, and crucially on the calorimeter system. 
The main driving parameters are its single hadron energy resolution, at low jet energies, and the charged hadron-neutral hadron separation power for more energetic jets with higher particle density. 
More precisely, this means the capability of properly assigning energy depositions in the calorimeters to either charged hadrons measured in the tracking system, or reconstructing them as neutral hadron showers. 
Mis-assignments, generally referred to as {\it confusion}, lead to non-gaussian degradations of the jet energy resolution, due to double-counting of charged shower fragments misinterpreted as neutral particles, or losses of neutral hadron energy erroneously assigned to a charged track. 
Photon-hadron confusion is also important, but simpler to resolve, due to the more compact photon showers and the more effective use of shower shape criteria for their identification. 
This is, for example, demonstrated in~\cite{Jeans:2012jj} that uses the GARLIC algorithm for photon identification.
The particle flow algorithm minimises confusion by a set of successively applied algorithms, each compensating for shortcomings of previous stages and refining the result, thereby making use of details of the shower topology. 
It is of high interest to test these algorithms on real data, because, even with the very good agreement between data and simulations presented in the previous sections, it is  in principle still possible that the reconstruction makes use of so far untested features of the shower, or of the detector modelling in the simulation. 

\begin{figure}[htb]
\includegraphics[width=.8\hsize]{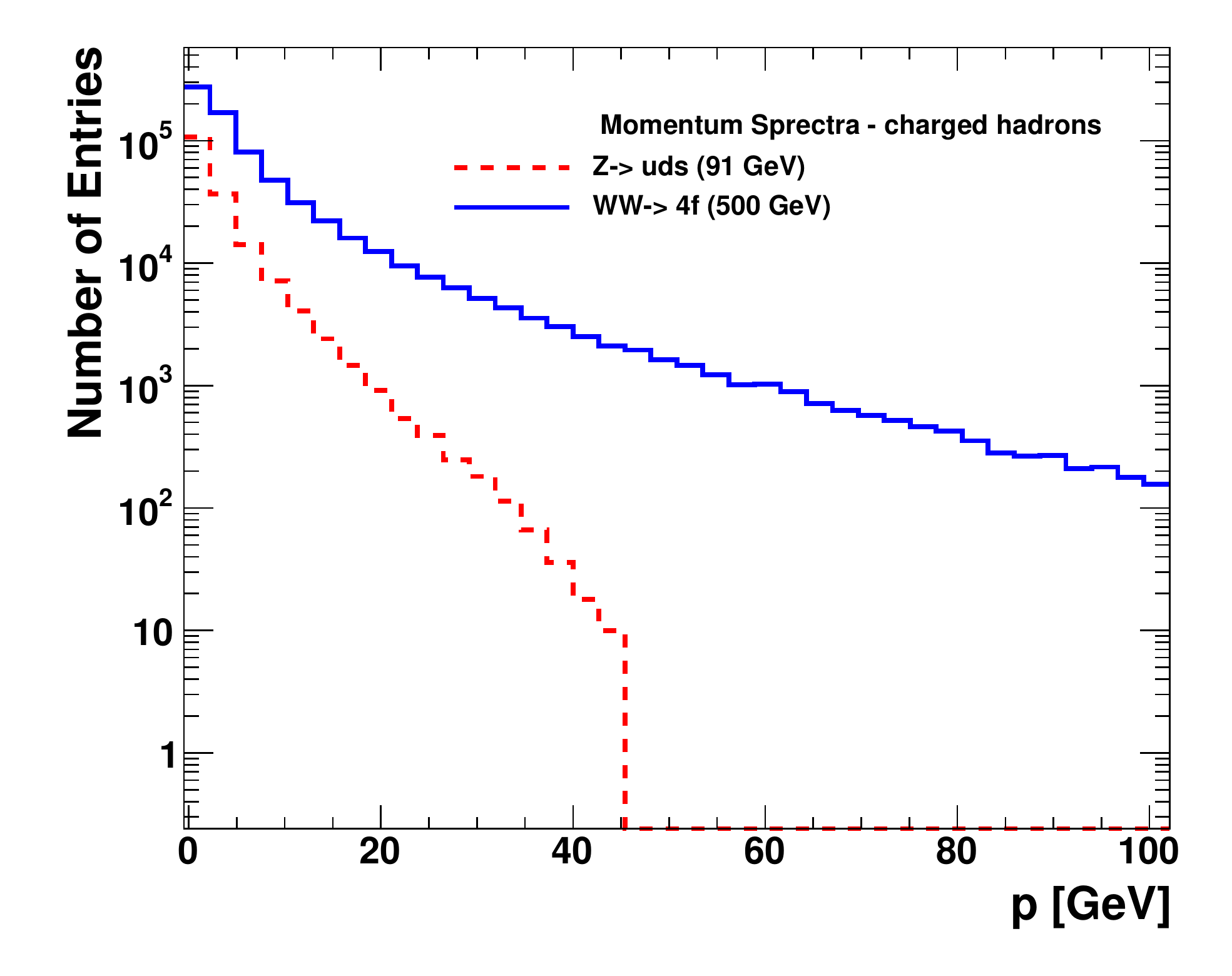}
\includegraphics[width=.8\hsize]{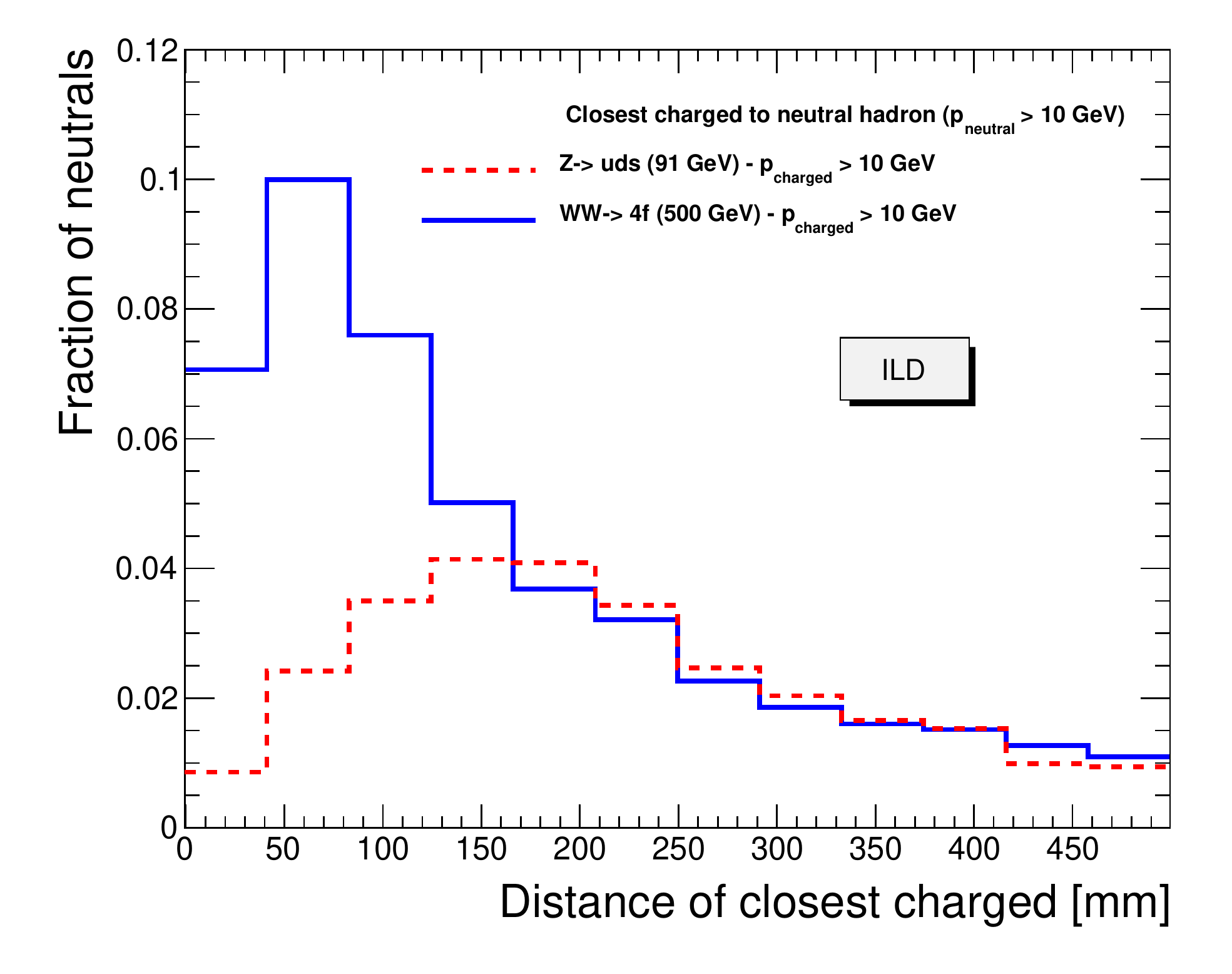}
\caption{\label{fig:Pflow:spectra} Top: Momentum distribution for charged particles in simulated ${\rm Z} \rightarrow uds$ and WW$\rightarrow 4f$ events. 
Bottom: Distribution of distances to closest charged track for neutrals produced in such events, measured at the front face of the electromagnetic calorimeter of the ILD detector.} 
\end{figure}

In order to illustrate the topological situation to resolve, Fig.~\ref{fig:Pflow:spectra} shows distributions of particle momenta and distances for two classes of  linear collider events: $e^+e^-\rightarrow {\rm Z} \rightarrow q\bar{q}$ with $q=u,d,s$ and $e^+e^-\rightarrow {\rm WW}\nu\bar{\nu}\rightarrow 4f$ where $f$ denotes quarks or leptons. 
The first is representative for the decay of heavy bosons produced near threshold, the second for a more complex event topology at higher energy and with more energetic jets. 
The spectra have been obtained~\cite{Brianne:2014} using simulations of the ILD detector~\cite{Behnke:2013lya} at the ILC, and the distances have been measured at the front face of the electromagnetic calorimeter, after propagating the particles in the 3.5T magnetic field. 
They show that typcial single particle energies are around 10~GeV, and that distances between neighbouring particles change considerably with centre-of-mass or jet energy. 
While for the decays of 
Z bosons at rest particles are well separated and showers will rarely overlap, at higher energies the density is higher, and distances of a few centimetres need to be resolved. 
These typical situations, which are relevant for the contribution of confusion to the jet energy resolution, have been studied using test beam events.

\subsection{Event overlay method} 
The method to apply particle flow algorithms to test beam data was first used~\cite{Morgunov:2004ed} in the framework of optimising the granularity of the CALICE AHCAL prototype with simulations. 
It uses event mixing techniques to overlay two test-beam events on top of each other and to study the reconstruction performance as a function  of the transverse separation between them.
This is possible since, due to the high granularity, the mean occupancy per event, and the fraction of calorimeter cells shared by two showers, is small.   
As it is possible to reconstruct the shower starting point, the charged track inside the calorimeter pointing to it can be removed from the event in software, such that a pseudo-neutral hadron shower is constructed from charged hadron beam data. 


For the study with real data~\cite{Adloff:2011ha}, data taken with the CALICE silicon tungsten ECAL and scintillator steel AHCAL taken at the CERN SPS have been used. 
Beam energies of 10 and 30~GeV have been chosen as being representative for particles in 100~GeV jets. 
Pure pion samples are selected using \u{C}erenkov counter information. 
In order not to dilute the energy measurement with leakage effects, events with more than  5\% of the visible energy in the TCMT are rejected. 
This suppresses showers starting late in the HCAL and thus reduces the average longitudinal separation of showers, which makes the separation harder and the results more conservative. 

Pseudo-neutral particles are constructed and laterally displaced by a distance between 5 and 30~cm, using the track in the upstream wire chamber as reference. 
The showers were mapped onto the ILD detector model with the same longitudinal segmentation as the ECAL and AHCAL test beam prototype, and 1x1~cm$^2$ and 3x3~cm$^2$ transverse cell size in ECAL and HCAL, respectively. 
Pairs consisting of one displaced pseudo-neutral and one charged shower were positioned such that the particles had normal  incidence to a barrel octant. 
Hit energies of the few shared cells were added. 
The possibility that two hits below threshold add up to one hit exceeding it was neglected, since in simulations it contributes only 0.1\% to the shower energy. 
The Pandora algorithm was slightly modified and adapted to this configuration, for example photon identification was skipped,  and the track helix was replaced by a straight line, since there was no magnetic field. 
Simulated test beam events were subjected to exactly the same procedure as data. 

\subsection{Shower reconstruction} 

\begin{figure*}[htb]
\includegraphics[width=.4\textwidth]{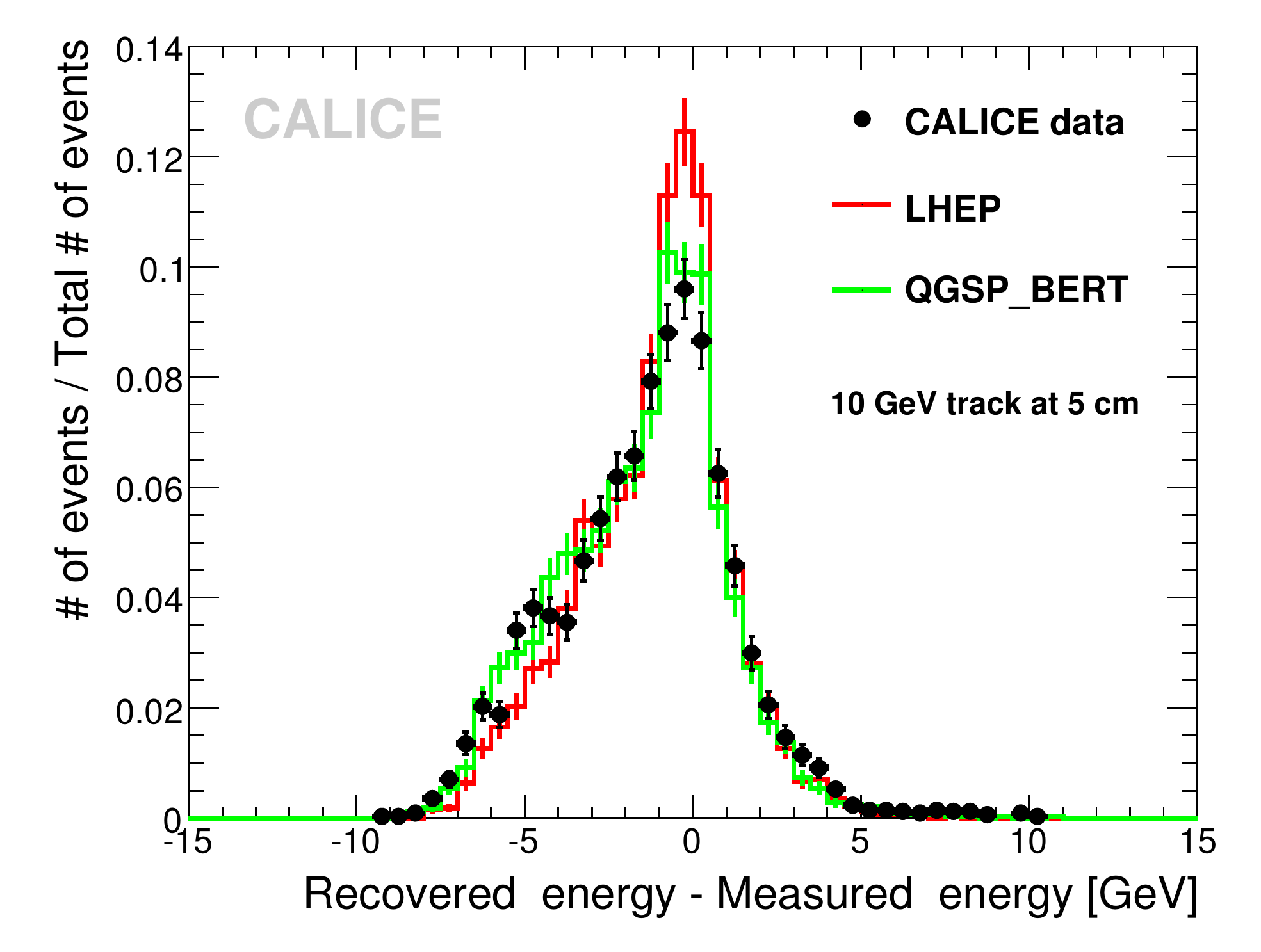}
\includegraphics[width=.4\textwidth]{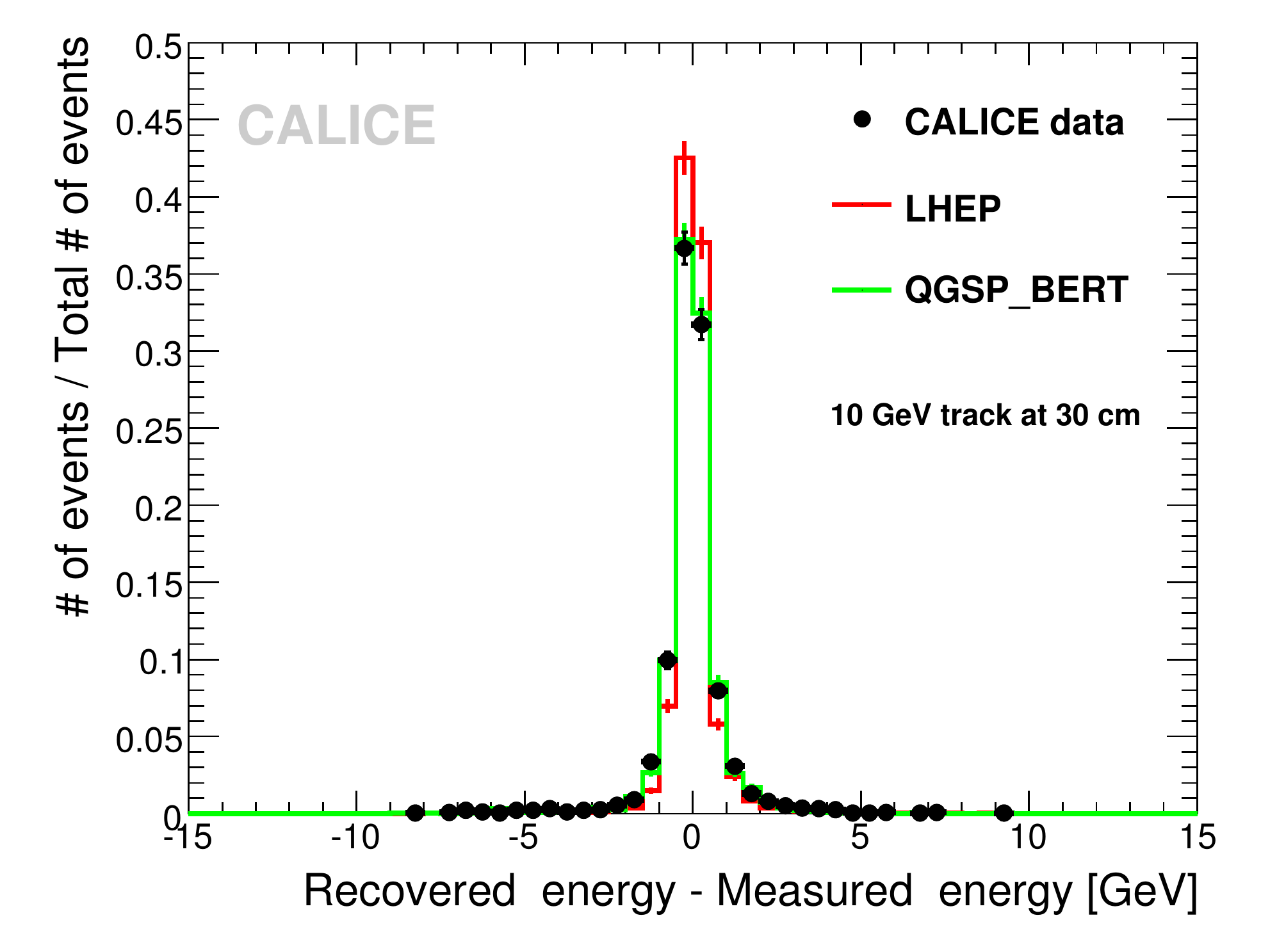}
\includegraphics[width=.4\textwidth]{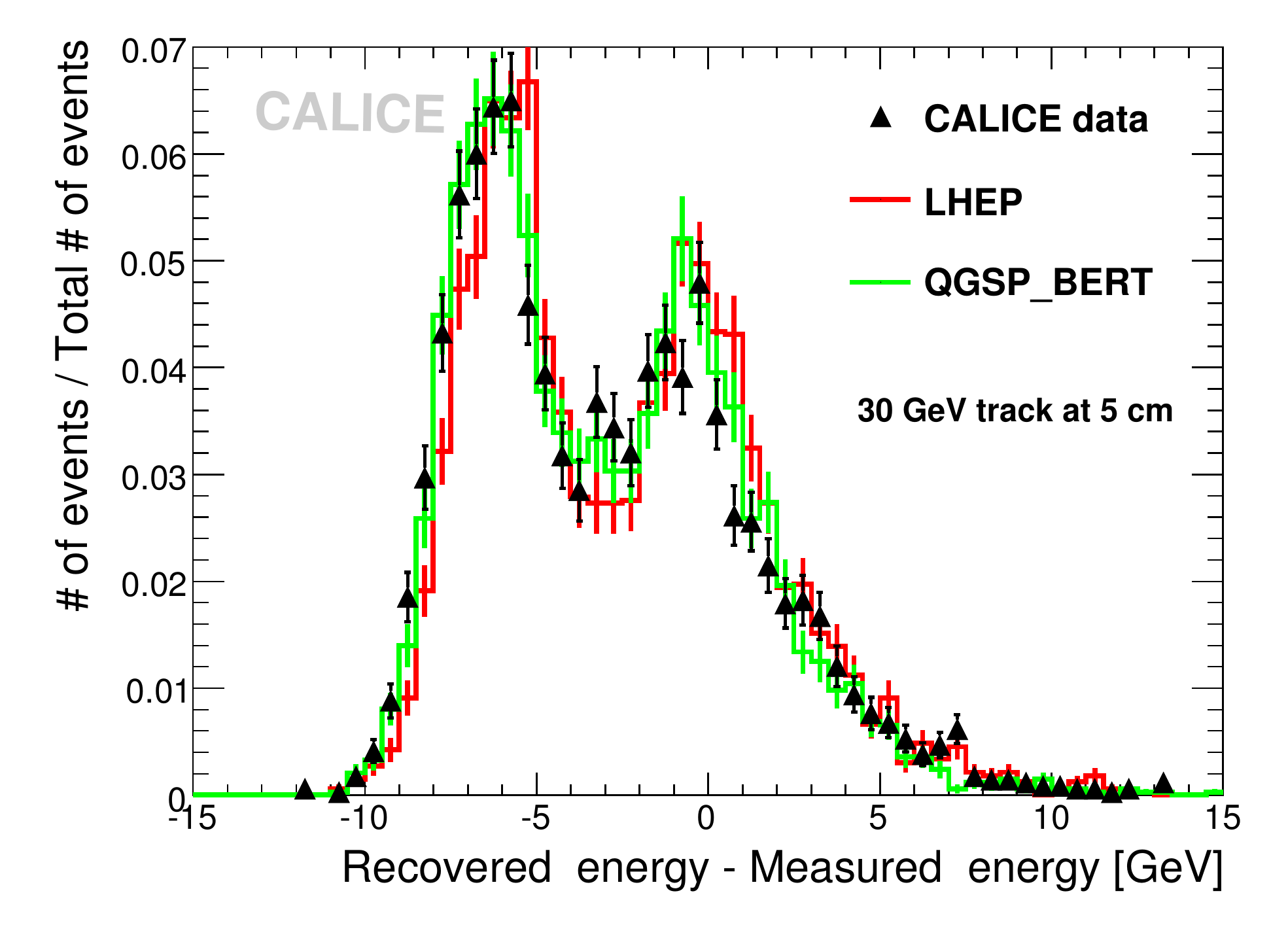}
\includegraphics[width=.4\textwidth]{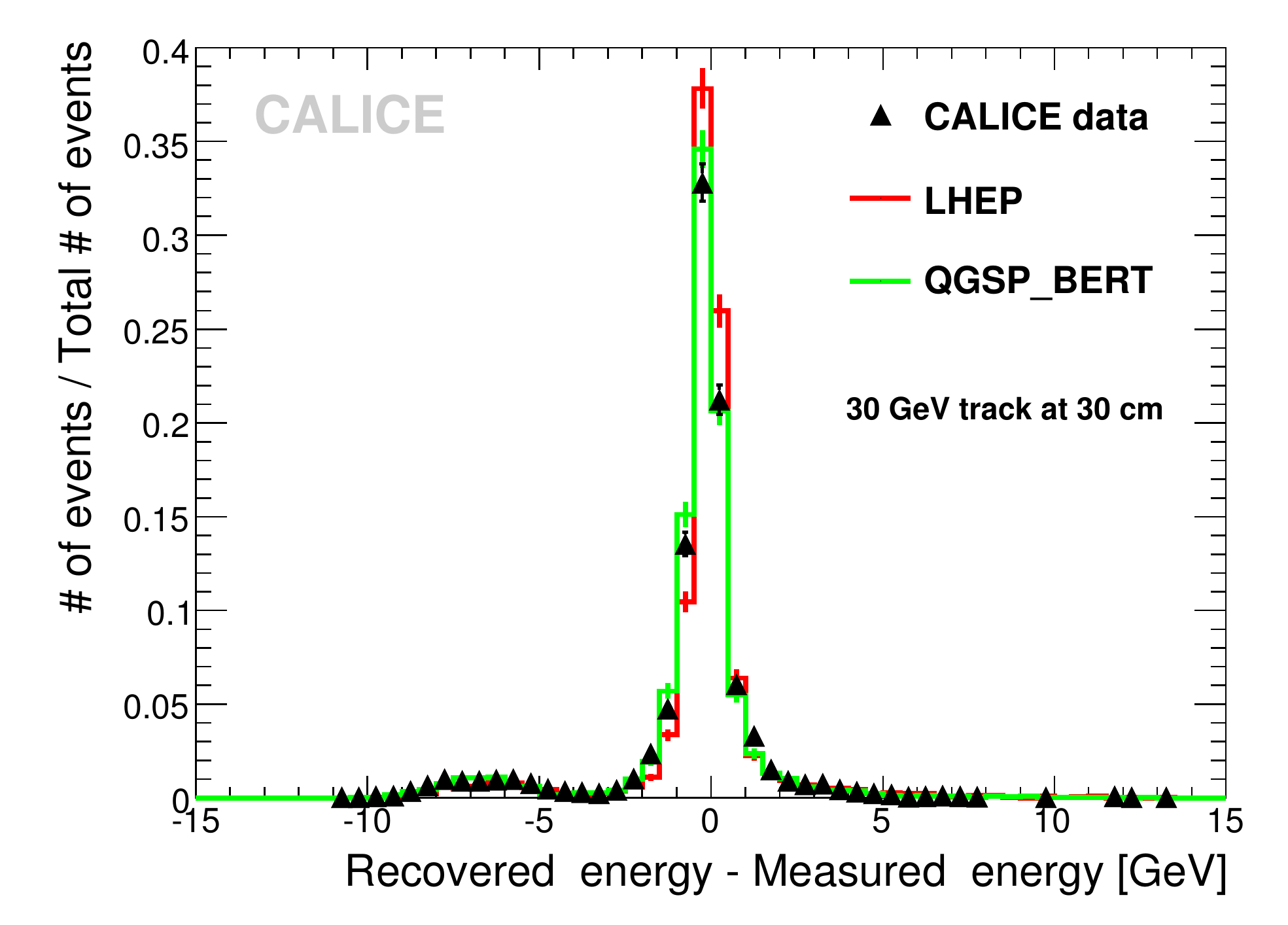}
\caption{\label{fig:Pflow:difference} Difference between the recovered energy and the measured energy
for the 10~GeV neutral hadron at 5 cm (left) and at 30 cm (right) from the 10~GeV (top) and 30~GeV (bottom) charged hadrons. Data taken with a combination of SiW ECAL and Fe-AHCAL (black) are compared to MC predictions for LHEP (red) and QGSP\_BERT (green) physics lists. From~\cite{Adloff:2011ha}.} 
\end{figure*}

The effect of confusion is quantified by studying the deterioration of the neutral particle measurement induced by the presence of a nearby charged particle shower.
This is measured by comparing the reconstructed energy of the neutral with the energy of the original undisturbed calorimetric measurement, before the overlay procedure. 
The neutral energy is the one obtained by Pandora in addition to that of the precisely known charged track. 
This deterioration is shown in Fig.~\ref{fig:Pflow:difference} for a neutral of 10~GeV, for two different energies of the charged particle, and two different distances between charged and neutral. 
Here the electromagnetic energy scale is used. 
On that scale a 10~GeV charged shower has a mean reconstructed energy of 8.2~GeV, and a neutral of 7.6~GeV, with the difference due to the energy deposited by the removed primary track. 
The worst deterioration occurs for a high energy (30~GeV) shower at close (5~cm) distance (bottom left). 
The left peak corresponds to the situation where most of the neutral energy is wrongly assigned to the charged track.
This is more likely to happen if the charged energy fluctuates downwards, because Pandora makes use of the expected shower energy, based on the tracking information, in addition to the topological criteria. 
Still, even at so close distance, there is still a large fraction of neutrals reconstructed with little deviation from the un-perturbed measurement. 
The case where the charged track showers in the ECAL and the neutral in the HCAL is much more difficult to resolve than in the reverse situation, where the primary track seen in the ECAL helps to disentangle the showers. 
The confusion for close distance (left plots) is considerably reduced for smaller charged particle energy, and largely disappears at larger distance. 

The simulated distributions are superimposed. 
The more recent QGSP-BERT physics list describes all four situations remarkably well. 
In contrast, the out-dated parameterisation-based LHEP list, which also tends to predict too narrow showers, underestimates the confusion effects. 

\subsection{Study of confusion effects with data}

\begin{figure}[t]
\includegraphics[width=.8\hsize]{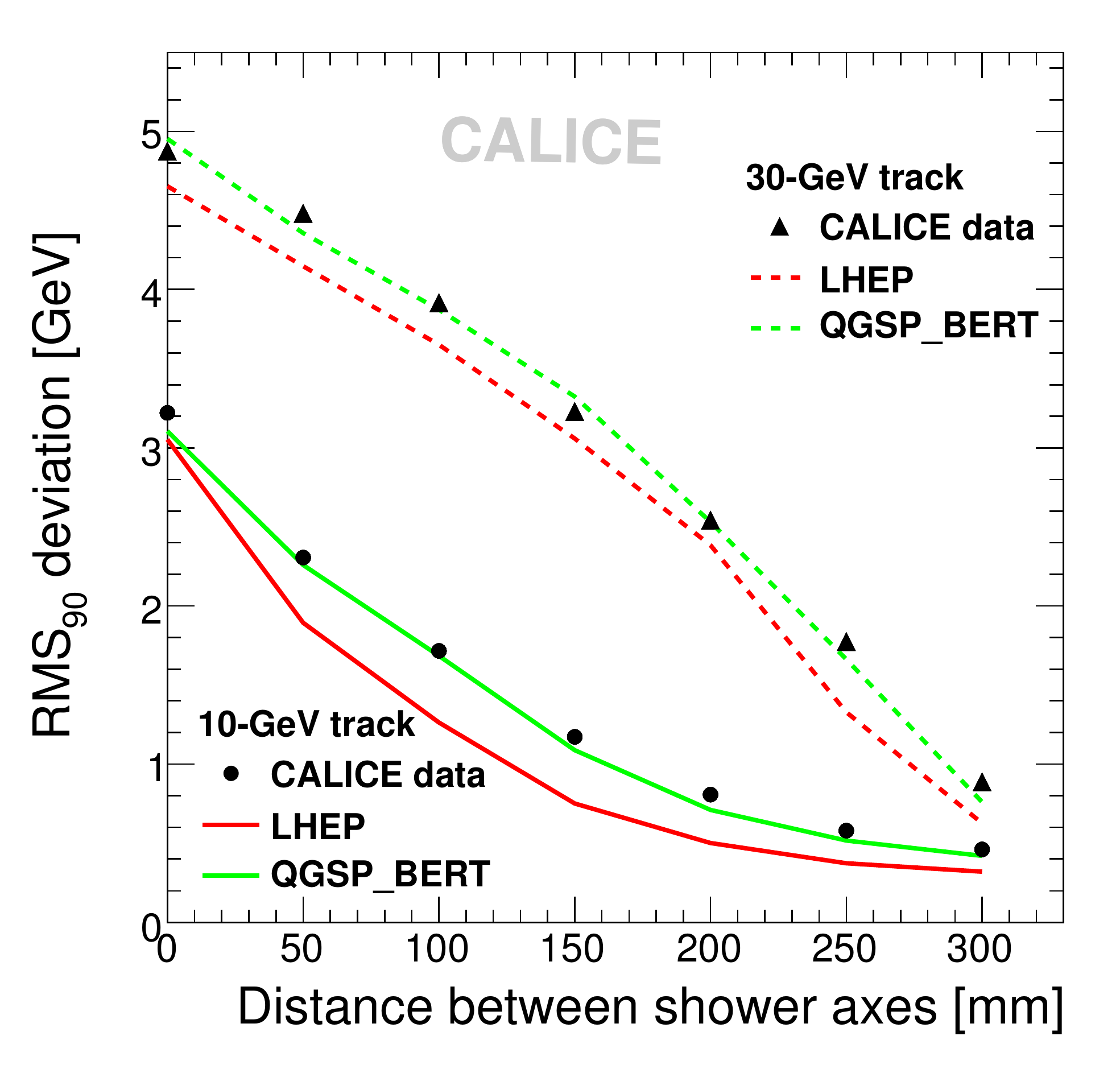}
\caption{\label{fig:Pflow:rms} $RMS_{90}$ deviations of the recovered energy of neutral 10~GeV
hadrons from its measured energy vs. the distance from 
charged 10~GeV (circles and continuous lines) and 30~GeV (triangles and dashed lines)
hadrons for beam data taken with a combination of SiW ECAL and Fe-AHCAL (black) and for Monte Carlo simulated data, for
both LHEP (red) and QGSP\_BERT (green) physics lists. From~\cite{Adloff:2011ha}.} 
\end{figure}

 
In order to quantify the effect of confusion as a function of particle separation, the root mean square (RMS) difference of recovered energy {\it versus} the original measurement, as plotted in Fig.~\ref{fig:Pflow:difference}, was calculated for different values of the distance between charged and neutral particles. 
(Note that this is not the RMS  width of the distribution, but the RMS deviation from zero, which is larger).
In order not to overemphasise the tails, mostly for large distance,  the RMS$_{90}$ -- the RMS of the central part containing 90\% of the entries -- is shown here; see Fig.~\ref{fig:Pflow:rms}.
The simulation based on QGSP-BERT physics also describes the dependency on distance very well, while the LHEP list generally underestimates the confusion. 
RMS variations of several GeV are non-negligible in relation to the aimed-at jet energy performance at a linear collider. However, the impact on the average jet resolution is moderate, since the abundance of 30~GeV particles in 100~GeV jets is small and the probability for very small distances is low. 

\begin{figure}[htb]
\includegraphics[width=.8\hsize]{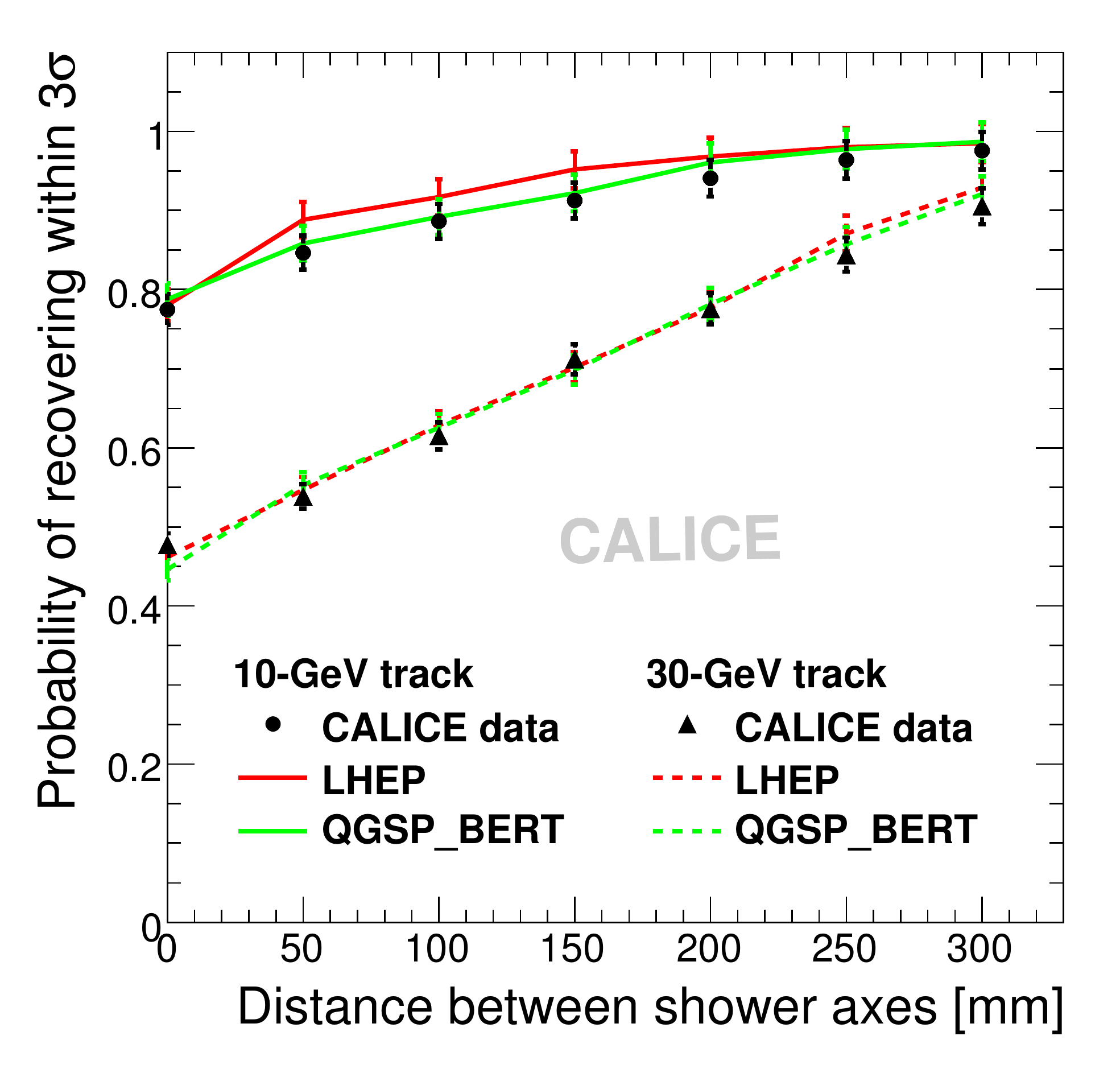}
\caption{\label{fig:Pflow:quality} Probability of  neutral 10~GeV hadrons energy recovering within
3 standard deviations from its real energy vs. the distance from charged 10~GeV
(circles and continuous lines) and 30~GeV (triangles and dashed lines) hadrons for beam data taken with a combination of SiW ECAL and Fe-AHCAL 
(black) and for Monte Carlo simulated data, for both LHEP (red) and
QGSP\_BERT (green) physics lists. From~\cite{Adloff:2011ha}} 
\end{figure}

Another way to quantify the effect is to calculate the fraction of neutral particles measured with an energy that is consistent with the nominal value within errors given by the calorimetric resolution $\sigma$, which in this study is 55\% 
times $\sqrt{7.6}$~GeV for data. 
This fraction, for a $\pm 3\sigma$ interval, is shown as a function of the distance to the nearby shower in 
Fig.~\ref{fig:Pflow:quality}. 
The absolute values of the probabilities at all distances are very well described by the recent physics list. 
The picture is qualitatively the same if a narrower $\pm 2\sigma$ window is chosen. 
In that case, the probabilities approach 55\% and 35\% at zero distance, for overlaid 10~GeV and 30~GeV charged showers, respectively.
The confusion effect decreases as the energy of the neutral particle increases. 
Further studies with the test beam data have shown that the RMS deviation shows little dependence on energy, and since the absolute energy resolution increases with $\sqrt{E}$, the probability to recover the shower energy 
within $\pm 3\sigma$ increases as well, i.e.\ the hadron energy resolution becomes the dominating effect. 

The size of the confusion effects measured here are a conservative approximation of those in a collider detector. 
The absence of a magnetic field, which on average separates the rear parts of the showers more from each other than the early parts,  the requirement of containment in the HCAL, the larger tiles in the outer parts of the prototype and the distortions due to mapping on the uniformly fine granularity of the ILD detector all make the task of shower separation more difficult. 
However, in this study they affect data and simulation in the same way and do not devalue the conclusion on the excellent agreement between them for the more recent physics list. 
The simulation studies on the jet energy performance of the ILD detector~\cite{Thomson:2009rp} have been performed using different physics lists and have shown that the jet energy resolution is rather robust against the choice of the model. 
The difference between the physics lists demonstrates that the performance of the algorithm and the study made here are indeed sensitive to details of the shower modelling, therefore the agreement with the QGSP-BERT model is a very significant and non-trivial result. 
It also shows that there are no hidden or not simulated effects in the detector hardware or its calibration, which would degrade the performance with respect to the simulated one. 


%

%
%
\section{Summary and outlook%
\label{sec:Summary}}


%

\begin{figure}[hb]
\includegraphics[width=0.95\hsize]{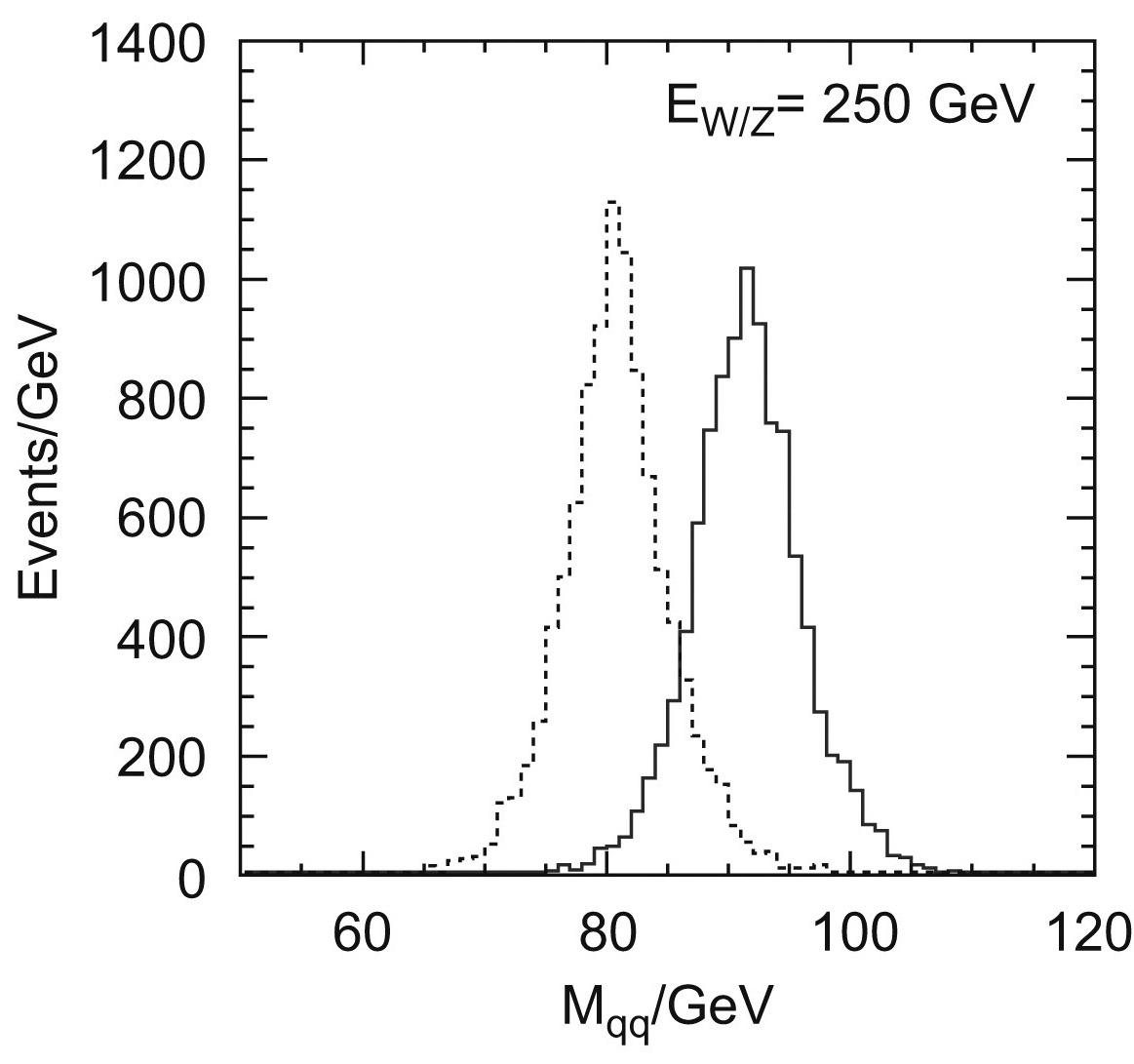}
\caption{\label{fig:Sum:WZ} Reconstructed invariant mass distribution for the hadronic system in ZZ$\rightarrow d\bar{d}\nu\bar{\nu}$ and 
WW$\rightarrow u\bar{d}\mu^-\bar{\nu}_{\mu}$ events simulated in the ILD detector. From~\cite{Thomson:2009rp}.}
\end{figure}

Research in particle flow calorimetry performed in the past decade has significantly enhanced our capabilities to precisely measure the energy of jets in high energy collider experiments. 
Detailed simulation studies in the framework of preparing detector concepts for future electron positron linear colliders such as ILC and CLIC have shown that  resolutions around 3.5-4\% can be realised for jet energies from 50 up to 1500~GeV. 
Such a resolution was shown to provide a separation of W and Z bosons decaying into pairs of jets on the basis of the di-jet invariant mass, as shown in 
Fig.~\ref{fig:Sum:WZ}.
This progress has come through a parallel development of sophisticated topological shower reconstruction algorithms like the Pandora package, and of novel detector technologies with which the required granularity can be realised in practice.  Examples of such technologies are silicon pad arrays, silicon photo-multipliers on small scintillator cells, or advanced gas amplification structures with readout segmented in two dimensions.

A broad experimental effort, driven by the need to develop calorimetry for the ILC, and mostly organised by the CALICE collaboration, has resulted in an experimental foundation for the particle flow approach. 
It complements the simulation studies with complete detectors and complex collider event topologies;  together they establish this novel approach to calorimetry.
Large-scale prototypes have been constructed, based on the novel technologies proposed for the future linear colliders and featuring the same fine granularity in three dimensions as used in the simulations of the jet energy performance. 
They have been exposed to test beams with muons, electrons and hadrons in major campaigns at CERN and Fermilab.
The accumulated data cover a large energy range from a few up to more than 100~GeV and allow testing the basic performance -- stability, linearity and resolution -- of the prototypes, the adequacy of shower model based simulations in great detail, and the performance of the algorithm in disentangling complex event topologies under the conditions of real data. 

The beam tests started with the silicon ECAL and the scintillator-SiPM HCAL technology, and consequently the analysis of these data is most advanced. 
However, all technologies have been successfully commissioned, have recorded large data sets and, in the results presented so far, demonstrated their basic performance. 
No show stoppers were found. 
Therefore the proof of principle based on the full analyses holds for the feasibility of particle-flow type reconstruction in fine-grained calorimeters in general. 
The data are available and ensure that final performance evaluations, and quantitative validations of detector and shower physics models can be expected also for the more recently tested technologies in the near future. 

The CALICE data confirm the progress made in recent years in improving the theory-motivated shower models implemented in the Geant4 framework, which had been driven by test beam studies with LHC calorimeter prototypes, and they provide a basis for further refinements. 
Electron beams have been used to validate the understanding of the detector response itself in terms of simulation, since electromagnetic showers can be precisely predicted. 
Overall, the more recent hadron shower models reproduce the pion and proton  data in the relevant energy range with an accuracy of a few percent. 
Shower shapes are well described in terms of average depth, while their width remains somewhat under-estimated as energies increase. 
Above around 50~GeV the shower core is too pronounced in simulations, both in longitudinal and radial extension.
Despite these remaining imperfections, the recent models describe the details of the showers remarkably well. 
The accuracy of the hadronic shower modelling has reached a level which qualifies them as a tool for quantitative detector optimisation. 
   
The studies probe the simulations in more detail than just average response and shapes, and the high granularity has been used extensively. 
Shower profiles were measured from the reconstructed location of the first hard interaction and thus provide more sensitivity to the different components relevant for the different phases of the evolution in ECAL and HCAL. 
The fine segmentation is an ideal basis for software compensation techniques and was quantitatively studied: it improves the resolution by up to 20\%, and a stochastic term of about 45\% for the hadronic energy resolution shows that highly granular calorimeters can have competitive energy measurement performance. 
The reconstructed starting point also enters into topology-based leakage estimation, a novel technique to further improve resolutions at higher energies, based on the fine spatial information. 

The CALICE results include a first demonstration that digital hadron calorimetry, based on hit counting only, works in principle, both conceptually and technologically, as demonstrated with prototypes with half a million channels.
The stochastic contributions to the achieved resolution is comparable to that obtained with analogue methods. 
The constant term appears to be affected by saturation effects at higher energies, and it was shown that the semi-digital method could mitigate these by combining the information from multiple thresholds. 
Quantitatively understanding the high-energy response in terms of simulations and disentangling instrumental effects from those due to high particle densities is a challenge which is being actively addressed in the analysis of the existing data. 
These studies will form a basis for the optimisation of read-out granularity and number of bits. 

Studies using tungsten as HCAL absorber extend the reach of the validation to applications at multi-TeV energies, such as at CLIC, where denser absorbers are considered. 
They probe the shower models in a regime where neutrons play a more important role than in steel, and the adequacy of the simulations is shown to be equally good as for the lighter material. 
Using a minimal but well conceived set-up with a few cells equipped with fast sampling electronics,  a first look into the time evolution of hadron showers became possible. 
The results show that this is well modelled but that in tungsten the detailed simulation of neutron transport -- so-called high performance versions of physics lists -- is necessary.
They support the simulated particle flow performance at CLIC energies, where both spatial and timing information are critical for pile-up rejection. 
The tungsten scintillator results are also relevant for the scintillator ECAL option of an ILC detector. 

Particle flow reconstruction makes use of the shower topology in greater detail than could be probed with earlier beam tests, for LHC detectors for example. 
The observed multiplicity of charged tracks visible in the shower evolution is quantitatively reproduced by simulations of gas and scintillator HCALs. 
A test of the particle flow reconstruction methods was performed by applying the Pandora algorithm to test beam events overlaid on each other, and quantifying confusion effects by measuring the degradation  of neutral hadron energy measurement through the presence of close-by charged hadron showers, and comparing results with real data to those with simulations. 
It adds to the realism of the study that it uses data taken with the combined set-up of ECAL and HCAL and makes use of the fine spatial resolution of both.
The excellent agreement represents important underlying evidence for the applicability of the method for jet measurements at future colliders. 
The results obtained so far form a cornerstone of the conceptual detector designs proposed for the ILC.

Further research will focus on completion of the validation through test beam data analysis. 
Important issues are the validation of hadronic response simulation in the gaseous HCALs and the scintillator ECAL, 
and the application of particle flow algorithms on data taken with alternative combinations of detectors. 
Data sets are available to probe the interplay of the the silicon ECAL and RPC DHCAL, both with 1~cm$^2$ segmentation, and the combination of ECAL and AHCAL both with scintillator read-out. 
In parallel, the particle flow reconstruction algorithms will be developed further. 
In particular the topological reconstruction in the (semi-) digital HCALs will be optimised, using real data. 

The effort to prepare options for detectors at the planned international linear collider is now addressing the technological challenges of integrating the enormous channel densities.
Engineering solutions are being developed that are scalable towards a full collider detector whilst maintaining the compactness and minimising the dead spaces needed for supports and services. 
In practice, this requires embedding the front end electronics into the detector volume and therefore keeping power consumption as low as possible, for example using power cycling according to the time structure of the accelerator.  
Data need to be digitised and zero-suppressed in the front end, and compressed on the detector.
Low material budget solutions need to be developed for power distribution and dissipation.
Such concepts have been developed for the different technologies, and small sets of detector modules with silicon and scintillator sensors have been successfully tested~\cite{Behnke:2013lya}, but power cycling must still be fully established, including operation in in high magnetic fields. 
The SDHCAL test beam prototype represents the largest system to test these concepts and has been routinely operated with cycled power to reduce cooling needs. 
However, none of the prototypes addresses all the challenges yet.
Full scale beam tests, as reported here, will also be necessary with scalable technological prototypes in order to 
re-establish performance and calibration and correction procedures at system level. 

The experience collected through these first steps towards realistic prototypes, together with the established confidence into the realism of the simulations in detail now provides the basis for a stringent cost-performance optimisation. 
In parallel, the large channel counts also require a level of industrialisation which goes beyond that of existing large calorimeter systems. 
This has barely begun, and will influence further developments, too. 

Recently, on the technology frontier, new options have appeared which may further enhance the potential of fine-grained calorimetry. 
Examples are monolithic active pixel sensors for ultra-finely segmented electromagnetic calorimeters, 
or digital silicon photomultipliers~\cite{Frach:2012zz} which may considerably simplify the front end readout of scintillator-based calorimeters. 
These will be evaluated and potentially explored in prototypes, simulations and new reconstruction algorithms. 

Particle flow methods are now successfully and routinely applied at the LHC, with decisive gains in jet and missing energy performance.
Yet, due to the much coarser spatial information in the ECAL and only minimal segmentation in the HCAL, when compared to the conceived linear collider detectors, the achieved resolutions are not yet reaching the stringent ILC goals.    
Encouraged by the demonstrated performance, and also by the successful experiments reported here, fine-grained calorimeter solutions are now being considered for the upgrades of LHC detectors for the high-luminosity phase. 
This brings additional challenges, both for the reconstruction software side, due to larger particle densities and energies, and pile-up, but also on the hardware side. 
Sensors need to meet higher radiation tolerance levels and rate capability needs, and the readout has to cope with much larger bandwidth, within tight power budget limitations. 

In all these future directions, algorithms and technologies need to be developed in parallel, and meet in test beam experiments.

\begin{acknowledgments}
We would like to thank our CALICE, ILD and SiD  colleagues for the excellent cooperation and fruitful discussions, and in particular
J.\ Brau, M.\ Breidenbrach, J.-C.\ Brient, M.\ Danilov, R.-D.\ Heuer, D.\ Jeans, Y.\ Karyotakis, I.\ Laktineh, L.\ Linssen, W.-D.\ Schlatter, F.\ Simon, T.\ Takeshita, D.\ Ward and J.\ Yu
for their leadership and support. 
We are deeply grateful to our late colleague Vasily Morgunov for his farsighted vision and continuous inspiration.

Many results quoted in this paper are based on test beam campaigns carried out by the CALICE collaboration, which would not have been possible without  the engineers and technicians who constructed the prototypes, 
and which were hosted and supported by DESY, CERN and Fermilab laboratories.
The work of the CALICE collaboration was supported by FNRS and FWO, Belgium; NSERC, Canada; NSFC, China; MPO CR and VSC CR, Czech Republic; Commission of the European Communities; IN2P3/CNRS, France; BMBF, DFG, Helmholtz Association and MPG, Germany; ISF, Israel; MEXT and JSPS, Japan; CRI(MST) and MOST/KOSEF, Korea; NFR, Norway; MES of Russia and ROSATOM, Russian Federation; MICINN-MINECO and CPAN, Spain; STFC, United Kingdom; DOE and NSF, United States of America.
\end{acknowledgments}

\bibliography{ExpPFA}

\end{document}